\documentclass[11pt,letterpaper]{article}

\usepackage[margin=1in]{geometry}
\usepackage{amsmath}
\usepackage{amssymb}
\usepackage{amsthm}
\usepackage{mathtools}
\usepackage{enumitem}
\usepackage{array}
\usepackage{etoolbox}
\usepackage{microtype}
\usepackage{placeins}
\usepackage{pgfplots}
\pgfplotsset{compat=1.16}
\usepgfplotslibrary{fillbetween}
\usepackage[colorlinks=true,allcolors=blue]{hyperref}
\usepackage[nameinlink,noabbrev]{cleveref}
\patchcmd{\abstract}{\small}{\normalsize}{}{}\patchcmd{\abstract}{\quotation}{}{}{}\patchcmd{\endabstract}{\endquotation}{}{}{}
\newtheorem{theorem}{Theorem}[section]
\newtheorem{lemma}[theorem]{Lemma}
\newtheorem{proposition}[theorem]{Proposition}
\newtheorem{corollary}[theorem]{Corollary}
\theoremstyle{definition}

\newtheorem*{remark*}{Remark}

\newcommand{\R}{\mathbb R}

\newcommand{\E}{\mathbb E}
\newcommand{\Prob}{\mathbb P}
\newcommand{\eps}{\varepsilon}
\newcommand{\one}{\mathbf 1}

\newcommand{\TopSet}{\operatorname{Top}}
\newcommand{\rank}{\operatorname{rank}}

\newcommand{\bstar}{\beta^*}
\newcommand{\gc}{\gamma_{\mathrm c}}
\newcommand{\Dstar}{\Delta^{\!*}}

\newcommand{\NN}[1]{N_{#1}}
\newcommand{\Work}{W}

\newcounter{step}
\newcounter{proofcnt}
\AtBeginEnvironment{proof}{\stepcounter{proofcnt}\setcounter{step}{0}}

\crefname{section}{Section}{Sections}
\Crefname{section}{Section}{Sections}
\crefname{appendix}{Appendix}{Appendices}
\crefname{theorem}{Theorem}{Theorems}
\crefname{proposition}{Proposition}{Propositions}
\crefname{lemma}{Lemma}{Lemmas}
\crefname{corollary}{Corollary}{Corollaries}
\crefname{remark}{Remark}{Remarks}
\crefname{step}{Step}{Steps}
\crefname{definition}{Definition}{Definitions}
\crefname{figure}{Figure}{Figures}
\crefname{table}{Table}{Tables}
\crefname{equation}{}{}

\title{Sharp Limits of Graph-Based\\
Nearest Neighbor Search on the Sphere}
\author{Kejing Lu\\Department of Electrical and Electronic Engineering\\University of Yamanashi, Japan\\\texttt{k.riku@yamanashi.ac.jp}}
\date{}

\begin{document}

\pagenumbering{roman}
\maketitle
\begin{abstract}
We study the degree needed for greedy graph search to find the exact nearest neighbor from every starting point on graphs of $n$ independent data points on the unit sphere in $\mathbb R^d$. For uniform data, we determine an explicit sublinear degree threshold when $d$ is proportional to $\log n$, and the threshold $2^{d/2+o(d)}$ in the dense regime $\sqrt{\log n}\ll d\ll\log n$. Above these thresholds by fixed exponent gaps, the nearest-neighbor graph succeeds from every entry for most uniform queries with high probability, with evaluation cost on the same exponential scale as its degree. Matching lower bounds hold for every graph designed after observing the sample.

For nonuniform data in the dense regime, local density variation shifts the critical degree. Under explicit local regularity conditions and within a specified range of density variation, we derive a sharp worst-case degree curve and show that the nearest-neighbor graph attains it. We construct sampling densities and query distributions for which, with high probability, every graph whose degree lies below this curve by a fixed exponent gap has failing entries for most queries.

Allowing fresh uniform random entries reduces the cost of exact recovery. For uniform data in the dense regime, we determine the exact exponent for the number of distinct distance evaluations over a range of degrees below the critical threshold. Independent restarts on a single graph joining points within a prescribed radius attain this exponent with high probability. Matching lower bounds hold for adaptive local exploration on threshold and nearest-neighbor graphs using entries independent of the data. Optimizing within this degree range lowers the exponent from $\frac12\log2\approx0.347$ for success from every entry to approximately $0.243$.
\end{abstract}

\thispagestyle{empty}
\clearpage
\tableofcontents
\thispagestyle{empty}
\clearpage
\pagenumbering{arabic}

\section{Introduction}\label{sec:intro}

\subsection{Graph navigation and the every-entry question}

Approximate nearest-neighbor search (ANNS) is a fundamental problem in data
mining and machine learning. To support high-performance search at scale, many
real-world systems currently build their ANNS indices from proximity graphs.
Among the most prominent examples are HNSW~\cite{MalkovYashunin2020HNSW},
NSG~\cite{FuEtAl2019NSG}, and DiskANN~\cite{SubramanyaEtAl2019DiskANN}.
Although these indices use different graph structures, their common navigation
principle has its algorithmic roots in Kleinberg's seminal work on the
small-world phenomenon~\cite{Kleinberg2000SmallWorld}. In Kleinberg's model,
decentralized routing moves a message toward a known target node by repeatedly
forwarding it to the available contact closest to that target. Proximity-graph search transfers this principle from a target node in the
network to an external query. Specifically, a navigation graph $G$ assigns a
short neighbor list $G(x)$ to each data point $x$. Its degree is
$\Delta(G)=\max_x |G(x)|$. We use $\Delta$ for a chosen list size or degree bound. Starting from an entry point, greedy search repeatedly
moves to the listed neighbor closest to the query and stops when none is closer.
We measure its evaluation cost by the number of distinct data points whose
distance to the query is evaluated.

Despite their widespread use and strong empirical performance, proximity-graph
indices are not yet supported by a comparably complete theory. Existing average-case analyses have made important progress by studying navigation from a
random or otherwise prescribed entry point on specific graph constructions. Laarhoven~\cite{Laarhoven2018GraphTradeoffs} analyzed random restarts on near-neighbor graphs over random spherical data and asked whether matching lower
bounds for graph search could be established. Building on this line of work, Prokhorenkova and Shekhovtsov~\cite{ProkhorenkovaShekhovtsov2020} proved, for $\log\log n\ll d\ll\log n$ and queries sampled near a designated nearest neighbor, that a threshold graph succeeds from their prescribed random start when its radius exceeds $\sqrt2$ times the typical nearest-neighbor distance. They also showed how adding edges between distant points and retaining several candidates during search can reduce its cost. Together,
these works provide the main quantitative foundations for graph navigation on random spherical data. These analyses, however, focus on specified graph constructions and prescribed entry rules, so they do not determine whether the degree scale they obtain is intrinsic to graph navigation or can be improved by a graph designed after
observing the sample. Moreover, existing analyses almost exclusively assume uniform sampling. Under nonuniform sampling, it therefore remains unknown how local density variation changes the degree required for exact navigation from every entry and whether the resulting degree is unavoidable for every graph.

Before discussing how to tackle these issues in more detail, we describe our setting. In this paper, we work on the unit sphere because cosine similarity is widely used in embedding-based retrieval to identify texts or images with similar semantic content. Moreover, the sphere is compact and has no boundary,
which allows density variation to be studied without additional tail or boundary
conditions. Within this domain, we draw the data independently from a probability
density rather than consider arbitrary point sets since, without assumptions on the
data, sublinear guarantees fail even for popular implementations. Indyk and
Xu~\cite{IndykXu2023WorstCase} exhibited inputs on which HNSW, NSG, and the
fast-preprocessing variant of DiskANN require empirical query time linear in the
data-set size to achieve reasonable accuracy. Random sampling therefore provides
a natural setting for determining when sublinear navigation is possible.

Notably, even within this distributional setting, the behavior of greedy search depends
on how its entry point is chosen. Laarhoven studied random restarts
\cite{Laarhoven2018GraphTradeoffs}. Prokhorenkova and Shekhovtsov considered a
single random entry \cite{ProkhorenkovaShekhovtsov2020}. However, practical
systems use other rules. For example, NSG starts every search from a Navigating
Node fixed during construction~\cite{FuEtAl2019NSG}, whereas HNSW enters its base
graph only after navigating its upper layers. The resulting base-layer entry is
thus query-dependent and may already lie close to the query. We therefore first study when greedy search succeeds from every entry, since an
every-entry guarantee covers all such entry rules at once and reveals what the stored graph alone can achieve. We later examine how the answer changes when either exact recovery or success
from every entry is relaxed.

In this setting, our central question is the following.

\begin{quote}
\emph{For a data set of $n$ random points on the unit sphere
$\mathbb S^{d-1}\subset\mathbb R^d$, can greedy search find the exact nearest neighbor from every entry with
sublinear evaluation cost? If so, what degree and evaluation cost are necessary
and sufficient, and can a navigation graph designed after observing the
complete data set improve either bound?}
\end{quote}

Before answering this question, we fix the asymptotic setting. We let $n\to\infty$ and allow $d=d(n)\to\infty$. A statement holds with high probability if its probability over the random data set is $1-o(1)$. It holds for most queries if the set of queries for which it holds has probability $1-o(1)$ under the query distribution. Unless stated otherwise, queries are uniform on $\mathbb S^{d-1}$. We distinguish two regimes according to how $d$ grows relative to $\log n$, where $\log$ denotes the natural logarithm. We call $d=\lfloor c\log n\rfloor$ for a fixed $c>0$ the \emph{logarithmic regime} and call $\sqrt{\log n}\ll d\ll\log n$ the \emph{dense regime}. For uniform data, the nearest-neighbor distance stays bounded away from zero in the logarithmic regime and tends to zero in the dense regime. Together, these regimes capture the transition from a constant nearest-neighbor distance to a vanishing one while the degree remains the leading scale of the evaluation cost. As $c$ grows, the degree threshold in the logarithmic regime approaches linear in $n$. At the other end, when $d=O(\sqrt{\log n})$, reaching the nearest neighbor from a distant entry already requires moves on the leading exponential scale if edges have length comparable to the nearest-neighbor distance. Therefore, we focus on the logarithmic and dense regimes in this paper.

With the conventions above, our answer to the central question is as follows. For uniform density, the
answer is yes, and the threshold is sharp. The degree and evaluation cost are
$n^{\beta^*(c)\pm o(1)}$ in the logarithmic regime, where
$\beta^*(c)=\frac{c}{2}\log(2-e^{-2/c})<1$, and
$2^{d/2\pm o(d)}$ in the dense regime. The graph that stores the $\Delta$
nearest neighbors of each point achieves these bounds from every entry for most
queries. Every graph of smaller degree has failing entries for most queries,
even if it is designed after observing the complete sample. In the dense
regime, the same dichotomy extends to a broad class of nonuniform densities
along a sharp worst-case curve determined by their local variation. Above the
curve, the nearest-neighbor graph still succeeds, with evaluation cost on the
same exponential scale as its degree. Below the curve, there exist a density
with the prescribed variation and a query distribution spread across the sphere
for which every graph of smaller degree fails in the same sense. By contrast, in
the logarithmic regime the answer depends on the shape of the density.
When the density is at most $e^{o(d)}$ times the uniform density, a suitable graph handles regions of arbitrarily low or zero density.
Certain concentrations of high density, however, raise the critical degree for every graph.

\subsection{Model and the critical curve}
\label{sec:model}
\label{sec:intro-curve}

We now specify the model and notation needed to explain the critical curve. Write $D_n=(X_1,\ldots,X_n)$, where the data points are independent with common distribution $P=f\,d\sigma_{d-1}$. Here $\sigma_{d-1}$ is normalized surface measure on $\mathbb S^{d-1}$ and $f$ is the sampling density. We use the sample indices as data labels and identify each label with its point, since the points are almost surely distinct. For a measurable set $A\subseteq\mathbb S^{d-1}$, the probability that a data point lies in $A$ is $P(A)=\int_A f\,d\sigma_{d-1}$. A query $q\sim Q$ is independent of $D_n$, and $Q$ is fixed before the data are sampled. Both the sampling and query distributions may depend on $n$. We compare points using geodesic distance $\delta(\cdot,\cdot)$. To describe neighborhoods on the sphere, write $C(x,\theta)$ for the closed spherical cap of radius $\theta$ centered at $x$. Its normalized surface measure is $F_{d-1}(\theta)=\sigma_{d-1}(C(x,\theta))$, independent of $x$ by rotational symmetry. For a fixed integer $k\ge1$, the target points form the set $\TopSet_k(q)$ of the $k$ data points nearest to $q$, so $k=1$ gives exact nearest-neighbor search.

After observing $D_n$, one may choose any navigation graph $G$, where each $G(x)$ is a neighbor list of distinct points in $D_n\setminus\{x\}$. We write $d_G(x)=|G(x)|$ for its length. A navigation run from an entry $x_0$ first evaluates $x_0$. At each visited point $x$, it expands $x$ by scanning the entire list $G(x)$ and evaluating the distance to $q$ of each point not already cached. Success is recorded after a completed expansion if the evaluated points include a member of $\TopSet_k(q)$, even when the entry itself is a target. Otherwise, the run moves to the listed point closest to $q$ if it is strictly closer than $x$, and stops if no listed point is closer. We write $\Work_G(q;x_0)$ for the number of distinct points evaluated through the first completed expansion after which a target has been evaluated, including the entry itself, and set $\Work_G(q;x_0)=\infty$ if the run stops without evaluating a target. The graph $G$ \emph{succeeds from every entry for $q$} if $\Work_G(q;x_0)<\infty$ for every $x_0\in D_n$. A guarantee from every entry for most queries means that, with probability $1-o(1)$ over the data, there is a set of queries of $Q$-measure $1-o(1)$ on which $G$ succeeds from every entry. We measure storage by $\sum_{x\in D_n}|G(x)|$, and average degree by $n^{-1}\sum_{x\in D_n}|G(x)|$. Notably, evaluation cost is different from the number of expansions because one expansion scans an entire neighbor list and different lists may overlap. Across restarts, the cost counts the union of all evaluated points, including evaluations made by unsuccessful runs, so repeated evaluations are charged only once. We apply these measures to two canonical graph families. The threshold graph $G_\rho$ lists all other data points within distance $\rho$ of each data point, while the nearest-neighbor graph $\NN{\Delta}$ lists the $\Delta$ nearest other data points. The uniform upper bound also applies to graphs that retain all edges within a slightly smaller radius and have degree at most $2\Delta$ (see \cref{thm:critical}(a)).

We now turn from the navigation model to the geometry that determines the critical degree. We begin with uniform data and give the geometric intuition behind the threshold. Let $\theta_1$ satisfy $nF_{d-1}(\theta_1)=1$. A cap of radius $\theta_1$ contains one data point in expectation, so $\theta_1$ gives the typical exact nearest-neighbor scale. When the current point $x$ has distance $\theta=\delta(q,x)$ from $q$, an improving neighbor on $G_\rho$ must lie in $C(x,\rho)\cap C(q,\theta)$. We call this intersection the \emph{lens}. Comparing the surface measure of the lens with $F_{d-1}(\theta)$ gives the fraction of improving locations available within radius $\rho$. For $0<\theta,\rho<\pi/2$, \cref{lem:lens} shows that the exponential loss in this fraction vanishes when $\cos\rho\le\cos^2\theta$. Exact recovery requires enough improving neighbors down to the scale $\theta_1$, so we evaluate the boundary at $\theta_1$. Since $nF_{d-1}(\rho)$ gives the leading scale of neighbor-list length on $G_\rho$, we define the critical radius $\rho^*$ and the corresponding critical degree $\Dstar$ by
\begin{equation}\label{eq:intro-critical-radius}
  \cos\rho^*=\cos^2\theta_1
  \qquad\text{and}\qquad
  \Dstar=nF_{d-1}(\rho^*).
\end{equation}
The first equality in \eqref{eq:intro-critical-radius} gives $\sin\rho^*/\sin\theta_1=\sqrt{1+\cos^2\theta_1}$. Since $\theta_1\to0$ in the dense regime, the critical radius satisfies $\rho^*=(\sqrt2+o(1))\theta_1$. To compute the critical degree, we use $nF_{d-1}(\theta_1)=1$ to write $\Dstar=F_{d-1}(\rho^*)/F_{d-1}(\theta_1)$. Applying the spherical cap asymptotics from \cref{lem:cap} gives exponent $\bstar(c)$ in the logarithmic regime and $\gc=\frac12\log2$ in the dense regime. In the two regimes, the critical degree therefore satisfies
\begin{equation}\label{eq:intro-uniform-curve}
  \Dstar=
  \begin{cases}
    n^{\bstar(c)+o(1)}, & d=\lfloor c\log n\rfloor,\\[1mm]
    e^{(\gc+o(1))d}=2^{d/2+o(d)}, & \sqrt{\log n}\ll d\ll\log n.
  \end{cases}
\end{equation}
Thus the constant $\sqrt2$ and the exponent $\bstar(c)$ both follow from the same spherical identity.

For nonuniform data in the dense regime, the nearest-neighbor scale and the probability of finding a point in a lens depend on location. For each $x\in\mathbb S^{d-1}$, let $\rho(x)$ be the smallest radius whose cap has probability at least $1/n$ under $P$. Continuity of cap probability in the radius gives $nP(C(x,\rho(x)))=1$, so $\rho(x)$ is the local nearest-neighbor scale. We also write $\rho_f(x)$ when the dependence on the density needs to be explicit. To keep the number of complete expansions subexponential in $d$, we assume that $\rho_{\min}:=\inf_{x\in\mathbb S^{d-1}}\rho(x)$ satisfies $\rho_{\min}=e^{-o(d)}$. For a candidate degree $\Delta=e^{\gamma d}$ with fixed $\gamma>0$, fix $0\le\kappa\le\bar\kappa<\gamma$ and choose a sufficiently large constant factor depending on $\gamma$, $\kappa$, and $\bar\kappa$. We consider queries $q$ for which the ratio between the essential upper and lower bounds of $f$ within the cap centered at $q$ whose radius is this factor times $\rho(q)$ is at most $e^{\kappa d}$. This condition controls the improving neighbors once the current point is near $q$. To guarantee progress from every entry before the run reaches this neighborhood, we require the corresponding density ratio to be at most $e^{\bar\kappa d}$ within the cap centered at $x$ whose radius is the same factor times $\rho(x)$, for every $x\in\mathbb S^{d-1}$. The essential bounds disregard sets of surface measure zero. The precise radius and density conditions appear in \cref{sec:prelim-nonuniform,loc:nn-sufficiency}.

To determine the worst-case critical degree, impose the density bound with parameter $\kappa$ at every center, with a sufficiently large fixed radius multiplier independent of the degree being tested. Write $\Dstar(\kappa)$ for the critical degree over these densities satisfying the bound on $\rho_{\min}$. For $0\le\kappa\le\frac12\log\frac32$, our matching upper and lower bounds give
\begin{equation}\label{eq:curve}
  \Dstar(\kappa)=\bigl(4e^{2\kappa}-2\bigr)^{d/2+o(d)}.
\end{equation}
At $\kappa=0$, \eqref{eq:curve} recovers the dense uniform scale in
\eqref{eq:intro-uniform-curve}. For arbitrarily fixed $\kappa$ in this range, let $\Delta=e^{\gamma d}$ and let $Q$ be any query distribution with a density on the sphere, fixed before sampling. If $e^{2\gamma}>4e^{2\kappa}-2$, then under the stated bound on $\rho_{\min}$ and the density bound with parameter $\bar\kappa$ at every center, with probability $1-e^{-\Omega(d)}$ over the sample, $\NN{\Delta}$ succeeds from every entry at every query satisfying the local density bound with parameter $\kappa$, outside an exceptional set of $Q$-probability $e^{-\Omega(d)}$. Conversely, if $e^{2\gamma}<4e^{2\kappa}-2$, we construct a density with the prescribed variation and a query distribution spread across the sphere for which, with probability $1-o(1)$ over the sample, every graph of degree at most $\Delta$ has failing entries for most queries. Taking $\bar\kappa=\kappa$ gives the density class in \eqref{eq:curve}, over which $\NN{\Delta}$ attains the worst-case critical degree.

This description of the critical degree in terms of $\kappa$ alone is specific
to the dense regime. In the logarithmic regime, variation in the magnitude of $f$
can shift the local nearest-neighbor angle by a constant amount. The sufficient condition for $\NN{\Delta}$ therefore depends on both the global upper bound on $f$ and the ratio between its essential upper and lower bounds over the sphere. The resulting conditions are stated in
\cref{thm:tolerance-moderate}. A different graph construction gives a stronger
guarantee for sparse regions that are not covered by these conditions. If
$f$ is bounded above by $e^{o(d)}$, then at every fixed degree exponent
strictly between $\bstar(c)$ and $1$, the constructed graph achieves exact
recovery from every entry for most uniform queries, even when $f$ is
arbitrarily small or zero on parts of the sphere. Its evaluation cost has the
same exponential scale as its degree. In the other direction, there are
densities that place sufficiently large probability mass in a small region
and force every navigation graph to use a degree exponent strictly larger
than $\bstar(c)$, even for uniform queries and even when the graph is designed
after observing the complete sample. We prove the navigation guarantee for this construction in
\cref{nat:navigation} and the lower bound in \cref{enr:obstruction}.

\subsection{Main results}\label{sec:results}

We now state the navigation guarantees and lower bounds associated with the curves in \cref{eq:intro-uniform-curve,eq:curve}. We then quantify the cost of fresh random starts and the quality of the point returned when exact recovery is relaxed. The degree and density exponents in these statements are fixed constants. The words \emph{above} and \emph{below} refer to a fixed positive margin between the degree exponent and the relevant threshold exponent.

\begin{theorem}[Exact navigation above the curve, informal]\label{thm:informal-above}
Fix $k\ge1$ and a query distribution $Q$ with a density on
$\mathbb S^{d-1}$, fixed before the data are sampled. The following
conclusions hold.
\begin{enumerate}[label=(\alph*)]
\item Suppose that the data are uniform. In the logarithmic regime, let
$\Delta=n^\beta$ with $\bstar(c)<\beta<1$. In the dense regime, let
$\Delta=e^{\gamma d}$ with $\gamma>\gc$. In either regime, choose $\rho$ so
that $nF_{d-1}(\rho)=\Delta$. Then, with high probability, greedy navigation
on $G_\rho$ and $\NN{\Delta}$ succeeds from every entry for every query
outside a set of $Q$-measure $o(1)$. The searches use $O(1)$ greedy moves in
the logarithmic regime and $e^{o(d)}$ moves in the dense regime, so their
evaluation cost is $\Delta^{1+o(1)}$.

\item In the dense regime, suppose that the data are sampled from
$P=f\,d\sigma_{d-1}$. Let
$0\le\kappa\le\frac12\log(3/2)$ and $\Delta=e^{\gamma d}$, and choose
$\bar\kappa$ with $\kappa\le\bar\kappa<\gamma$. Assume
$\rho_{\min}=e^{-o(d)}$, and fix a sufficiently large constant depending on
$\gamma$, $\kappa$, and $\bar\kappa$. Suppose that, for every
$x\in\mathbb S^{d-1}$, the ratio between the essential upper and lower
bounds of $f$ in the cap centered at $x$ whose radius is this constant times $\rho(x)$ is at
most $e^{\bar\kappa d}$. If $e^{2\gamma}>4e^{2\kappa}-2$ with a fixed
margin, then, with high probability, greedy navigation on $\NN{\Delta}$
succeeds from every entry for every query $q$ satisfying the density ratio
bound $e^{\kappa d}$ in the cap centered at $q$ whose radius is the same constant times
$\rho(q)$, except for queries in a set of $Q$-measure $o(1)$. The search uses $e^{o(d)}$ greedy moves and has evaluation
cost $\Delta^{1+o(1)}$.
\end{enumerate}
\end{theorem}

We next turn to the complementary behavior below the curve. For uniform data, the lower bound is measured under the uniform query distribution. For nonuniform data, the lower-bound construction distributes the query mass among a growing number of separated regions across the sphere. We call a sequence $Q_n$ \emph{globally spread} if, for every fixed $0<\theta<\pi/2$, $\sup_{p\in\mathbb S^{d-1}}\Prob_{q\sim Q_n}(q\in C(p,\theta))\longrightarrow0$.

\begin{theorem}[Failure of exact navigation below the curve, informal]\label{thm:informal-below}
Fix $k\ge1$.
\begin{enumerate}[label=(\alph*)]
\item Suppose that the data are uniform. In the logarithmic regime, let
$\Delta=n^\beta$ with $0<\beta<\bstar(c)$. In the dense regime, let
$\Delta=e^{\gamma d}$ with $0<\gamma<\gc$. Then there is a constant $a>0$
such that, with high probability over the data, every $G$ admits a
measurable query set of surface measure $1-o(1)$ on which every query $q$
has at least $(1-o(1))n^a$ entries in the logarithmic regime and at least
$(1-o(1))e^{ad}$ entries in the dense regime satisfying
$\Work_G(q;x_0)\ge\Delta$. If $\Delta(G)\le\Delta$, every query in this set
has at least $(1-o(1))n^a$ failing entries in the logarithmic regime and at
least $(1-o(1))e^{ad}$ failing entries in the dense regime. Moreover, with
high probability over the data, any $G$ that succeeds from every entry on a
query set of surface measure $1-o(1)$ has average degree at least
$(\Dstar)^{1-o(1)}$, and every query in its success set outside a subset of
surface measure $o(1)$ has evaluation cost at least
$(\Dstar)^{1-o(1)}$ from a $1-o(1)$ fraction of entries.

\item In the dense regime, fix $\kappa\ge0$ and let
$\Delta=e^{\gamma d}$, where $\gc<\gamma\le\log2$ and
$e^{2\gamma}<4e^{2\kappa}-2$. There exist a positive sampling density $f$
whose values across $\mathbb S^{d-1}$ vary by exactly a factor
$e^{\kappa d}$ after sets of surface measure zero are ignored, and a
globally spread query distribution $Q_n$ with a density on
$\mathbb S^{d-1}$. The sampling density and query distribution are fixed
before the data are sampled, and $\rho_{\min}=e^{-o(d)}$. With high
probability over the data, every $G$ of degree at most $\Delta$ has an
exceptional query set of $Q_n$-measure $o(1)$, possibly depending on $G$,
outside which greedy navigation fails from at least $e^{\Omega(d)}$
entries.
\end{enumerate}
\end{theorem}

Together, \cref{thm:informal-above,thm:informal-below} establish the critical curve for exact navigation from every entry. For uniform data, the curve gives the sharp degree and evaluation-cost exponents. For nonuniform data in the dense regime, under the density bound with parameter $\kappa$ at every center and for $0\le\kappa\le\frac12\log(3/2)$, $\NN{\Delta}$ attains the optimal worst-case degree threshold in \eqref{eq:curve} among all navigation graphs, including those designed after observing the complete sample.

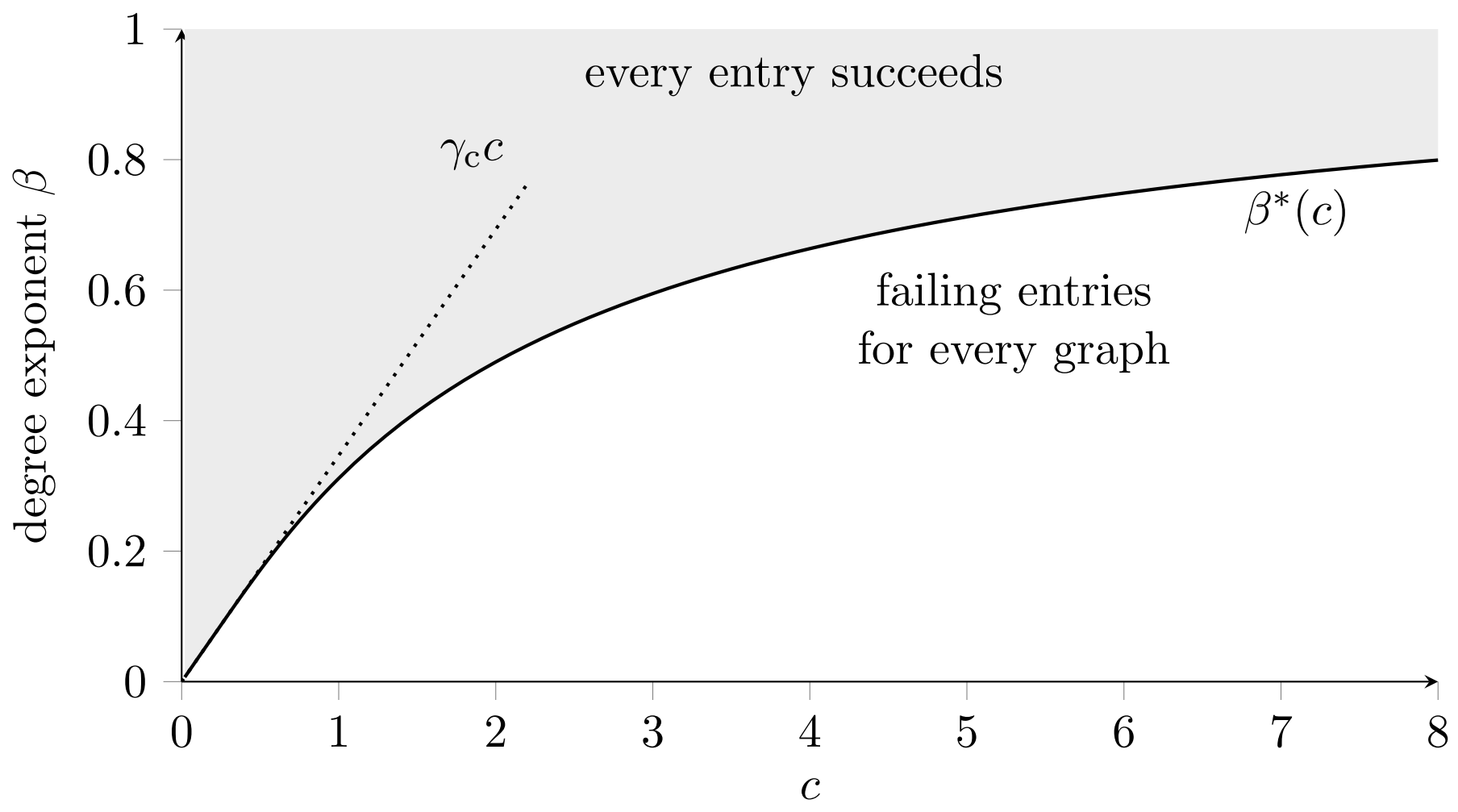
\begin{figure}[htb]
\centering
\begin{tikzpicture}
\begin{axis}[width=0.72\textwidth,height=0.42\textwidth,
  xmin=0,xmax=8,ymin=0,ymax=1,
  xlabel={$c$},ylabel={degree exponent $\beta$},
  axis lines=left,tick align=outside,
  every axis plot/.append style={thick}]
\addplot[name path=star,black] coordinates {(0.020,0.0069) (0.050,0.0173) (0.100,0.0347) (0.150,0.0520) (0.200,0.0693) (0.250,0.0866) (0.300,0.1039) (0.350,0.1210) (0.400,0.1380) (0.450,0.1546) (0.500,0.1710) (0.550,0.1870) (0.600,0.2025) (0.650,0.2177) (0.700,0.2324) (0.750,0.2467) (0.800,0.2605) (0.850,0.2739) (0.900,0.2868) (0.950,0.2994) (1.000,0.3115) (1.100,0.3347) (1.200,0.3564) (1.300,0.3767) (1.400,0.3959) (1.500,0.4139) (1.600,0.4308) (1.700,0.4468) (1.800,0.4620) (1.900,0.4763) (2.000,0.4899) (2.100,0.5028) (2.200,0.5150) (2.300,0.5267) (2.400,0.5378) (2.500,0.5484) (2.600,0.5585) (2.700,0.5681) (2.800,0.5774) (2.900,0.5862) (3.000,0.5947) (3.100,0.6029) (3.200,0.6107) (3.300,0.6182) (3.400,0.6254) (3.500,0.6324) (3.600,0.6391) (3.700,0.6455) (3.800,0.6518) (3.900,0.6578) (4.000,0.6636) (4.100,0.6692) (4.200,0.6746) (4.300,0.6799) (4.400,0.6850) (4.500,0.6899) (4.600,0.6947) (4.700,0.6993) (4.800,0.7038) (4.900,0.7081) (5.000,0.7123) (5.100,0.7165) (5.200,0.7204) (5.300,0.7243) (5.400,0.7281) (5.500,0.7318) (5.600,0.7353) (5.700,0.7388) (5.800,0.7422) (5.900,0.7455) (6.000,0.7487) (6.100,0.7518) (6.200,0.7549) (6.300,0.7579) (6.400,0.7608) (6.500,0.7636) (6.600,0.7664) (6.700,0.7691) (6.800,0.7717) (6.900,0.7743) (7.000,0.7769) (7.100,0.7793) (7.200,0.7818) (7.300,0.7841) (7.400,0.7864) (7.500,0.7887) (7.600,0.7909) (7.700,0.7931) (7.800,0.7952) (7.900,0.7973) (8.000,0.7993)};
\node[font=\normalsize,anchor=north] at (axis cs:7.1,0.78) {$\bstar(c)$};
\addplot[black,dotted,domain=0:2.2] {0.34657*x};
\node[font=\normalsize,anchor=south east] at (axis cs:2.15,0.76) {$\gc c$};
\addplot[name path=top,draw=none] coordinates {(0.02,1) (8,1)};
\addplot[gray!15] fill between[of=star and top];
\node[font=\normalsize,align=center] at (axis cs:3.9,0.93) {every entry succeeds};
\node[font=\normalsize,align=center] at (axis cs:5.3,0.55) {failing entries\\for every graph};
\end{axis}
\end{tikzpicture}
\caption{The uniform critical curve. For uniform data with $d=\lfloor c\log n\rfloor$ and degree $\Delta=n^\beta$, the solid curve is $\beta=\bstar(c)$. In the shaded region above it, $\NN{\Delta}$ succeeds from every entry for most uniform queries. Below it, every graph of degree at most $\Delta$ has failing entries for most uniform queries. Both statements hold with probability $1-o(1)$ over the data when $\beta$ stays a fixed positive distance from $\bstar(c)$ (\cref{thm:informal-above,thm:informal-below}). The dotted line $\beta=\gc c$ approximates the critical curve for small $c$, since $\bstar(c)\sim\gc c$ as $c\to0$.}
\label{fig:phase}
\end{figure}

Allowing fresh starting points offers a way to retain exact recovery at degrees below the threshold for success from every entry. A \emph{uniform random entry} is a data label chosen uniformly and independently of the query and the sampled coordinates. Independent restarts repeat greedy navigation on the same sampled graph from fresh uniform random entries. To compare restarts with more flexible searches, we introduce \emph{local exploration algorithms}. At each step, such an algorithm may request an entry or expand an already evaluated point whose neighbor list has not yet been scanned. Requested labels either follow a sequence fixed before sampling or are drawn afresh uniformly and independently of the sampled coordinates and the exploration history. Requesting an entry evaluates its distance to the query, and expanding a point scans its entire neighbor list. Success is recorded after the first entry request or complete expansion whose evaluated points include a member of $\TopSet_k(q)$. The cost counts all distinct points evaluated through that operation, across all earlier entry requests and complete expansions. The decision to request an entry and the choice of a point to expand may depend on the query, the algorithm's internal randomness, and the complete exploration history, consisting of the requested and expanded labels, scanned lists, and positions of evaluated points. All random entry choices are included in the algorithm's randomness. Beam search belongs to this class when started from such entries because it retains several evaluated candidates and selects which to expand.

\begin{theorem}[The exact cost of relaxing the entry guarantee, informal]\label{thm:informal-restarts}
Fix $k\ge1$. For uniform data in the dense regime, let $\Delta=e^{\gamma d}$, where $\frac12\log\frac43\le\gamma<\gc$, and choose $\rho$ so that $nF_{d-1}(\rho)=\Delta$. For the explicit exponent $\eta^*(\gamma)>0$ in \eqref{eq:dense-fresh-start-exponent}, and for every query fixed before sampling, one greedy run from a uniform random entry succeeds with probability $e^{-(\eta^*(\gamma)+o(1))d}$ over the data and the entry on both $G_\rho$ and $\NN{\Delta}$. On either graph, every local exploration algorithm requires at least $e^{(\gamma+\eta^*(\gamma)-o(1))d}$ distinct evaluations to achieve constant success probability. This lower bound is attained on $G_\rho$ by independent restarts, which succeed with probability $1-o(1)$ over the data and the restart entries using at most $e^{(\gamma+\eta^*(\gamma)+o(1))d}$ distinct evaluations. Optimizing the degree over this range gives the exponent $\min_{\frac12\log(4/3)\le\gamma<\gc}\{\gamma+\eta^*(\gamma)\}\approx0.243$, below the every-entry exponent $\gc\approx0.347$.
\end{theorem}

The preceding theorem quantifies the saving in evaluation cost obtained by allowing fresh starts while retaining exact recovery. A run that fails to find the exact nearest neighbor, however, may still return a point close to the query. At the degrees used for fresh starts, how close to the query can greedy navigation get from every entry? We answer this question by bounding the rank of the stopping point of a \emph{full greedy run}, which continues until no listed neighbor is closer. Here rank refers to the ordering of the data by distance to $q$, with rank $1$ denoting the exact nearest neighbor. For uniform data in the dense regime, let $\Delta=e^{\gamma d}$ with $0<\gamma<\gc$ and choose $\rho$ so that $nF_{d-1}(\rho)=\Delta$. The exponent governing the rank bound is $\nu_\gamma=\frac12\log\bigl(e^{4\gamma}/(4(e^{2\gamma}-1))\bigr)$. For every fixed $\varepsilon>0$, with high probability every full greedy run on $G_\rho$ or $\NN{\Delta}$ from every entry stops at a point of rank at most $e^{(\nu_\gamma+\varepsilon)d}$ for every query outside a set whose measure under $\sigma_{d-1}$ is $e^{-\Omega(d)}$.

For nonuniform data, relaxing exact recovery to a bound on rank also permits greater density variation while preserving the guarantee from every entry. Fix $\varepsilon>0$ and a query distribution $Q$ with a density on $\mathbb S^{d-1}$, chosen before the data are sampled. In the dense regime, let $\Delta=e^{\gamma d}$ with $\gc<\gamma\le\log2$ and $0\le\kappa<\gamma-\gc$, and assume the regularity and minimum-radius conditions of \cref{thm:approximate}, with density ratio at most $e^{\kappa d}$ in the prescribed cap at every center. Then, with high probability, every full greedy run on $\NN{\Delta}$ from every entry stops within $e^{o(d)}$ moves at a point of rank at most $2e^{(\kappa+\varepsilon)d}$ for every query outside a set of probability $o(1)$ under $Q$.

\subsection{Technical overview}\label{sec:overview}

The lens geometry determines whether a single greedy step can improve the current
point, whereas the results in \cref{sec:results} concern the outcome and cost
of the entire search. Turning the estimate for one step into these results
creates three separate proof problems. Above the curve, we must force progress
from every entry and rule out stopping near the query before the nearest
neighbor is found. For nonuniform data, this argument must also account for
changes in the direction distributions of closer points and in the distance covered by a
neighbor list of fixed length. Below the curve, the lower bound must hold even
when all neighbor lists are chosen after the complete sample is known. When fresh random entries are allowed, we must show that adaptive local exploration cannot improve on independent restarts per operation at the exponential scale. We must also show that the success probability of one run reflects the fraction of successful entries in most sampled data sets. Above the curve, we average over possible stopping points and compare directional distributions. Below the curve, we combine sampling and geometric estimates that hold for all graphs chosen from the sample. For fresh starts, we compare all adaptive routes and then resample the data to analyze two entries on the same graph.

\subsubsection{Navigation above the curve}

We begin with uniform data and $k=1$, since finding the nearest neighbor also gives the result for every larger fixed $k$. Near the nearest-neighbor scale, only a constant number of closer points are expected. An empty improving lens can therefore mean either that navigation has failed or that the current point is already the nearest neighbor. Choose an auxiliary neighbor radius $r$ for which $nF_{d-1}(r)$ lies above the critical degree and below $\Delta$ by fixed exponential margins. We distinguish the two cases through false local minima, points that have a closer data point but no closer neighbor in $G_r$. For a data point $x$ at distance $\theta$ from $q$, let $t=F_{d-1}(\theta)$ be the surface mass of closer locations and let $\ell=\sigma_{d-1}(C(x,r)\cap C(q,\theta))$ be the mass of the improving lens. Independence gives
\[
  \Prob\{\text{$x$ is a false local minimum}\mid x\}
  =(1-\ell)^{n-1}-(1-t)^{n-1}.
\]

The difference of powers is crucial because it remains small after summing over all possible stopping points. We choose a final query cap slightly larger than $C(q,\theta_1)$ and small enough relative to $r$ that \cref{lem:inside}(a) gives $\ell\ge(1-e^{-\Omega(d)})t$ throughout it. The mean value theorem bounds the difference of powers by $(n-1)(t-\ell)(1-\ell)^{n-2}$. For uniform $x$ and fixed $q$, the variable $t$ is uniform on $[0,1]$. Using $t-\ell\le e^{-\Omega(d)}t$ together with this lower bound on $\ell$, summing over the $n$ possible choices of $x$ leaves an exponentially small factor times the cancellation
\[
  n(n-1)\int_0^1 t(1-t)^{n-2}\,dt=1.
\]
The expected number of false local minima in the final cap is therefore exponentially small. Averaging over queries and applying Markov's inequality shows that, with probability $1-o(1)$ over the sample, every run entering the cap finds the nearest neighbor for most queries.

This conclusion inside the final cap gives an every-entry guarantee only after we show that every run starting outside the cap reaches it. At the cap boundary, almost all of a slightly smaller query cap lies within distance $r$ of the current point and gives a prescribed decrease in query distance. This smaller cap contains exponentially many data points in expectation, so the probability that it contains no such neighbor is doubly exponentially small. For a current point farther from $q$, the set of locations giving the same decrease within distance $r$ has at least as much surface measure. Hence every move outside the final cap decreases the distance to $q$ by a fixed fraction of $\theta_1$, and the run enters the cap after $O(1/\theta_1)$ moves.

It remains to control the additional moves inside the final cap, where each improvement may be much smaller. Any two visited points separated by at least two moves must be more than $r$ apart, since otherwise the later point would have been a better successor at the earlier point. The points in the even and odd positions therefore form two sets whose distinct members are more than $r$ apart. Summing the pairwise squared distances within either set and using that all its points lie in the final cap bounds its size by a constant independent of $d$. Together with the moves before entering the final cap, this gives $O(1)$ moves in the logarithmic regime and $e^{o(d)}$ moves in the dense regime. Finally, concentration and the exponential gap between $nF_{d-1}(r)$ and $\Delta$ show that $G_\rho$ and $\NN{\Delta}$ contain every edge of $G_r$ and have neighbor lists of length $O(\Delta)$. Since each expansion scans a complete neighbor list, the evaluation cost is at most $\Delta^{1+o(1)}$.

The same separation between reaching the final cap and completing the search inside it applies to nonuniform data in the dense regime, but density variation changes both the mass of an improving lens and the distance covered by a nearest-neighbor list. Inside the final cap, we choose one auxiliary radius no larger than the distance from any data point in that cap to the farthest point in its list, so every such list contains all data points within this radius. Fixing the current point's distance from $q$, we represent the current point and a point drawn from $P$, conditional on being closer to $q$, by the directions from $q$ toward them, represented by their normalized projections onto the subspace orthogonal to $q$. Replacing the closer direction by its opposite turns failure to lie in the auxiliary neighbor cap into a lower bound on the inner product of these two directions. The density bound near $q$ with parameter $\kappa$ controls both direction densities relative to uniform measure. The false local minimum probability is a convex nondecreasing function of the closer $P$-mass outside the neighbor cap, so we can maximize its average by concentrating the two direction densities on caps with a common axis (see \cref{lem:rearrangement}). This replaces arbitrary direction densities by an explicit configuration whose exponential rate can be calculated.

Keeping this directional comparison sharp requires averaging over the current point's distance before applying an upper bound to its direction density. At a fixed distance from $q$, the direction density is normalized by the mean of $f$ over all directions at that distance. The density of the current point's distance contains the same mean, so the two factors cancel upon integration. The remaining cap constraints require the final query cap to contain exponentially many data points at the density lower bound and the auxiliary neighbor cap to remain within the degree budget at the density upper bound. Cap measures scale with the $(d-1)$st power of the sine of the radius. To identify the critical degree, we examine these constraints as the fixed auxiliary margins shrink. Near the boundary, the ratio of the sine of the final query cap radius to the sine of the auxiliary neighbor radius approaches $e^{\kappa-\gamma}$ from above, while the exponent in the upper bound on the expected number of possible stopping points approaches $\kappa$.

The rearranged configuration then determines the critical curve. After replacing both direction densities by caps with a common axis, we split each cap according to projection onto that axis. The projection threshold for a closer point is chosen so that the uniform probability beyond it offsets the factor $e^{\kappa d}$ from the density bound. The threshold for the current point similarly offsets the upper bound on the expected number of possible stopping points. In the limiting comparison, the smallest product of thresholds that meets both requirements approaches $1-e^{-2\kappa}$. The spherical cosine identity gives the inner product required for the closer point to miss the auxiliary neighbor cap, which approaches $e^{2(\gamma-\kappa)}/2-1$. When this value exceeds the product of the two projection thresholds, the normalized components orthogonal to the common axis must have inner product bounded below by a positive constant, an event with exponentially small probability. The required inequality is
\[
  \frac{e^{2(\gamma-\kappa)}}{2}-1>1-e^{-2\kappa}
  \quad\Longleftrightarrow\quad
  e^{2\gamma}>4e^{2\kappa}-2.
\]
Under this strict inequality, the directional comparison and the averaging over the current point's distance make the expected number of false local minima exponentially small. To obtain a guarantee from every entry, we must also force runs starting outside the final cap to reach it. Just outside the cap, the density bound with parameter $\kappa$ provides exponentially many candidate neighbors that reduce the query distance by a fixed factor. At points $x$ farther from $q$, the density bound with parameter $\bar\kappa<\gamma$ at every center provides exponentially many candidate neighbors that reduce the distance by at least a constant times $\rho_f(x)$. Combining this progress toward the cap with $\rho_f(x)\ge\rho_{\min}=e^{-o(d)}$ and the bound on moves inside the cap gives $e^{o(d)}$ moves from every entry and evaluation cost $\Delta^{1+o(1)}$.

The above dense regime argument uses the fact that the nearest-neighbor angle tends to zero. In the logarithmic regime, this angle remains bounded away from zero, so the sufficient condition for $\NN{\Delta}$ involves both the density level and a density ratio bound requiring a positive lower bound. To allow arbitrarily low or zero density, we use a different graph construction. When $f\le e^{o(d)}$, the first task is to show that most query caps have enough $P$-mass even when the density at their centers is arbitrarily small or zero. We reduce the cap radius slightly and average the masses of caps of the reduced radius over their centers. Since no cap has more than $e^{o(d)}$ times its uniform mass, the centers whose reduced caps have at least half their uniform mass form a set of surface measure $e^{-o(d)}$. Spherical isoperimetry shows that enlarging this set by any fixed positive angle covers almost the whole sphere. For most queries, a reduced cap about one of these centers lies inside the original query cap, giving the required lower bound on its $P$-mass at an arbitrarily small cost in the exponent.

The resulting lower bound on the $P$-mass of most query caps supplies the sampled points needed by the graph construction. We use a fixed sequence of query caps shrinking toward the nearest-neighbor scale and subsamples whose sizes increase as the caps shrink. Every neighbor list contains a common small subsample, which lets distant entries reach the largest query cap, together with nearby points from the larger subsamples. From a point in one query cap, the intersection of its neighbor cap with the next smaller query cap has enough $P$-mass to contain a point from the larger subsample assigned to that next query cap, except with doubly exponentially small probability. Increasing the subsample sizes compensates for the shrinking query caps, while the density upper bound keeps the neighbor lists within the degree budget. At the final scale, using the full sample in the neighbor lists supplies the remaining closer points, and the false local minimum estimate prevents the run from stopping before it reaches the nearest neighbor. The resulting graph gives exact recovery from every entry for most uniform queries whenever $\bstar(c)<\beta<1$, using $O(1)$ moves and evaluation cost $O(\Delta)$.

\subsubsection{Failure below the curve}

To show that the degree curve is sharp, the lower bound must cover graphs whose neighbor lists are chosen after the complete sample is observed. Such a graph can combine nearby edges, which exploit local clusters in the sample, with farther edges chosen to help particular query directions. We separate the stored neighbors according to their distance from the current point. Random sampling bounds the number of nearby points that improve the query distance, while spherical geometry bounds the fraction of query directions for which any fixed farther point is improving. The second estimate holds for every potential stored edge, so the two bounds can be made simultaneous for all graphs constructed from the sample.

For uniform data, we select points whose ranks lie between an exponentially growing threshold above $k$ and twice that threshold. These points provide many possible entries outside $\TopSet_k(q)$. We split potential neighbors at a distance below the critical radius, calling points within this distance nearby and the remaining points farther. A nearby improving point must lie in a lens whose surface measure is exponentially smaller than the cap of points closer to $q$. Since a selected point has rank on the order of the threshold, only the same order of data points lie in this closer cap. Choosing the threshold so that its growth rate is smaller than the exponential loss of the lens therefore makes nearby improving points unlikely. Each farther point improves the distance for only an exponentially small fraction of query directions. Summing this directional bound over the allowed number of stored neighbors controls every possible neighbor list. Averaging both estimates over uniform queries shows that, with probability $1-o(1)$ over the sample, every graph whose degree is below the critical scale has the following property. For most queries, only an exponentially small fraction of the selected points have an improving listed neighbor. All remaining selected points are failing entries.

Averaging over uniform queries also converts the estimate for the fraction of selected points with an improving neighbor into lower bounds on average degree and evaluation cost. As a uniform query varies, every sample point belongs to the selected rank interval for nearly the same surface measure of queries. If a graph succeeds from every entry for most queries, all but a vanishing fraction of its points must therefore have neighbor lists at the critical scale. Otherwise, too many selected entries would have short lists containing no improving neighbor. Since expanding an entry scans its complete neighbor list, starting from a point with such a long list already incurs the corresponding evaluation cost. For an arbitrary graph, the same argument uses a tested degree $\Delta$ below the critical scale. For most uniform queries, all but a vanishing fraction of the selected entries either have no improving listed neighbor and fail or evaluate more than $\Delta$ points in their first expansion. These alternatives give $\Work_G(q;x_0)\ge\Delta$ for the selected entries in \cref{thm:informal-below}(a).

The nonuniform lower bound also separates nearby improving edges from farther ones, but matching the shifted curve requires a sampling density that suppresses both kinds of improvement. We place density close to $1$ in an inner ball and outside a surrounding annulus, while lowering the annulus density by a factor $e^{\kappa d}$. The inner ball supplies many entries, and the exterior supplies the points in $\TopSet_k(q)$. We place the queries at a prescribed distance from the common center and select inner points within a prescribed distance of the query. These selected entries and closer exterior points lie on opposite sides of the query, as in the rearranged configuration of the upper bound. The geometry is chosen so that, at degree $e^{\gamma d}$, farther exterior points improve selected entries for only an exponentially small fraction of the query locations, even after summing over all selected entries and their neighbor lists. If the annulus instead had density close to $1$, the expected number of points there that are closer to $q$ than a selected entry would have exponent approaching $\frac12\log((e^{2\gamma}+2)/4)$ (see \eqref{win:boundary-identities}). Lowering the annulus density by $e^{\kappa d}$ makes such an intermediate point exponentially unlikely when $\kappa$ exceeds $\frac12\log((e^{2\gamma}+2)/4)$ by a fixed amount. Random sampling also rules out nearby exterior points that improve the query distance. For all remaining query locations, every improving move from a selected inner point remains among the selected inner points, and no complete neighbor list scanned there contains a point of $\TopSet_k(q)$. Placing separated copies of this construction across the sphere gives the globally spread query distribution in \cref{thm:informal-below}(b).

The logarithmic regime requires a different obstruction because the nearest-neighbor angle stays bounded away from zero. We therefore concentrate most of the sampling mass in a cap. For most uniform queries, $\TopSet_k(q)$ lies inside the cap while many possible entries lie outside it. Success from every entry would require the graph to provide improving edges from exterior points into the cap for most query directions. Random sampling rules out nearby improving edges. For farther edges, the directional estimate gives a probability that is exponentially smaller than one over the product of the degree and the number of exterior points. Summing over all exterior points and their at most $\Delta$ stored neighbors therefore still leaves only an exponentially small fraction of queries with an improving edge into the cap. For every graph of degree at most $\Delta$, most queries consequently have failing exterior entries. Concentrating most of the sampling mass in a cap of exponentially small surface measure raises the necessary degree exponent above the uniform value $\bstar(c)$ (see \cref{enr:obstruction}).

\subsubsection{Relaxing the navigation guarantee}

For uniform data, failure from many entries below the critical degree leaves open how many fresh starts exact recovery requires. The difficulty is to compare a direct jump toward the nearest neighbor with routes through arbitrarily many intermediate points. A successful search must cross a range of query distances where improving neighbors are exponentially rare. Write $\theta_{\mathrm{trap}}$ for its outer boundary, where their expected number has exponent zero. Let $g^\rho(\varphi\to\psi)$ be the exponent governing the expected number of neighbors in $G_\rho$ near query distance $\psi$ from a point at distance $\varphi$. This combines an edge probability, symmetric in its endpoints, with the number of candidate points at the destination. With $\chi^\rho(\varphi,\psi)=\tfrac12(g^\rho(\varphi\to\psi)+g^\rho(\psi\to\varphi))$, the rate separates into a symmetric term and a logarithmic correction as
\[
 g^\rho(\varphi\to\psi)
 =\chi^\rho(\varphi,\psi)+\frac12\log\frac{\sin\psi}{\sin\varphi}.
\]
The logarithmic terms telescope along a route. In the degree range of \cref{thm:informal-restarts}, the symmetric term is largest on the diagonal for a fixed mean query distance, negative inside the trap boundary, and increasing in each distance. Intermediate steps can therefore only lower the total exponent. The best route jumps directly from the outer boundary to the nearest-neighbor scale, with exponent $-\eta^*(\gamma)$.

To turn this route comparison into a lower bound for adaptive exploration, we must account for choices made after earlier neighbor lists have been read. Revealing a cap containing each scanned list leaves the unrevealed points independent and uniform outside the exposed regions. With at most $e^{O(d)}$ entry requests and expansions, these regions have total measure $o(1)$ in the dense regime. Dividing query distances into short intervals then lets us sum over all possible routes without fixing the algorithm's choices in advance. Fresh entries hit $C(q,\theta_{\mathrm{trap}})$ with probability $e^{-\omega(d)}$, so constant success probability requires $e^{(\eta^*(\gamma)-o(1))d}$ expansions. Caching creates a further difficulty, since many expansions could evaluate the same points. Every set of expanded points of size below $e^{\eta^*(\gamma)d}$ by a fixed exponent margin contains a subset of at least a $1/d$ fraction whose points are separated by a fixed fraction of $\theta_1$. The total pairwise overlap of these neighbor lists is exponentially smaller than the sum of their lengths. The union therefore has at least the degree times the number of expansions, up to a subexponential factor, giving the distinct-evaluation exponent $\gamma+\eta^*(\gamma)$.

To attain the preceding evaluation cost lower bound by independent restarts, we show that, for each query fixed before sampling and most sampled threshold graphs, greedy search succeeds from a fraction at least $e^{-(\eta^*(\gamma)+o(1))d}$ of entries. With the nearest neighbor removed and its distance fixed, the greedy path on the remaining points is independent of the nearest neighbor's hidden direction. The probability that this direction places the nearest neighbor in a neighbor cap along the path gives the single-run exponent. An estimate up to $e^{o(d)}$ factors, however, cannot establish concentration of the successful-entry fraction by a direct second-moment calculation. We retain paths of subexponential length that stay outside a cap slightly smaller than $C(q,\theta_{\mathrm{trap}})$, preserving the single-run exponent. Resampling after revealing one such path produces a second path independent of the first, and rotational symmetry makes their joint success probability against the hidden direction an exact product. We then compare the second path on the original and resampled data. For the retained paths, the probability that replaced points change the second path has decay exponent strictly larger than $\eta^*(\gamma)$. To affect joint success, such a change must also occur with a hidden direction that makes the first path succeed, contributing a further factor $e^{-(\eta^*(\gamma)-o(1))d}$. The resulting error is smaller than the product of the success probabilities, so the fraction of successful entries concentrates. Independent restarts on the same graph therefore attain the matching exponent. Minimizing $\gamma+\eta^*(\gamma)$ over the degree range in \cref{thm:informal-restarts} gives approximately $0.243$.

Even when some entries do not reach the nearest neighbor, the subcritical lens estimates force every full greedy run to stop inside a specified query cap. The number of sampled points in that cap therefore bounds the stopping rank. For nonuniform data, the progress estimates outside the final cap in the upper-bound proof remain valid over a wider density range than exact recovery, yielding the rank guarantee (see \cref{thm:approximate}).
\subsection{Related work}\label{sec:related}

Random input models have enabled quantitative analyses of greedy search on specific proximity graphs. Laarhoven~\cite{Laarhoven2018GraphTradeoffs} derived tradeoffs between query time and space for greedy walks with random restarts, matching the hashing and filtering tradeoff of~\cite{AndoniEtAl2017Tradeoffs} as the approximation factor approaches one and storage approaches linear. These methods build on locality-sensitive hashing and filtering~\cite{AndoniRazenshteyn2015,BeckerEtAl2016Sieving}. For uniform spherical data, Prokhorenkova and Shekhovtsov~\cite{ProkhorenkovaShekhovtsov2020} proved exact navigation from the prescribed random entry for $\log\log n\ll d\ll\log n$ when the query is sampled near a designated nearest neighbor and the neighbor radius exceeds $\sqrt2$ times the nearest-neighbor scale. With beam search and added long edges, their query time exponent, normalized by dimension using natural logarithms, can approach $\frac12\log(27/16)\approx0.262$~\cite[Corollary~3]{ProkhorenkovaShekhovtsov2020}. Subsequent analyses address missing edges~\cite{ShrivastavaSongXu2023}, the degree and expected path length of sparse neighborhood graphs~\cite{MaEtAl2025SNG}, and the transition from power-law to subpolynomial growth in search cost as the data set grows relative to its intrinsic dimension~\cite{MaghrebiEtAl2026PowerLaw}. Random input analyses also cover hierarchical graphs on a torus, including models with densities bounded above and below by fixed positive constants~\cite{BhamidiEtAl2026Geometry}. For $d=o(\log n/\log\log n)$, that hierarchical graph analysis gives approximate recovery from every entry for all queries and expected move bounds for a fixed query.

Alongside analyses of prescribed graphs, navigable graph theory asks how many edges are needed to provide an improving neighbor from every data point toward every other data point. \cite{DiwanEtAl2024Navigable} gave sparse constructions for arbitrary point sets and average-degree lower bounds for random points in $O(\log n)$ dimensions. Khanna, Padaki, and Waingarten~\cite{KhannaEtAl2026Sparse} related sparse graph construction to set cover and gave approximation algorithms and hardness results, while \cite{ConwayEtAl2026Constructing} developed faster construction algorithms. Allowing each point a small fraction of destinations for which no improving edge is required permits sparser graphs~\cite{AviMusco2026AlmostNavigable}. Waizer and Danieli~\cite{WaizerDanieli2026Price} proved matching sparsity lower bounds, including a lower bound on every vertex's degree for random spherical data, by bounding how many ordered pairs of data points a stored edge can serve. Adaptive beam search connects these guarantees for destinations in the data set to external queries, giving approximate recovery and exact recovery for a particular parameter choice~\cite{AlJazzaziEtAl2025AdaptiveBeam}.

External queries can also be handled by imposing geometric conditions on an otherwise arbitrary data set. Indyk and Xu~\cite{IndykXu2023WorstCase} proved approximate search guarantees for a variant of DiskANN with slower preprocessing on data of bounded intrinsic dimension. Its query cost is polylogarithmic in the ratio of the largest to the smallest pairwise distance. They also constructed inputs on which the published HNSW, NSG, and faster preprocessing DiskANN implementations empirically take linear query time to achieve reasonable accuracy. \cite{GollapudiEtAl2025SortPrune} strengthened the DiskANN guarantees and analyzed beam search. Sparse graphs for approximate search have also been constructed in doubling metrics and Euclidean space~\cite{AryaMount1993,LuTao2025Proximity,HarPeledRaichelRobson2025}. Using a sparse graph together with an auxiliary data structure, \cite{GilibertiEtAl2026WithoutSpread} obtained logarithmic query time in the data set size for fixed dimension and a fixed approximation factor. Pruning constructions give exact recovery when the query lies within a prescribed distance of the data~\cite{HarwoodDrummond2016FANNG,PengEtAl2023Tau,LiYanLu2026FastConvergent}.

Even on a fixed graph, the entry rule changes both the probability of success and the cost of search. Adaptive entry selection has been compared with fixed central entries through theoretical bounds and experiments~\cite{OguriMatsui2024Entry}. Laarhoven~\cite[full version, Appendix~A.3]{Laarhoven2018GraphTradeoffs} analyzed random restarts. For restarts on one sampled graph, Prokhorenkova and Shekhovtsov~\cite[Supplementary Section~E]{ProkhorenkovaShekhovtsov2020} identified dependence between attempts as a technical obstacle. Since caching charges repeated evaluations only once, the total cost also depends on overlap between scanned neighbor lists. \cite{WangEtAl2024SteinerHardness} studied graph search hardness using the size of the union of scanned neighbor lists on a given graph.

The analyses of prescribed proximity graphs establish when those constructions succeed, while our degree comparison asks whether another graph chosen after observing the sample could use a smaller degree for success from every entry. The navigable graph lower bounds already allow arbitrary graph choices, but concern destinations in the data set rather than exact nearest neighbors of external queries. For nonuniform data, the hierarchical graph analysis uses fixed density bounds, whereas we allow density ratios exponential in dimension and determine their effect on the required degree in the dense regime over the variation range in \eqref{eq:curve}. For restarts, we address the dependence between attempts identified by Prokhorenkova and Shekhovtsov and compare independent restarts with local exploration algorithms that cache evaluated points, in the graph models and degree range of \cref{thm:informal-restarts}.

\section{Geometric foundations of the critical curve}\label{sec:prelim}

In this section, we develop geometric and probability estimates that connect neighbor radii, improving steps, and ranks under uniform and nonuniform sampling.

\subsection{Conventions}

For a query $q$ and $x\in D_n$, define $\rank_q(x)=1+\#\{y\in D_n:\delta(q,y)<\delta(q,x)\}$. We exclude queries on the bisectors $\{q\in\mathbb S^{d-1}:\delta(q,x)=\delta(q,y)\}$ of distinct data points $x,y\in D_n$, since each bisector has surface measure zero. For a query fixed before sampling, distance ties also have data probability zero. For a nonempty neighbor list $G(x)$, the \emph{successor} of $x$ is its listed point closest to $q$. Write $x_0,x_1,\ldots$ for the successive points visited by a greedy run, whose distances to $q$ strictly decrease. A run that succeeds within $H$ complete expansions has evaluation cost at most $1+H\Delta(G)$, including the entry and the entire final expansion.

For the asymptotic bounds, we assume throughout that $d\to\infty$, $d=O(\log n)$, and $\log n=o(d^2)$. Unless stated otherwise, the constants implicit in $O(\cdot)$ and $\Omega(\cdot)$ may depend on $k$, on $c$ in the logarithmic regime, and on the fixed exponents, multipliers, and margins in the hypotheses, including $\beta$, $\gamma$, $\kappa$ and $\eps$ where present. All bounds hold for sufficiently large $n$, with the threshold allowed to depend on the sequence $d=d(n)$.

\subsection{Spherical caps and nearest-neighbor scales}

To convert neighbor radii into degree scales for uniform data, we estimate the following spherical cap probabilities, writing $F=F_{d-1}$.

\begin{lemma}\label{lem:cap}
For every fixed $\eps>0$, uniformly for $0<\theta\le\pi/2-\eps$,
\[
  F(\theta)=\frac{(\sin\theta)^{d-1}}{\sqrt{2\pi(d-1)}\,\cos\theta}\bigl(1+O(d^{-1})\bigr).
\]
The angular density is $F'(\theta)=\alpha_{d-1}(\sin\theta)^{d-2}$, where the normalizing constant satisfies
\[
  \alpha_{d-1}=\frac{\Gamma(d/2)}{\sqrt\pi\,\Gamma((d-1)/2)}
  =\sqrt{\frac{d-1}{2\pi}}\bigl(1+O(d^{-1})\bigr).
\]
For every $0<\theta<\pi/2$, the exact remainder can be written using a number $\xi$ depending on $d$ and $\theta$ as
\[
  F(\theta)=\frac{\alpha_{d-1}(\sin\theta)^{d-1}}{(d-1)\cos\theta}(1+\xi)^{-1}
  \qquad\text{and}\qquad
  0\le\xi\le\frac{\tan^2\theta}{d-1}.
\]
\end{lemma}

\begin{proof}
The distance from a fixed point to a uniform point of $\mathbb S^{d-1}$ has density $\alpha_{d-1}(\sin t)^{d-2}$ on $[0,\pi]$. Substituting $u=\sin t$ and integrating by parts gives
\[
  \int_0^\theta(\sin t)^{d-2}\,dt
  =\frac{(\sin\theta)^{d-1}}{(d-1)\cos\theta}
  -\frac1{d-1}\int_0^{\sin\theta}\frac{u^d}{(1-u^2)^{3/2}}\,du.
\]
The subtracted term is nonnegative and at most $\tan^2\theta/(d-1)$ times the integral on the left. This gives the stated bound on $\xi$, while Stirling's formula gives the estimate for $\alpha_{d-1}$.
\end{proof}

The cap estimate applies at $\theta_1$ because comparing $nF(\theta)$ with $1$ at fixed radii shows that $\theta_1$ stays bounded away from $\pi/2$ in both regimes. Substituting $nF(\theta_1)=1$ into \cref{lem:cap} gives $\sin\theta_1=e^{-1/c}(1+O(\log d/d))$ in the logarithmic regime. In the dense regime, $\theta_1\to0$ and the nearest-neighbor scale satisfies
\begin{equation}\label{eq:theta1-dense}
  \frac1{\sin\theta_1}
  =n^{1/(d-1)}\bigl(1+O(\log d/d)\bigr)
  \qquad\text{and}\qquad
  \log\frac1{\theta_1}
  =(1+o(1))\frac{\log n}{d-1}=o(d).
\end{equation}

To compare other cap radii with $\theta_1$, we normalize the estimate in \cref{lem:cap} by $F(\theta_1)=1/n$. For every fixed $\eps>0$, the resulting comparison is uniform for $0<\theta\le\pi/2-\eps$ and gives
\begin{equation}\label{eq:cap-ratio}
  nF(\theta)=\Bigl(\frac{\sin\theta}{\sin\theta_1}\Bigr)^{d-1}
  \frac{\cos\theta_1}{\cos\theta}\bigl(1+O(d^{-1})\bigr).
\end{equation}

To express the two degree choices as neighbor radii, fix $0<\beta<1$ and $\gamma>0$, with $\gamma<1/c$ in the logarithmic regime. Define $\rho_\beta$ and $r_\gamma$ by $nF(\rho_\beta)=n^\beta$ and $nF(r_\gamma)=e^{\gamma d}$, respectively. Both radii stay bounded away from $\pi/2$, so \eqref{eq:cap-ratio} gives
\begin{equation}\label{eq:radii}
  \frac{\sin\rho_\beta}{\sin\theta_1}
  =n^{\beta/(d-1)}\bigl(1+O(d^{-1})\bigr)
  \qquad\text{and}\qquad
  \frac{\sin r_\gamma}{\sin\theta_1}
  =e^\gamma\bigl(1+O(d^{-1})\bigr).
\end{equation}

\subsection{Improving geometry and the uniform critical scale}

To determine which neighbor radii in \eqref{eq:radii} permit greedy progress, we estimate the fraction of closer locations available within the neighbor radius. For $\theta=\delta(x,q)$, this fraction is the surface measure of the lens $C(x,\rho)\cap C(q,\theta)$ divided by $F(\theta)$. The next lemma quantifies the exponential loss.

\begin{lemma}[Lens rate]\label{lem:lens}
Let $\delta(x,q)=\theta\in(0,\pi/2)$ and $0<\rho<\pi/2$. Put $a=\cos\rho$ and $b=\cos\theta$, and define $J_\theta(\rho)$ to measure the exponential loss by
\[
  J_\theta(\rho)=
  \begin{cases}
    0, & a\le b^2,\\[1mm]
    -\dfrac12\log\dfrac{(1-a)(a-\cos2\theta)}{\sin^4\theta}, & a>b^2.
  \end{cases}
\]
For every fixed $\eps>0$, the following statements hold.
\begin{enumerate}[label=\textup{(\alph*)},leftmargin=*]
\item Uniformly for $0<\theta\le\pi/2-\eps$ and $0<\rho\le\pi/2-\eps$,
\[
  \frac{\sigma_{d-1}\bigl(C(x,\rho)\cap C(q,\theta)\bigr)}{F(\theta)}
  =\exp\bigl\{-(d-1)J_\theta(\rho)+O(\log (d-1)+\log(1/\sin\theta)+\log(1/\sin\rho))\bigr\}.
\]
If $\rho\ge\eps\theta$, the error simplifies to $O(\log (d-1)+\log(1/\sin\theta))$.
\item Uniformly for $0<\theta\le\pi/2-\eps$ and $0<\rho<\pi/2$,
\[
  \frac{\sigma_{d-1}\bigl(C(x,\rho)\cap C(q,\theta)\bigr)}{F(\theta)}
  \le e^{-(d-1)J_\theta(\rho)}\bigl(1+O(d^{-1})\bigr).
\]
\end{enumerate}
\end{lemma}

\begin{proof}
The lens is determined by the projections of a point onto the plane spanned by $x$ and $q$. By rotational invariance, choose coordinates in this plane so that $x=e_1$ and $q=be_1+se_2$, where $s=\sin\theta$. If $Y$ is uniform on $\mathbb S^{d-1}$ and $z=(Y_1,Y_2)$, then $z$ has density
\[
  f_{d-1}(z)=\frac{d-2}{2\pi}(1-\|z\|^2)^{(d-4)/2}
\]
on the unit disk. The remaining coordinates have already been integrated out in this marginal density. The two cap conditions give the planar region $\mathcal D=\{z\in\mathbb R^2:\|z\|\le1,\ z_1\ge a,\ bz_1+sz_2\ge b\}$, so the lens mass is $\int_{\mathcal D} f_{d-1}(z)\,dz$.

Because $f_{d-1}(z)$ decreases with $\|z\|$, the exponential scale of this integral is determined by the point $p\in\mathcal D$ of smallest Euclidean norm. If the constraint $z_1\ge a$ is omitted, the closest point satisfying $bz_1+sz_2\ge b$ is $b(b,s)$, whose first coordinate is $b^2$. It remains feasible when $a\le b^2$, so in this case $p=b(b,s)$. When $a>b^2$, $z_1\ge a$ excludes that point, and both linear constraints must be tight. Solving them gives $p=(a,b(1-a)/s)$. Therefore, with $u_p=1-\|p\|^2$,
\[
  u_p=
  \begin{cases}
    s^2, & a\le b^2,\\[1mm]
    \dfrac{(1-a)(a-\cos2\theta)}{s^2}, & a>b^2.
  \end{cases}
\]
In the second case, $u_p$ is positive because $a>b^2>\cos2\theta$.

The density at $p$ immediately gives an upper bound on the integral, since $\mathcal D$ has area at most $\pi$. To obtain a lower bound with the same leading exponent, we need a region of positive area on which the density remains within a constant factor of its value at $p$. Set $\eta=u_p/(5(d-1))$ and use the parallelogram $\mathcal R=p+\{\alpha e_1+\alpha'(b,s):0\le\alpha,\alpha'\le\eta\}$. Moving in either generating direction can only increase the two linear forms that define $\mathcal D$, so $\mathcal R$ remains inside both half-planes. Every $z\in\mathcal R$ also satisfies $\|z\|^2\le\|p\|^2+5\eta$, and hence $1-\|z\|^2\ge(1-1/(d-1))u_p$. This keeps $\mathcal R$ inside the unit disk and its density within a constant factor of the density at $p$. The area $s\eta^2$ therefore gives a lower bound that differs from the upper bound by at most a constant multiple of $(d-1)^2/(s u_p^2)$. The lens mass has logarithm
\[
  \log\sigma_{d-1}\bigl(C(x,\rho)\cap C(q,\theta)\bigr)
  =\frac {d-1}2\log u_p+O\bigl(\log (d-1)+\log(1/s)+\log(1/u_p)\bigr).
\]

It remains to express the error and the leading term in the parameters of the lemma. If $a\le b^2$, then $u_p=s^2$. If $a>b^2$, then $a-\cos2\theta=2s^2-(1-a)\ge s^2$, so $u_p\ge1-a\ge2\rho^2/\pi^2$. Consequently, $u_p\ge c_0\min\{s^2,\rho^2\}$ for an absolute constant $c_0>0$. Since $\rho$ and $\sin\rho$ are comparable on $(0,\pi/2)$, this gives the error in part~(a). If $\rho\ge\eps\theta$, there is a constant $c_\eps>0$, depending only on $\eps$, such that $u_p\ge c_\eps s^2$. This removes the term $\log(1/\sin\rho)$. By \cref{lem:cap}, $\log F(\theta)=(d-1)\log s+O(\log(d-1))$ uniformly in the stated range. Dividing by $F(\theta)$ therefore replaces the leading factor $u_p^{(d-1)/2}$ by $(u_p/s^2)^{(d-1)/2}$. Since $u_p/s^2=e^{-2J_\theta(\rho)}$, this proves part~(a).

The area estimate in part~(a) introduces an error depending on $\rho$. For the sharper upper bound in part~(b), we enclose the entire lens in one spherical cap. Convexity of $\mathcal D$ and minimality of $p$ give $z\cdot p\ge\|p\|^2$ for every $z\in\mathcal D$, since moving from $p$ toward $z$ cannot decrease the squared norm. After lifting back to $\mathbb S^{d-1}$, this places the lens in the cap of angular radius $\arccos\|p\|$ centered at $p/\|p\|$. The constraint $p\cdot(b,s)\ge b$ also gives $\|p\|\ge b$, so this cap has radius at most $\theta$. We may therefore apply \cref{lem:cap} to both caps to obtain
\[
  \frac{\sigma_{d-1}\bigl(C(x,\rho)\cap C(q,\theta)\bigr)}{F(\theta)}
  \le\left(\frac{u_p}{s^2}\right)^{(d-1)/2}\frac{b}{\|p\|}\bigl(1+O(d^{-1})\bigr)
  \le e^{-(d-1)J_\theta(\rho)}\bigl(1+O(d^{-1})\bigr),
\]
which proves part~(b).
\end{proof}

The boundary where $J_\theta(\rho)$ vanishes will be used at several query distances. Write $r_*(\theta)$ for this radius, defined by $\cos r_*(\theta)=\cos^2\theta$. Then $J_\theta(\rho)=0$ for $\rho\ge r_*(\theta)$ and is positive for $\rho<r_*(\theta)$. The lens estimate varies the neighbor for a fixed query, whereas a bound for one stored edge must fix its endpoints $x,y$ and vary the query direction. The next lemma estimates the fraction of these directions for which $y$ is closer to the query than $x$.

\begin{lemma}[Improving directions]\label{lem:directions}
Let $x,y\in \mathbb S^{d-1}$ with $\delta(x,y)=t\in(0,\pi)$, and let $0<\theta<\pi/2$. Among the queries $q$ satisfying $\delta(q,x)=\theta$, those for which $\delta(q,y)<\theta$ form a spherical cap in the $(d-2)$-sphere of directions at $x$. Its cosine threshold is $h=\cot\theta\tan(t/2)$. The cap is empty when $h\ge1$. When $h<1$, its normalized measure is at most $\sqrt {d-1}\,(1-h^2)^{(d-4)/2}$ and decreases as $t$ increases. Moreover, for $d\ge5$ and $0\le h<1$, this measure is at least $c_0(d-1)^{-1/2}(1-h^2)^{(d-2)/2}$ for an absolute constant $c_0>0$. Consequently, for every fixed $\eps>0$, the measure is $(1-h^2)^{(d-1)/2}e^{O_\eps(\log (d-1))}$ uniformly for $0\le h\le1-\eps$.
\end{lemma}

\begin{proof}
Write $q=\cos\theta\,x+\sin\theta\,v$, where $v$ is a unit vector in $x^\perp$. Under normalized measure on the queries at distance $\theta$ from $x$, the direction $v$ is uniform on this unit sphere. Similarly, write $y=\cos t\,x+\sin t\,w$ for a unit vector $w\in x^\perp$. Substituting these representations into the condition $q\cdot y>\cos\theta$ gives
\[
\delta(q,y)<\theta
\quad\Longleftrightarrow\quad
v\cdot w>\cot\theta\tan(t/2)=h.
\]
Thus the improving directions form a spherical cap with cosine threshold $h$. Because $\tan(t/2)$ increases with $t$, the threshold increases and the cap measure decreases as $t$ increases. When $h\ge1$, no unit vector $v$ can satisfy $v\cdot w>h$, so the cap is empty.

It remains to estimate the measure of this cap. For uniform $v$ on the unit sphere in $x^\perp$, let $Z=v\cdot w$. Then $Z$ has density $\alpha_{d-2}(1-z^2)^{(d-4)/2}$ on $[-1,1]$, where $\alpha_{d-2}=\Gamma((d-1)/2)/(\sqrt\pi\,\Gamma((d-2)/2))$. Stirling's formula gives $\alpha_{d-2}=\sqrt{(d-2)/(2\pi)}(1+O(d^{-1}))$, so $\alpha_{d-2}\le\sqrt {d-1}$ for all sufficiently large $d$. Since the density decreases on $[0,1]$, we have
\[
\Prob(Z>h)
=\alpha_{d-2}\int_h^1(1-z^2)^{(d-4)/2}\,dz
\le\sqrt {d-1}\,(1-h^2)^{(d-4)/2}.
\]

For a matching lower bound, we integrate over $[h,h+\ell_{\mathrm{int}}]$, where $\ell_{\mathrm{int}}=(1-h^2)/(d-1)$. When $d\ge5$, this interval lies in $[h,1]$, and every $z$ in it satisfies $1-z^2\ge(1-h^2)(1-3/(d-1))$. The density throughout the interval is therefore at least a constant multiple of its value at $h$. Integrating over its length and using the lower bound on $\alpha_{d-2}$ gives $\Prob(Z>h)\ge c_0(d-1)^{-1/2}(1-h^2)^{(d-2)/2}$ after adjusting the absolute constant $c_0$. The two bounds also give the stated estimate uniformly for $0\le h\le1-\eps$.
\end{proof}

The probability estimate in \cref{lem:directions} also controls the fraction of a query cap lying outside a prescribed neighbor radius. Conditioning a uniform point in that cap on its distance from the query leaves a uniform direction, so the containment question again reduces to one coordinate.

\begin{lemma}[Cap containment]\label{lem:inside}
Let $\delta(x,q)=\theta$ and let $Y$ be uniform on $C(q,\psi_0)$, where $0<\theta,\psi_0<\pi/2$. Fix a neighbor radius $0<\rho<\pi$.
\begin{enumerate}[label=\textup{(\alph*)},leftmargin=*]
\item Suppose that $\cos\theta\cos\psi_0-\cos\rho\ge\tau\sin\theta\sin\psi_0$ for some $\tau>0$. If $0<\tau<1$, then $\Prob(\delta(x,Y)>\rho)\le\sqrt {d-1}\,(1-\tau^2)^{(d-4)/2}$. If $\tau\ge1$, this probability is zero.
\item If $\cos\theta\cos\psi_0\ge\cos\rho$, then $\Prob(\delta(x,Y)\le\rho)\ge\frac12$. Consequently, if $\cos\rho\le\cos^2\theta$, then $C(x,\rho)\cap C(q,\theta)$ has at least half the measure of $C(q,\theta)$.
\end{enumerate}
\end{lemma}

\begin{proof}
Choose a unit vector $w\perp q$ such that $x=\cos\theta\,q+\sin\theta\,w$. Conditional on $\psi=\delta(q,Y)$, write $Y=\cos\psi\,q+\sin\psi\,u$, where $0\le\psi\le\psi_0$ and $u$ is uniform on the unit sphere in $q^\perp$. Setting $Z=w\cdot u$ gives $x\cdot Y=\cos\theta\cos\psi+\sin\theta\sin\psi\,Z$, so $Z$ determines membership in the cap about $x$.

For part~(a), the event $\delta(x,Y)>\rho$ implies $x\cdot Y<\cos\rho$. For $\psi>0$, the preceding identity and the inequalities $\cos\psi\ge\cos\psi_0$ and $\sin\psi\le\sin\psi_0$ give
\[
Z<\frac{\cos\rho-\cos\theta\cos\psi}{\sin\theta\sin\psi}
\le-\tau\frac{\sin\psi_0}{\sin\psi}
\le-\tau.
\]
By symmetry, the coordinate tail bound in the proof of \cref{lem:directions} gives $\Prob(\delta(x,Y)>\rho)\le\Prob(Z<-\tau)\le\sqrt {d-1}\,(1-\tau^2)^{(d-4)/2}$ when $0<\tau<1$. When $\tau\ge1$, the event is impossible because $Z\ge-1$.

For part~(b), conditional on any fixed value of $\psi$, symmetry gives $\Prob(Z\ge0\mid\psi)=1/2$. Whenever $Z\ge0$, we have $x\cdot Y\ge\cos\theta\cos\psi\ge\cos\theta\cos\psi_0\ge\cos\rho$, and hence $\delta(x,Y)\le\rho$. Therefore, $\Prob(\delta(x,Y)\le\rho\mid\psi)\ge1/2$ for every $\psi$. Averaging over $\psi$ proves $\Prob(\delta(x,Y)\le\rho)\ge1/2$. Finally, taking $\psi_0=\theta$ shows that if $\cos\rho\le\cos^2\theta$, then $\sigma_{d-1}(C(x,\rho)\cap C(q,\theta))/F(\theta)\ge1/2$, which proves the lens claim.
\end{proof}

To pass from these geometric estimates to the critical degree, we evaluate the boundary at $\theta_1$. Here $r_*(\theta_1)=\rho^*$ as in \eqref{eq:intro-critical-radius}, and \cref{lem:inside}(b) guarantees at least half the measure of $C(q,\theta_1)$ in the lens. Substituting $\sin\rho^*/\sin\theta_1=\sqrt{1+\cos^2\theta_1}$ into \eqref{eq:cap-ratio}, we obtain the critical degree in both regimes as
\begin{equation}\label{eq:uniform-critical-degree}
  \Dstar=
  \begin{cases}
    n^{\bstar(c)+o(1)}, & d=\lfloor c\log n\rfloor,\\
    e^{(\gc+o(1))d}, & \sqrt{\log n}\ll d\ll\log n.
  \end{cases}
\end{equation}

\subsection{Ranks and concentration}

The geometric estimates give cap probabilities, while navigation depends on the points present in the sample. To relate these probabilities to rank, suppose that the data are uniform, fix a label $i$ and a query $q$ independent of $D_n$, and write $x=X_i$. Conditional on $x$ and $q$, the number of other data points closer to $q$ is binomial, so
\begin{equation}\label{eq:rank-law}
  \rank_q(x)-1\sim\mathrm{Bin}\bigl(n-1,F(\delta(q,x))\bigr).
\end{equation}
To control deviations from the mean in \eqref{eq:rank-law}, we use the standard Chernoff bounds. For $B\sim\mathrm{Bin}(N,p)$ with mean $\mu=Np>0$ and $s>0$, these bounds are
\begin{equation}\label{eq:chernoff}
  \Prob(B\ge\mu+s)\le\exp\Bigl(-\frac{s^2}{2(\mu+s/3)}\Bigr)
  \qquad\text{and}\qquad
  \Prob(B\le\mu-s)\le\exp\Bigl(-\frac{s^2}{2\mu}\Bigr).
\end{equation}

When averaging these estimates over a point or a query, we also use the fact that $F(\delta(q,x))$ is uniform on $[0,1]$ for fixed $q$ and uniform $x$, or for fixed $x$ and uniform $q$.

\subsection{Nonuniform local scales}\label{sec:prelim-nonuniform}

\phantomsection\label{eq:local-nn-radius}
Under nonuniform sampling, the number of points in a cap is governed by its $P$-probability rather than its surface measure. To use the geometric estimates in this setting, we control the density on neighborhoods scaled by the local radii from \cref{sec:model}. We retain $\rho_f(x)=\rho(x)$ for the smallest radius whose cap has $P$-probability at least $1/n$, so $nP(C(x,\rho_f(x)))=1$ and $\rho_{\min}=\inf_{x\in\mathbb S^{d-1}}\rho_f(x)$.

For degree $\Delta=\lceil e^{\gamma d}\rceil$ with fixed $\gamma>0$ in the dense regime, \eqref{eq:radii} gives neighbor radius $(e^\gamma+o(1))\theta_1$ under uniform sampling. We therefore control the nonuniform density on $C(x,Ke^\gamma\rho_f(x))$, where the fixed constant $K>0$ specifies the size of the neighborhood. For $\kappa\ge0$, we call $f$ \emph{$(\kappa,K)$-regular} if $\log(1/\rho_{\min})=o(d)$ and, at every center $x$, the essential oscillation of $\log f$ on $C(x,Ke^\gamma\rho_f(x))$ is at most $\kappa d$. For a measurable set $A$, this means $\operatorname*{ess\,osc}_A\log f=\operatorname*{ess\,sup}_A\log f-\operatorname*{ess\,inf}_A\log f$. The density thus varies by at most a factor $e^{\kappa d}$ on each such cap. The condition on $\rho_{\min}$ keeps the later expansion bound $O(1+\rho_{\min}^{-1})$ subexponential in $d$. Uniform data satisfy these conditions with $\kappa=0$ and $\rho_f(x)=\theta_1$.

The density bound transfers the surface measures in the lens estimates to probabilities under $P$. If measurable sets $A$ and $B$ of positive surface measure lie in one cap $C(x,Ke^\gamma\rho_f(x))$, integrating the upper and lower density bounds over them gives
\begin{equation}\label{eq:local-mass-comparison}
  e^{-\kappa d}\frac{\sigma_{d-1}(B)}{\sigma_{d-1}(A)}
  \le \frac{P(B)}{P(A)}
  \le e^{\kappa d}\frac{\sigma_{d-1}(B)}{\sigma_{d-1}(A)}.
\end{equation}

To express the dense critical curve in \eqref{eq:curve} as a bound on density variation at a fixed degree exponent, define
\begin{equation}\label{eq:tolerance-function}
  h(\gamma)=\frac12\log\frac{e^{2\gamma}+2}{4}.
\end{equation}
For $\gc\le\gamma\le\log2$, the boundary of the dense curve in \eqref{eq:curve} is $\kappa=h(\gamma)$.

In the logarithmic regime, the local density level also changes the leading exponent. Write $\lambda=\lim_{n\to\infty}(\log n)/d=1/c$. If $\log f=\mu d+o(d)$ uniformly in a region, the expected number of points in a cap contained there is its surface measure multiplied by $e^{(\lambda+\mu+o(1))d}$. Replacing $\lambda$ by $\lambda+\mu$ changes both the local nearest-neighbor radius and the radius reached by a neighbor list of prescribed length. The logarithmic nonuniform bounds in \cref{sec:tolerance-moderate,sec:logarithmic-lower} therefore depend on density level and variation.

\section{The uniform critical curve}\label{sec:upper}

We now prove that \eqref{eq:uniform-critical-degree} gives the critical degree scale for uniform data. The upper bound applies to graphs that retain every edge of a threshold graph. We say that $G$ \emph{contains} $G_r$ if $G_r(x)\subseteq G(x)$ for every $x\in D_n$. The following theorem states the matching upper and lower bounds.

\begin{theorem}[Uniform critical degree]\label{thm:critical}
For uniform data on $\mathbb S^{d-1}$, fix $k\ge1$. In the logarithmic regime, let $\Delta=\lceil n^\beta\rceil$ with fixed $0<\beta<1$. In the dense regime, let $\Delta=\lceil e^{\gamma d}\rceil$ with fixed $\gamma>0$.
\begin{enumerate}[label=\textup{(\alph*)},leftmargin=*]
\item \emph{(Above the critical degree.)} Fix an absolutely continuous query distribution $Q$ before drawing the data. In the logarithmic regime, suppose that $\beta>\bstar(c)$, choose $\beta'\in(\bstar(c),\beta)$, and set $\rho'=\rho_{\beta'}$. There is a constant $\eta>0$ such that, with probability at least $1-n^{-\eta}$ over $D_n$, there is a measurable set $Q_D\subseteq\mathbb S^{d-1}$ with $\Prob_{q\sim Q}(q\in Q_D)\ge1-n^{-\eta}$ on which every graph $G$ containing $G_{\rho'}$ and satisfying $\Delta(G)\le2\Delta$ succeeds from every entry. Each search uses $O(1)$ moves and has evaluation cost at most $\Delta^{1+o(1)}$. In the dense regime, suppose that $\gamma>\gc$, choose $\gamma'\in(\gc,\gamma)$, and set $\rho'=r_{\gamma'}$. There are positive constants $c_1$ and $c_2$ such that, with probability at least $1-e^{-c_1d}$ over $D_n$, there is a measurable set $Q_D\subseteq\mathbb S^{d-1}$ with $\Prob_{q\sim Q}(q\in Q_D)\ge1-e^{-c_2d}$ on which every graph $G$ containing $G_{\rho'}$ and satisfying $\Delta(G)\le2\Delta$ succeeds from every entry. Each search uses $O(1/\theta_1)=e^{o(d)}$ moves and has evaluation cost at most $\Delta^{1+o(1)}$. These data events can be chosen so that the same query set, success guarantee, move bound, and evaluation cost bound also apply to $G_{\rho_\beta}$ in the logarithmic regime, $G_{r_\gamma}$ in the dense regime, and $\NN{\Delta}$ in both regimes.
\item \emph{(Below the critical degree.)} In the logarithmic regime, suppose that $\beta<\bstar(c)$. There are positive constants $a$ and $\eta$ such that, with probability at least $1-n^{-\eta}$ over $D_n$, every navigation graph $G$ constructed from the sample admits a measurable set $Q_G\subseteq \mathbb S^{d-1}$ with $\sigma_{d-1}(Q_G)\ge1-n^{-\eta}$ such that, for every $q\in Q_G$, at least $(1-o(1))n^a$ entries $x_0$ satisfy $\Work_G(q;x_0)\ge\Delta$. In the dense regime, suppose that $\gamma<\gc$. There are positive constants $a$, $c_1$, and $c_2$ such that, with probability at least $1-e^{-c_1d}$ over $D_n$, every navigation graph $G$ constructed from the sample admits a measurable set $Q_G\subseteq \mathbb S^{d-1}$ with $\sigma_{d-1}(Q_G)\ge1-e^{-c_2d}$ such that, for every $q\in Q_G$, at least $(1-o(1))e^{ad}$ entries $x_0$ satisfy $\Work_G(q;x_0)\ge\Delta$. If $\Delta(G)\le\Delta$, then for every $q\in Q_G$, greedy navigation fails from at least $(1-o(1))n^a$ entries in the logarithmic regime and at least $(1-o(1))e^{ad}$ entries in the dense regime.
\end{enumerate}
\end{theorem}

To prove part~(a), we first find conditions on a general radius $r$ that force progress on every graph containing $G_r$. We then choose $r=\rho'$ in \cref{sec:instances} and check that the threshold and nearest-neighbor graphs retain these edges within the required degree bound.

\subsection{Uniform progress from short edges}

\phantomsection\label{eq:core-conditions}
Short edges must force progress toward a cap around the query and prevent unsuccessful stopping inside that cap. Fix $r\in(0,\pi/2)$ and choose the cap radius $A$ so that $\theta_1<A<\pi/2$ and $\cos^2A>\cos r$.
To prevent unsuccessful stopping inside $C(q,A)$, we need almost all locations closer to $q$ than the current point to lie within its neighbor radius $r$. By \cref{lem:inside}(a), the fraction of closer locations beyond radius $r$ is exponentially small whenever $\tau=(\cos^2A-\cos r)/\sin^2A$ has a fixed positive lower bound. The inequality $\cos^2A>\cos r$ will also bound the number of visited points inside the cap. For progress outside the cap, put $h=(A-\theta_1)/2$. We will require each move there to decrease the distance to $q$ by at least $h$. From a point at distance $A$ from $q$, every point of $C(q,A-h)$ gives this decrease, and $A-h>\theta_1$. Write $c_{\mathrm{pop}}=d^{-1}\log(nF(A-h))$ for the exponent of its expected number of data points. A fixed positive lower bound on $c_{\mathrm{pop}}$ makes this expectation exponential in $d$. We choose $r$ and $A$ in \cref{sec:instances} so that $\tau$ and $c_{\mathrm{pop}}$ are bounded below by positive constants.

For a fixed query $q$, let $\mathcal P(q)$ be the event that every data point $x$ outside $C(q,A)$ has a data point $y$ with $\delta(x,y)\le r$ and $\delta(q,y)\le\delta(q,x)-h$. On any graph containing $G_r$, this forces a decrease of at least $h$ whenever an expansion outside the cap does not already succeed. After entering the cap, a run stays there because its distances decrease. To exclude unsuccessful stopping inside it, we consider false local minima relative to $\TopSet_k(q)$, namely data points outside $\TopSet_k(q)$ with no closer neighbor in $G_r$. Let $\mathcal C(q)$ be the event that $C(q,A)$ contains no such point. The next proposition establishes both properties for most queries on one sample and bounds the search from every entry.

\begin{proposition}[Uniform greedy progress]\label{prop:certificate}
Let $r,A,h,\tau,c_{\mathrm{pop}},\mathcal C(q)$, and $\mathcal P(q)$ be as above, with $\theta_1<A<\pi/2$ and $\cos^2A>\cos r$. Assume that $\tau\ge\tau_0$ for a fixed $\tau_0>0$ and that $c_{\mathrm{pop}}$ is bounded below by a positive constant independent of $n$. Fix an absolutely continuous query distribution $Q$ before sampling. For each sample $D_n$, let $Q_D\subseteq\mathbb S^{d-1}$ be the measurable set of queries for which both $\mathcal C(q)$ and $\mathcal P(q)$ hold, with bisectors excluded as in \cref{sec:prelim}. There is a constant $\zeta=\zeta(\tau_0)>0$ such that $\Prob_{q\sim Q}(q\in Q_D)\ge1-e^{-\zeta d/2}$ with probability at least $1-6e^{-\zeta d/2}$ over $D_n$, for all sufficiently large $n$. Let $K=(1-\cos r)/(\cos^2A-\cos r)>1$ and $H=\lceil(\pi-A)/h\rceil+2\lceil K\rceil+2$. For every $q\in Q_D$, greedy navigation on every graph $G$ containing $G_r$ succeeds from every entry within $H$ complete expansions, using at most $H$ moves and evaluation cost at most $1+H\Delta(G)$.
\end{proposition}

\begin{proof}
Fix a query $q$. To bound $\Prob(\mathcal C(q)^c)$, we compare the mass of all closer locations with the mass of those within distance $r$ of a given data point. Condition on the location $x$ of a fixed data label with $0<\theta=\delta(q,x)\le A$. Write $t=F(\theta)$ for the mass of closer locations and $\ell=\sigma_{d-1}(C(x,r)\cap C(q,\theta))$ for the mass accessible within radius $r$. Since $\cos^2\theta-\cos r\ge\tau\sin^2\theta$, \cref{lem:inside}(a) applies with $\psi_0=\theta$. Choose $\zeta=\zeta(\tau_0)>0$ sufficiently small that the bound in \cref{lem:inside}(a) on the fraction of $C(q,\theta)$ outside $C(x,r)$ is at most $e^{-\zeta d}$ for all sufficiently large $n$ and all $\tau\ge\tau_0$. For $\tau\ge1$, the omitted fraction is zero. Setting $\eps_d=e^{-\zeta d}$ gives $\ell\ge(1-\eps_d)t$.

For $x$ to be a false local minimum, at least $k$ other data points must lie in $C(q,\theta)$, but none may lie in the lens. Since $k\ge1$, independence bounds this probability, conditional on $x$, by
\[
  (1-\ell)^{n-1}-(1-t)^{n-1}
  \le(n-1)\eps_{d} t\bigl(1-(1-\eps_{d})t\bigr)^{n-2},
\]
using the mean value theorem and $t-\ell\le\eps_d t$. Taking the worst case over $x$ would lose a factor $n$ when summed over the sample. Instead, we retain the dependence on $t$ and average over $x$. Let $N_q$ be the number of false local minima in $C(q,A)$. Over all possible locations of $x$, the variable $t=F(\delta(q,x))$ is uniform on $[0,1]$, as recorded after \eqref{eq:chernoff}. Restricting to $x\in C(q,A)$ gives the integral over $t\le F(A)$, which we extend to $[0,1]$ to obtain
\begin{equation}\label{step:core}
\begin{aligned}
  \E N_q
  &\le n(n-1)\eps_{d}\int_0^1t\bigl(1-(1-\eps_{d})t\bigr)^{n-2}\,dt\\
  &\le\frac{\eps_{d}}{(1-\eps_{d})^2}\le4\eps_{d}.
\end{aligned}
\end{equation}
After rescaling the integration variable by $1-\eps_d$, the identity $\int_0^1t(1-t)^{n-2}\,dt=1/(n(n-1))$ cancels the factor $n(n-1)$. Since $\mathcal C(q)$ fails only if $N_q\ge1$, we obtain $\Prob(\mathcal C(q)^c)\le\E N_q\le4\eps_d$.

To bound $\Prob(\mathcal P(q)^c)$, consider a point $x$ with $\theta=\delta(q,x)\ge A$. Neighbors within distance $r$ that reduce the query distance by at least $h$ lie in $C(x,r)\cap C(q,\theta-h)$. We compare the mass of this region with $F(A-h)$, the surface measure of $C(q,A-h)$ at the boundary $\theta=A$, and claim that
\begin{equation}\label{step:progress}
  \sigma_{d-1}\bigl(C(x,r)\cap C(q,\theta-h)\bigr)
  \ge(1-\eps_{d})F(A-h).
\end{equation}
At $\theta=A$, shrinking the query cap from radius $A$ to $A-h$ preserves the containment margin $\tau$, since $\cos(A-h)\ge\cos A$ and $\sin(A-h)\le\sin A$. Hence $\cos A\cos(A-h)-\cos r\ge\tau\sin A\sin(A-h)$, so \cref{lem:inside}(a) with $\psi_0=A-h$ proves \eqref{step:progress} at the boundary.

For larger $\theta$, we compare the available directions in polar coordinates around $x$. A possible neighbor $y$ has distance $\psi=\delta(x,y)\in(0,r]$. At each fixed $\psi$, surface measure makes the direction of $y$ uniform. Let $Z$ be the cosine of the angle between that direction and the direction from $x$ toward $q$. For $A\le\theta<\pi$, the spherical cosine identity gives
\[
  \delta(q,y)\le\theta-h
  \quad\Longleftrightarrow\quad
  Z\ge\frac{\sin h}{\sin\psi}
  +\cot\theta\,\frac{\cos h-\cos\psi}{\sin\psi}.
\]
The condition is impossible for $\psi<h$ by the triangle inequality. For $\psi\ge h$, the coefficient of $\cot\theta$ is nonnegative. Increasing $\theta$ therefore lowers the threshold on $Z$, while rotational invariance leaves its distribution unchanged. The fraction of allowed directions cannot decrease at any fixed $\psi$. Since the radial surface measure around $x$ is also independent of $\theta$, integration over $0<\psi\le r$ proves \eqref{step:progress} for $A\le\theta<\pi$. At $\theta=\pi$, the distance condition becomes $\psi\ge h$, and continuity of the measure of $C(x,r)\cap C(q,\theta-h)$ extends the bound to the endpoint.

By \eqref{step:progress}, conditional on a data point $x$ outside $C(q,A)$, the probability that none of the other $n-1$ points lies in $C(x,r)\cap C(q,\theta-h)$ is at most $\exp(-(n-1)(1-\eps_d)F(A-h))\le\exp(-e^{c_{\mathrm{pop}}d}/2)$. A union bound over the $n$ labels gives $\Prob(\mathcal P(q)^c)\le n\exp(-e^{c_{\mathrm{pop}}d}/2)$.

The estimates for $\Prob(\mathcal C(q)^c)$ and $\Prob(\mathcal P(q)^c)$ hold for each fixed query. To control the measure of queries outside $Q_D$ on one sample, we integrate the failure probabilities against the prescribed distribution $Q$. Membership in $Q_D$ is measurable jointly in the data and query because $\mathcal C(q)$ and $\mathcal P(q)$ use finitely many comparisons of distances and ranks. Tonelli's theorem therefore gives
\[
  \E\,\Prob_{q\sim Q}(q\notin Q_D)
  \le4e^{-\zeta d}+n\exp(-e^{c_{\mathrm{pop}}d}/2)
  \le5e^{-\zeta d}
\]
for all sufficiently large $n$, using the fixed positive lower bound on $c_{\mathrm{pop}}$ and the asymptotic assumptions of \cref{sec:prelim}. The bisectors excluded from $Q_D$ have zero probability under the absolutely continuous law $Q$. Applying Markov's inequality to $\Prob_{q\sim Q}(q\notin Q_D)$ yields $\Prob_{q\sim Q}(q\notin Q_D)\le e^{-\zeta d/2}$ with probability at least $1-5e^{-\zeta d/2}$ over the sample. This proves the required measure bound for $Q_D$.

Fix a sample, a query $q\in Q_D$, and a graph $G$ containing $G_r$. While a run remains outside $C(q,A)$, $\mathcal P(q)$ ensures that every expansion either succeeds or is followed by a decrease of at least $h$. Since the entry is at distance at most $\pi$ from $q$, there are at most $\lceil(\pi-A)/h\rceil+1$ expansions outside the cap. This includes a possible successful final expansion with no following move. After a run enters $C(q,A)$, its decreasing distances keep it there and $\mathcal C(q)$ prevents unsuccessful stopping.

Inside the cap, individual improvements may be small, but any two points visited at least two moves apart have distance greater than $r$. Otherwise, the later point would have been a listed neighbor closer to $q$ than the successor chosen at the earlier point. The points expanded inside the cap at even and odd positions along the run therefore form two subsequences with pairwise distances greater than $r$. Write either subsequence as $z_1,\ldots,z_N$. Their pairwise squared Euclidean distances exceed $2(1-\cos r)$, whereas $q\cdot z_i\ge\cos A$ places all of them in the Euclidean ball of radius $\sin A$ centered at $\cos A\,q$. Here $\cos^2A>\cos r$ gives $1-\cos r>\sin^2A$, so the pairwise squared distances exceed twice the squared ball radius. For $N\ge2$, let $\bar z=N^{-1}\sum_i z_i$ be their mean. The comparison of squared distances gives
\begin{equation}\label{step:moves}
  2\binom N2(1-\cos r)
  <\sum_{i<j}\|z_i-z_j\|^2
  =N\sum_i\|z_i-\bar z\|^2
  \le N^2\sin^2A.
\end{equation}
The final inequality in \eqref{step:moves} holds because the sum of squared distances to the mean is at most the sum of squared distances to $\cos A\,q$. Rearranging \eqref{step:moves} gives $N<(1-\cos r)/(\cos^2A-\cos r)=K$, so each subsequence has at most $\lceil K\rceil$ points. There are therefore at most $2\lceil K\rceil$ expansions inside $C(q,A)$. Adding the expansions outside $C(q,A)$ gives at most $H$ complete expansions in total. Each expansion evaluates at most $\Delta(G)$ new points. Including the entry, the evaluation cost is at most $1+H\Delta(G)$, with the successful final expansion counted in full.
\end{proof}

\subsection{Completing the uniform upper bound}\label{sec:instances}\label{sec:retention}

To prove \cref{thm:critical}(a), we choose $r$ and $A$ using the gap above the critical degree. We verify the hypotheses of \cref{prop:certificate} in each regime, then check that the threshold and nearest-neighbor graphs contain $G_r$ and have degree at most $2\Delta$.

\begin{proof}[Proof of \cref{thm:critical}(a)]
In the logarithmic regime, take $r=\rho_{\beta'}$ with $\bstar(c)<\beta'<\beta<1$. The estimate for $\theta_1$ after \cref{lem:cap}, together with \eqref{eq:radii}, gives $\theta_1\to\theta_\infty=\arcsin(e^{-1/c})$ and $r\to r_\infty=\arcsin(e^{-(1-\beta')/c})$. The inequality $\beta'>\bstar(c)$ gives $\cos r_\infty<\cos^2\theta_\infty$, leaving room to choose a fixed $A\in(\theta_\infty,\pi/2)$ with $\cos^2A>\cos r_\infty$. For all sufficiently large $n$, this choice gives $\theta_1<A<\pi/2$ and $\cos^2A>\cos r$, as required by \cref{prop:certificate}. The definitions of $\tau$ and $c_{\mathrm{pop}}$, together with \eqref{eq:cap-ratio}, give $\tau\to(\cos^2A-\cos r_\infty)/\sin^2A>0$ and $c_{\mathrm{pop}}\to\log(\sin((A+\theta_\infty)/2)/\sin\theta_\infty)>0$, so \cref{prop:certificate} applies. Since $h\to(A-\theta_\infty)/2>0$ and $K$ has a finite limit, it gives $H=O(1)$ complete expansions. The probability of failure over the data and the excluded query probability are both $e^{-\Omega(d)}=n^{-\Omega(1)}$.

In the dense regime, take $r=r_{\gamma'}$ with $\gc<\gamma'<\gamma$. Now \eqref{eq:radii} gives $r/\theta_1\to e^{\gamma'}>\sqrt2$, so we can choose the cap radius as $A=a'r$, where $a'$ is a fixed constant satisfying $e^{-\gamma'}<a'<1/\sqrt2$. The bound $a'>e^{-\gamma'}$ ensures $A>\theta_1$, while $a'<1/\sqrt2$ ensures $\cos^2A>\cos r$, for all sufficiently large $n$. To verify the remaining hypotheses of \cref{prop:certificate} and bound $K$, we use \eqref{eq:cap-ratio} and expand sine and cosine at zero. This gives $\tau\to1/(2a'^2)-1>0$ and $c_{\mathrm{pop}}\to\log((1+a'e^{\gamma'})/2)>0$, while $K\to1/(1-2a'^2)$. Since $h=(\tfrac12(a'e^{\gamma'}-1)+o(1))\theta_1$, the proposition gives $H=O(1/\theta_1)=e^{o(d)}$ complete expansions. The probability of failure over the data and the excluded query probability are both $e^{-\Omega(d)}$. In either regime, the move bound is $H$, and every graph containing $G_r$ with degree at most $2\Delta$ has evaluation cost at most $1+2H\Delta=\Delta^{1+o(1)}$ on the query set of \cref{prop:certificate}.

It remains to verify the required edge inclusion and degree bound for the threshold and nearest-neighbor graphs. Write $s=\rho_\beta$ in the logarithmic regime and $s=r_\gamma$ in the dense regime, so $s$ is the radius of the threshold graph in \cref{thm:critical}(a). Conditional on the position of a data point $x$, the lengths $|G_s(x)|$ and $|G_r(x)|$ are binomial with respective means $(n-1)F(s)$ and $(n-1)F(r)$. Applying \eqref{eq:chernoff} at the respective thresholds $2nF(s)$ and $2nF(r)$ and taking a union bound over the $n$ labels shows that, with probability at least $1-2n\exp(-3nF(r)/8)$, every $x\in D_n$ satisfies $|G_s(x)|\le2nF(s)$ and $|G_r(x)|\le2nF(r)$. The strict gap between the degree exponents gives $nF(r)/\Delta\to0$, so $|G_r(x)|<\Delta$ for every $x$. Thus $G_s$ contains $G_r$ and has degree at most $2\Delta$, and the $\Delta$ nearest neighbors of every point include its entire list in $G_r$. Hence $\NN{\Delta}$ also contains $G_r$ and has degree $\Delta$. Adding $2n\exp(-3nF(r)/8)$ to the data failure bound in \cref{prop:certificate} preserves the error rates stated in \cref{thm:critical}(a).
\end{proof}

\subsection{Traps for arbitrary navigation graphs}\label{sec:lower}

The upper bound uses short edges to force an improving move. For the lower bound, we show that a list shorter than the critical degree cannot supply such a move at many points, regardless of how the graph is chosen after observing the data. We examine points with ranks from $T$ through $2T-1$, where $T>k$ is an integer, and call the interval $[T,2T)$ the \emph{rank window}. A cutoff radius $\rho$ separates possible neighbors into two groups. An improving point within distance $\rho$ of the current point $x$ is unlikely because the lens $C(x,\rho)\cap C(q,\delta(q,x))$ has small mass. A point farther than $\rho$ from $x$ improves the distance for only a small fraction of query directions. Bounding these groups separately lets us control every data-dependent neighbor list through a property of the data alone.

Let $L$ be a positive integer. We seek entries that either fail or require at least $L$ evaluations, and call them traps. It suffices first to study lists of length at most $L$, since a longer list already costs more than $L$ at the first complete expansion. To locate the chosen ranks geometrically, define $\vartheta$ by $nF(\vartheta)=T$, so a cap of radius $\vartheta$ contains $T$ data points in expectation. For $0<\theta<\pi/2$ with $\cot\theta\tan(\rho/2)<1$, write $I_\theta(\rho)=-\frac12\log(1-\cot^2\theta\tan^2(\rho/2))$. By \cref{lem:directions}, a fixed point at distance at least $\rho$ from $x$ is closer to the query than $x$ for at most an $e^{-dI_\theta(\rho)+O(\log d)}$ fraction of the directions with $\delta(q,x)=\theta$, provided $\cot\theta\tan(\rho/2)$ stays a fixed distance below one. Write $I=I_\vartheta(\rho)$ and $J=J_\vartheta(\rho)$, where $J_\theta$ is the lens rate from \cref{lem:lens}. The factor $e^{-dJ+o(d)}$ bounds the lens mass relative to $C(q,\vartheta)$ when $\delta(q,x)=\vartheta$. To obtain exponentially many entries and keep the rank fluctuations negligible after summing over labels, we require $T$ to grow exponentially in $d$. The additional requirements that $Te^{-dJ}$ and $Le^{-dI}$ be exponentially small control nearby improving points and farther stored neighbors, respectively. \phantomsection\label{eq:trap-margins}For a fixed $c_{\mathrm{trap}}>0$, we assume that $T\ge e^{{c_{\mathrm{trap}}}d}$, $dJ\ge\log T+{c_{\mathrm{trap}}}d$, and $dI\ge\log L+{c_{\mathrm{trap}}}d$.
To keep the geometric estimates uniform, assume also that $\vartheta\le\pi/2-\eps$, $\eps\vartheta\le\rho<r_*(\vartheta)$, and $\cot\vartheta\tan(\rho/2)\le1-\eps$ for a fixed $\eps>0$. Here $r_*(\vartheta)$ is the lens boundary defined by $\cos r_*(\vartheta)=\cos^2\vartheta$ in \cref{sec:prelim}. We first prove the lower bound under these conditions and then choose $T$, $L$, and $\rho$ below the critical degree in each regime.

\begin{theorem}[Traps in a rank window]\label{thm:traps}
Let $T>k$ and $L$ be positive integers, let $\rho\in(0,\pi/2)$ be the distance cutoff, and retain $\vartheta$, $I=I_\vartheta(\rho)$, and $J=J_\vartheta(\rho)$ as defined above. Suppose that $T\ge e^{c_{\mathrm{trap}}d}$, $dJ\ge\log T+c_{\mathrm{trap}}d$, and $dI\ge\log L+c_{\mathrm{trap}}d$. Assume also that $\vartheta\le\pi/2-\eps$, $\eps\vartheta\le\rho<r_*(\vartheta)$, and $\cot\vartheta\tan(\rho/2)\le1-\eps$, for fixed $c_{\mathrm{trap}},\eps>0$. For all sufficiently large $n$, with probability at least $1-e^{-{c_{\mathrm{trap}}} d/4}$ over the uniform data set $D_n$, simultaneously for every navigation graph $G$ constructed from the data, there is a measurable query set $Q_G$ with $\sigma_{d-1}(Q_G)\ge1-e^{-{c_{\mathrm{trap}}} d/8}$ such that, for every $q\in Q_G$, at least $(1-e^{-{c_{\mathrm{trap}}} d/8})T$ entries $x_0$ of rank in $[T,2T)$ satisfy $\Work_G(q;x_0)\ge L$. The same entries can be chosen so that each either has $d_G(x_0)>L$ or is a failing entry with $\Work_G(q;x_0)=\infty$.
\end{theorem}

\begin{proof}
We first consider a graph $H$ of degree at most $L$ and bound how many points in the rank window have a listed neighbor closer to the query. For each data point $x$, let $p_x(q)=F(\delta(q,x))$ be the surface measure of the cap of points closer to $q$ than $x$. Use the indicator $R_x(q)=\one\{T\le\rank_q(x)<2T\}$ to record whether the rank of $x$ lies in $[T,2T)$. To apply the geometric estimates, we compare $p_x(q)$ with the interval $\mathcal W=[T/(4n),4T/n]$ of cap masses. We separately bound points with rank in $[T,2T)$ but $p_x(q)\notin\mathcal W$, nearby improving points with $p_x(q)\in\mathcal W$, and farther listed neighbors. The first two bounds use the randomness of the data. The farther-neighbor bound holds for every fixed data set and every list of length at most $L$.

To apply the rates $J$ and $I$ throughout $\mathcal W$, we first compare the corresponding angles with $\vartheta$. If $p_x(q)\in\mathcal W$ and $\theta=\delta(q,x)$, then $nF(\theta)\in[T/4,4T]$. Since $\vartheta\le\pi/2-\eps$ and $F(\pi/2-\eps/2)/F(\pi/2-\eps)=e^{\Omega(d)}$, we have $\theta\le\pi/2-\eps/2$, and \eqref{eq:cap-ratio} gives $\sin\theta=\sin\vartheta(1+O(d^{-1}))$. To control the resulting change in the rates, write $h_\theta=\cot\theta\tan(\rho/2)$ for the directional cosine threshold. Differentiation at fixed $\rho$ gives $\partial I_\theta(\rho)/\partial\log(\sin\theta)=-h_\theta^2/((1-h_\theta^2)\cos^2\theta)$ and $0\le\partial J_\theta(\rho)/\partial\log(\sin\theta)\le1$, with $J_\theta(\rho)$ continuous across its zero boundary. The bounds $h_\theta\le1-\eps/2$ and $\theta\le\pi/2-\eps/2$ keep these derivatives uniformly bounded. Thus using the reference angle $\vartheta$ changes $dI_\theta(\rho)$ and $dJ_\theta(\rho)$ by only $O(1)$.

We also need to absorb the errors in the geometric estimates. The bound $h_\theta\le1-\eps/2$ keeps $I_\theta(\rho)$ bounded, while $\rho\ge\eps\vartheta$ and $\sin\theta/\sin\vartheta=1+O(d^{-1})$ keep $J_\theta(\rho)$ bounded. Replacing the geometric factor $d-1$ by $d$ therefore adds only $O(1)$ to the exponents. Moreover, $\rho<r_*(\vartheta)$ keeps $\rho$ bounded away from $\pi/2$, so \cref{lem:lens,lem:directions} apply uniformly with a common error $\omega_n=O(\log d+\log(1/\sin\vartheta)+\log(1/\sin\rho))=o(d)$. Here $\omega_n=o(d)$ follows from $\vartheta>\theta_1$, $\rho\ge\eps\vartheta$, and \eqref{eq:theta1-dense}, with the logarithmic cap estimate used in the logarithmic regime. The bound $\vartheta\le\pi/2-\eps$ also keeps $\mathcal W$ within the range of cap probabilities for large $n$.

We next bound the points with rank in $[T,2T)$ but cap mass $p_x(q)\notin\mathcal W$. Let $M_x(q)=R_x(q)\one\{p_x(q)\notin\mathcal W\}$. Conditional on $x$ and $q$, \eqref{eq:rank-law} makes $\rank_q(x)-1$ a binomial variable of mean $(n-1)p_x(q)$. If $p_x(q)<T/(4n)$, a rank of at least $T$ requires at least $T-1$ other data points closer to $q$ than $x$, far more than the expected number. If $p_x(q)>4T/n$, a rank below $2T$ requires fewer than $2T-1$ such points, far fewer than the expected number. In both cases, \eqref{eq:chernoff} bounds the probability by $e^{-c_1T}$ for an absolute constant $c_1>0$. Averaging over $q$ and $x$ therefore gives $\E\int M_x(q)\,d\sigma_{d-1}(q)\le e^{-c_1T}$.

For the nearby points, let $N_x(q)$ indicate that some other data point $y$ satisfies $\delta(x,y)<\rho$ and $\delta(q,y)<\delta(q,x)$. Such a point lies in the lens $C(x,\rho)\cap C(q,\delta(q,x))$. Conditional on $x$ and $q$ with $p_x(q)\in\mathcal W$, a union bound over the other data labels and \cref{lem:lens} give $\Prob(N_x(q)=1\mid x,q)\le np_x(q)e^{-dJ+\omega_n}$. For fixed $x$, the variable $p_x(q)$ is uniform on $[0,1]$ under $q\sim\sigma_{d-1}$. Integrating this conditional bound yields
\[
  \E\int\one\{p_x(q)\in\mathcal W\}N_x(q)\,d\sigma_{d-1}(q)
  \le\int_{T/(4n)}^{4T/n}np\,e^{-dJ+\omega_n}\,dp
  \le\frac{8T^2}{n}e^{-dJ+\omega_n}.
\]
Because it considers all nearby data points, this bound applies to every graph.

For the farther neighbors, fix the entire data set and any graph $H$ of degree at most $L$. Let $P_x^{H,\mathrm{far}}(q)$ indicate that $H(x)$ contains an improving point at distance at least $\rho$ from $x$. At each fixed value of $\delta(q,x)$, a uniform query has a uniform direction around $x$. By \cref{lem:directions} and the rate comparison throughout $\mathcal W$, each such listed point is closer to $q$ than $x$ for at most an $e^{-dI+\omega_n}$ fraction of these directions. A union bound over the list and integration over $p_x(q)\in\mathcal W$ give
\[
  \int\one\{p_x(q)\in\mathcal W\}P_x^{H,\mathrm{far}}(q)\,d\sigma_{d-1}(q)
  \le\frac{4T}{n}L e^{-dI+\omega_n}.
\]
On each fixed data set, the bound covers every list chosen after observing the data.

We combine the classification of improving neighbors with the deterministic bound for farther neighbors in a quantity that depends only on the sample. Let $P_x^H(q)$ indicate that $H(x)$ contains a point closer to $q$ than $x$. A point in the rank window with an improving listed neighbor either contributes to $M_x(q)$ or has $p_x(q)\in\mathcal W$ and an improving point accounted for by $N_x(q)$ or $P_x^{H,\mathrm{far}}(q)$. Off the bisectors, the window contains exactly $T$ points, so $T^{-1}\sum_xR_x(q)P_x^H(q)$ is the fraction with an improving listed neighbor. To bound its average over uniform queries, define
\begin{gather}
  Y=\frac1T\sum_x\int\bigl(M_x(q)+\one\{p_x(q)\in\mathcal W\}N_x(q)\bigr)\,d\sigma_{d-1}(q)+4Le^{-dI+\omega_n}.\notag\\
\intertext{For every graph $H$ of degree at most $L$, this classification and the farther-neighbor bound give}
  \frac1T\int\sum_xR_x(q)P_x^H(q)\,d\sigma_{d-1}(q)\le Y.\label{eq:Y}
\end{gather}
Taking expectations in the definition of $Y$ and using the estimates for $M_x(q)$ and $N_x(q)$ gives
\[
  \E Y\le\frac nT e^{-c_1T}
  +8Te^{-dJ+\omega_n}
  +4Le^{-dI+\omega_n}.
\]
The assumptions of \cref{thm:traps} make $T$ exponentially large and both $Te^{-dJ}$ and $Le^{-dI}$ exponentially small. Together with $\omega_n=o(d)$ and $\log\log n=o(d)$, this gives $\E Y\le e^{-{c_{\mathrm{trap}}} d/2}$ for all sufficiently large $n$. Applying Markov's inequality over the random sample gives $Y\le e^{-{c_{\mathrm{trap}}} d/4}$ with probability at least $1-e^{-{c_{\mathrm{trap}}} d/4}$. Fix a data set satisfying this bound. For each graph $H$ of degree at most $L$, apply Markov's inequality to \eqref{eq:Y} over uniform queries. Outside a set of surface measure at most $e^{-{c_{\mathrm{trap}}} d/8}$, the fraction $T^{-1}\sum_xR_x(q)P_x^H(q)$ is at most $e^{-{c_{\mathrm{trap}}} d/8}$. After excluding this set and the bisectors, at least $(1-e^{-{c_{\mathrm{trap}}} d/8})T$ points in the rank window have no improving neighbor in $H$ for each remaining query.

To pass from bounded degree to an arbitrary navigation graph $G$, retain $G(x)$ in $H$ when $d_G(x)\le L$ and replace it by the empty list otherwise. Since $H$ has degree at most $L$, the preceding query bound applies to $H$ on the fixed data set. Let $Q_G$ be the resulting set of queries, with the bisectors excluded, so $\sigma_{d-1}(Q_G)\ge1-e^{-{c_{\mathrm{trap}}} d/8}$. For every $q\in Q_G$, at least $(1-e^{-{c_{\mathrm{trap}}} d/8})T$ entries $x_0$ of rank in $[T,2T)$ have no improving neighbor in $H$. For each such entry with $d_G(x_0)\le L$, the list $G(x_0)=H(x_0)$ has no improving point. Since $\rank_q(x_0)\ge T>k$, neither the entry nor any listed neighbor is a target, so greedy navigation stops unsuccessfully at $x_0$. If instead $d_G(x_0)>L$, its first complete expansion evaluates all $d_G(x_0)$ distinct neighbors before success is recorded, giving evaluation cost greater than $L$. Thus $\Work_G(q;x_0)\ge L$ in both cases, proving the theorem.
\end{proof}

\subsection{Completing the uniform lower bound}\label{sec:uniform-lower-parameters}

To complete \cref{thm:critical}(b), we choose $T$, $L$, and $\rho$ so that \cref{thm:traps} applies below the critical degree. In particular, $dJ$ must exceed $\log T$ and $dI$ must exceed $\log L$ by fixed positive multiples of $d$.

\begin{proof}[Proof of \cref{thm:critical}(b)]
The identity $\cos r_*(\theta)=\cos^2\theta$ gives $\cot^2\theta\tan^2(r_*(\theta)/2)=\cos^2\theta/(1+\cos^2\theta)$. Hence $I_\theta(r_*(\theta))=\frac12\log(1+\cos^2\theta)$ and $J_\theta(r_*(\theta))=0$. For fixed $\theta$, the rate $I_\theta(\rho)$ increases with $\rho$, whereas $J_\theta(\rho)$ decreases to zero as $\rho$ increases to $r_*(\theta)$. Both rates are continuous on the parameter ranges used below.

In the logarithmic regime, set $L=\lceil n^\beta\rceil$ with $\beta<\bstar(c)$. For the limiting angle $\theta_\infty=\arcsin(e^{-1/c})$ computed in the proof of \cref{thm:critical}(a), we have $cI_{\theta_\infty}(r_*(\theta_\infty))=\bstar(c)>\beta$. We can therefore choose a fixed $\rho<r_*(\theta_\infty)$ close enough to the boundary that $cI_{\theta_\infty}(\rho)>\beta$. This choice also gives $J_{\theta_\infty}(\rho)>0$. For $T=\lfloor n^a\rfloor$ with $0<a<1$, the cap estimate gives $\vartheta\to\arcsin e^{-(1-a)/c}$. As $a\downarrow0$, this limit approaches $\theta_\infty$. By continuity, we can fix a sufficiently small $a>0$ so that $I>\beta/c$ and $J>a/c$ with fixed positive margins for all large $n$. Taking a sufficiently small fixed ${c_{\mathrm{trap}}}>0$ below these margins and below $a/c$ verifies the assumptions on $T$, $I$, and $J$ in \cref{thm:traps}. The remaining angular conditions hold because the angles converge to positive constants below $\pi/2$ and $\rho<r_*(\vartheta)$.

In the dense regime, set $L=\lceil e^{\gamma d}\rceil$ with $\gamma<\gc$. We write $\rho=u\vartheta$, where the fixed ratio $0<u<\sqrt2$ will be chosen to make $I>\gamma$ and $J>0$. As $\vartheta\to0$, we have
\begin{equation}\label{eq:uniform-dense-trap-rates}
  I_\vartheta(u\vartheta)\longrightarrow-\frac12\log(1-u^2/4)\qquad\text{and}\qquad
  J_\vartheta(u\vartheta)\longrightarrow-\frac12\log\bigl(u^2(1-u^2/4)\bigr).
\end{equation}
Choose $u$ so that $2\sqrt{1-e^{-2\gamma}}<u<\sqrt2$. By \eqref{eq:uniform-dense-trap-rates}, $I$ tends to a value larger than $\gamma$ and $J$ tends to a positive value. Fix $a>0$ smaller than the limit of $J_\vartheta(u\vartheta)$ and set $T=\lfloor e^{ad}\rfloor$. By \eqref{eq:cap-ratio}, $\sin\vartheta=e^a\sin\theta_1(1+O(d^{-1}))$, so $\vartheta\to0$. The errors in \eqref{eq:uniform-dense-trap-rates} are $O(\vartheta^2)=o(1)$. A sufficiently small fixed ${c_{\mathrm{trap}}}>0$ therefore verifies the assumptions on $T$, $I$, and $J$ in \cref{thm:traps}, and $\rho=u\vartheta<r_*(\vartheta)$ for large $n$. Since $u<\sqrt2$, $\cot\vartheta\tan(\rho/2)$ stays a fixed distance below one, as required by \cref{thm:traps}.

With the hypotheses verified in both regimes, \cref{thm:traps} applies simultaneously to every navigation graph with probability at least $1-e^{-{c_{\mathrm{trap}}} d/4}$ over $D_n$. For each graph $G$, it gives a measurable query set $Q_G$ with $\sigma_{d-1}(Q_G)\ge1-e^{-{c_{\mathrm{trap}}} d/8}$. For every $q\in Q_G$, at least $(1-e^{-{c_{\mathrm{trap}}} d/8})T$ entries $x_0$ satisfy $\Work_G(q;x_0)\ge L=\Delta$. If $\Delta(G)\le\Delta$, greedy navigation fails from at least $(1-e^{-{c_{\mathrm{trap}}} d/8})T$ entries for every $q\in Q_G$. The data failure probability and the excluded surface measure are $n^{-\Omega(1)}$ in the logarithmic regime and $e^{-\Omega(d)}$ in the dense regime. Substituting $T=\lfloor n^a\rfloor$ or $T=\lfloor e^{ad}\rfloor$, respectively, and absorbing integer rounding into the $o(1)$ factors proves part~(b).
\end{proof}

\subsection{Average degree and evaluation cost}\label{sec:uniform-cost}

To strengthen \cref{thm:critical}(b) from many traps to a lower bound on average degree, we relate each point's degree to the queries for which its rank lies in $[T,2T)$. We show that, with probability $1-o(1)$ over the data, every point belongs to this rank window for nearly the same surface measure of queries. Combined with \eqref{eq:Y}, this balance forces most neighbor lists to be long whenever navigation succeeds from every entry for most queries.

\begin{theorem}[Degree and evaluation-cost lower bounds]\label{thm:uniform-cost}
Fix $k\ge1$ and $0\le\eps<1$. For uniform data in either regime, with probability $1-o(1)$ over $D_n$, every navigation graph $G$ that succeeds from every entry on a measurable query set $S_G$ with $\sigma_{d-1}(S_G)\ge1-\eps$ has all but $(\eps+o(1))n$ points of degree at least $(\Dstar)^{1-o(1)}$. In particular, its average degree is at least $(1-\eps-o(1))(\Dstar)^{1-o(1)}$. For every query in $S_G$ outside a set of surface measure $o(1)$, all but an $(\eps+o(1))$ fraction of entries $x_0$ satisfy $\Work_G(q;x_0)\ge(\Dstar)^{1-o(1)}$. The conclusion also holds for $\eps=\eps_n\to0$.
\end{theorem}

\begin{proof}
To treat both regimes in one argument, write $(b_n,\chi_*)=(\log n,\bstar(c))$ in the logarithmic regime and $(b_n,\chi_*)=(d,\gc)$ in the dense regime. Thus $b_n$ is the exponent scale and $\chi_*$ is the corresponding critical exponent, with $\Dstar=e^{(\chi_*+o(1))b_n}$. Fix a degree exponent $0<t<\chi_*$ and let $L=\lceil e^{tb_n}\rceil$. Choose $T$ and $\rho$ as in \cref{sec:uniform-lower-parameters}, using $t$ in place of $\beta$ or $\gamma$. We prove the bounds for this fixed $t$, then let $t$ approach $\chi_*$.

To convert \cref{thm:traps} into a degree bound for most data points, we estimate the surface measure of queries for which each point has rank in $[T,2T)$. For $x\in D_n$, let $W_x=\{q:T\le\rank_q(x)<2T\}$ and $w_x=\sigma_{d-1}(W_x)$. For any fixed $A_0>0$, we will bound all $w_x$ simultaneously with failure probability at most $n^{-A_0}$. Use ${\eps_{\mathrm{rank}}}=\sqrt{32(A_0+4)\log n/T}$ as the relative slack when comparing ranks with their binomial means. The choices of $T$ give $T\gg\log n$ and $T\le n/3$, so ${\eps_{\mathrm{rank}}}=o(1)$. With $p_x(q)=F(\delta(q,x))$ as in the proof of \cref{thm:traps}, define the query sets approximating $W_x$ from inside and outside by
\[
\begin{aligned}
  \mathcal I_x&=\{q:(1+{\eps_{\mathrm{rank}}})T\le np_x(q)\le(2-{\eps_{\mathrm{rank}}})T\},\\
  \mathcal O_x&=\{q:(1-{\eps_{\mathrm{rank}}})T\le np_x(q)\le(2+{\eps_{\mathrm{rank}}})T\}.
\end{aligned}
\]
For fixed $x$, the variable $p_x(q)$ is uniform under $q\sim\sigma_{d-1}$, giving $\sigma_{d-1}(\mathcal I_x)=(1-2{\eps_{\mathrm{rank}}})T/n$ and $\sigma_{d-1}(\mathcal O_x)=(1+2{\eps_{\mathrm{rank}}})T/n$. Conditional on $x$ and $q$, \eqref{eq:rank-law} gives a binomial mean $(n-1)p_x(q)$ for $\rank_q(x)-1$. For $q\in\mathcal I_x$, the event $q\notin W_x$ requires a deviation of at least ${\eps_{\mathrm{rank}}}T/2$ from this mean for all large $n$. For $q\notin\mathcal O_x$, the event $q\in W_x$ requires an upper deviation when $np_x(q)<(1-{\eps_{\mathrm{rank}}})T$ and a lower deviation when $np_x(q)>(2+{\eps_{\mathrm{rank}}})T$. Monotonicity bounds the upper tail by its value at the lower endpoint and the lower tail by its value at the upper endpoint. At both endpoints, the required deviation is again at least ${\eps_{\mathrm{rank}}}T/2$. Thus \eqref{eq:chernoff} bounds each error probability by $2e^{-{\eps_{\mathrm{rank}}}^2T/32}$.

Integrating these error bounds over $q$ by Tonelli's theorem, with $x$ fixed, shows that the expectation over the remaining $n-1$ data points of $\sigma_{d-1}(\mathcal I_x\setminus W_x)+\sigma_{d-1}(W_x\setminus\mathcal O_x)$ is at most $4e^{-{\eps_{\mathrm{rank}}}^2T/32}$. Markov's inequality and a union bound over the data labels show that each such sum is at most ${\eps_{\mathrm{rank}}}T/n$ simultaneously, except with probability at most $4n^{-A_0-2}/({\eps_{\mathrm{rank}}}T)\le n^{-A_0}$. Comparing $W_x$ with $\mathcal I_x$ and $\mathcal O_x$ therefore gives, with probability at least $1-n^{-A_0}$, simultaneously for every $x\in D_n$, the bounds
\begin{equation}\label{eq:uniform-window-weights}
  (1-3{\eps_{\mathrm{rank}}})\frac Tn\le w_x\le(1+3{\eps_{\mathrm{rank}}})\frac Tn.
\end{equation}

We now combine \eqref{eq:uniform-window-weights} with the bound $Y\le e^{-{c_{\mathrm{trap}}} d/4}$ from the proof of \cref{thm:traps}. Both bounds hold with probability at least $1-e^{-\Omega(b_n)}$. Fix a data set satisfying both bounds and any graph $G$ that succeeds from every entry on $S_G$. Let $\mathcal D_G=\{x\in D_n:d_G(x)\le L\}$ be the points with short lists, and form $H$ by retaining $G(x)$ for $x\in\mathcal D_G$ and using an empty list elsewhere. For $q\in S_G$ and $x\in\mathcal D_G$ with $q\in W_x$, the list $H(x)=G(x)$ must contain an improving point. Otherwise navigation from $x$ would fail, since $T>k$ puts $x$ outside $\TopSet_k(q)$. Applying \eqref{eq:Y} to $H$ therefore bounds the integral over $S_G$ of the number of data points $x\in\mathcal D_G$ with $q\in W_x$ by $TY$. Outside $S_G$, the contribution is at most $\eps T$, because every query outside the bisectors has exactly $T$ points in the rank window. The integral over the whole query sphere equals $\sum_{x\in\mathcal D_G}w_x$ and is consequently at most $(Y+\eps)T$. Combining this upper bound with \eqref{eq:uniform-window-weights} gives
\[
  \frac{|\mathcal D_G|}{n}\le\frac{\eps+Y}{1-3{\eps_{\mathrm{rank}}}}=\eps+o(1).
\]
Every point outside $\mathcal D_G$ has degree greater than $L$, so the average degree is at least $(1-\eps-o(1))e^{tb_n}$.

The same long lists force an evaluation cost greater than $L$ at the first expansion. For every query $q\in S_G$ and every entry $x\notin\mathcal D_G$, complete scanning evaluates all $d_G(x)>L$ distinct neighbors before success is recorded. Hence $\Work_G(q;x)>L$ for all but an $(\eps+o(1))$ fraction of entries.

For each fixed $t<\chi_*$, the lower bounds on degree and evaluation cost hold simultaneously for every graph and uniformly in $0\le\eps<1$. We may let $t$ approach $\chi_*$ slowly enough that $T/\log n\to\infty$ and the probability that either \eqref{eq:uniform-window-weights} or $Y\le e^{-{c_{\mathrm{trap}}} d/4}$ fails still tends to zero. This gives the scale $(\Dstar)^{1-o(1)}$ in both the degree and evaluation-cost bounds. Uniformity in $\eps$ also permits $\eps_n\to0$.
\end{proof}

Taking $\eps=o(1)$ in \cref{thm:uniform-cost} gives the critical exponent for both average degree and evaluation cost averaged over entries whenever a graph succeeds from every entry for most uniform queries. Together with \cref{thm:critical}(a), this completes the uniform critical curve.

\section{Nonuniform data and the critical degree}\label{sec:nonuniform}

We now extend the uniform results of \cref{sec:upper} to nonuniform sampling. In the dense regime, the uniform nearest-neighbor scale tends to zero, and local density variation determines a sharp worst-case degree curve over the range in \eqref{eq:curve}. \Cref{sec:tolerance} states this curve and gives the directional comparison used in its upper bound. \Cref{loc:localized-tolerance} proves the upper bound with separate density conditions near and far from the query, while \cref{win:subsection} proves the matching obstruction for every graph. In the logarithmic regime, the uniform nearest-neighbor angle stays bounded away from zero, so the density level also changes the local nearest-neighbor angle and the radius needed for a neighbor list of prescribed length. We therefore treat its upper and lower bounds separately in \cref{sec:tolerance-moderate,sec:logarithmic-lower}. The upper bounds include a graph construction that allows arbitrarily deep low-density regions, while the lower bounds show how sufficiently strong high-density concentration defeats every graph of the prescribed degree.

\subsection{The dense nonuniform critical curve}\label{sec:tolerance}

To determine the dense curve, we compare the effects of density variation on the probability of an improving lens and on the radius needed to collect a fixed number of neighbors. Under the $(\kappa,K)$-regularity conditions of \cref{sec:prelim-nonuniform}, the comparison gives the boundary $\kappa=h(\gamma)$, equivalently $e^{2\gamma}=4e^{2\kappa}-2$ by \eqref{eq:tolerance-function}. The following theorem gives navigation on $\NN{\Delta}$ above the curve and a matching obstruction for every navigation graph below it.

\begin{theorem}[The dense nonuniform critical curve]\label{thm:tolerance}\label{thm:tolerance-upper}
Write $\kappa_0=\frac12\log(3/2)$ for the density endpoint in \eqref{eq:curve}. In the dense regime, fix $k\ge1$, $\gamma>\gc$, and $\kappa\ge0$, and let $\Delta=\lceil e^{\gamma d}\rceil$.
\begin{enumerate}[label=\textup{(\alph*)},leftmargin=*]
\item \emph{(Above the curve.)} If $0\le\kappa\le\kappa_0$ and $\kappa<h(\gamma)$, then for all sufficiently large fixed $K$ depending on $\gamma$ and $\kappa$, all $(\kappa,K)$-regular sampling densities, and all absolutely continuous query distributions $Q$ fixed before the data, with probability $1-e^{-\Omega(d)}$ over $D_n$, there is a Borel query set $Q_D$ with $\Prob_{q\sim Q}(q\in Q_D)\ge1-e^{-\Omega(d)}$ such that $\NN{\Delta}$ succeeds from every entry for every $q\in Q_D$. For every $q\in Q_D$, every run uses at most $H=O(1+\rho_{\min}^{-1})=e^{o(d)}$ complete expansions and has evaluation cost at most $1+\Delta H=e^{(\gamma+o(1))d}$. The implicit constants depend only on $k$, $K$, $\gamma$, and $\kappa$.
\item \emph{(Below the curve.)} If $\gamma\le\log2$ and $\kappa>h(\gamma)$, there exist a positive sampling density $f$ for which the essential supremum of $\log f$ over the sphere minus its essential infimum is exactly $\kappa d$ and an absolutely continuous query distribution $Q_n$, both fixed before the data, such that the density is $(\kappa,K)$-regular for every fixed $K>0$ and, with probability $1-e^{-\Omega(d)}$ over $D_n$, every navigation graph of degree at most $\Delta$ has failing entries for all queries outside a graph-dependent set of $Q_n$-probability $e^{-\Omega(d)}$. For each such query, at least $e^{\Omega(d)}$ entries fail. The query distributions satisfy $\sup_{p\in \mathbb S^{d-1}}\Prob_{q\sim Q_n}(q\in C(p,\theta))\to0$ for every fixed $0<\theta<\pi/2$.
\end{enumerate}
\end{theorem}

The upper bound is the specialization $\bar\kappa=\kappa$ of the more general navigation theorem in \cref{loc:localized-tolerance}. We give its full proof there, derive part~(a), and then choose a degree sequence approaching the curve on one fixed family of density neighborhoods. Part~(b) is proved in \cref{win:subsection}. The directional comparison used for the upper bound and in \cref{thm:tolerance-moderate} compares the tangent direction of a sampled point about $q$ with the opposite tangent direction of a point drawn from $P$ conditional on being closer to $q$ than the sampled point. Taking the opposite direction turns the condition that the closer point misses the neighbor cap into a lower bound on the inner product of the two directions.

To state this comparison, let $\omega$ be normalized surface measure on the sphere of directions $\mathbb S^{d-2}$. Write $B_s,B_t\ge1$ for upper bounds on two probability densities of directions relative to $\omega$. In the application, $B_s$ bounds the density of a sampled point's tangent direction at $q$, conditional on its distance from $q$. The bound $B_t$ applies to the opposite of the tangent direction at $q$ of a point drawn from $P$ conditional on being closer to $q$ than the sampled point. To construct the comparison densities, fix a center $e\in\mathbb S^{d-2}$. For any density upper bound $B\ge1$ relative to $\omega$, let $E_B$ be the cap centered at $e$ with $\omega(E_B)=1/B$, and define $h_B=B\,\one_{E_B}$. Taking $B=B_s$ and $B=B_t$ gives the two comparison densities, supported on caps with the same center. To bound the probability that a closer point misses the current point's neighbor cap, for a direction density $g$ and $0<\tau<1$ define $K_\tau g(u)=\int\one\{u\cdot v\ge\tau\}\,g(v)\,d\omega(v)$. This is the probability that a direction drawn from $g$ has inner product at least $\tau$ with a fixed direction $u$. In the application, $u$ is the tangent direction of the current point at $q$, and $g$ describes the opposite directions of closer points.

\begin{lemma}[Rearrangement with unequal bounds]\label{lem:rearrangement}
Let $f_s$ and $g$ be probability densities on $\mathbb S^{d-2}$ relative to $\omega$, with $f_s\le B_s$ and $g\le B_t$ almost everywhere, where $B_s,B_t\ge1$. If $\Phi:[0,1]\to[0,\infty)$ is continuous, nondecreasing, and convex with $\Phi(0)=0$, then
\[
  \int f_s\,\Phi(K_\tau g)\,d\omega\le\int h_{B_s}\,\Phi(K_\tau h_{B_t})\,d\omega,
\]
where the two caps on the right have the same center.
\end{lemma}

\begin{proof}
We first compare the integral over an arbitrary measurable set of directions. For a measurable set $E$ of measure $0<s\le1$, the integral $\int_E K_\tau g\,d\omega$ is linear in $g$. Subject to $g\le B_t$ and $\int g\,d\omega=1$, it is maximized by $g=B_t\one_F$ for a set $F$ of measure $1/B_t$. The spherical rearrangement inequality of Baernstein and Taylor~\cite[Theorem~2]{BaernsteinTaylor1976}, applied to the nondecreasing kernel $\one\{u\cdot v\ge\tau\}$, moves $E$ and $F$ to caps with a common center. Thus
\begin{equation}\label{eq:rearr-sets}
  \int_EK_\tau g\,d\omega\le\int_{E_{1/s}}K_\tau h_{B_t}\,d\omega.
\end{equation}
We also need that $K_\tau h_{B_t}(u)$ increases as $u$ approaches $e$, so that $E_{1/s}$ selects its largest values. To see this, take distinct $u,u'$ with $u'\cdot e\ge u\cdot e$, and let $\mathcal R$ be reflection with normal $w=(u'-u)/\|u'-u\|$. This reflection exchanges $u$ and $u'$. For $v\cdot w\ge0$, both $\one_{E_{B_t}}(v)-\one_{E_{B_t}}(\mathcal Rv)$ and $\one\{u'\cdot v\ge\tau\}-\one\{u\cdot v\ge\tau\}$ are nonnegative. Pairing $v$ with $\mathcal Rv$ expresses $K_\tau h_{B_t}(u')-K_\tau h_{B_t}(u)$ as $B_t$ times the integral of their product over the hemisphere $v\cdot w\ge0$, proving the required monotonicity.

To pass from this setwise comparison to the convex function $\Phi$, let $H=K_\tau g$ and $G=K_\tau h_{B_t}$. Sort their values in decreasing order, obtaining functions $H^*,G^*$ on $(0,1)$ with the same value distributions as $H,G$. Equation~\eqref{eq:rearr-sets} and the preceding monotonicity give $\int_0^sH^*\le\int_0^sG^*$ for every $0\le s\le1$. Writing $t_+=\max\{t,0\}$, the identity $\int_0^s(H^*-z)_+=\sup_{0\le t\le s}\{\int_0^tH^*-zt\}$ gives the same comparison for the positive parts above each level $z\ge0$. A continuous nondecreasing convex $\Phi$ with $\Phi(0)=0$ is a nonnegative multiple of the identity plus a nonnegative weighted integral of these positive-part functions. Consequently, $\int_0^s\Phi(H^*)\le\int_0^s\Phi(G^*)$. Finally, $\int f_s\Phi(H)\,d\omega$ is maximized over $f_s\le B_s$ by placing mass where $\Phi(H)$ is largest, on a set of $\omega$-measure $1/B_s$. Taking $s=1/B_s$ and multiplying by $B_s$ proves the lemma.
\end{proof}

For the angular estimates below, write $C_{d-1}(t)$ for the normalized measure of $\{u\in\mathbb S^{d-2}:u\cdot e\ge t\}$ and $C_{d-2}(t)$ for the corresponding cap probability on $\mathbb S^{d-3}$. Their rate is $I(t)=-\frac12\log(1-t^2)$ for $0<t<1$. By \cref{lem:directions}, $C_{d-1}(t)=\exp\{-dI(t)+O(\log d)\}$ uniformly on compact subsets of $(0,1)$, and the same estimate holds for $C_{d-2}(t)$.

\subsection{Navigation with a sharper density bound near the query}\label{loc:localized-tolerance}\label{sec:scope}

To prove \cref{thm:tolerance}(a), we separate the density condition near the query from the one needed to reach it. Density variation at most $e^{\kappa d}$ near $q$, with $\kappa<h(\gamma)$, rules out false local minima. Variation at most $e^{\bar\kappa d}$ around every possible current point, with $\bar\kappa<\gamma$, supplies progress from distant entries. The next theorem gives this separation. Taking $\bar\kappa=\kappa$ will yield \cref{thm:tolerance}(a).

\begin{theorem}[Navigation with a sharper density bound near the query]\label{loc:nn-sufficiency}
In the dense regime fix $k\ge1$, $\gamma>\gc$, $0\le\kappa\le\kappa_0$ with $\kappa<h(\gamma)$, and $\kappa\le\bar\kappa<\gamma$. There is a finite $K_0=K_0(\gamma,\kappa,\bar\kappa)$ such that the following holds for every fixed $K\ge K_0$. Let $\Delta=\lceil e^{\gamma d}\rceil$ and let $P=f\,d\sigma_{d-1}$ be fixed before the data, with $f>0$ almost everywhere. Suppose that
\[
 \log(1/\rho_{\min})=o(d)\qquad\text{and}\qquad
 \operatorname*{ess\,osc}_{C(x,Ke^\gamma\rho(x))}\log f\le\bar\kappa d
 \quad\text{for every }x\in\mathbb S^{d-1}.
\]
Define the set of queries satisfying the sharper local bound by
\[
 \mathcal Q_\kappa=\{q:\operatorname*{ess\,osc}_{C(q,Ke^\gamma\rho(q))}\log f\le\kappa d\}.
\]
The set $\mathcal Q_\kappa$ is closed, hence Borel. For every absolutely continuous query distribution $Q$ fixed before the data, with probability at least $1-e^{-c_*d}$ over $D_n$ there is a Borel set $\mathcal B$ with $\Prob_{q\sim Q}(q\in\mathcal B)\le e^{-c_*d}$ such that greedy navigation on $\NN{\Delta}$ from every entry succeeds for every $q\in\mathcal Q_\kappa\setminus\mathcal B$. For every $q\in\mathcal Q_\kappa\setminus\mathcal B$ and every entry, the number of complete expansions is at most $H=O(1+\rho_{\min}^{-1})=e^{o(d)}$, and the evaluation cost is at most $1+\Delta H=e^{(\gamma+o(1))d}$. Here $c_*>0$ and the implicit constant in $H$ depend only on $(\gamma,\kappa,\bar\kappa)$. The threshold for sufficiently large $n$ is independent of $Q$.
\end{theorem}

\begin{proof}
We prove success for $k=1$ and then use the same bounds for every fixed $k\ge1$. The proof first chooses auxiliary neighbor radii whose caps fit in the stored lists. Near $q$, we choose one radius whose edges appear in the lists of all points in the final query cap. Those edges support the rearrangement estimate inside that cap and a contraction toward it. Farther away, each current point uses its own auxiliary radius to make progress.

Choose an auxiliary degree exponent $g$ below $\gamma$ but above both $\bar\kappa$ and $\frac12\log(4e^{2\kappa}-2)$, the exponent in \eqref{eq:curve}. Take $g=\tfrac12(\gamma+\max\{\bar\kappa,\tfrac12\log(4e^{2\kappa}-2)\})$. We enlarge the local variation parameter to $\kappa_1=\tfrac12(\kappa+h(g))$ to leave fixed margins in the angular comparisons.
Then $\bar\kappa<g<\gamma$ and $\kappa<\kappa_1<h(g)<g-\gc$. Define $r(x)$ by $nP(C(x,r(x)))=e^{gd}$, so its cap contains $e^{gd}$ points in expectation. The gap $g<\gamma$ will ensure that all its sampled points fit in the neighbor list of $x$.

Choose $a$ as the ratio of the sine of the inner-cap radius to the sine of the common neighbor radius. Put $A=g+\log a$ for the exponent in the upper bound on its expected number of points, and $\tau=1/(2a^2)-1$ for the cosine threshold after taking the opposite direction of a closer point. Start at $a_*=\max\{1/2,e^{\kappa_1-g}\}<1/\sqrt2$, and choose $a$ slightly larger with $a<1/\sqrt2$ and $A>\kappa_1$, so that $0<\tau<1$ and $\tau^2>(1-e^{-2\kappa_1})(1-e^{-2A})$.
Such a choice exists throughout the stated range, including $\gamma>\log2$ and $\kappa=\kappa_0$. At $a=a_*>1/2$, we have $A=\kappa_1$, and the strict inequality is equivalent to $\kappa_1<h(g)$. At $a=a_*=1/2$, we have $\tau=1$, while the right side is strictly smaller than one. Continuity therefore permits the required increase of $a$. Choose $t_0,v_0\in(0,1)$ with $I(t_0)>A$, $I(v_0)>\kappa_1$ and $t_0v_0<\tau$. The infimum of $t_0v_0$ under the two rate constraints is $\sqrt{(1-e^{-2\kappa_1})(1-e^{-2A})}$, which is smaller than $\tau$, so the choices exist. In the rearrangement estimate, $t_0$ limits the projection of the current direction onto $e$, and $v_0$ limits the projection of the opposite direction of a closer point.

The bound $\bar\kappa$ determines the comparison between the auxiliary neighbor radii and the local nearest-neighbor radii. Fix $B_0=1/10$ and $c_0=\cos B_0$ for the cap derivative estimates. We will show that the auxiliary neighbor radius at $x$ is at most $e^{(g+\bar\kappa+1)/c_0}$ times $\rho(x)$ and keep the auxiliary radii below $B_0$ by choosing the multiplier $K$ in $C(x,Ke^\gamma\rho(x))$ sufficiently large.

The neighborhood size must also support progress from distant entries when the density can vary by a factor $e^{\bar\kappa d}$. Choose $w=\min\{1/4,\sqrt{1-e^{-(g-\bar\kappa)}}\}$, so that the probability $C_{d-1}(w)$ of a uniform direction having cosine at least $w$ toward $q$ has rate $I(w)\le(g-\bar\kappa)/2<g-\bar\kappa$. Put $J=4/w$ and $c_p=c_0w/8$. At query distance at least $J$ times the current point's auxiliary radius, a point at distance between $0.9r(x)$ and $r(x)$ from $x$, with directional cosine at least $w$ toward $q$, will decrease the query distance by at least $c_pr(x)$. Set $D=2J\exp((\bar\kappa+\kappa_1)/c_0)$ to express this distance requirement using the common radius near the query. Write $L=Ke^\gamma$ for the multiplier in the assumed cap $C(x,L\rho(x))$. All the caps used in the radius comparisons lie there when $K\ge K_0$, where we take
\[
 K_0=e^{(g+\bar\kappa+1)/c_0-\gamma}\max\{\pi(D+1)/B_0,\ 2J+1,\ D+1\}.
\]

We now verify the radius comparisons using the bound $\bar\kappa$ at every center. Positive density makes cap probability continuous and strictly increasing with its radius, so both $\rho(x)$ and $r(x)$ are uniquely determined. Cap containment gives $|\rho(x)-\rho(y)|\le\delta(x,y)$ and $|r(x)-r(y)|\le\delta(x,y)$. Write $\rho_{\max}=\max_{x\in\mathbb S^{d-1}}\rho(x)$ and $r_{\max}=\max_{x\in\mathbb S^{d-1}}r(x)$, which are attained by continuity. Eventually $\rho_{\max}<\pi/L$. Otherwise $C(x,L\rho(x))$ at a maximizer of $\rho$ would cover the sphere, giving $f\ge e^{-\bar\kappa d}$ almost everywhere and $1/n\ge e^{-\bar\kappa d}F(\pi/L)$. This contradicts \cref{lem:cap} at $\pi/L$, because $\log n/d\to\infty$.

Integrating the cap density and bounding the cosine by $c_0$ gives $(d-1)c_0\le tF'(t)/F(t)\le d-1$ for $0<t\le B_0$. Since $\rho(x)<\pi/L$, the choice of $K_0$ gives $e^{(g+\bar\kappa+1)/c_0}\rho(x)<B_0/(D+1)$ and places both $C(x,\rho(x))$ and $C(x,e^{(g+\bar\kappa+1)/c_0}\rho(x))$ inside $C(x,L\rho(x))$. Integrating the lower derivative bound and applying the density comparison gives
\[
 nP\bigl(C(x,e^{(g+\bar\kappa+1)/c_0}\rho(x))\bigr)
 \ge e^{(d-1)(g+\bar\kappa+1)-\bar\kappa d}.
\]
For all sufficiently large $d$, this lower bound exceeds $e^{gd}$. Consequently, uniformly in $x$, we have
\begin{equation}\label{loc:radii}
 \rho(x)\le r(x)\le e^{(g+\bar\kappa+1)/c_0}\rho(x)
 \qquad\text{and}\qquad r_{\max}<B_0/(D+1).
\end{equation}
On this smaller radius range, the logarithmic derivative $tF'(t)/F(t)$ is at least $(d-1)\cos(B_0/(D+1))\ge dc_0$ for all sufficiently large $d$.

Near a query, we use its sharper density bound to obtain a common radius retained at all nearby points. For every $q\in\mathbb S^{d-1}$, let $M_q=\operatorname*{ess\,sup}_{C(q,L\rho(q))}f$. Fix $q\in\mathcal Q_\kappa$ and define $R_q$ by $nM_qF(R_q)=e^{gd}$. Since $r(q)\le e^{(g+\bar\kappa+1)/c_0}\rho(q)<L\rho(q)$, the cap $C(q,r(q))$ lies where the density is between $M_qe^{-\kappa_1d}$ and $M_q$ almost everywhere. Comparing its probability $P(C(q,r(q)))=M_qF(R_q)$ with these density bounds and integrating the cap derivative estimate gives
\begin{equation}\label{loc:query-radius}
 R_q\le r(q)\qquad\text{and}\qquad r(q)/R_q\le e^{\kappa_1/c_0}.
\end{equation}
For $\delta(q,x)\le DR_q$, combining $R_q\le r(q)\le e^{(g+\bar\kappa+1)/c_0}\rho(q)$ with the choice of $K_0$ gives $C(x,R_q)\subseteq C(q,(D+1)R_q)\subseteq C(q,L\rho(q))$. Therefore $nP(C(x,R_q))\le e^{gd}$ and $R_q\le r(x)$. Also $DR_q<B_0<\pi/2$ by \eqref{loc:radii}.

Outside $C(q,DR_q)$, the progress estimate uses the current point's own radius $r(x)$ rather than $R_q$. To apply the direction threshold $w$, we need the query distance to be at least $Jr(x)$. We therefore prove
\begin{equation}\label{loc:outer-dichotomy}
 \delta(q,x)\ge DR_q\quad\Longrightarrow\quad\delta(q,x)\ge Jr(x).
\end{equation}
If $\theta=\delta(q,x)<Jr(x)$, the Lipschitz property gives $r(q)<(J+1)r(x)$. Both caps $C(x,r(x))$ and $C(q,r(q))$ then lie in $C(x,(2J+1)r(x))\subseteq C(x,L\rho(x))$. They have equal $P$-probability, and their average densities differ by at most $e^{\bar\kappa d}$. The cap derivative bound therefore gives $r(x)/r(q)\le e^{\bar\kappa/c_0}$. Combining this with \eqref{loc:query-radius} yields $\theta<J e^{(\bar\kappa+\kappa_1)/c_0}R_q=DR_q/2$, proving \eqref{loc:outer-dichotomy}.

These radius comparisons concern the underlying density. To use them for navigation, we also need the sampled neighbor lists to retain the points within $r(x)$. Conditional on a data point at $x$, the number of other data points in $C(x,r(x))$ is binomial with mean at most $e^{gd}$. A multiplicative Chernoff bound and a union over labels give probability at least $1-n\exp(-(2\log2-1)e^{gd})=1-e^{-\omega(d)}$ that all these numbers are at most $2e^{gd}<\Delta$. Every neighbor list of $\NN{\Delta}$ then contains all other data points within $r(x)$. For each fixed query, we next bound progress failures using independence of the sampled points, then add the probability of missing required edges.

Write $R=R_q$ and define the inner query cap $C(q,\Theta)$ by $\Theta=\arcsin(a\sin R)$, using the ratio $a$ chosen above. We first bound the expected number of sampled points in this cap. The direction comparison will then bound the expected number of false local minima relative to the radius-$R$ edges among them. Substituting $u=\sin t$ in the cap integral gives $F(\theta)=\alpha_{d-1}\int_0^{\sin\theta}u^{d-2}(1-u^2)^{-1/2}du$ for $\theta<\pi/2$. Scaling $u\mapsto au$ and using $R<B_0$ yields $c_0a^{d-1}F(R)\le F(\Theta)\le a^{d-1}F(R)$, and hence $nM_qF(\Theta)\le a^{-1}\exp(Ad)$. Let $Z_{\mathrm{core}}(q)$ count the points $x\in D_n\cap C(q,\Theta)$ for which another sampled point is closer to $q$, but no sampled point within distance $R$ of $x$ is closer to $q$. Once the auxiliary neighbors are retained, $R\le r(x)$ ensures that every unsuccessful stopping point in the inner cap is among these points. We will prove
\[
  \E Z_{\mathrm{core}}(q)\le12\eps_{d}+a^{-1}\exp(Ad)C_{d-1}(t_0),
\]
where $\eps_d=\exp(\kappa_1d)C_{d-1}(v_0)+C_{d-2}(\tau-t_0v_0)$ bounds $K_\tau h_{B_t}(u)$ whenever $u\cdot e\le t_0$, with $B_t=\exp(\kappa_1d)$. The inequalities $I(v_0)>\kappa_1$, $I(t_0)>A$, and $t_0v_0<\tau$ make both terms in the bound on $\E Z_{\mathrm{core}}(q)$ exponentially small in $d$.

The relation $\sin\Theta=a\sin R$ also restricts the directions of closer points that lie outside the neighbor cap. Let $\delta(q,x)=\theta\le\Theta$ and $0<\delta(q,z)=t\le\theta$. By the spherical cosine identity, $\delta(x,z)>R$ requires the cosine between their directions about $q$ to be less than $(\cos R-\cos\theta\cos t)/(\sin\theta\sin t)$. This threshold increases with $t$, since its derivative has the sign of $\cos\theta-\cos R\cos t>0$. At $t=\theta$, it is $1-(1-\cos R)/\sin^2\theta\le1-1/(a^2(1+\cos R))\le-\tau$. Thus every closer point outside $C(x,R)$ has direction with inner product below $-\tau$ with the direction of $x$.

To average this condition over a sampled data point, use polar coordinates about $q$. For a unit direction $u$ orthogonal to $q$, set $f_\theta(u)=f(\cos\theta\,q+\sin\theta\,u)$ and let $\mu(\theta)=\int f_\theta\,d\omega$ be its angular average. For almost every $0<\theta\le\Theta$, we have $0<\mu(\theta)<\infty$, and the conditional direction density is $s_\theta=f_\theta/\mu(\theta)$, bounded by $B_s(\theta)=M_q/\mu(\theta)\ge1$. A point drawn from $P$ conditional on $C(q,\theta)$ has direction density at most $B_t$. Indeed, the bound with parameter $\kappa$ on $C(q,L\rho(q))$ gives $s_t\le e^{\kappa d}\le B_t$ for almost every radius $t$, and the mixture over radii preserves this bound. Here $B_t\ge2$ for all sufficiently large $n$. Write $p_\theta=P(C(q,\theta))$ and $\ell_x=P(C(q,\theta)\cap C(x,R))$. Independence of the other $n-1$ points makes the probability that $x$ contributes to $Z_{\mathrm{core}}(q)$ exactly $(1-\ell_x)^{n-1}-(1-p_\theta)^{n-1}$. Taking the opposites of the closer-point directions changes the preceding threshold from $-\tau$ to $\tau$, so the missing fraction $(p_\theta-\ell_x)/p_\theta$ is bounded by the cap operator $K_\tau$ applied to the density of the opposite directions, evaluated at the direction of $x$ about $q$. That density is bounded by $B_t$.

With $N=n-1$ and $p=p_\theta$, define $\Phi(z)=(1-p+pz)^N-(1-p)^N$. The probability that $x$ contributes to $Z_{\mathrm{core}}(q)$ is $\Phi((p_\theta-\ell_x)/p_\theta)$, so it is a function of the fraction of the probability mass of $C(q,\theta)$ under $P$ that lies outside $C(x,R)$. This function is nondecreasing and convex and satisfies $\Phi(0)=0$. Apply \cref{lem:rearrangement} with $f_s=s_\theta$ and the density of the opposite direction of a point drawn from $P$ conditional on $C(q,\theta)$ as the second density. The lemma bounds the expected contribution to $Z_{\mathrm{core}}(q)$ conditional on $\delta(q,x)=\theta$ by the integral with cap densities $h_{B_s(\theta)}$ and $h_{B_t}$. For a direction $u$ with $u\cdot e\le t_0$, let $V$ be uniform on $E_{B_t}$. The probability of $V\cdot e>v_0$ is at most $B_tC_{d-1}(v_0)$. When $V\cdot e\le v_0$, we also have $V\cdot e\ge0$ because $B_t\ge2$. Hence $(u\cdot e)(V\cdot e)\le t_0v_0$. If $u=-e$, then $u\cdot V\ge\tau$ is impossible. Otherwise the component of $u$ orthogonal to $e$ is nonzero, and $u\cdot V\ge\tau$ requires the inner product of the normalized components orthogonal to $e$ to be at least $\tau-t_0v_0$. Conditional on $V\cdot e$, the normalized component of $V$ is uniform on $\mathbb S^{d-3}$ in $e^\perp$. Its inner product with the fixed normalized component of $u$ therefore exceeds $\tau-t_0v_0$ with probability at most $C_{d-2}(\tau-t_0v_0)$. Thus $K_\tau h_{B_t}(u)\le\eps_{d}$ whenever $u\cdot e\le t_0$.

We now use this operator bound in the failure probability. For $z\le\eps_{d}$, we have $\Phi(z)\le Npz(1-p+pz)^{N-1}\le Np\eps_{d}\exp\{-(N-1)p(1-\eps_{d})\}$. Directions with $u\cdot e>t_0$ have probability at most $B_s(\theta)C_{d-1}(t_0)$ under $h_{B_s(\theta)}$, and there we use $\Phi\le1$. Together these bounds give
\[
  \E\bigl[(1-\ell_x)^{n-1}-(1-p_\theta)^{n-1}\,\big|\,\theta\bigr]
  \le Np_\theta\eps_{d}\exp\{-(N-1)p_\theta(1-\eps_{d})\}+B_s(\theta)C_{d-1}(t_0).
\]
To obtain the expected number of such points, integrate the conditional bound against the distance law $\mu(\theta)\,dF(\theta)$ of one sampled point, then sum over the $n$ labels. In the first term, $dp_\theta=\mu(\theta)\,dF(\theta)$ allows us to substitute the cap probability itself. Extending the resulting integral to $[0,1]$ bounds this contribution by
\[
  nN\eps_{d}\int_0^1p\exp\{-(N-1)p(1-\eps_{d})\}\,dp
  \le\frac{nN\eps_{d}}{(N-1)^2(1-\eps_{d})^2}\le12\eps_{d}.
\]
In the second term, $\mu(\theta)B_s(\theta)=M_q$ cancels the angular average, giving $nM_qF(\Theta)C_{d-1}(t_0)\le a^{-1}\exp(Ad)C_{d-1}(t_0)$. This proves the bound on $\E Z_{\mathrm{core}}(q)$ and excludes false local minima in the inner cap with probability $1-\exp\{-\Omega(d)\}$ for each fixed query.

For $\Theta\le\delta(q,x)\le DR$, the density bound with parameter $\kappa$ near $q$ must leave enough candidates for a fixed contraction of the query distance. Let $d_*=\min\{A-\kappa_1,g-\kappa_1-\gc\}>0$ be the smaller of the two available exponent margins. Choose $1-b=\min\{1/(4D),d_*/(4(D+1)),1/4\}$ and $s=1-(1-b)D$, where $b$ is the contraction factor and $s$ is the fraction of the neighbor radius left before contraction. These choices give $b,s\ge3/4$ and $\log b+\log s\ge-d_*/2$, so $g-\kappa_1+\log a+\log b>0$ and $g-\kappa_1-\gc+\log s+\log b>0$. The contraction parameters depend only on $(\gamma,\kappa,\bar\kappa)$.

With this contraction fixed, we find a neighbor that contracts the query distance by $b$ whenever $\Theta\le\theta=\delta(q,x)\le DR$. Start with the lens $C(x,sR)\cap C(q,\theta)$, leaving the radius difference $R-sR$ for the contraction. If $\cos^2\theta\ge\cos(sR)$, \cref{lem:inside}(b) gives surface mass at least $F(\theta)/2\ge c_0a^{d-1}F(R)/2$. Otherwise \cref{lem:directions} bounds the improving-direction threshold by $\cot\theta\tan(sR/2)$. The inequality $\cos^2\theta<\cos(sR)$ bounds this in turn by $\sqrt{\cos(sR)/(1+\cos(sR))}<1/\sqrt2$. All directions with cosine at least $1/\sqrt2$ toward $q$ therefore improve throughout the cap of radius $sR$. The direction estimate and the upper bound on $tF'(t)/F(t)$ give lens mass at least $F(sR)e^{-\gc d-O(\log d)}\ge F(R)\exp\{(\log s-\gc)d-O(\log d)\}$.

Contract each point of this lens toward $q$ by multiplying its distance to $q$ by $b$. The point moves by at most $(1-b)\theta\le(1-b)DR$, so $sR+(1-b)DR=R$ places its image in $V=C(x,R)\cap C(q,b\theta)$. The contraction has Jacobian at least $b^{d-1}$, since $\sin(bt)\ge b\sin t$. Moreover, $V\subseteq C(q,L\rho(q))$, where $f\ge M_qe^{-\kappa_1 d}$. Using the defining equality for $R_q$, the two lens bounds therefore give
\[
  nP(V)\ge\exp\bigl\{[g-\kappa_1+\log b+\min\{\log a,\log s-\gc\}]d-O(\log d)\bigr\}=e^{\Omega(d)},
\]
with a positive exponent because $g-\kappa_1+\log a+\log b>0$ and $g-\kappa_1-\gc+\log s+\log b>0$. Conditional on $x$, independence of the other $n-1$ points makes the probability that $V$ contains none of them at most $\exp(-e^{\eta'd})$ for a fixed $\eta'>0$. Summing over the $n$ labels bounds the probability that some data point in the middle region has no other data point in its corresponding set $V$ by $n\exp(-e^{\eta'd})=e^{-\omega(d)}$, using $\log n=o(d^2)$. Once the auxiliary neighbors are retained, every data point in $V$ appears in the list of $x$, because $R\le r(x)$. Each greedy move therefore multiplies the query distance by at most $b$ in the middle region.

In the outer region $\delta(q,x)\ge DR$, we use the direction threshold chosen for the density bound with parameter $\bar\kappa$. Suppose $\theta=\delta(q,x)\ge DR$. By \eqref{loc:outer-dichotomy}, $\theta\ge Jr(x)$. Write $r=r(x)$ and let $V_x$ consist of points at distance $t\in[0.9r,r]$ from $x$ whose directional cosine toward $q$ is at least $w$. At $x=-q$, use the whole annulus. For $\theta<\pi$, the spherical cosine rule gives a decrease of at least $c_pr$ whenever the directional cosine is at least
\[
 \frac{\sin(c_pr)}{\sin t}+\cot\theta\,\frac{\cos(c_pr)-\cos t}{\sin t}.
\]
For $Jr\le\theta\le\pi/2$, this threshold is at most $c_p/(0.9c_0)+\tan(r/2)/(Jr)\le7w/18<w$, using $r\le B_0$, $\sin t\ge c_0t$, $\cot\theta\le1/\theta$ and $\tan(r/2)\le r$ on $[0,B_0]$. For $\theta\ge\pi/2$, the cotangent term is nonpositive. At the antipode the decrease is exactly $t\ge0.9r>c_pr$. Thus every point in $V_x$ supplies the required decrease. The cap derivative bound gives an annular fraction at least $1-0.9^{c_0d}$. Both the annulus and the cap $C(x,r)$ lie in $C(x,L\rho(x))$, so \eqref{eq:local-mass-comparison}, with $\bar\kappa$ in place of $\kappa$, gives
\[
 nP(V_x)\ge e^{(g-\bar\kappa)d}(1-0.9^{c_0d})C_{d-1}(w)
 =e^{(g-\bar\kappa-I(w))d-O(\log d)}=e^{\Omega(d)}.
\]
Conditioning on each data point's position separately and taking a union over labels bounds by $n\exp(-e^{c_3d})$ the probability that any point in the outer region has no other data point in its corresponding set $V_x$, where $c_3=(g-\bar\kappa-I(w))/2>0$. Retention places these improving points in the current point's neighbor list. Greedy navigation therefore decreases the query distance by at least $c_pr(x)$ at each move in the outer region, unless that expansion has already succeeded.

To integrate the three progress estimates over queries, we first verify that their failure sets are Borel. Put $W_q=C(q,L\rho(q))$ and $m_q=\operatorname*{ess\,inf}_{W_q}f$. The finite density variation on $W_q$ gives $0<m_q\le M_q<\infty$. For every real $t>0$, consider the maps
\[
 q\longmapsto\int_{W_q}(f-t)_+\,d\sigma_{d-1}
 \qquad\text{and}\qquad q\longmapsto\int_{W_q}(t-f)_+\,d\sigma_{d-1}.
\]
Continuity of $\rho$, the zero surface measure of cap boundaries and domination by $f+t$ make both maps continuous. Their zero sets are respectively $\{M_q\le t\}$ and $\{m_q\ge t\}$, so $M_q$ is lower semicontinuous and $m_q$ is upper semicontinuous. Hence $\mathcal Q_\kappa=\{q:M_q-e^{\kappa d}m_q\le0\}$ is closed, and $R_q$, determined from $M_q$, is Borel. The property $Z_{\mathrm{core}}(q)=0$ and the existence of other sampled points in each set $V$ and $V_x$ at its corresponding middle or outer data point are finite distance comparisons. Their failure indicators are therefore measurable.

The estimates for each fixed $q\in\mathcal Q_\kappa$ can now be integrated against $Q$. Let $\chi_*=\min\{I(v_0)-\kappa_1,I(\tau-t_0v_0),I(t_0)-A\}>0$ and $c_4=\chi_*/8$. The inner cap estimate bounds its failure probability by $e^{-\chi_*d+O(\log d)}$, while the middle and outer failure probabilities are $e^{-\omega(d)}$. Thus the probability that any of the three properties fails is at most $e^{-c_4d}$ for all sufficiently large $n$. Let $\mathcal B$ consist of the queries in $\mathcal Q_\kappa$ for which $Z_{\mathrm{core}}(q)>0$ or one of those sets $V$ or $V_x$ contains no other sampled point, together with the data bisectors. Tonelli and Markov at threshold $e^{-c_4d/2}$ give, with probability at least $1-e^{-c_4d/2}$, an exceptional set $\mathcal B$ of $Q$-probability at most $e^{-c_4d/2}$, since the bisectors have $Q$-probability zero. Combining this conclusion with retention and taking $c_*=\chi_*/32$ gives the stated data probability and exceptional query probability for all sufficiently large $n$.

Fix a data set for which retention holds and $\mathcal B$ has the stated probability bound. For any $q\in\mathcal Q_\kappa\setminus\mathcal B$ and any entry, outer moves decrease the query distance by at least $c_p\rho_{\min}$. Middle moves multiply it by at most $b$ over a distance ratio $DR/\Theta\le D/(ac_0)$. These two regions therefore require $O(1+\rho_{\min}^{-1})$ moves.

Inside the inner cap, every list contains the radius-$R$ edges. Any two path vertices separated by at least two moves must be more than $R$ apart, since the later vertex would otherwise be a listed neighbor closer to $q$ than the immediate successor actually chosen. In each parity subsequence, the Euclidean separation exceeds $2\sin(R/2)$, while all vertices lie within distance $2\sin(\Theta/2)$ of $q$. The squared ratio of these radii is
\[
 \frac{1-\cos\Theta}{1-\cos R}
 =a^2\frac{1+\cos R}{1+\cos\Theta}\le a^2<\frac12.
\]
The pairwise squared-distance identity in the proof of \cref{prop:certificate} bounds each subsequence by $\lceil(1-2a^2)^{-1}\rceil$ vertices. The absence of false local minima forces success, so the total number of complete expansions is $H=O(1+\rho_{\min}^{-1})=e^{o(d)}$, including the final expansion and the expansion of an entry that is already the nearest neighbor. The evaluation cost is at most $1+\Delta H$. For fixed $k\ge1$, success occurs no later than for $k=1$, because the nearest neighbor belongs to $\TopSet_k(q)$. The same complete-expansion and cost bounds therefore apply.
\end{proof}

\begin{proof}[Proof of \cref{thm:tolerance}(a)]
Finite log-density oscillation on $C(x,Ke^\gamma\rho(x))$, a cap containing positive $P$-mass, makes its essential density bounds finite and positive. A finite subcover of these caps gives $f>0$ almost everywhere on the sphere. Take $\bar\kappa=\kappa$ in \cref{loc:nn-sufficiency} and choose $K\ge K_0(\gamma,\kappa,\kappa)$. Every query then belongs to $\mathcal Q_\kappa$, giving the stated bounds.
\end{proof}

To obtain the degree scale in \eqref{eq:curve}, the density condition must remain on one fixed neighborhood while the degree approaches the curve. For fixed $0\le\kappa\le\kappa_0$, take $\gamma_j=\frac12\log(4e^{2\kappa}-2)+1/j$. In the proof of \cref{loc:nn-sufficiency} with $\bar\kappa=\kappa$, we have $g-\kappa>\gc$, so $w=1/4$ and $J=16$. The parameters $g,\kappa_1,D$ and $e^{\gamma_j}K_0(\gamma_j,\kappa,\kappa)$ remain bounded as $j\to\infty$, including at $\kappa=\kappa_0$. Choose one fixed multiplier $L$ larger than all the required values of $e^{\gamma_j}K_0$, and assume the density bound on every cap $C(x,L\rho(x))$. For exponent $\gamma_j$, apply \cref{loc:nn-sufficiency} with $K=Le^{-\gamma_j}$, so the assumed caps are unchanged.

For each fixed $j$, the radius comparisons and probability estimates depend only on $\kappa,\gamma_j,L$ and $n,d$, uniformly over the density class and the query distribution. In particular, the contradiction establishing $\rho_{\max}<\pi/L$ uses only $1/n\ge e^{-\kappa d}F(\pi/L)$, so its threshold is independent of the density. Choose increasing deterministic thresholds $N_j$ so that, for all $n\ge N_j$, both error probabilities are at most $1/j$ and the constant multiplying $1+\rho_{\min}^{-1}$ in the expansion bound is at most $e^{d/j}$. Use degree $\lceil e^{\gamma_jd}\rceil$ for $N_j\le n<N_{j+1}$. This gives one degree sequence for the entire density class. For each density sequence satisfying $\log(1/\rho_{\min})=o(d)$, its degree and evaluation cost are $e^{(\frac12\log(4e^{2\kappa}-2)+o(1))d}$. The resulting sample probability and query probability are $1-o(1)$. Together with \cref{thm:tolerance}(b) and \cref{thm:critical}(b) at $\kappa=0$, this proves the sharp worst-case scale throughout \eqref{eq:curve}.

\subsection{Below the dense curve}\label{win:subsection}\label{sec:curve-lower}\label{sec:dense-obstructions}

To match the upper bound with an obstruction for every graph, we prove \cref{thm:tolerance}(b) by constructing many possible entries in balls surrounded by annuli of lower density. Few improving points lie in those annuli. For stored edges longer than a distance cutoff specified below and ending outside the ball and its surrounding annulus, only a small fraction of query directions allow improvement. The sample estimates will hold before any graph is chosen. For each graph of degree at most $\Delta$, we then bound the exceptional query directions for its stored edges. Repeating the construction across the sphere gives the query distribution.

\begin{proof}[Proof of \cref{thm:tolerance}(b)]
We first choose a geometric configuration that balances the number of possible entries against the fraction of query directions for which a fixed stored edge can improve an entry. We then perturb it to obtain strict margins. Use the directional rate $I$ and cap probability $C_{d-1}$ from \cref{sec:tolerance}. Choose $\gamma_1>\gamma$ sufficiently close that $h(\gamma_1)<\kappa$, using the formula in \eqref{eq:tolerance-function} also when $\gamma_1>\log2$. The gaps $\gamma_1-\gamma$ and $\kappa-h(\gamma_1)$ will leave positive margins in the edge and density estimates.

The configuration consists of a ball containing possible entries, an annulus of lower density around it, and closer points beyond the annulus. We first use Euclidean coordinates with the outer annulus radius normalized to one, then scale to the sphere. Fix an axis through the center on which the query will lie. A representative entry and a representative closer point have axial coordinates $x$ and $B$, each at perpendicular distance $L$ from that axis. To specify the configuration, put $E=e^{2\gamma_1}>2$ and $Z_0=E^2-E-1$, and take $x=Z_0^{-1/2}$, $L=\sqrt{(E-2)/Z_0}$, ${r_{\mathrm{src}}}=\sqrt{(E-1)/Z_0}$, and $B=(E-1)/\sqrt{Z_0}$. Here ${r_{\mathrm{src}}}$ is the radius of the ball containing the entries. Place the query at axial coordinate $t_0=(B+x)/2$. Its common distance to the two representative points is $D_0=\sqrt{L^2+(B-x)^2/4}$, and the typical distance between such points with independent perpendicular directions is $\ell_0=\sqrt{(B-x)^2+2L^2}$.
These choices satisfy ${r_{\mathrm{src}}}^2=x^2+L^2<1$, $B^2+L^2=1$, $0<x<t_0<B<1$, $0<L<{r_{\mathrm{src}}}$, and $t_0+D_0<2$.

To compare the numbers of points and stored neighbors, set $a=-\log L$ and $\eps=\log({r_{\mathrm{src}}}/L)$, and choose the common angular scale $R$ by $nF(R)=e^{ad}$. The dense regime and \cref{lem:cap} give $R\to0$ and $\log(1/R)=o(d)$. For fixed positive radius multipliers, the same estimate gives
\begin{equation}\label{win:cap-scaling}
 F(vR)/F(R)=e^{d\log v+o(d)}
\end{equation}
uniformly on compact ranges of $v>0$. After scaling to the sphere, the outer cap has angular radius $R$ and the source cap has angular radius ${r_{\mathrm{src}}}R$. Under uniform sampling their expected numbers of points are $e^{ad}$ and $e^{(\eps+o(1))d}$, respectively. The directional estimate below has exponent $J=\frac12\log((1+x^2)/L^2)$. The two relations that connect these exponents to the density and degree are
\begin{equation}\label{win:boundary-identities}
 a+\log D_0=\tfrac12\log\frac{E+2}{4}=h(\gamma_1)
 \qquad\text{and}\qquad J-\eps=\gamma_1.
\end{equation}
The inequality $a+\log D_0<\kappa$ will make the expected number of sampled points in the low density part of the improving region exponentially small. For candidate neighbors outside the outer cap and separated from the entry by at least the distance cutoff, we bound the fraction of query directions placing both endpoints of a stored edge within the query cap. The inequality $J-\eps>\gamma$ will keep this bound exponentially small after summing over the possible entries and their neighbor lists. We first establish the direction bound for one fixed edge.

First fix normalized vectors $X,Y\in\R^{d-1}$ for a source and a candidate neighbor outside the unit ball, with $|X|={r_{\mathrm{src}}}$, $|Y|=r\ge1$, and $|X-Y|\ge\ell_0$. A unit query direction $u$ that places both points within distance $D_0$ of the query satisfies $|t_0u-X|\le D_0$ and $|t_0u-Y|\le D_0$. These bounds give $u\cdot X\ge x$ and $u\cdot Y\ge B+(r^2-1)/(B+x)$. To combine the projection requirements into one cap condition, put $V=xX+BY$, $s_0=x^2+B^2$, and $z=r^2-1$. The projection of $V$ and its squared norm satisfy
\begin{align*}
 u\cdot V&\ge s_0+\frac{B}{B+x}z=:S(z),\\
  |V|^2&\le x^2{r_{\mathrm{src}}}^2+B^2r^2+xB({r_{\mathrm{src}}}^2+r^2-\ell_0^2)
 =s_0(1+x^2)+B(B+x)z=:N(z).
\end{align*}
For $z\ge0$, both $S(z)$ and $N(z)$ are positive. Thus $S(z)^2/N(z)$ is a lower bound on the squared cosine threshold of the cap about $V/|V|$. The derivative of $S(z)^2/N(z)$ has the sign of $2BN(z)/(B+x)-B(B+x)S(z)$, an affine function with positive slope and constant term $s_0B\ell_0^2/(B+x)>0$. The minimum is therefore attained at $z=0$, giving squared threshold at least $s_0/(1+x^2)=1-e^{-2J}$.

We need this direction bound both at the initial parameters and when the source radius and query distance vary slightly. Write $s=|X|$ for the normalized radius of the source and $r=|Y|$ for that of the candidate neighbor, $t'$ for the normalized query distance from the center, and $D$ for their common distance bound from the query. Let $\ell$ be the distance cutoff between nearby and farther pairs, to be chosen below $\ell_0$, and reserve an exponent margin $\zeta=(\gamma_1-\gamma)/8$. The corresponding bounds on the projection and squared norm are
\[
\begin{aligned}
 \mathcal S(s,r,t',D)&=\frac{x(t'^2+s^2-D^2)+B(t'^2+r^2-D^2)}{2t'},\\
 \mathcal N(s,r,\ell)&=x^2s^2+B^2r^2+xB(s^2+r^2-\ell^2).
\end{aligned}
\]
At $(s,t',D,\ell)=({r_{\mathrm{src}}},t_0,D_0,\ell_0)$, the preceding calculation holds uniformly for $1\le r\le2$. By continuity and compactness, choose a fixed neighborhood of this tuple on which
\begin{equation}\label{win:far-neighborhood}
 \mathcal S,\mathcal N>0,\qquad \mathcal S^2/\mathcal N>1-e^{-2(J-\zeta/2)},\qquad\text{and}\qquad t'+D<2
\end{equation}
uniformly for $1\le r\le2$. Choose a distance cutoff $\ell<\ell_0$ in this neighborhood. We will also restrict relevant sources to a thin interval of radii. Choose $\eta_s>0$ sufficiently small that $s\in[{r_{\mathrm{src}}}-\eta_s,{r_{\mathrm{src}}}]$ stays in the same neighborhood, and put $r_s={r_{\mathrm{src}}}-\eta_s>0$. To exclude sources farther inside the source ball, set $s_s=(t_0^2+r_s^2-D_0^2)/(2t_0)$ and $L_s=\sqrt{r_s^2-s_s^2}$, requiring $0<s_s<\min\{r_s,t_0\}$ and $0<L_s<L$. These conditions hold for sufficiently small $\eta_s$, since $0<s_s<\min\{r_s,t_0\}$ holds strictly at $r_s={r_{\mathrm{src}}}$, while $L_s=L$ there. The derivative with respect to the squared radius is positive at this endpoint, since
\[
 \left.\frac{d}{dv}\left[v-\left(\frac{t_0^2+v-D_0^2}{2t_0}\right)^2\right]\right|_{v={r_{\mathrm{src}}}^2}=1-x/t_0>0.
\]
The strict inequality $L_s<L=e^{-a}$ persists when $t_0,D_0$ in the expression for $L_s$ are replaced by nearby $t',D$. Fix a neighborhood of $(t_0,D_0)$ on which $L_s$ remains below $e^{-a}$ by a fixed positive margin. For a center $p$ with $\delta(p,q)=t'R$, the cap containment calculation below will then make the expected number of points in $C(p,r_sR)\cap C(q,DR)$ exponentially small.

To apply the direction estimate on the sphere, identify the tangent space at a center $p$ with $\R^{d-1}$. For a tangent vector $X$, $\exp_p(X)$ denotes travel from $p$ by distance $|X|$ in direction $X/|X|$, with $\exp_p(0)=p$. Here $|\cdot|$ is Euclidean norm. Write the query as $q=\exp_p(t'Ru)$ and the two points as $\exp_p(RX)$ and $\exp_p(RY)$. If the second point is outside $C(p,R)$ and within distance $DR$ of $q$, then $1\le |Y|\le t'+D<2$. Expanding the spherical cosine identity for the two points at $R=0$ gives, uniformly for $|X|,|Y|\le2$, the metric comparison
\begin{equation}\label{win:metric-expansion}
 R^{-2}\delta(\exp_p(RX),\exp_p(RY))^2=|X-Y|^2+O(R^2).
\end{equation}
For pairs at mutual distance at least $\ell R$, \eqref{win:far-neighborhood} therefore gives a direction cap with squared threshold at least $1-e^{-2(J-\zeta/2)}+O(R^2)$. Since $R\to0$ in the dense regime, this threshold is at least $1-e^{-2(J-\zeta)}$ for large $n$. The fraction of query directions placing both points within distance $DR$ of $q$ is consequently at most
\begin{equation}\label{win:far-cap}
 \exp\{-(J-\zeta+o(1))d\}.
\end{equation}
A threshold at least one gives zero direction measure. The errors are uniform over the fixed parameter neighborhoods and all such pairs, so \eqref{win:far-cap} applies after the data and graph have been chosen.

For pairs closer than $\ell R$, we bound the probability that such a pair appears in the sample. We first enclose the region containing an entry and the region containing a closer exterior point in small caps. Suppose $\delta(p,q)=t'R$. Let $rR$ be the radius of the cap about $p$ and $DR$ the permitted distance from $q$, with $r,D>0$ fixed. In this containment calculation, $s=(t'^2+r^2-D^2)/(2t')$ is the axial coordinate of the containing cap center in units of $R$, and $L_S=\sqrt{r^2-s^2}$ is its leading radius multiplier. At $r=r_s$, $t'=t_0$, and $D=D_0$, these are the quantities $s_s$ and $L_s$ defined above. When $0<s<\min\{r,t'\}$, the intersection $C(p,rR)\cap C(q,DR)$ lies in a cap with center at distance $sR+O(R^3)$ from $p$ toward $q$ and radius $(L_S+O(R^2))R$. To see this, combine the two cosine inequalities with weights $1-s/t'$ and $s/t'$, then normalize the vector $(1-s/t')p+(s/t')q$. The resulting cosine threshold is
\[
 \frac{(1-s/t')\cos(rR)+(s/t')\cos(DR)}{\sqrt{1-2(s/t')(1-s/t')(1-\cos(t'R))}}
 =1-\tfrac12(r^2-s^2)R^2+O(R^4).
\]
Expanding the normalized vector along the geodesic from $p$ to $q$ gives the cap center's location. Both expansions are uniform on compact parameter sets satisfying the strict conditions.

The second point lies in the exterior region $C(q,DR)\setminus C(p,R)$. For this region, put $b'=(t'^2+1-D^2)/(2t')$ and $L_T=\sqrt{1-b'^2}$, and suppose $t'<b'<1$. The region $C(q,DR)\setminus C(p,R)$ is contained in a cap centered at distance $b'R$ from $p$ toward $q$, with radius $(L_T+O(R^2))R$. Indeed, the unit center is $z=\alpha q-\beta p$, where $\alpha=\sin(b'R)/\sin(t'R)>0$ and $\beta=\alpha\cos(t'R)-\cos(b'R)>0$. Every $y\in C(q,DR)\setminus C(p,R)$ satisfies $z\cdot y\ge\alpha\cos(DR)-\beta\cos R$. The right side is $1-\frac12(1-b'^2)R^2+O(R^4)$, using $D^2-t'^2-1=-2t'b'$, which proves the radius bound.

At $(r,t',D)=({r_{\mathrm{src}}},t_0,D_0)$, the two containing caps have radii $(L+O(R^2))R$ and centers separated by $(B-x+O(R^2))R$. Independent uniform points in them typically have squared distance close to $\ell_0^2R^2$, so pairs within distance $\ell R$ have exponentially small probability. For the needed uniform bound, choose $\tau\in(0,1/4)$ sufficiently small that
\begin{equation}\label{win:pair-gap}
 (B-x)^2+2(1-\tau)^2L^2-2\tau\{2(B-x)L+L^2\}>\ell^2.
\end{equation}
This inequality remains strict when the two radii and the distance between their centers vary in a sufficiently small compact neighborhood. In either cap, the probability that a uniform point's radial distance is less than $1-\tau$ times the cap radius is $\exp\{(\log(1-\tau)+o(1))d\}$. Conditional on the radii, the unit tangent directions are independent and uniform in their respective tangent spheres. Regard both as vectors in $\mathbb R^d$ and take their projections onto the unit direction along the chord joining the two cap centers. Each projection has absolute value at most $\tau$, except with probability $2C_{d-1}(\tau)$. The same probability bound holds for the inner product of the two tangent directions in $\mathbb R^d$, conditional on the first direction, because its projection onto the second tangent space has norm at most one. When both points have radial distance at least $1-\tau$ times their respective cap radii and all three scalar products have absolute value at most $\tau$, expanding the radial representations bounds squared Euclidean distance divided by $R^2$ from below by the left side of \eqref{win:pair-gap}, up to $O(R^2)$. The cross terms involving the line joining the centers decrease this lower bound by at most $4\tau(B-x)L$, and the cross term between the two directions decreases it by at most $2\tau L^2$. These bounds vary continuously in the chosen neighborhood. Geodesic distance is at least Euclidean distance. Thus the probability of distance at most $\ell R$ is at most $e^{-c_pd}$ for some fixed $c_p>0$, uniformly in that neighborhood. One may take $c_p=\frac1{16}\min\{-\log(1-\tau),I(\tau)\}$ for sufficiently large $n$.

\phantomsection\label{win:pair-intensity}
To transfer this pair bound to the sample, shrink the parameter neighborhood until $a+\log L_S<c_p/8$ and $a+\log L_T<c_p/8$, which is possible because both left sides vanish at $(r,t',D)=({r_{\mathrm{src}}},t_0,D_0)$. Let $\widehat C_S$ and $\widehat C_T$ be the caps containing $C(p,{r_{\mathrm{src}}}R)\cap C(q,DR)$ and $C(q,DR)\setminus C(p,R)$, respectively. Their angular radii are $(L_S+O(R^2))R$ and $(L_T+O(R^2))R$. For a density $f\le2$, the product density of two distinct labels is at most $4$. The expected number of ordered pairs of distinct sample labels in the two regions at distance at most $\ell R$ is the integral below. Enlarging its domains to the containing caps and applying the pair probability bound gives
\begin{align}
 &n(n-1)\iint_{[C(p,{r_{\mathrm{src}}}R)\cap C(q,DR)]\times[C(q,DR)\setminus C(p,R)]}
 \one_{\{\delta(y,z)\le\ell R\}}f(y)f(z)\,d\sigma_{d-1}(y)d\sigma_{d-1}(z)\nonumber\\
 &\hspace{12mm}\le4n^2\sigma_{d-1}(\widehat C_S)\sigma_{d-1}(\widehat C_T)e^{-c_pd}
 \le e^{-c_pd/2}.\label{win:near-pairs}
\end{align}
Here $nF(R)=e^{ad}$ and the bounds $a+\log L_S<c_p/8$ and $a+\log L_T<c_p/8$ control the expected cap sizes, with $o(d)$ errors in their exponents. The original regions are disjoint because ${r_{\mathrm{src}}}<1$, so only distinct labels contribute.

The direction and pair bounds leave room to move the query slightly so that many source points and at least $k$ closer points are present. Keep the neighborhoods, $\ell$, $\eta_s$, and $c_p$ fixed. For a small $\eta>0$, put $t=t_0+\eta$, $d_-=D_0$, and $d_+=\sqrt{L^2+(t-x)^2+\eta^2}$. Choose $\eta$ sufficiently small that $(t,d_-)$ and $(t,d_+)$ stay in all the chosen neighborhoods, $t<B$, and $a-\kappa+\log d_+<0$. The last inequality follows for small $\eta$ from \eqref{win:boundary-identities} and $h(\gamma_1)<\kappa$. To describe the intersections with the two query caps, define
\[
 s_\pm(t')=\frac{t'^2+{r_{\mathrm{src}}}^2-d_\pm^2}{2t'},\quad L_\pm(t')=\sqrt{{r_{\mathrm{src}}}^2-s_\pm(t')^2},\quad\text{and}\quad b_-(t')=\frac{t'^2+1-d_-^2}{2t'}.
\]
The quantities $s_\pm$ give the center positions along the query axis, after division by $R$, for caps containing $C(p,{r_{\mathrm{src}}}R)\cap C(q,d_\pm R)$. Their leading radii divided by $R$ are $L_\pm$. The quantity $b_-$ gives the center position of a cap containing $C(q,d_-R)\setminus C(p,R)$. At $t'=t$, they satisfy
\begin{equation}\label{win:strict-counts}
 s_+(t)=x-\frac{\eta^2}{2t}<x<s_-(t),\qquad L_-(t)<L<L_+(t),\qquad\text{and}\qquad b_-(t)<B.
\end{equation}
The inequality $L_-(t)<L=e^{-a}$ will make the expected number of sources in the smaller query cap exponentially small, whereas $s_+(t)<x$ allows exponentially many sources in the larger cap. The inequality $b_-(t)<B$ allows points outside the annulus to be closer than those sources. For small $\eta$, $0<s<\min\{r,t'\}$ still holds for the cap containing $C(p,rR)\cap C(q,DR)$, while $t'<b'<1$ still holds for the cap containing $C(q,DR)\setminus C(p,R)$.

To obtain exponentially many sampled points of both kinds, choose $z_s$ strictly between $s_+(t)/{r_{\mathrm{src}}}$ and $x/{r_{\mathrm{src}}}$, and $z_t$ strictly between $b_-(t)$ and $B$. Since $I(x/{r_{\mathrm{src}}})=\eps$ and $I(B)=a$, the exponents $d_s=\eps-I(z_s)$ and $d_t=a-I(z_t)$ are positive. Choose the half width $w>0$ so that $t-w>0$ and, for every $t'\in[t-w,t+w]$, both $(t',d_-)$ and $(t',d_+)$ remain in the fixed parameter neighborhoods used for \eqref{win:far-cap}, the containing caps, and \eqref{win:near-pairs}. Throughout this interval, keep the strict comparisons in \eqref{win:strict-counts} and the inequalities $s_+(t')/{r_{\mathrm{src}}}<z_s$ and $b_-(t')<z_t$. Also keep the radius of the cap containing $C(p,r_sR)\cap C(q,d_+R)$ smaller than $e^{-a}R$ by a fixed proportion, as ensured by the choice following the definition of $L_s$. We can then choose radial thicknesses $\nu,\xi\in(0,1)$ sufficiently small that the normalized squared query distances, for radial coordinate $r$ and directional cosine at least $z_s$ or $z_t$, satisfy
\begin{align}
 t'^2+r^2-2t'rz_s&<d_+^2\quad({r_{\mathrm{src}}}(1-\nu)\le r\le{r_{\mathrm{src}}}),\label{win:source-witness}\\
  t'^2+r^2-2t'rz_t&<d_-^2\quad(1\le r\le1+\xi).\label{win:target-witness}
\end{align}
The strict inequalities at $r={r_{\mathrm{src}}}$ and $r=1$ ensure these bounds by continuity. The choices are made before $n$ varies, in the order $\gamma_1$, then $\ell,\eta_s$, then $\tau,c_p$, then $\eta$, $z_s,z_t$, $w$, and finally $\nu,\xi$.

We now place the configurations across the sphere at the scale $R$ already chosen. The $O(R^2)$ errors in \eqref{win:metric-expansion} and in the containing cap radii contribute $O(dR^2)$ to the logarithms of the probability bounds. Together with the cap estimates, the uniform error in those logarithms is $O(dR^2+\log d+\log(1/R))=o(d)$, without requiring $dR^2\to0$. Take a deterministic maximal $4R$-separated set of centers $p_1,\ldots,p_M$. Assign each point to its nearest center, breaking ties deterministically, to obtain Borel cells $V_i$, and put $w_i=\sigma_{d-1}(V_i)$. The $2R$-caps are disjoint up to null boundaries, and the $4R$-caps cover the sphere.

At each center, let $S_i=C(p_i,{r_{\mathrm{src}}}R)$ be the source cap. Before normalization, set the density to one there and outside all annuli, and to $e^{-\kappa d}$ in each surrounding annulus. Write $L_n=\bigcup_i(C(p_i,R)\setminus S_i)$ for their union and $Z_n=1-(1-e^{-\kappa d})\sigma_{d-1}(L_n)$ for the normalizing constant. Define $f_n=Z_n^{-1}e^{-\kappa d}$ on $L_n$ and $f_n=Z_n^{-1}$ elsewhere. The annuli lie in the $R$-caps, so $\sigma_{d-1}(L_n)\le MF(R)$. Disjointness of the $2R$-caps gives $MF(2R)\le1$, and hence $\sigma_{d-1}(L_n)\le F(R)/F(2R)=e^{-(\log2+o(1))d}$.
Consequently $Z_n=1-e^{-\Omega(d)}$ and $f_n\le2$ for large $n$, as required in \eqref{win:near-pairs}. Both density levels occur on sets of positive measure, so the global log-density oscillation is exactly $\kappa d$ and the oscillation on every cap is at most $\kappa d$. The bound $f_n\le2$ also gives $\rho_{\min}\ge F^{-1}(1/(2n))$, hence $\log(1/\rho_{\min})=o(d)$ by \cref{lem:cap}. Thus $f_n$ satisfies the required regularity conditions.

The query distribution $Q_n$ uses all these configurations. It chooses $i$ with weights $w_i$, chooses $t'$ uniformly in $[t-w,t+w]$ and an independent uniform tangent direction $u$ at $p_i$, and sets $q=\exp_{p_i}(t'Ru)$. The condition $t'+d_+<2$ puts this query in the interior of $C(p_i,2R)$. The $2R$-caps about distinct $p_i$ have disjoint interiors, so the generated query $q$ determines $i$. Conditional on $i$, the density relative to surface measure is proportional to $\sin^{-(d-2)}(t'R)$ on that radial interval, so $Q_n$ is absolutely continuous. To show that it remains spread across the sphere, couple $q$ to $z\sim\sigma_{d-1}$ by choosing the cell containing $z$ and then sampling $q$ about its center. Their distance is at most $r_n=(4+t+w)R\to0$. The triangle inequality and \cref{lem:cap} therefore give, for each fixed $0<\theta<\pi/2$,
\[
 \sup_{p\in\mathbb S^{d-1}}\Prob_{q\sim Q_n}(q\in C(p,\theta))
 \le F(\theta+r_n)=\exp\{d(\log\sin\theta+o(1))\}\longrightarrow0.
\]

It remains to show that most queries under $Q_n$ have many failing entries on one sample, simultaneously for every graph. First fix a query generated about $p_i$. Let $N_i$ be the number of sample points in $S_i$, with mean $nZ_n^{-1}F({r_{\mathrm{src}}}R)=e^{(\eps+o(1))d}$. Chernoff gives
\begin{equation}\label{win:total-sources}
 \Prob\{N_i>e^{(\eps+\zeta)d}\}\le\exp\{-e^{\Omega(d)}\}.
\end{equation}
The possible failing entries are the selected sources $S_i(q)=D_n\cap S_i\cap C(q,d_+R)$. By \eqref{win:source-witness}, points at distances in $[{r_{\mathrm{src}}}(1-\nu)R,{r_{\mathrm{src}}}R]$ from $p_i$, whose tangent directions at $p_i$ have inner product at least $z_s$ with the query direction $u$, lie in $S_i\cap C(q,d_+R)$ for large $n$. The expected number of sampled points satisfying these distance and direction conditions is
\[
 nZ_n^{-1}[F({r_{\mathrm{src}}}R)-F({r_{\mathrm{src}}}(1-\nu)R)]C_{d-1}(z_s)=e^{(d_s+o(1))d}.
\]
Another application of Chernoff shows that $|S_i(q)|\ge e^{d_sd/2}$ except with probability $\exp\{-e^{d_sd/2}\}$ for sufficiently large $n$.

To separate these entries from closer points, use the cap containing $S_i\cap C(q,d_-R)$. By \eqref{win:strict-counts},
\begin{equation}\label{win:no-close-source}
 \Prob\{D_n\cap S_i\cap C(q,d_-R)\ne\varnothing\}
 \le e^{(a+\log L_-(t')+o(1))d}=e^{-\Omega(d)}.
\end{equation}
For the inner part $C(p_i,r_sR)\cap C(q,d_+R)$, use the same containment with $r=r_s$ and $D=d_+$. The parameter choice following the definition of $L_s$ makes the radius of this containing cap smaller than $e^{-a}R$ by a fixed proportion, uniformly in $t'$. Thus
\begin{equation}\label{win:radial-exclusion}
 \Prob\{D_n\cap C(p_i,r_sR)\cap C(q,d_+R)\ne\varnothing\}\le e^{-\Omega(d)}.
\end{equation}
The set $L_n\cap C(q,d_+R)$ contains no sampled point except with probability at most
\begin{equation}\label{win:low-void}
 nZ_n^{-1}e^{-\kappa d}F(d_+R)=e^{(a-\kappa+\log d_++o(1))d}=e^{-\Omega(d)}.
\end{equation}
Finally, \eqref{win:near-pairs} and Markov bound by $e^{-c_pd/2}$ the probability of a selected source and a sample point in $C(q,d_+R)\setminus C(p_i,R)$ at mutual distance at most $\ell R$. These probability bounds apply to the same fixed query and can be combined by a union bound.

We also need at least $k$ sampled points closer than every selected source. For $y$ with $\phi=\delta(p_i,y)\in(0,\pi)$, write $v_y=(y-\cos\phi\,p_i)/\sin\phi$ for its unit tangent direction at $p_i$. With $u$ the query direction at this center, use the region $T_i(q)=\{y:R<\delta(p_i,y)\le(1+\xi)R,\ u\cdot v_y\ge z_t\}$.
This region lies outside all the low density annuli because $1+\xi<2$ and distinct centers are $4R$-separated. Its expected number of sampled points is at least
\[
 nZ_n^{-1}[F((1+\xi)R)-F(R)]C_{d-1}(z_t)\ge e^{(d_t+o(1))d}.
\]
It therefore contains at least $k$ sampled points except with probability $\exp\{-e^{\Omega(d)}\}$. By \eqref{win:target-witness}, every point in $T_i(q)$ is strictly within distance $d_-R$ of $q$. When $D_n\cap S_i\cap C(q,d_-R)=\varnothing$, these sampled points are closer than every sampled point in $S_i$. In particular, $S_i(q)$ contains no member of $\TopSet_k(q)$.

We now combine the upper bound on $N_i$, the lower bound on $|S_i(q)|$, the exclusions in \cref{win:no-close-source,win:radial-exclusion,win:low-void,win:near-pairs}, and the existence of $k$ closer sampled points. For a fixed data set, let $A(D_n)$ be the $Q_n$-probability of queries violating at least one of these requirements. The requirements defining $A(D_n)$ are finite distance and cardinality comparisons, so their failure set is Borel. The fixed-query bounds are uniform in $i,t',u$. Integrating over $t',u$ and summing with weights $w_i$ gives $\E A(D_n)\le e^{-\Omega(d)}$ by Tonelli and $\sum_iw_i=1$. Markov yields a data event of probability $1-e^{-\Omega(d)}$ on which $A(D_n)\le e^{-\Omega(d)}$. This event is determined before any graph is chosen.

Fix a data set with $A(D_n)\le e^{-\Omega(d)}$, and let $G$ be any graph of degree at most $\Delta$. The preceding sample estimates exclude improving neighbors outside $S_i(q)$ unless the listed point lies outside $C(p_i,R)$ and is at distance at least $\ell R$ from its source. We now exclude queries for which such a listed point lies closer to $q$. For each center with $N_i\le e^{(\eps+\zeta)d}$, the number of listed neighbors across all data points in $S_i$ is at most $N_i\Delta\le e^{(\eps+\zeta+\gamma+o(1))d}$. Consider a pair whose source has distance in $[r_sR,{r_{\mathrm{src}}}R]$ from $p_i$, whose listed neighbor lies outside $C(p_i,R)$, and whose mutual distance is at least $\ell R$. For each fixed $t'$, \eqref{win:far-cap} bounds the fraction of directions placing both endpoints within distance $d_+R$ of the query. A union bound over the listed neighbors gives at most
\[
 e^{(\eps+\gamma+2\zeta-J+o(1))d}=e^{(-6\zeta+o(1))d}.
\]
Integrating this direction fraction over $t'$ and summing with weights $w_i$ gives the same bound on the probability of the additional excluded queries. That query set may depend on $G$, but its bound holds for every $G$ on the fixed data event. The selected sources at radius below $r_sR$ are already absent by \eqref{win:radial-exclusion}, and nearby pairs leaving $C(p_i,R)$ are already excluded by \eqref{win:near-pairs}. Every improving listed neighbor of a point in $S_i(q)$ lies within distance $d_+R$ of the query, so these exclusions cover all ways to reach a listed neighbor outside $C(p_i,R)$.

For a query satisfying the sample requirements used to define $A(D_n)$ and outside the additional excluded set for $G$, every improving stored edge from $S_i(q)$ ends in $S_i(q)$. A sampled point of $S_i$ outside $S_i(q)$ is farther than $d_+R$ from the query. By \eqref{win:low-void}, $C(q,d_+R)\cap(C(p_i,R)\setminus S_i)$ contains no sampled point, and the pair estimates exclude improving listed neighbors outside $C(p_i,R)$. Thus the finite set $S_i(q)$, of size at least $e^{d_sd/2}$, cannot be left by strictly improving moves and contains no member of $\TopSet_k(q)$. Every run starting there eventually stops without success. Scanning entire lists cannot reveal a target either, because every member of $\TopSet_k(q)$ is closer than every sampled point of $S_i$ and would therefore be an improving listed neighbor.

The bounds on $N_i$ and $|S_i(q)|$, the exclusions in \cref{win:no-close-source,win:radial-exclusion,win:low-void,win:near-pairs}, the existence of $k$ closer points in $T_i(q)$, and the final union bound for listed neighbors all have fixed positive exponential margins, uniformly for $t'\in[t-w,t+w]$. Choose $c_*>0$ smaller than one hundredth of these margins, including the exponent $d_s/2$ for the number of selected sources. For sufficiently large $n$, the uniform lower order errors and the factors from Markov and the final union bound are absorbed into the margins. The data event then has probability at least $1-e^{-c_*d}$, and on it every graph of degree at most $\Delta$ has at least $e^{c_*d}$ failing entries for every query outside a set of $Q_n$-probability at most $e^{-c_*d}$. This proves \cref{thm:tolerance}(b).
\end{proof}

\subsection{Upper bounds in the logarithmic regime}\label{sec:tolerance-moderate}

We now consider the logarithmic regime, where the uniform nearest-neighbor angle converges to a positive constant. The upper bound for $\NN{\Delta}$ depends on both the density level and its variation. We first quantify that dependence, then construct a different graph that, for densities bounded above by $e^{o(d)}$, succeeds from every entry for most uniform queries even with arbitrarily deep low-density regions.

Let $d=\lfloor c\log n\rfloor$, $\beta^*(c)<\beta<1$, $\Delta=\lceil n^\beta\rceil$, $\lambda=1/c$ and $\gamma=\beta/c$, so that $n=e^{\lambda d+O(1)}$. To describe a region where $f$ has exponential level $e^{\mu d}$, write $\nu=\lambda+\mu>0$. In the cap estimate of \cref{lem:cap}, this level multiplies the expected number of sampled points by $e^{\mu d}$. For $0\le g<\nu$, let $R(g,\nu)$ be the limiting radius of a cap containing $e^{gd}$ points in expectation at this density level. Thus $R(0,\nu)$ is the limiting nearest-neighbor angle. We use the ratio $X(g,\nu)=(1-\cos R(g,\nu))/\sin^2R(0,\nu)$ to compare the neighbor radius with that nearest-neighbor scale. The cap estimate gives the following formulas for $R$ and $X$, and we define the permitted density variation $h$ by
\[
  R(g,\nu)=\arcsin e^{g-\nu},\qquad X(g,\nu)=\frac{e^{2g}}{1+\cos R(g,\nu)},\qquad\text{and}\qquad
  h(g,\nu)=\frac12\log\frac{1+X(g,\nu)}2 .
\]
For $\nu=\lambda$, $R(\gamma,\lambda)=\lim\rho_\beta$ and $R(0,\lambda)=\lim\theta_1$. The condition $X(\gamma,\lambda)>1$ is equivalent to $\cos R(\gamma,\lambda)<\cos^2R(0,\lambda)$, the limiting critical-radius inequality in \eqref{eq:intro-critical-radius}. Hence $h(\gamma,\lambda)>0$ exactly when $\beta>\bstar(c)$. As the density level $\mu$ increases, both the local nearest-neighbor angle and the local neighbor-list radius decrease, and $X(\gamma,\nu)$ decreases in $\nu$, so denser regions are harder to navigate at the same degree. Fix a sequence $\vartheta_{d}=o(d)$ with $\vartheta_{d}\ge\log4$. For $\mu_+,\mu_-\ge0$, let $\mathcal D(\mu_+,\mu_-)$ consist of probability densities satisfying $-\mu_-d-\vartheta_d\le\log f\le\mu_+d+\vartheta_d$ almost everywhere.
The term $\vartheta_d$ absorbs subexponential factors in the density bounds. The next theorem expresses the sufficient condition using the largest allowed density level and the global density ratio.

\begin{theorem}[Nearest-neighbor navigation in the logarithmic regime]\label{thm:tolerance-moderate}
Fix $c>0$, $\bstar(c)<\beta<1$, $k\ge1$, and $\bar\mu,\kappa\ge0$. Suppose that $\kappa<h(\gamma,\lambda+\bar\mu)$. Let the sampling density satisfy $Me^{-\kappa d-2\vartheta_d}\le f\le M$ almost everywhere for some $M\le e^{\bar\mu d+\vartheta_d}$. For every absolutely continuous query distribution $Q$ fixed before sampling, with probability $1-e^{-\Omega(d)}$ over the data, $\NN{\Delta}$ succeeds from every entry for every query outside a set of $Q$-probability $e^{-\Omega(d)}$. Each run uses $O(1)$ complete expansions and has evaluation cost $O(\Delta)$. The constants depend only on $(c,\beta,\bar\mu,\kappa)$, and the threshold for sufficiently large $n$ may also depend on the fixed sequence $\vartheta_d=o(d)$.
\end{theorem}

\begin{proof}
Put $\nu=\lambda+\bar\mu$. We choose the auxiliary degree exponent below $\gamma$ and the inner cap using this largest density level. By continuity of $h(\cdot,\nu)$ and $\kappa<h(\gamma,\nu)$, choose $0<g<\gamma$ sufficiently close to $\gamma$ that $\kappa<h(g,\nu)$, choose $\kappa<\kappa'<h(g,\nu)<g$, and put $R_\infty=R(g,\nu)$ and $X_g=X(g,\nu)>1$. The fixed gap $\kappa'-\kappa$ absorbs the subexponential density factors. The constant $a$ will set the inner-cap radius, while $A=g+\log a$ bounds its expected number of points on the exponential scale. As $a$ decreases to $e^{\kappa'-g}$, the limiting directional threshold becomes $X_ge^{-2\kappa'}-1$. The inequality $e^{2\kappa'}<(1+X_g)/2$ makes this larger than $1-e^{-2\kappa'}$, so continuity permits a choice $e^{\kappa'-g}<a<1$ for which $A>\kappa'$, $\tau_\infty=1/(a^2(1+\cos R_\infty))-1>0$, and $\tau_\infty^2>(1-e^{-2\kappa'})(1-e^{-2A})$. Choose $0<\tau<\min\{\tau_\infty,1\}$ so that $\tau^2>(1-e^{-2\kappa'})(1-e^{-2A})$. With the directional rate $I$ and direction-cap probabilities $C_{d-1},C_{d-2}$ defined in \cref{sec:tolerance}, we can choose projection thresholds $t_0,v_0$ as in the proof of \cref{loc:nn-sufficiency} so that $I(v_0)>\kappa'$, $I(t_0)>A$, and $t_0v_0<\tau$.

Outside the inner cap we will contract the distance to the query by a fixed factor $b<1$. Choose $b$ sufficiently close to $1$ that $\zeta_b=1-(1-b)\pi/R_\infty>0$, $g-\kappa'+\log a+\log b>0$, and $g-\kappa'+\log\zeta_b+\log b-\tfrac12\log(1+\cos(\zeta_bR_\infty))>0$. The radius $\zeta_bR_\infty=R_\infty-(1-b)\pi$ leaves room for contraction inside the neighbor cap. For a current distance $\theta=\delta(q,x)$, the last two inequalities keep the expected numbers of candidates exponentially large in the cases $\theta\le\theta_c$ and $\theta>\theta_c$, respectively, with $\theta_c$ specified below. Such a choice is possible because $A>\kappa'$ and $\kappa'<h(g,\nu)<\tfrac12\log X_g=g-\tfrac12\log(1+\cos R_\infty)$.

The upper density bound determines a radius whose edges are retained by every list. For the given $M$, we have $Me^{-\kappa'd}\le f\le M$ for large $d$. Define $R_n$ by $nMF(R_n)=e^{gd}$. Normalization gives $M\ge1$, while \cref{lem:cap} and $M\le e^{\bar\mu d+\vartheta_d}$ place $R_n$ between $R_\infty-o(1)$ and $R(g,\lambda)+o(1)<\pi/2$. Thus $\cos R_n$ has a uniform positive lower bound. Every cap $C(x,R_n)$ has $P$-mass at most $e^{gd}/n$. Since $g<\gamma$, the retention argument in the proof of \cref{loc:nn-sufficiency} gives probability $1-e^{-\omega(d)}$ that every list of $\NN{\Delta}$ contains all other data points within distance $R_n$ of its center. The same radius works at every center because $M$ bounds the density globally.

The retained edges rule out unsuccessful stopping points in the inner cap $C(q,\Theta_n)$, where $\Theta_n=\arcsin(a\sin R_n)$. We apply the inner-cap argument in the proof of \cref{loc:nn-sufficiency}, keeping $1+\cos R_n$ instead of bounding it by $2$. If $\delta(q,x)=\theta\le\Theta_n$, a closer point at distance $0<t\le\theta$ from $q$ can lie outside $C(x,R_n)$ only if the cosine between its tangent direction at $q$ and the direction from $q$ toward $x$ is at most $1-(1-\cos R_n)/\sin^2\Theta_n=-\tau_n$, where $\tau_n=1/(a^2(1+\cos R_n))-1\ge\tau_\infty-o(1)>\tau$. Thus the same rearrangement estimate applies with $\tau$ and the density ratio $\exp\{\kappa'd\}$. In the rearranged comparison, $K_\tau h_{B_t}(u)\le\eps_d$ whenever $u\cdot e\le t_0$, with $B_t=\exp\{\kappa'd\}$ and $\eps_d=\exp\{\kappa'd\}C_{d-1}(v_0)+C_{d-2}(\tau-t_0v_0)$. Since $nMF(\Theta_n)\le a^{d-1}nMF(R_n)=a^{-1}\exp\{Ad\}$, the expected number of false local minima in the inner cap is at most $12\eps_d+a^{-1}\exp\{Ad\}C_{d-1}(t_0)=\exp\{-\Omega(d)\}$ by the choices of $t_0,v_0$.

Outside the inner cap, we need a retained neighbor whose distance to the query is at most $b\theta$, where $\theta=\delta(q,x)\ge\Theta_n$. We first find candidate points in a smaller neighbor cap, then contract them toward $q$. Let $\zeta_n=1-(1-b)\pi/R_n\ge\zeta_b-o(1)$ and $\theta_c=\arccos\sqrt{\cos(\zeta_nR_n)}$. If $\theta\le\theta_c$, let $Y_x$ be the half of $C(q,\theta)$ whose tangent directions at $q$ have nonnegative inner product with the direction from $q$ toward $x$. The cosine identity puts this set in $C(x,\zeta_nR_n)$, and its measure satisfies $\sigma_{d-1}(Y_x)=F(\theta)/2\ge\cos R_n\,a^{d-1}F(R_n)/2$. If $\theta>\theta_c$, take $Y_x$ to be the points in $C(x,\zeta_nR_n)$ whose tangent directions at $x$ have cosine with the direction from $x$ toward $q$ at least $w_c=\sqrt{\cos(\zeta_nR_n)/(1+\cos(\zeta_nR_n))}$. For $\theta_c<\theta<\pi/2$, \cref{lem:directions} puts these points in $C(q,\theta)$ because $\cot\theta\tan(\zeta_nR_n/2)\le w_c$. For $\pi/2\le\theta<\pi$ and $y\in Y_x$, the term $\cos\theta(\cos\delta(x,y)-1)$ and the directional term in the spherical cosine identity are both nonnegative, so again $\cos\delta(q,y)\ge\cos\theta$. At $\theta=\pi$, every other point in the neighbor cap is closer, so any direction can be used to define $Y_x$. For $\theta>\theta_c$, \cref{lem:directions} and cap scaling give $\sigma_{d-1}(Y_x)\ge\zeta_n^{d-1}F(R_n)(1+\cos(\zeta_nR_n))^{-(d-1)/2}e^{-O(\log d)}$. Keeping the exact threshold $w_c$, rather than replacing it by $1/\sqrt2$, preserves the positive exponents used to choose $b$.

Contract each point of $Y_x$ along its geodesic toward $q$ by multiplying its distance to $q$ by $b$, and denote the image by $V_x$. This moves points by at most $(1-b)\pi=(1-\zeta_n)R_n$, so $V_x\subseteq C(x,R_n)\cap C(q,b\theta)$. The contraction has Jacobian at least $b^{d-1}$. Combining this factor with either lower bound on $\sigma_{d-1}(Y_x)$, the density bound $f\ge Me^{-\kappa'd}$, and $nMF(R_n)=e^{gd}$ gives $(n-1)P(V_x)\ge e^{\Omega(d)}$ by the two positive exponents in the choice of $b$. The first candidate bound uses the uniform positive lower bound on $\cos R_n$. For the second, $\zeta_nR_n=R_n-(1-b)\pi$ makes $\log\zeta_n-\tfrac12\log(1+\cos(\zeta_nR_n))$ nondecreasing in $R_n$. Conditional on $x$, the other sampled points are independent. Taking a union bound over the $n$ choices of the current data point $x$ therefore bounds the probability that some $V_x$ contains no other data point by $n\exp(-e^{\Omega(d)})$.

We now combine the fixed-query estimates and bound the length of the run. Integrating the fixed-query failure bounds against $Q$ and applying Markov's inequality, as in the proof of \cref{loc:nn-sufficiency}, gives, with probability $1-e^{-\Omega(d)}$ over the data, the required progress for all queries outside a set of $Q$-measure $e^{-\Omega(d)}$. The search reaches the inner cap after at most $\lceil\log(\pi/\Theta_n)/\log(1/b)\rceil$ contractions. Inside, vertices separated by at least two moves are more than $R_n$ apart, and they lie in the Euclidean ball of radius $\sin\Theta_n$ about $\cos\Theta_n\,q$. The pairwise squared-distance identity from \eqref{step:moves} gives fewer than $2(1-\cos R_n)/(2(1-\cos R_n)-2\sin^2\Theta_n)=(1+\tau_n)/\tau_n$ points in each of the even-indexed and odd-indexed subsequences. Thus the number of moves and complete expansions is $O(1)$, and the evaluation cost is $O(\Delta)$.

\end{proof}

For the class $\mathcal D(\mu_+,\mu_-)$, take $M=e^{\mu_+d+\vartheta_d}$, $\bar\mu=\mu_+$ and $\kappa=\mu_++\mu_-$. The sufficient condition is then $\mu_++\mu_-<h(\gamma,\lambda+\mu_+)$. When the global density ratio is $e^{o(d)}$, normalization permits $\bar\mu=\kappa=0$ after absorbing that variation into $\vartheta_d$.

\phantomsection\label{nat:section}

For densities with $\operatorname{ess\,sup}f/\operatorname{ess\,inf}f=n^{o(1)}$, the theorem gives the same upper degree exponent $\bstar(c)$ as for uniform data. This guarantee for $\NN{\Delta}$ uses both upper and lower density bounds. If only the upper density is bounded by a subexponential factor, changing the graph still gives the uniform upper degree exponent, even when the density vanishes in some regions. The construction uses a fixed number of nested subsamples. Every neighbor list contains a common small subsample and nearby points from the larger subsamples. To make progress through these subsamples, we need a lower bound on the probability mass of query caps despite the absence of a lower density bound. The next lemma supplies it when the query density is also bounded by a subexponential factor, including uniform queries and queries drawn from $P$.

\begin{lemma}[Sampling mass of query caps]\label{nat:cap-population}
Let $P_{d}=f_{d}\sigma_{d-1}$ be a probability law with $0\le f_{d}\le e^{\vartheta_{d}}$ a.e., where $\vartheta_d$ is any fixed sequence satisfying $\vartheta_d=o(d)$. Fix $0<a<b<\pi/2$ and $\xi>0$. Let $Q_{d}$ be a deterministic query distribution with a density satisfying $dQ_{d}/d\sigma_{d-1}\le e^{\omega_{d}}$ a.e., where $\omega_{d}=o(d)$.
There is $b_0>0$ such that, for all sufficiently large $d$, there exists a Borel set $E_d$, depending only on $P_d$, with $\Prob_{q\sim Q_{d}}(q\in E_{d})\le e^{-b_0d}$ such that
\begin{equation}\label{nat:cap-lower}
 P_{d}(C(q,t))\ge e^{-\xi d}F(t)
 \quad(q\notin E_{d}, a\le t\le b).
\end{equation}
The constants may depend on $a,b,\xi$. The threshold for sufficiently large $d$ may also depend on the sequences in the hypotheses.
\end{lemma}

\begin{proof}
We first obtain the bound for uniform queries at one fixed radius $D\in(0,\pi/2)$, allowing an exponent loss $\zeta>0$. Choose $u\in(0,D)$ sufficiently small that $\log\sin(D-u)>\log\sin D-\zeta/2$. Let $g(z)=P_d(C(z,D-u))$ and $A_d=\{z:g(z)\ge F(D-u)/2\}$. Fubini gives $\int g\,d\sigma_{d-1}=F(D-u)$, while $g\le e^{\vartheta_d}F(D-u)$, so $\sigma_{d-1}(A_d)\ge\tfrac12e^{-\vartheta_d}=e^{-o(d)}$. The function $g$ is continuous by dominated convergence, since cap boundaries have $P_d$-measure zero. Thus $A_d$ is compact. Spherical isoperimetry says that, among closed sets of a given surface measure, a spherical cap minimizes the measure of points at distance at most $u$ from the set, for every fixed $u>0$~\cite[Theorem~2.1]{FigielLindenstraussMilman1977}. By \cref{lem:cap}, a cap of measure at least $e^{-o(d)}$ has radius at least $\pi/2-u/2$ for all large $d$. Consequently the set of points at distance greater than $u$ from every point of $A_d$ has measure at most
\[
 F(\pi/2-u/2)=\exp\bigl(d\log\cos(u/2)+O(\log d)\bigr).
\]
For every remaining $q$, choose $z\in A_d$ with $\delta(q,z)\le u$. The inclusion $C(z,D-u)\subset C(q,D)$ gives
\[
 P_{d}(C(q,D))\ge F(D-u)/2\ge e^{-\zeta d}F(D)
\]
for large $d$. We have proved an exponentially small exceptional surface measure for every fixed radius and every fixed positive exponent loss.

A finite grid extends the estimate to all radii in the prescribed interval. Choose a finite grid $a=t_0<\cdots<t_L=b$ with $\log\sin t_{i+1}-\log\sin t_i<\xi/8$. Apply the preceding step with exponent loss $\xi/8$ to its finitely many radii. For any $t_i\le t\le t_{i+1}$, monotonicity of cap mass and \cref{lem:cap} give
\[
 P_{d}(C(q,t))\ge e^{-\xi d/8}F(t_i)\ge e^{-\xi d}F(t)
\]
outside a set of exponentially small surface measure, uniformly in $t$. The density bound on $Q_d$ multiplies this exceptional measure by at most $e^{\omega_d}=e^{o(d)}$, preserving exponential decay.

To obtain the required Borel exceptional set, note that $(q,t)\mapsto P_d(C(q,t))$ is continuous on $\mathbb S^{d-1}\times[a,b]$ by dominated convergence. The set where \eqref{nat:cap-lower} fails for some $t$ is Borel, since one may test a countable dense set of radii. Taking this failure set as $E_d$ completes the proof.
\end{proof}

\begin{theorem}[Navigation by a graph built from nested subsamples]\label{nat:navigation}
Fix $c>0$, $\bstar(c)<\beta<1$, and $k\ge1$. Let $d=\lfloor c\log n\rfloor$ and $\Delta=\lceil n^\beta\rceil$. For every sequence of probability laws $P_n=f_n\sigma_{d-1}$ with $0\le f_n\le e^{\vartheta_n}$ a.e., where $\vartheta_n=o(d)$, there is a navigation graph $G$ on the data, with neighbor lists of at most $\Delta$ entries and built without using the query distribution, with the following property. For every query distribution $Q_n$ fixed before the data are drawn and satisfying $dQ_n/d\sigma_{d-1}\le e^{o(d)}$ almost everywhere, with probability $1-e^{-\Omega(d)}$ over the data the graph $G$ succeeds from every entry for every query outside a Borel set of $Q_n$-measure $e^{-\Omega(d)}$, within $O(1)$ moves and with evaluation cost $O(\Delta)$. In particular, this holds on $\mathcal D(0,\kappa)$ for every fixed $\kappa\ge0$.
\end{theorem}

\begin{proof}
We choose the subsample sizes so that each level supplies enough neighbors to reach a smaller query cap. Since $\beta>\bstar(c)$ gives $\gamma>\bstar(c)/c=\tfrac12\log(2-e^{-2\lambda})$, choose $\tfrac12\log(2-e^{-2\lambda})<g<\gamma$, reserving degree exponent $g$ for the construction. This choice gives $e^{2g}>2-e^{-2\lambda}$, equivalently $e^{2g}-1>\sqrt{1-e^{2(g-\lambda)}}$. Hence the gap used to space the subsample exponents is $H_*=g-\tfrac12\log(1+\sqrt{1-e^{2(g-\lambda)}})>0$.
To let the subsamples grow while the query caps shrink, set $J=\max\{1,\lceil8(\lambda-g)/H_*\rceil\}$ and $\lambda_j=g+j(\lambda-g)/J$ for $0\le j\le J$, dividing the sample-size exponent from $g$ to $\lambda$ into increments at most $H_*/8$.
For each level $1\le j\le J$, the radius $r_j$ is chosen so that, for $e^{(\lambda_j+o(1))d}$ points sampled uniformly on the sphere, the expected number of neighbors has exponent $g$. The query-cap radius $A_j$ is chosen to contain exponentially many points from the preceding subsample. To verify $\cos^2A_j>\cos r_j$, we introduce $H_j$ below to write this condition as $\lambda_j+\log\sin A_j<H_j$. Define
\begin{equation}\label{nat:levels}
 r_j=\arcsin e^{g-\lambda_j},\quad A_j=\arcsin e^{-\lambda_{j-1}+H_*/4},\quad\text{and}\quad H_j=g-\tfrac12\log\left(1+\sqrt{1-e^{2(g-\lambda_j)}}\right).
\end{equation}
The angles lie in $(0,\pi/2)$, the $A_j$ decrease strictly, and $H_j\ge H_*$. The identity $1-\cos r_j=e^{2(H_j-\lambda_j)}$ and the inequality $\lambda_j+\log\sin A_j=(\lambda-g)/J+H_*/4\le3H_*/8<H_j$ give the required cap containment. The choices also leave exponentially many points from the preceding subsample in each query cap and make the final query cap radius $A_J$ larger than the limiting nearest-neighbor angle for uniform data. In particular, we have
\begin{equation}\label{nat:margins}
 \cos^2A_j>\cos r_j,\qquad
 \lambda_{j-1}+\log\sin A_j=H_*/4,\qquad\text{and}\qquad
 A_J>\arcsin e^{-\lambda}.
\end{equation}
Let $\tau=\min_{1\le j\le J}(\cos^2A_j-\cos r_j)/\sin^2A_j>0$ be the common containment margin and set $\eta=\tfrac14\log(1/(1-\min\{\tau,1/2\}^2))>0$. By \cref{lem:inside}(a), if $\delta(q,x)\le A_j$, then $C(x,r_j)$ omits at most an $e^{-\eta d}$ fraction of the surface measure of $C(q,A_j)$. This bound holds uniformly in $x$ and $1\le j\le J$ for all large $d$. Choose $\xi=\min\{H_*/16,\eta/16\}>0$ as the exponent loss in \cref{nat:cap-population}, small enough to preserve both the lower bounds on the expected numbers of points and the containment estimates. All these choices are fixed before $n$ varies.

Those levels now define one graph using only the sampled data. For $0\le j<J$, let $N_j=\min\{n,\lceil e^{\lambda_jd}\rceil\}$, and set $N_J=n$. Let $S_j$ consist of the first $N_j$ sample labels. The proposed neighbor list of a label $x$ is
\begin{equation}\label{nat:graph}
 (S_0\setminus\{x\})\ \cup\
 \bigcup_{j=1}^J\{y\in S_j\setminus\{x\}:\delta(x,y)\le r_j\}.
\end{equation}
If it has more than $\Delta$ points, truncate it in label order to $\Delta$ points. This defines $G$ with the asserted degree for every sample. For large $n$, we have $N_j=\lceil e^{\lambda_jd}\rceil$ for $j<J$.

Conditionally on one source location $X_i=x$, the number of other $S_j$ labels in $C(x,r_j)$ is binomial, with either $N_j$ or $N_j-1$ trials and mean at most
\[
 N_j e^{\vartheta_n}F(r_j)
 =\exp\bigl((\lambda_j+\log\sin r_j+o(1))d\bigr)
 =e^{(g+o(1))d}.
\]
This bound is uniform in $x$ and the finite index $j$, including $j=J$. Since $g<\gamma$, a binomial upper-tail bound and a union over $nJ$ pairs show, with failure probability $\exp(-e^{\Omega(d)})$, that the number of subsample points in each cap is at most $\Delta/(2J)$. Also $N_0<\Delta/2$ eventually. Hence on this single data event no neighbor list is truncated and, in particular, the entire radius-$r_J$ graph is retained.

To show that the retained edges provide progress, we first lower-bound the probability mass available in each query cap. Set $a=\tfrac12\arcsin e^{-\lambda}$ and apply \cref{nat:cap-population} on $[a,A_1]$ with exponent loss $\xi$. Let $E_n$ be its deterministic exceptional set, which depends only on $P_n$ and not on $Q_n$. For every $q\notin E_n$, we have
\begin{equation}\label{nat:all-cap-lower}
 P_n(C(q,t))\ge e^{-\xi d}F(t)\quad(a\le t\le A_1).
\end{equation}
This bound applies to every $A_j$, and $E_n$ has exponentially small $Q_n$-measure. We now fix $q\notin E_n$ and estimate the sampling probabilities directly, combining them with the event that no neighbor list is truncated only after these estimates are established.

We show that the retained subsample points move the search through the successive query caps. The common subsample $S_0$ contains a point of $C(q,A_1)$ except with probability
\[
 (1-P_n(C(q,A_1)))^{N_0}
 \le\exp\bigl(-N_0P_n(C(q,A_1))\bigr)
 \le\exp\bigl(-e^{H_*d/8}\bigr),
\]
by \eqref{nat:margins}, \eqref{nat:all-cap-lower} and $\xi\le H_*/16$. When this point in $S_0$ exists and neighbor lists are untruncated, every source outside $C(q,A_1)$ has a greedy successor inside that cap.

Now fix $1\le j<J$ and a source $x$ with $\theta=\delta(q,x)\le A_j$. For a point at distance $0\le\psi\le A_{j+1}$ from $q$, the containment margin gives
\[
 \cos\theta\cos\psi-\cos r_j
 \ge\cos^2 A_j-\cos r_j
 \ge\tau\sin\theta\sin\psi.
\]
Applying \cref{lem:inside}(a) with $\psi_0=A_{j+1}$ shows that $C(x,r_j)$ omits at most an $e^{-\eta d}$ fraction of the surface measure of $C(q,A_{j+1})$. The upper density bound and \eqref{nat:all-cap-lower} therefore give, for all large $d$,
\begin{equation}\label{nat:band-lens}
 \begin{split}
 P_n(C(x,r_j)\cap C(q,A_{j+1}))
 &\ge P_n(C(q,A_{j+1}))-e^{\vartheta_n-\eta d}F(A_{j+1})\\
 &\ge\tfrac12e^{-\xi d}F(A_{j+1}).
 \end{split}
\end{equation}
Conditional only on $X_i=x$, at least $N_j-1$ other labels in $S_j$ remain independent with law $P_n$. By $\lambda_j+\log\sin A_{j+1}=H_*/4$, the probability that none belongs to the lens in \eqref{nat:band-lens} is at most $\exp(-e^{H_*d/8})$, uniformly in this source location. A union over all $n$ source labels and $J-1$ levels is still doubly exponentially small. On the resulting event, every source at distance from $q$ in $(A_{j+1},A_j]$ has a greedy successor in $C(q,A_{j+1})$. Thus every run reaches $C(q,A_J)$ in at most $J$ moves. Extra neighbor-list entries can only make the successor closer.

Once the search reaches the smallest query cap, the edges of $G_{r_J}$ prevent unsuccessful termination. We bound false local minima for $k=1$, which suffices for every fixed $k$. For a point $x$ at distance $\theta\in[a,A_J]$ from $q$, let $T=P_n(C(q,\theta))$ and $L=P_n(C(x,r_J)\cap C(q,\theta))$. By \cref{lem:inside}(a), the fraction of the surface measure of $C(q,\theta)$ outside $C(x,r_J)$ is at most $e^{-\eta d}$. Combining this with the upper density bound and \eqref{nat:all-cap-lower} bounds the missed $P_n$-probability fraction by $u_d=e^{-\eta d/2}$. For all large $d$, we have
\[
 0\le T-L\le e^{\vartheta_n-\eta d}F(\theta)
 \le e^{\vartheta_n-\eta d+\xi d}T\le u_{d}T.
\]
Conditional on $x$, the probability that it is a false local minimum for $G_{r_J}$ is
\[
 (1-L)^{n-1}-(1-T)^{n-1}
 \le(n-1)u_{d}T\bigl(1-(1-u_{d})T\bigr)^{n-2}.
\]
The radial distribution $t\mapsto P_n(C(q,t))$ is continuous, because $P_n$ gives every cap boundary measure zero. Its probability integral transform is therefore uniform on $[0,1]$, even if this distribution function has flat intervals. Integrating over source positions and summing over all $n$ labels bounds the expected number of false local minima in this band by
\[
 n(n-1)u_{d}\int_0^1t\bigl(1-(1-u_{d})t\bigr)^{n-2}\,dt
 \le\frac{u_{d}}{(1-u_{d})^2}\le4u_{d}.
\]
The expected number of sampled points in the remaining cap $C(q,a)$ satisfies
\[
 nP_n(C(q,a))\le ne^{\vartheta_n}F(a)
 =\exp\bigl((\lambda+\log\sin a+o(1))d\bigr)=e^{-\Omega(d)}.
\]
Consequently the probability of any false local minimum in the smallest query cap $C(q,A_J)$ is exponentially small, uniformly for $q\notin E_n$.

The existence of suitable subsample points and the absence of false local minima together give progress for each fixed query. To obtain one sample on which these properties hold for most queries, let $Z(D_n,q)$ be the indicator that at least one of them fails, either for $S_0$, for a successive distance interval, or for the smallest query cap. This indicator is jointly Borel, since the properties use finitely many comparisons of distances. The fixed-query bounds give $\E Z(D_n,q)\le e^{-b_1d}$ for some fixed $b_1>0$, uniformly over $q\notin E_n$ for all large $d$. Tonelli and Markov give
\[
 \Prob\left\{\int_{E_n^c}Z(D_n,q)\,dQ_n(q)>e^{-b_1d/2}\right\}
 \le e^{-b_1d/2}.
\]
Intersect the event $\int_{E_n^c}Z(D_n,q)\,dQ_n(q)\le e^{-b_1d/2}$ with the event that no neighbor list is truncated, and add $E_n$ to the exceptional queries. Removing the data bisectors, which have $Q_n$-measure zero by absolute continuity, gives the claimed sample probability and Borel exceptional query set.

For every remaining query, every run either succeeds earlier or reaches $C(q,A_J)$ within $J$ moves. Inside this cap, every point outside $\TopSet_k(q)$ has an improving neighbor, so the run cannot stop unsuccessfully. To bound the path length inside $C(q,A_J)$, take every second visited point inside $C(q,A_J)$. Any two of these vertices are more than $r_J$ apart, otherwise the later, closer vertex was in the earlier retained neighbor list, contrary to greedy choice. All lie in the Euclidean ball of radius $\sin A_J$ centered at $\cos A_J\,q$. Applying \eqref{step:moves} with $r=r_J$ and $A=A_J$ bounds the number $N$ of points in either the even-indexed or odd-indexed subsequence by $N<K:=(1-\cos r_J)/(\cos^2 A_J-\cos r_J)$ when $N\ge2$. Since $K>1$, the same bound holds for $N=0,1$. There are at most $2\lceil K\rceil+2$ moves inside $C(q,A_J)$. The total move bound $J+2\lceil K\rceil+2$ is independent of $n$, and the evaluation cost is at most $1+(J+2\lceil K\rceil+3)\Delta$.
\end{proof}

The degree exponent is sharp over $\mathcal D(0,\kappa)$ under uniform queries for every fixed $\kappa\ge0$, since the class includes uniform sampling and \cref{thm:critical}(b) applies below $\bstar(c)$.

\subsection{Lower bounds in the logarithmic regime}\label{sec:logarithmic-lower}

The preceding construction handles arbitrarily low density by changing the graph. We now ask when the density instead forces every graph to use long neighbor lists. We first show that subexponential global density variation preserves the uniform lower exponent, then construct high density concentrations that force a strictly larger exponent.

\begin{corollary}[The uniform lower exponent under small global variation]\label{thm:nonuniform-logarithmic-small}
Fix $k\ge1$, $c>0$, and $0<\beta<\bstar(c)$ in the logarithmic regime, and let $\Delta=\lceil n^\beta\rceil$. Suppose that $K_n=\operatorname{ess\,sup}f/\operatorname{ess\,inf}f=n^{o(1)}$ and that the ratio of the query density to $f$ is at most $n^{o(1)}$, with the query distribution fixed before sampling. There are constants $a,\eta>0$ such that, with probability at least $1-n^{-\eta}$ over $D_n$, every navigation graph $G$ has a measurable query set of $Q$-probability at least $1-n^{-\eta}$ on which at least $(1-o(1))\lfloor n^a\rfloor$ entries $x_0$ satisfy $\Work_G(q;x_0)\ge\Delta$. From each such entry whose neighbor list has length at most $\Delta$, greedy navigation fails to evaluate a member of $\TopSet_k(q)$.
\end{corollary}

\begin{proof}
Use the logarithmic-regime choices of $T=\lfloor n^a\rfloor$, $L=\lceil n^\beta\rceil$, cutoff radius $\rho$, and rates $I,J$ from the proof of \cref{thm:critical}(b). Normalization gives $K_n^{-1}\le f\le K_n$. Choose $J_n=n^{o(1)}$ with $dQ/dP\le J_n$, so integration against $Q$ is bounded by $J_nK_n$ times surface integration. By the binomial rank estimate in the proof of \cref{thm:traps}, the rank interval $[T,2T)$ corresponds, up to an error $e^{-c_1T}$ for an absolute $c_1>0$, to caps with $P$-probability in $[T/(4n),4T/n]$. Their surface probabilities therefore lie in $\mathcal W_*=[T/(4nK_n),4K_nT/n]$. By \cref{lem:cap}, enlarging the interval to $\mathcal W_*$ changes the logarithms of its sine radii by $O((1+\log K_n)/d)$. The derivative bounds in the proof of \cref{thm:traps} consequently change $dI$ and $dJ$ by only $O(1+\log K_n)=o(d)$.

Replacing surface probabilities by $P$-probabilities and surface integration by $Q$-integration contributes factors $e^{o(d)}$, since $J_nK_n^4=e^{o(d)}$. The analogue of $Y$ in \eqref{eq:Y}, with cap probabilities taken under $P$ and query integrals under $Q$, thus has expectation at most
\[
 \frac nT e^{-c_1T}+e^{o(d)}\bigl(Te^{-dJ}+Le^{-dI}\bigr)=n^{-\Omega(1)}.
\]
The fixed margins in $dJ-\log T$ and $dI-\log L$ absorb these factors. Markov's inequality over the sample gives one bound valid for all graphs with lists of length at most $L$, with probability $1-n^{-\Omega(1)}$. For each such graph, Markov's inequality over queries leaves $(1-o(1))T$ points in the rank interval with no improving listed neighbor outside a graph-dependent query set of $Q$-probability $n^{-\Omega(1)}$. Their ranks exceed $k$, and no point of $\TopSet_k(q)$ can be in such a list, so the run stops without success after its first complete expansion. For an unrestricted graph, retain its lists of length at most $L$ and replace longer lists by empty ones. Each resulting failing entry either still fails in the original graph or has a first complete expansion costing more than $L$. This proves both assertions.
\end{proof}

\phantomsection\label{enr:section}

The preceding lower bound matches the uniform exponent when the global density variation is subexponential. To force a larger degree, we concentrate most of the sampling mass in a cap while leaving exponentially many data points outside it. For every graph of the degree allowed below, most uniform queries have many exterior entries with no improving route into the cap. The construction is governed by a degree boundary $\Gamma(\mu,\lambda)$ and a required density variation $K_{\rm cap}(\mu,\lambda)$, where $\mu>0$ is the exponent of the density inside the cap. For $\lambda>0$, define

\[
\begin{aligned}
 \Gamma(\mu,\lambda)&=\lambda+\frac12\log\!\left(1-e^{-2\mu}(1-e^{-2\lambda})^2\right),\\
  K_{\rm cap}(\mu,\lambda)&=\lambda+\mu+\frac12\log\!\left(1-e^{-2\mu}(1-e^{-2\lambda})\right).
\end{aligned}
\]
A degree $\Delta=e^{\gamma d}$ with $\gamma<\Gamma(\mu,\lambda)$ leaves too few stored edges to reach the enriched cap from all exterior entries. The condition $\kappa>K_{\rm cap}(\mu,\lambda)$ allows enough exterior points while keeping the log-density oscillation below $\kappa d$. These are the two conditions in the following theorem, and $K_{\rm cap}(\mu,\lambda)>\mu$ ensures that the stated density class is defined.

\begin{theorem}[Enrichment and arbitrary navigation graphs]\label{enr:obstruction}
Fix $c>0$, $\bstar(c)<\beta<1$, and $k\ge1$, and put $d=\lfloor c\log n\rfloor$, $\lambda=1/c$, $\gamma=\beta/c$, and $\Delta=\lceil n^\beta\rceil$. Fix $\mu>0$ and $\kappa>0$. If $\gamma<\Gamma(\mu,\lambda)$ and $\kappa>K_{\rm cap}(\mu,\lambda)$, there are deterministic, everywhere positive two-level densities $f_n$ in $\mathcal D(\mu,\kappa-\mu)$ with global log-density oscillation at most $\kappa d$ and $\operatorname{ess\,sup}f_n=e^{(\mu+o(1))d}$ such that the following holds for $Q_n=\sigma_{d-1}$. With probability $1-e^{-\Omega(d)}$ over $D_n$, simultaneously for every navigation graph with neighbor lists of at most $\Delta$ points, navigation fails from $e^{\Omega(d)}$ entries for every query outside a Borel set of query measure $e^{-\Omega(d)}$ that may depend on the graph. For each such query, a set of $e^{\Omega(d)}$ failing entries is disjoint from $\TopSet_k(q)$ and closed under strictly improving greedy moves.
\end{theorem}

\begin{proof}
We construct a cap containing most sampling mass and select exterior entries that are farther from $q$ than at least $k$ points in that cap. To make those entries fail for every graph of degree at most $\Delta$, we exclude nearby improving pairs by a sampling estimate and bound, for each farther stored edge, the measure of queries for which it improves the distance. The number of exterior points will be small enough to sum this directional bound over all their stored edges. We use the directional rate $I$ and direction-cap probability $C_{d-1}$ defined in \cref{sec:tolerance} for these estimates.

Fix the cap center $p$ and write $T=e^{-\mu}$ for the sine of its limiting angular radius $B=\arcsin T$. Set $b=\sqrt{1-e^{-2\lambda}}$ and $s=Tb$. For queries orthogonal to $p$, $s$ is the limiting largest query inner product among sampled points in the cap: among $e^{(\lambda+o(1))d}$ independent tangent directions, the largest cosine with a fixed direction approaches $b$, while the cap boundary has sine $T$. We will choose the exterior-entry thresholds just below $s$. By \cref{lem:cap}, the surface measure of $\{x:q\cdot x\ge s\}$ has rate $\epsilon_0=I(s)\in(0,\lambda)$. Let $0<\epsilon<\epsilon_0$ be an exponent to be chosen below for the number of exterior points. For large $d$, define the radius $B_n\in(0,\pi/2)$ of the enriched cap by $F(B_n)=e^{-\mu d}$, so $B_n\to B$ by \cref{lem:cap}. With the center $p$ fixed above, define the background density $p_n$ and the sampling density $f_n$ by
\begin{equation}\label{enr:eq6}
p_n=e^{-(\lambda-\epsilon)d},\qquad\text{and}\qquad
 f_n(x)=p_n+(1-p_n)e^{\mu d}\one_{\{\delta(p,x)\le B_n\}}.
\end{equation}
This integrates to one and satisfies $p_n\le f_n\le e^{\mu d}$. Its upper level is $e^{\mu d}(1-o(1))$, and its log-density oscillation is at most $(\lambda+\mu-\epsilon)d$. Choosing $\lambda+\mu-\epsilon<\kappa$ will give $p_n>e^{-(\kappa-\mu)d}$ and place this density in $\mathcal D(\mu,\kappa-\mu)$ without the subexponential slack. Let $A_n=C(p,B_n)$ be the enriched cap and $A_n^c$ the source region. The density on $A_n^c$ is exactly $p_n$, while the density on $A_n$ lies between $\frac12e^{\mu d}$ and $e^{\mu d}$ for all large $n$.

We first derive the geometric estimates for queries orthogonal to $p$ and exterior points $x$ satisfying $p\cdot x=0$. The inequalities on $\epsilon$ needed for the density class will be verified when it is chosen below. For a point $y\in C(p,B)$ with $x\cdot y\le s^2$, an improving edge from $x$ to $y$ when $q\cdot x\ge s$ must also satisfy $q\cdot y\ge s$. These two constraints can be combined with positive weights $a=T^2-s^2$ and $1-s^2$. Their weighted vector is $ax+(1-s^2)(y-(p\cdot y)p)$, whose projection onto $q$ is at least $(a+1-s^2)s$. Since $|y-(p\cdot y)p|\le T$, its squared norm is at most $N_0$ below. The allowed query directions therefore lie in a cap with positive cosine threshold $z_0$, where
\[
\begin{gathered}
 N_0=a^2+(1-s^2)^2T^2+2a(1-s^2)s^2=(1+T^2-2s^2)(T^2-s^4),\quad
 \text{and}\quad z_0^2=\frac{(a+(1-s^2))^2s^2}{N_0}.
\end{gathered}
\]
The identity $1-z_0^2=(1-s^2)(T^2-s^2)/(T^2-s^4)$ gives the directional rate
\[
 J:=I(z_0)=\epsilon_0+\lambda+\tfrac12\log(1-T^2b^4)
 =\epsilon_0+\Gamma(\mu,\lambda).
\]
To sum the directional bound over stored edges, we will choose the expected number of exterior points to have an exponent strictly smaller than $\epsilon_0$. Since $J-\epsilon_0=\Gamma(\mu,\lambda)>\gamma$, choose $\zeta>0$ with $16\zeta<J-\epsilon_0-\gamma$ to leave a fixed margin after including the degree exponent. We use this estimate for edges beyond a distance cutoff. When the two endpoints are very close, their query constraints nearly coincide and the directional rate alone is insufficient to sum over all edges. We therefore exclude such sampled pairs directly. For $\tau>0$, call a pair nearby when $x\cdot y\ge s^2+\tau$, with $\tau$ chosen sufficiently small below. At the boundary configuration $p\cdot q=p\cdot x=0$ and $q\cdot x=s$, combine this condition with $p\cdot y\ge\cos B$ and $q\cdot y\ge s$. Put $b_0=\cos B$ and $B_0=b_0^2+s^2$. The vector $b_0p+sq+\tau x$ has squared norm $N_1$ and projection at least $L_0$ onto every such $y$, giving cap threshold $Z_N$. These values are $L_0=B_0+\tau s^2+\tau^2$, $N_1=B_0+2\tau s^2+\tau^2$, and $Z_N=L_0/\sqrt{N_1}$.
The identity $L_0^2-B_0N_1=\tau^2\{B_0+(s^2+\tau)^2\}>0$ shows $I(Z_N)>I(\sqrt{B_0})=\lambda+\mu$ whenever $Z_N<1$. Thus the containing cap is small enough to offset both the exponent $\lambda$ of the sample size and the density exponent $\mu$ inside the enriched cap. We will extend this configuration to sources in a strip $|p\cdot x|\le\eta$, with $\eta>0$ chosen after $\tau$. Choose $\tau>0$ sufficiently small that $s^2+\tau<1$, $Z_N<1$, and the strict rate bound in \eqref{enr:eq1} holds at $\eta=0$. Let $\chi=(I(Z_N)-\lambda-\mu)/4>0$ record the margin for nearby pairs. Choose $0<\eta<\tfrac14\min\{\cos B,s,\sqrt{1-s^2}\}$ sufficiently small that the directional rate for farther edges remains above $J-\zeta$, namely
\begin{equation}\label{enr:eq1}
 \frac{(a+(1-s^2))^2s^2}
 {a^2+(1-s^2)^2T^2+2a(1-s^2)(s^2+\tau+\eta)}
 >1-e^{-2(J-\zeta)}.
\end{equation}
For nearby pairs, the same strip changes the bound on the squared norm by at most $2b_0\tau\eta$. Require $I(L_0/\sqrt{N_1+2b_0\tau\eta})>\lambda+\mu+3\chi$.
The bound for farther edges follows by continuity from the definition of $J$ as $\tau,\eta\to0$. With $\tau$ fixed, the bound for nearby pairs follows by continuity in $\eta$ from $I(Z_N)=\lambda+\mu+4\chi$.

To keep the possible exterior entries in this strip when queries move off the equator, we bound the two regions outside it. Write $z=p\cdot q$. For a threshold $h>0$ and $|z|<1$, let $D_\pm(h,z)>0$ satisfy $D_\pm(h,z)^2=(h^2+\eta^2\mp2zh\eta)/(1-z^2)$. These are the cosine thresholds of caps containing $\{x:q\cdot x\ge h,\ \pm p\cdot x\ge\eta\}$ when the coefficients below are positive. For the plus sign, set $A=(\eta-zh)/(1-z^2)$, $C=(h-z\eta)/(1-z^2)$, and $V=Ap+Cq$. Then $p\cdot V=\eta$, $q\cdot V=h$, and $V\cdot x\ge A\eta+Ch=|V|^2$ on that region. The cap centered at $V/|V|$ with cosine threshold $|V|=D_+(h,z)$ therefore contains it. Replacing $p$ by $-p$ gives the minus sign. At $h=s,z=0$, the strict inequality $I(D_\pm(s,0))>\epsilon_0$ will make selected sources outside the strip exponentially unlikely.

The bound for nearby pairs must also survive the change in query and source positions. Let $b'$ vary near $\cos B$, representing $\cos B_n$. The thresholds $0<r<u<s$, to be chosen below, will select exterior entries by $q\cdot x\ge r$. A sampling estimate will rule out exterior points with $q\cdot x\ge u$. Consider a source $x$ with $|p\cdot x|\le\eta$ and $r\le q\cdot x\le u$, and a point $y$ in the enriched cap satisfying $p\cdot y\ge b'$, $q\cdot y\ge r$, and $x\cdot y\ge s^2+\tau$. The vector $b_0p+sq+\tau x$ has projection at least $\mathcal L(b',r)$ onto $y$ and squared norm at most $\mathcal N(z,u)$. These bounds give the cap threshold $\mathcal Z(b',r,u,z)$, where
\[
 \begin{aligned}
 \mathcal L(b',r)&=b_0b'+sr+\tau(s^2+\tau),\\
 \mathcal N(z,u)&=b_0^2+s^2+\tau^2+
             2b_0s|z|+2b_0\tau\eta+2s\tau u,\\
 \mathcal Z(b',r,u,z)&=\frac{\mathcal L(b',r)}
                            {\sqrt{\mathcal N(z,u)}}.
 \end{aligned}
\]
We now choose the density outside the enriched cap so that exponentially many sources meet the lower threshold $r$, but none meets the upper threshold $u$. The expected total number of exterior points will have exponent $\epsilon<\epsilon_0$. The identity $K_{\rm cap}(\mu,\lambda)=\lambda+\mu-\epsilon_0$ shows that $\kappa>K_{\rm cap}(\mu,\lambda)$ permits $\epsilon$ sufficiently close to $\epsilon_0$ from below. \phantomsection\label{enr:eq2}Choose $0<\epsilon<\epsilon_0$ with $\lambda+\mu-\epsilon<\kappa$ and thresholds $0<r<\sqrt{1-e^{-2\epsilon}}<u<s$, with $r,u$ sufficiently close to $s$. Then $\epsilon-I(r)>0$ supplies many sources, whereas $I(u)>\epsilon$ excludes closer source points. Choose a fixed $0<w_{\mathrm{qry}}<r/2$ for the query belt $|p\cdot q|\le w_{\mathrm{qry}}$, sufficiently small that the bounds for nearby pairs and the containing-cap thresholds hold for every $z$ in this belt and every $b'$ near $\cos B$. The additional conditions needed below are imposed at the same time. \phantomsection\label{enr:eq3}In particular, require $\epsilon-I(D_\pm(r,z))<0$ to exclude selected sources outside the strip. To make nearby pairs rare even after summing over selected sources, require
\begin{equation}\label{enr:eq4}
 0<\epsilon-I(r)<\chi,\quad 0<\mathcal Z(b',r,u,z)<1,\quad\text{and}\quad I(\mathcal Z(b',r,u,z))>\lambda+\mu+2\chi.
\end{equation}
Require also that the weights $A,C$ used to define the cap with threshold $D_+(h,z)$ remain positive at $h=r$, as do the weights obtained by replacing $p$ by $-p$. Require $D_\pm(r,z)\in(0,1)$. For stored edges beyond the cutoff for nearby pairs, preserve the directional rate $J-2\zeta$ by requiring
\begin{equation}\label{enr:eq5}
\frac{(a+(1-s^2))^2(r-{w_{\mathrm{qry}}})^2}
 {a^2+(1-s^2)^2(T')^2+2a(1-s^2)(s^2+\tau+\eta)}
 >1-e^{-2(J-2\zeta)}
\end{equation}
for every $T'$ sufficiently close to $T$, where $T'$ represents the sine of the perturbed cap radius. The parameter choices are compatible in the following order. As $\epsilon\uparrow\epsilon_0$ and $r,u\to s$, the positive difference $\epsilon-I(r)$ can be made arbitrarily small, while the choice of $\eta$ keeps $I(\mathcal Z)$ above $\lambda+\mu+3\chi$ at $b'=\cos B$, $r=u=s$, and $z=0$. This gives \eqref{enr:eq4}. The strict bound in \eqref{enr:eq1} gives \eqref{enr:eq5} when the changes in $r$, $T'$, and $w_{\mathrm{qry}}$ are sufficiently small. Thus we choose $\zeta,\tau,\chi,\eta$ first, then $\epsilon,r,u$, and finally $w_{\mathrm{qry}}$. With $\zeta,\tau,\chi,\eta,\epsilon,r,u,w_{\mathrm{qry}}$ fixed, choose closed neighborhoods of $b'=\cos B$ and $T'=T$ on which all strict bounds hold uniformly for $|z|\le w_{\mathrm{qry}}$. These parameter choices complete the specification of the density in \eqref{enr:eq6}. We now apply the geometric bounds to the number of sources, nearby pairs, and stored edges.

For this density, we verify that most queries have many exterior entries and at least $k$ closer points in the enriched cap. Consider the query belt $\mathcal B=\{q:|p\cdot q|\le w_{\mathrm{qry}}\}$, whose complement has surface measure $e^{-\Omega(d)}$. All sampling estimates that follow are uniform over fixed $q\in\mathcal B$. Let $S(q)=D_n\cap A_n^c\cap\{x:q\cdot x\ge r\}$ be the selected exterior entries.

\phantomsection\label{enr:eq7}The total number $N_S=|D_n\cap A_n^c|$ of sampled points outside the enriched cap has mean $np_n(1-e^{-\mu d})=e^{\epsilon d+O(1)}$. Chernoff bounds it by $N_S\le e^{(\epsilon+\zeta)d}$ except with doubly exponentially small probability. For a fixed query, the two containing caps with thresholds $D_\pm(r,z)$ and the strict bounds $I(D_\pm(r,z))>\epsilon$ give
\begin{equation}\label{enr:eq8}
np_n\,\sigma_{d-1}\{q\cdot x\ge r,\ |p\cdot x|>\eta\}
 \le e^{-d_rd}
\end{equation}
for a fixed $d_r>0$, after increasing the dimension threshold. Markov therefore ensures $|p\cdot x|\le\eta$ for every selected source, except with probability $e^{-d_rd}$.

To obtain many entries in $S(q)$, put $d_s=\epsilon-I(r)>0$. For large $n$, the region $\{x:q\cdot x\ge r,\ |p\cdot x|\le\eta\}$ lies outside $A_n$ because $\cos B_n>2\eta$, so its density is $p_n$. By \eqref{enr:eq8}, its expected number of sampled points is at least $np_nF(\arccos r)-e^{-d_rd}\ge\tfrac12e^{(d_s+o(1))d}$. Every sampled point in this region belongs to $S(q)$. \phantomsection\label{enr:eq9}Chernoff therefore gives $|S(q)|\ge e^{d_sd/2}$ except with doubly exponentially small probability. Since $I(u)>\epsilon$, a first-moment bound also shows that no sampled point outside $A_n$ satisfies $q\cdot x\ge u$, except with probability $e^{-(I(u)-\epsilon+o(1))d}$.

To separate these entries from $\TopSet_k(q)$, we show that at least $k$ sampled points in $A_n$ satisfy $q\cdot y>u$. Choose $z_t$ with $u/T<z_t<b$, so $I(z_t)<\lambda$. Choose $a_t\in(0,B/4)$ so small that $\sin(B-2a_t)z_t>u$, and if necessary decrease $w_{\mathrm{qry}}$ until $-{w_{\mathrm{qry}}}+\sqrt{1-{w_{\mathrm{qry}}}^2}\sin(B-2a_t)z_t>u$. For large $n$, $B_n\ge B-a_t$. Every point $y$ with $\delta(p,y)\in[B_n-a_t,B_n]$, whose tangent direction at $p$ has cosine at least $z_t$ with the direction from $p$ toward $q$, therefore satisfies $q\cdot y>u$, uniformly in the query belt. Decreasing ${w_{\mathrm{qry}}}$ preserves every previous condition. Fix its final value after this choice. The mean number of sampled points in this region is at least
\begin{equation}\label{enr:eq10}
\tfrac12ne^{\mu d}
 [F(B_n)-F(B_n-a_t)]\,C_{d-1}(z_t)
 =e^{(\lambda-I(z_t)+o(1))d}.
\end{equation}
Here $F(B_n-a_t)/F(B_n)=e^{-\Omega(d)}$ uniformly, and $F(B_n)=e^{-\mu d}$ exactly. Since $\lambda-I(z_t)>0$, the Chernoff bound shows that this region contains at least $k$ sampled points except with probability $\exp\{-e^{\Omega(d)}\}$, uniformly for queries in the belt.

Since $q\cdot x\ge r$ for $x\in S(q)$, any improving edge from $x$ to $y\in A_n$ also requires $q\cdot y\ge r$. We first exclude pairs satisfying these thresholds at distance at most $\ell=\arccos(s^2+\tau)$, independently of the graph. Consider $x$ with $|p\cdot x|\le\eta$ and $r\le q\cdot x\le u$, and $y\in A_n$ with $q\cdot y\ge r$ and $x\cdot y\ge s^2+\tau$. For each fixed $x$, put $W=b_0p+sq+\tau x$ and use $z=p\cdot q$ and $b'=\cos B_n$ in $\mathcal L$ and $\mathcal N$. The conditions on $y$ give $W\cdot y\ge b_0b'+sr+\tau(s^2+\tau)=\mathcal L(b',r)$, while $p\cdot x\le\eta$, $q\cdot x\le u$, and $p\cdot q=z$ give
\[
 |W|^2=b_0^2+s^2+\tau^2+
       2b_0sz+2b_0\tau(p\cdot x)+2s\tau(q\cdot x)
 \le\mathcal N(z,u).
\]
If $W=0$ there are no such points $y$, since $\mathcal L(b',r)>0$. Otherwise all such points $y$ lie in a cap of cosine threshold at least $\mathcal Z(b',r,u,z)$. The density on $A_n$ is at most $e^{\mu d}$. The possible source locations $A_n^c\cap\{x:q\cdot x\ge r\}$ have density $p_n$ and surface measure at most $F(\arccos r)$. Consequently the expected number of pairs with distinct data labels is at most
\begin{equation}\label{enr:eq11}
n(n-1)p_ne^{\mu d}
 F(\arccos r)\,F(\arccos\mathcal Z(\cos B_n,r,u,p\cdot q))
 \le e^{-\chi d/2}.
\end{equation}
For large $d$, \eqref{enr:eq4} bounds the exponent by $(\epsilon-I(r))+(\lambda+\mu-I(\mathcal Z))<-\chi$. Markov's inequality therefore excludes these pairs except with probability $e^{-\chi d/2}$. Together with \eqref{enr:eq8} and the bound $q\cdot x<u$ for sources, this excludes every improving edge from $S(q)$ into $A_n$ of length at most $\ell$, independently of the graph.

For a fixed edge of length at least $\ell$ from an exterior point into $A_n$, we now bound the measure of queries for which both endpoints have inner product at least $r$ with $q$. The perturbed weighted-vector estimate in \eqref{enr:eq5} gives a bound independent of how the edge was chosen. Fix a source $x$ with $|p\cdot x|\le\eta$ and a point $y\in A_n$, and suppose $\delta(x,y)\ge\ell$. Write $X=x-(p\cdot x)p$ and $Y=y-(p\cdot y)p$. These projections satisfy $|X|\le1$, $|Y|\le\sin B_n$, and $X\cdot Y\le s^2+\tau+\eta$.
At a fixed value of $z=p\cdot q$ with $|z|\le{w_{\mathrm{qry}}}$, write $q=zp+\sqrt{1-z^2}\,w$, where, under uniform query measure conditional on $p\cdot q=z$, $w$ is uniform on the unit sphere in $p^\perp$. If both points satisfy $q\cdot x,q\cdot y\ge r$, then $w\cdot X,w\cdot Y\ge r-{w_{\mathrm{qry}}}$. Consequently, for $V=aX+(1-s^2)Y$,
\[
\begin{gathered}
 w\cdot V\ge(a+(1-s^2))(r-{w_{\mathrm{qry}}}),\quad
 \text{and}\quad |V|^2\le a^2+(1-s^2)^2\sin^2B_n+2a(1-s^2)(s^2+\tau+\eta).
\end{gathered}
 \]
By \eqref{enr:eq5}, the possible $w$ lie in a direction cap with measure at most
\begin{equation}\label{enr:eq12}
\exp\{-(J-2\zeta+o(1))d\}.
\end{equation}
The bound is uniform in every listed neighbor and every $z\in[-{w_{\mathrm{qry}}},{w_{\mathrm{qry}}}]$. If $V=0$ the conditions $q\cdot x,q\cdot y\ge r$ are impossible. If the derived cap threshold exceeds one the event is empty. Thus \eqref{enr:eq12} holds after the complete data set and any graph have been fixed.

For each fixed $q\in\mathcal B$, the sampling estimates give exponentially many entries in $S(q)\subseteq A_n^c$ and at least $k$ points in $A_n$ closer to $q$ than every entry in $S(q)$. They also exclude every improving edge of length at most $\ell$ from $S(q)$ into $A_n$. Their combined failure probability is $e^{-\Omega(d)}$, uniformly in that query. The failure indicator is jointly Borel, since it uses finitely many comparisons of distances and inner products. Tonelli's theorem therefore bounds the expected surface measure of queries in $\mathcal B$ where any of these properties fails by $e^{-\Omega(d)}$. Markov's inequality over the sample bounds this exceptional query measure by $e^{-\Omega(d)}$ on one data event of probability $1-e^{-\Omega(d)}$. Intersect that event with the bound $N_S\le e^{(\epsilon+\zeta)d}$, and add $\mathcal B^c$ to the exceptional queries. The data event and the initial exceptional query set are defined without choosing a graph.

Fix such a data set and any graph with degree at most $\Delta$. The neighbor lists of all $N_S$ sampled points outside $A_n$ contain at most $N_S\Delta\le e^{(\epsilon+\zeta+\gamma+o(1))d}$ listed neighbors. Apply \eqref{enr:eq12} to every stored edge from $A_n^c$ into $A_n$ with $|p\cdot x|\le\eta$ and distance at least $\ell$. Outside the initial query exception, every point of $S(q)$ satisfies $|p\cdot x|\le\eta$. For each fixed graph, exclude the queries admitting a stored edge of length at least $\ell$ from $S(q)$ into $A_n$ at a point $y$ with $q\cdot y\ge r$. Integrating \eqref{enr:eq12} over $z=p\cdot q$ and summing over all eligible stored edges bounds this additional set's surface measure by
\[
 e^{(\epsilon+\gamma+3\zeta-J+o(1))d}
 =e^{-\Omega(d)}.
 \]
The exponent is negative because $\epsilon<\epsilon_0$ and $16\zeta<J-\epsilon_0-\gamma$. Both the initial and the additional exceptional query sets have surface measure $e^{-\Omega(d)}$.

For every query outside the combined exception, the lower bound $|S(q)|\ge e^{d_sd/2}$ supplies exponentially many selected sources. An improving neighbor in $A_n^c$ remains in $S(q)$ because its query inner product remains at least $r$. An improving neighbor $y\in A_n$ would also satisfy $q\cdot y\ge r$. The bound \eqref{enr:eq11} for nearby pairs excludes such edges of length at most $\ell$, and \eqref{enr:eq12} excludes longer stored edges outside the additional query exception. Thus $S(q)$ is closed under strictly improving greedy moves. Every sampled point outside $A_n$ satisfies $q\cdot x<u$, while \eqref{enr:eq10} supplies at least $k$ sampled points in $A_n$ with $q\cdot y>u$. Hence $\TopSet_k(q)\subseteq A_n$ and $S(q)$ contains no point of $\TopSet_k(q)$. A complete expansion at any point of $S(q)$ cannot evaluate a top-$k$ point either, since that point would be an improving neighbor in $A_n$. Every run starting in $S(q)$ therefore fails. Exclude also the finite union of bisectors, which has surface measure zero. The geometric estimates are uniform over the fixed closed parameter neighborhoods, and $B_n\to B$, so all stated probability bounds hold for sufficiently large $n$. This proves the theorem.
\end{proof}

To express the obstruction directly in terms of degree and density variation, write $\mu_{\min}$ for the lower bound on the density exponent $\mu$ required by this construction. The condition $\gamma<\Gamma(\mu,\lambda)$ is equivalent to $\mu>\mu_{\min}$, where
\[
 \mu_{\min}=\frac12\log
 \frac{(1-e^{-2\lambda})^2}{1-e^{-2(\lambda-\gamma)}}.
\]
The inequality $\gamma>\frac12\log(2-e^{-2\lambda})$ gives $\mu_{\min}>0$. The identity $K_{\rm cap}(\mu,\lambda)=\frac12\log\{1+e^{2\lambda}(e^{2\mu}-1)\}$ shows that $K_{\rm cap}$ is strictly increasing in $\mu$. Substituting $\mu_{\min}$ therefore shows that an enrichment level satisfying both theorem conditions exists whenever
\[
 \kappa>K_{\rm cap}(\mu_{\min},\lambda)
 =\frac12\log\frac{(e^{2\lambda}-1)(e^{2\gamma}-1)}{e^{2\lambda}-e^{2\gamma}}.
\]
Since $\Gamma(\mu,\lambda)\uparrow\lambda$ as $\mu\to\infty$, this construction gives an obstruction for every $\bstar(c)<\beta<1$ when the permitted density variation is sufficiently large, with uniform queries throughout.

\section{Relaxing the navigation guarantee}\label{sec:relaxed}\label{sec:approximate-navigation}
\Cref{sec:upper,sec:nonuniform} establish bounds for greedy navigation to find a member of $\TopSet_k(q)$ from every entry. On threshold and nearest-neighbor graphs, we now ask how much evaluation cost can be saved by allowing fresh random entries, and whether adaptive local exploration with caching can improve on independent restarts. The main result of this section answers these questions for uniform data in the dense regime. After proving it, we return to a guarantee from every entry and relax exact recovery instead, bounding the stopping rank in \cref{sec:stopping-rank}.

For uniform data in the dense regime, let $\Delta=\lceil e^{\gamma d}\rceil$ with $\frac12\log(4/3)\le\gamma<\gc$. The success probability of one greedy run from a uniform random entry is governed by the exponent
\begin{equation}\label{eq:dense-fresh-start-exponent}
  \eta^*(\gamma)=-\frac12\log\left(1-\frac{(2-e^{2\gamma})^2(3e^{2\gamma}-2)^2}{16e^{4\gamma}(e^{2\gamma}-1)}\right).
\end{equation}
For any query $q$ fixed before sampling, one greedy run on either $G_{r_\gamma}$ or $\NN{\Delta}$ succeeds with probability $e^{-(\eta^*(\gamma)+o(1))d}$ when we draw the uniform data and independently choose a uniform entry label (see \cref{thm:exact-exponent}). To determine the cost of attaining constant success probability, we compare independent restarts with the local exploration algorithms introduced before \cref{thm:informal-restarts}. The following theorem gives matching upper and lower bounds on the evaluation cost of fresh starts.

\begin{theorem}[The evaluation cost of fresh starts]\label{thm:fresh-start-cost}
For uniform data in the dense regime, fix $k\ge1$ and $\frac12\log(4/3)\le\gamma<\gc$, and let $\Delta=\lceil e^{\gamma d}\rceil$. For each query $q$ fixed before sampling, we average success probabilities over the random sample and the algorithm's random choices, including any random entries. On either $G_{r_\gamma}$ or $\NN{\Delta}$, every local exploration algorithm with data-independent entries requires at least $e^{(\gamma+\eta^*(\gamma)-o(1))d}$ distinct evaluations to succeed with probability bounded away from zero. Independent restarts from uniform random entries on $G_{r_\gamma}$, all using the same sample, succeed with probability tending to one using at most $e^{(\gamma+\eta^*(\gamma)+o(1))d}$ distinct evaluations. The minimum of $\gamma+\eta^*(\gamma)$ over the stated degree range is attained at $e^{2\gamma}=(12+2\sqrt3)/11$ and is approximately $0.243$, compared with the every-entry exponent $\gc\approx0.347$.
\end{theorem}

To prove \cref{thm:fresh-start-cost}, we first analyze the improving regions and bound the number of expansions in a run from a random entry in \cref{sec:random-entry}. We use these estimates to determine the single-run success exponent and bound adaptive local exploration in \cref{sec:local}. \Cref{sec:restarts} proves that repeated random entries succeed on one sampled threshold graph with probability tending to one. Finally, \cref{sec:tradeoff} accounts for complete scans and list overlaps to obtain matching bounds on distinct evaluations, even with a shared cache.

\subsection{Failure from a random entry}\label{sec:random-entry}

We begin the restart analysis with a bound on the success probability of one run from a uniform random entry. The following theorem shows that this probability is exponentially small below the critical degree in both regimes. Its proof controls the number of expansions and the probability of finding a point of $\TopSet_k(q)$ during a scan, providing the estimates used to determine the exact exponent in \cref{sec:local}. Throughout this subsection the data are uniform.

\begin{theorem}[Random-entry failure]\label{thm:random-entry}
Fix $k\ge1$ and suppose that the data are uniform.
\begin{enumerate}[label=\textup{(\alph*)},leftmargin=*]
\item In the logarithmic regime, fix $0<\beta<\bstar(c)$ and let $\Delta=\lceil n^\beta\rceil$. There is $\eta>0$ such that, with probability at least $1-n^{-\eta}$ over the data, a query set of surface measure at least $1-n^{-\eta}$ has the following property. For every query in this set, greedy navigation from a uniform random entry succeeds with probability at most $n^{-\eta}$ on either $G_{\rho_\beta}$ or $\NN{\Delta}$ conditional on the sampled data, with probability taken over the entry label.
\item In the dense regime, fix $0<\gamma<\gc$ and let $\Delta=\lceil e^{\gamma d}\rceil$. There is $\zeta>0$ such that, with probability at least $1-e^{-\zeta d}$ over the data, a query set of surface measure at least $1-e^{-\zeta d}$ has the following property. For every query in this set, greedy navigation from a uniform random entry succeeds with probability at most $e^{-\zeta d}$ on either $G_{r_\gamma}$ or $\NN{\Delta}$ conditional on the sampled data, with probability taken over the entry label.
\end{enumerate}
\end{theorem}

The proof isolates a small cap around the query that contains the targets but has no improving edges. A run entering this cap at a nontarget point stops unsuccessfully. Success therefore requires either starting at a point of $\TopSet_k(q)$ or revealing such a point in an expansion outside the cap. We bound the starting probability directly and control the probability of such an expansion by exposing only the caps scanned along the path and bounding the number of expansions. We first prove the bounds for $G_\rho$, then transfer them to $\NN{\Delta}$ by comparing its neighbor-list radii with $\rho$.

Let $\rho=\rho_\beta$ with $0<\beta<\bstar(c)$ in the logarithmic regime, and $\rho=r_\gamma$ with
$0<\gamma<\gc$ in the dense regime. In both cases there is a fixed ${c_{\mathrm{sub}}}>0$ with
\begin{equation}\label{eq:subcritical}
  \sin\rho\ge e^{{c_{\mathrm{sub}}}}\sin\theta_1\qquad\text{and}\qquad 1-\cos\rho\le e^{-{c_{\mathrm{sub}}}}\sin^2\theta_1
\end{equation}
for all large $n$. In the logarithmic regime $\sin\rho/\sin\theta_1\to e^{\beta/c}$ by \eqref{eq:radii},
and $1-\cos\rho<\sin^2\theta_1$ in the limit is equivalent to $\rho<\rho^*$, or $\beta<\bstar(c)$. In the dense regime $\rho/\theta_1\to e^\gamma\in(1,\sqrt2)$. For a point
$x$ with $\delta(q,x)=\theta$, let $U(x)$ be its improving region and $u(\theta)$ its surface measure. We also define a positive function $\Lambda$ for $\theta_1\le\theta\le\pi/2$ by
\[
  \begin{gathered}
  U(x)=C(x,\rho)\cap C(q,\theta),\quad u(\theta)=\sigma_{d-1}(U(x)),\quad
  \text{and}\quad\Lambda(\theta)^2=2(1-\cos\rho)-\frac{(1-\cos\rho)^2}{\sin^2\theta}.
  \end{gathered}
\]
For every fixed $\eps>0$ and $\theta_1\le\theta\le\pi/2-\eps$, the ratio $\Lambda(\theta)/\sin\theta_1$ determines the exponential scale of $nu(\theta)$ in \eqref{eq:u-rate}. The expected number of improving neighbors is $(n-1)u(\theta)$. We use $L_n=(\log n)^2$ as a cutoff for $nu(\theta)$, bounding the probability of no improving neighbor above the cutoff and of very small moves below it.

The search is analyzed through the following auxiliary exposure procedure. Fix the query $q$ and
choose the entry $x_0$ uniformly at random. The exploration reveals the position of
$x_0$ and then repeatedly expands the current point $x_i$ by revealing the unrevealed data points in $C(x_i,\rho)$ and their positions. It moves to the sampled point of $C(x_i,\rho)$ closest to $q$ if that point is closer to $q$ than $x_i$. It stops before expanding a current point in the trap cap specified after \eqref{eq:trap-cap}. Outside that cap, it stops after a complete expansion if it cannot move, or after $H_n$ expansions, with $H_n$ defined below. It ignores whether a
point of $\TopSet_k(q)$ has been revealed, and its stopping rule depends only on revealed
positions. Until the exploration stops or the original greedy run succeeds, whichever occurs first, the two procedures visit the same points from $x_0$, and
every point evaluated by the original run up to that time has been revealed. The trap estimate below will rule out later success, while the path-length estimate will make truncation at $H_n$ unlikely. Let $E_i$ be the union of
the caps revealed before the $i$-th expansion and $\mathcal H_i$ the revealed information at
that time.

\begin{proof}[Proof of \cref{thm:random-entry}]

\phantomsection\label{step:u-props}
We show that $u$ is continuous and nondecreasing on $(0,\pi)$, that for every fixed $\eps>0$ and
$\theta_1\le\theta\le\pi/2-\eps$
\begin{equation}\label{eq:u-rate}
  nu(\theta)=\Bigl(\frac{\Lambda(\theta)}{\sin\theta_1}\Bigr)^{d-1}e^{O(\log d+\log(1/\theta_1))},
\end{equation}
and that $\Lambda$ increases on $[\theta_1,\pi/2]$ from
$\Lambda(\theta_1)<\sin\theta_1$ to $\Lambda(\pi/2)=\sin\rho$.

In polar coordinates around $x$, write a point as $y=\cos t\,x+\sin t\,w$ with $t=\delta(x,y)$ and a
unit vector $w\perp x$, and let $v$ be the unit vector at $x$ pointing toward $q$.
The spherical cosine identity gives $y\in C(q,\theta)$ if and only if $w\cdot v\ge\cot\theta\tan(t/2)$. For fixed
$(t,w)$ the right side decreases in $\theta$, so $U(x)$ grows when $q$ moves away from $x$ along
a fixed geodesic, and $u$ is nondecreasing. Continuity follows from dominated
convergence. For \eqref{eq:u-rate}, if $\theta\ge\theta_1$, then $\sin^2\theta\ge\sin^2\theta_1>1-\cos\rho$ by \eqref{eq:subcritical}, so $\cos\rho>\cos^2\theta$, and \cref{lem:lens}
gives $u(\theta)=F(\theta)e^{-(d-1)J_\theta(\rho)+O(\log d+\log(1/\theta_1))}$, the error term of
\cref{lem:lens}(a) with $\log(1/\sin\rho)=\log(1/\theta_1)+O(1)$ and $\sin\theta\ge\sin\theta_1$. The
ratio $\rho/\theta$ need not be bounded below, so we use the form of \cref{lem:lens}(a)
that allows this ratio to tend to zero. Substituting the formula for $J_\theta(\rho)$ gives
$\sin^2\theta\,e^{-2J_\theta(\rho)}=(1-\cos\rho)(\cos\rho-\cos2\theta)/\sin^2\theta=\Lambda(\theta)^2$. Together with
\cref{lem:cap}, this yields $u(\theta)=\Lambda(\theta)^{d-1}e^{O(\log d+\log(1/\theta_1))}$. Since $nF(\theta_1)=1$,
\cref{lem:cap} gives $n=(\sin\theta_1)^{-(d-1)}e^{O(\log d)}$, which proves \eqref{eq:u-rate}. Finally, $\Lambda^2$ increases because $(1-\cos\rho)^2/\sin^2\theta$ decreases. Its endpoint value is $\Lambda(\pi/2)^2=2(1-\cos\rho)-(1-\cos\rho)^2=\sin^2\rho$. At the nearest-neighbor scale, we have
\[
  \Lambda(\theta_1)^2=\sin^2\theta_1-\frac{(\sin^2\theta_1-(1-\cos\rho))^2}{\sin^2\theta_1}<\sin^2\theta_1,
\]
with a gap bounded below in units of $\sin^2\theta_1$ by \eqref{eq:subcritical}.

\phantomsection\label{step:trap-target}
At $a=\theta_1$, we have $\sin a\,\Lambda(a)<\sin^2\theta_1$. By continuity, this strict inequality persists when $a$ is slightly larger. We can therefore choose angles $\theta_1<b<a<\rho$, with $a/\theta_1$ and
$b/\theta_1$ fixed and close to $1$ in the dense regime and $a,b$ fixed in the logarithmic regime,
such that for a fixed ${c_{\mathrm{fail}}}>0$ and all large $n$
\begin{equation}\label{eq:trap-cap}
  \frac{\sin a\,\Lambda(a)}{\sin^2\theta_1}\le e^{-2{c_{\mathrm{fail}}}}.
\end{equation}
Indeed, the left side equals $\Lambda(\theta_1)/\sin\theta_1<1$ at $a=\theta_1$. Call $C(q,a)$ the \emph{trap cap}
and $C(q,b)$ the \emph{target cap}. Since $\sin a>\sin\theta_1$, \eqref{eq:trap-cap} gives
$\Lambda(a)\le e^{-2{c_{\mathrm{fail}}}}\sin\theta_1$ and hence $nu(a)\le e^{-2{c_{\mathrm{fail}}} d+o(d)}$. We show three facts for every fixed
query $q$ and all large $n$.

(i) \emph{With probability at least $1-e^{-{c_{\mathrm{fail}}} d}$, no data point in $C(q,a)$ has another data
point within distance $\rho$ that is closer to $q$.} Let $N$ be the number of ordered pairs
$(x,y)$ of distinct data points with $\delta(q,x)\le a$ and $y\in U(x)$. Given $x$ with $\delta(q,x)\le a$, the
expected number of such $y$ is at most $nu(a)$, and the expected number of data points in
$C(q,a)$ is $nF(a)$. By \eqref{eq:cap-ratio} and \eqref{eq:u-rate},
\[
  \E N\le nF(a)\cdot nu(a)=\Bigl(\frac{\sin a\,\Lambda(a)}{\sin^2\theta_1}\Bigr)^{d-1}e^{o(d)}\le e^{-{c_{\mathrm{fail}}} d},
\]
and Markov's inequality gives the claim.

(ii) \emph{With probability at least $1-\exp(-e^{\Omega(d)})$, the target cap contains at least $k$
data points, so $\TopSet_k(q)\subseteq C(q,b)$.} The number of data points in $C(q,b)$ is binomial with
mean $nF(b)=(\sin b/\sin\theta_1)^{d-1}e^{O(1)}=e^{\Omega(d)}$. Apply \eqref{eq:chernoff}.

(iii) \emph{Every point $x$ with $\delta(q,x)\ge a$ satisfies $n\sigma_{d-1}(C(x,\rho)\cap C(q,b))\le e^{-{c_{\mathrm{fail}}} d}$.} Let
$\theta=\delta(q,x)$. For $\theta<\pi$, write a point $y$ with $0<\delta(q,y)=s\le b$ as $y=\cos s\,q+\sin s\,w$, and
$x=\cos\theta\,q+\sin\theta\,v$, with unit vectors $v,w\perp q$. Then $\delta(x,y)\le\rho$ if and only if
$w\cdot v\ge(\cos\rho-\cos\theta\cos s)/(\sin\theta\sin s)$. The derivative of the right side in $\theta$ is
$(\cos s-\cos\rho\cos\theta)/(\sin^2\theta\sin s)$, which is positive because $s\le b<\rho$. Hence the
mass of $C(x,\rho)\cap C(q,b)$ does not increase with $\theta$, and for $\theta\ge a$ it is at most its value
at $\theta=a$, which is at most $u(a)$ because $C(q,b)\subseteq C(q,a)$. The point $y=q$ has measure zero and does not affect this comparison. At $\theta=\pi$, the intersection is empty because $b+\rho<\pi$. Hence
$n\sigma_{d-1}(C(x,\rho)\cap C(q,b))\le nu(a)\le e^{-{c_{\mathrm{fail}}} d}$.

On the event of (i), a search at a point in the trap cap outside $\TopSet_k(q)$ stops without success. Its list has no closer sampled point and therefore contains no member of $\TopSet_k(q)$.

\phantomsection\label{step:strip}
To bound the number of expansions before stopping, we first control the mass of neighbors that offer only a small improvement. Let $A_0=\frac14\min\{\rho,a\}$, a fixed multiple of $\theta_1$ in the dense regime and a constant
in the logarithmic regime. For every fixed $\eps>0$, we show that for $a\le\theta\le\pi/2-\eps$, every point $x$ with $\delta(q,x)=\theta$,
and every $0<h<A_0$,
\begin{equation}\label{eq:strip}
  \sigma_{d-1}\bigl\{y\in C(x,\rho):\ \theta-h<\delta(q,y)<\theta\bigr\}\le\frac{dh}{A_0}\,u(\theta).
\end{equation}
Let $A=\min\{\rho,\theta\}/2\ge A_0$, and let $z$ be the point at distance $A$ from $x$ on the geodesic
from $x$ to $q$. Thus $\delta(q,z)=\theta-A$ and $z\in C(x,\rho)$. For $y\in U(x)$ let $\phi(y)$ be the point of
the geodesic from $z$ to $y$ at distance $(1-h/A)\,\delta(z,y)$ from $z$. The image
$\phi(y)$ remains in $C(x,\rho)$ because this cap is geodesically convex ($\rho<\pi/2$) and contains $z$ and $y$.
To bound its distance to $q$, we use the convexity of $\delta_q=\delta(q,\cdot)$ along geodesics inside the
open hemisphere around $q$. Along a unit-speed geodesic away from $q$, its second derivative is $\cot(\delta_q)(1-(\delta_q')^2)\ge0$ when $\delta_q<\pi/2$, and convexity extends through $q$ by continuity. Since $z,y\in C(q,\theta)$ with $\theta<\pi/2$,
$\delta_q(\phi(y))\le(h/A)(\theta-A)+(1-h/A)\theta=\theta-h$. To compare surface measures, use polar coordinates around $z$. The map multiplies the radial coordinate $t$ by $(1-h/A)$ and leaves the direction unchanged. Since the volume element is proportional to $(\sin t)^{d-2}dt\,d\omega$, its Jacobian $(1-h/A)(\sin((1-h/A) t)/\sin t)^{d-2}$ is at least $(1-h/A)^{d-1}$ because
$\sin((1-h/A) t)\ge(1-h/A)\sin t$ on $[0,\pi]$. Hence $C(x,\rho)\cap C(q,\theta-h)\supseteq\phi(U(x))$ has mass at least
$(1-h/A)^{d-1}u(\theta)$, and the strip in \eqref{eq:strip}, which lies in $U(x)\setminus C(q,\theta-h)$, has mass at
most $(1-(1-h/A)^{d-1})u(\theta)\le d(h/A)u(\theta)$.

\phantomsection\label{step:far-phase}
We now use the lens rate to bound progress away from the trap cap. The function $nu$ is continuous and nondecreasing on $[a,\pi/2-\eps]$ for a small fixed
$\eps>0$, with $nu(a)<2L_n$, while $nu(\pi/2-\eps)$ is exponentially large because
$\Lambda(\pi/2-\eps)\ge e^{{c_{\mathrm{sub}}}/2}\sin\theta_1$. Let $\theta_n\in(a,\pi/2-\eps)$ solve $nu(\theta_n)=2L_n$. Since
$\log L_n=o(d)$, \eqref{eq:u-rate} gives $\Lambda(\theta_n)=e^{o(1)}\sin\theta_1$. In the dense regime $\theta_n/\theta_1$ is bounded. For moves starting at distance at least $\theta_n$, set $h_{\mathrm{far}}=A_0/(4d)$. For moves starting at distance in $[a,\theta_n)$, set $h_{\mathrm{near}}=A_0/(64dL_n)$. Set $B_n=\lceil\theta_n/h_{\mathrm{near}}\rceil+2$ and use expansion budget $H_n=\lceil\pi/h_{\mathrm{far}}\rceil+2B_n+3$.
Since $A_0$ and $\theta_n$ are within constant factors of $\theta_1$ in the dense regime and constants in
the logarithmic regime, $H_n=O(d/\theta_1+dL_n)$. By \eqref{eq:theta1-dense} and $\log\log n=o(d)$ this is
$e^{o(d)}$ in both regimes. We show that, with probability at
least $1-ne^{-L_n/2}$, every data point $x$ with $\delta(q,x)\ge\theta_n$ has a data point $y$ with
$\delta(x,y)\le\rho$ and $\delta(q,y)\le \delta(q,x)-h_{\mathrm{far}}$. Let $V(x)=C(x,\rho)\cap C(q,\delta(q,x)-h_{\mathrm{far}})$. At
$\delta(q,x)=\theta_n$, the contraction $\phi$ used in the strip estimate with $h=h_{\mathrm{far}}$ sends $U(x)$ into $V(x)$ with
Jacobian at least $(1-1/(4d))^{d-1}\ge3/4$, so $\sigma_{d-1}(V(x))\ge\frac32L_n/n$. For
$\theta=\delta(q,x)\in(\theta_n,\pi)$, this mass cannot decrease. Indeed, a point at distance $t\ge
h_{\mathrm{far}}$ from $x$ lies in $V(x)$ if and only if its directional cosine toward $q$ is at least $\sin
h_{\mathrm{far}}/\sin t+\cot\theta\,(\cos h_{\mathrm{far}}-\cos t)/\sin t$. This threshold is nonincreasing
in $\theta$ because $\cos h_{\mathrm{far}}-\cos t\ge0$. Points with $t<h_{\mathrm{far}}$ cannot lie in
$V(x)$ by the triangle inequality, and $\theta=\pi$ follows by continuity of the intersection measure. Given a
data point $x$, the other $n-1$ points are independent and uniform, so the probability that none lies in $V(x)$
is at most $(1-\frac32L_n/n)^{n-1}\le e^{-L_n}$. A union bound over all $n$ points proves the claim.

\phantomsection\label{step:deferred}
To estimate success during the adaptive trajectory, we next determine the conditional law of the points not yet seen. Conditionally on $\mathcal H_i$, the unrevealed data points are independent and
uniform on $\mathbb S^{d-1}\setminus E_i$. Initially only $x_0$ is revealed and the other $n-1$ points are
independent and uniform. If the claim holds before the $i$-th expansion, then the center
$x_i$ is a function of $\mathcal H_i$, and revealing which unrevealed points lie in $C(x_i,\rho)$,
with their positions, conditions each remaining unrevealed point, independently, on
lying in $\mathbb S^{d-1}\setminus(E_i\cup C(x_i,\rho))=\mathbb S^{d-1}\setminus E_{i+1}$. Restricting a uniform law to a subset gives
the uniform law on that subset. During the first $H_n$ expansions,
$\sigma_{d-1}(E_i)\le H_nF(\rho)$, which is $n^{\beta-1+o(1)}$ or $e^{\gamma d+o(d)}/n$, and in
particular at most $1/2$ because $d=o(\log n)$ in the dense regime. Hence for every Borel set $W$, conditionally on $\mathcal H_i$, the
probability that some unrevealed point lies in $W$ is at most
$n\,\sigma_{d-1}(W)/(1-\sigma_{d-1}(E_i))\le2n\,\sigma_{d-1}(W)$.

Every point to which the exploration moves was unrevealed before that expansion. To see this,
maintain the property that before expanding $x_i$, every other revealed point is at least as
far from $q$ as $x_i$. Initially only $x_0$ is revealed. If the exploration moves to $x_{i+1}$,
its distance is smaller than that of $x_i$, so it cannot be an earlier revealed point. Since
$x_{i+1}$ is the sampled point in the scanned cap closest to $q$, every point newly revealed in that
expansion is at least as far from $q$ as $x_{i+1}$. Thus the property holds before the next expansion.

\phantomsection\label{step:inner-phase}
In the remaining interval $[a,\theta_n]$, there are few improving points, but \eqref{eq:strip} prevents many very small moves. We show that the exploration makes more than $2B_n$ moves from points $x$ with
$a\le \delta(q,x)\le\theta_n$ with probability at most $4^{-B_n}$. Call a move from $x_i$ \emph{slow} if
$\delta(q,x_i)\in[a,\theta_n]$ and $\delta(q,x_{i+1})>\delta(q,x_i)-h_{\mathrm{near}}$. By the conditional exposure argument, a slow move
requires an unrevealed point in the strip of \eqref{eq:strip} with $h=h_{\mathrm{near}}$, whose mass is
at most $dh_{\mathrm{near}}A_0^{-1}u(\theta_n)=1/(32n)$. So, conditionally on $\mathcal H_i$, the move is slow with
probability at most $1/16$. Let $\xi_j$ indicate that the $j$-th move made from a point at
distance in $[a,\theta_n]$ is slow, with $\xi_j=0$ if there is no such move. For every set $S$ of
$B_n$ indices, the chain rule and the conditional bound give
$\Prob(\xi_j=1\ \forall j\in S)\le16^{-B_n}$. Moves in this range that are not slow decrease the
distance by at least $h_{\mathrm{near}}$, so at most $\theta_n/h_{\mathrm{near}}+1\le B_n$ of them occur. More than $2B_n$
moves in the range thus force $\sum_{j\le2B_n}\xi_j\ge B_n$, which has probability at most
$\binom{2B_n}{B_n}16^{-B_n}\le4^{-B_n}$.

Fix a query $q$. Combine (i) and (ii) with the existence of a neighbor improving by $h_{\mathrm{far}}$ from every distance at least $\theta_n$, and the bound of $2B_n$ moves from $[a,\theta_n]$. Their failure probabilities sum to at most $e^{-{c_{\mathrm{fail}}}d}+\exp(-e^{\Omega(d)})+ne^{-L_n/2}+4^{-B_n}=e^{-\Omega(d)}$. When all these properties hold, the
exploration makes at most $\lceil\pi/h_{\mathrm{far}}\rceil$ moves from points at distance at least $\theta_n$
from $q$ and at most $2B_n$ moves from points in $[a,\theta_n]$, so it stops before exhausting its
budget, either in the trap cap or because it cannot move. Suppose greedy navigation from $x_0$ succeeds. If
$x_0\in\TopSet_k(q)$, call this a \emph{start in the target}. Otherwise let $x_i\notin\TopSet_k(q)$ be the point
whose complete expansion first succeeds, and choose any $y\in G_\rho(x_i)\cap\TopSet_k(q)$. By (ii), $y$ lies in
the target cap. Since $y$ is closer to $q$ than $x_i$ and within distance $\rho$, property (i) puts $x_i$ outside
the trap cap. Thus success from a nontarget entry requires an expansion outside the trap cap to reveal a point of
the target cap, which we call a \emph{direct jump}. A start in the target is covered by $x_0\in C(q,b)$, whose
probability is $F(b)=e^{-\Omega(d)}$. By (iii) and the conditional exposure argument, each
expansion of a point outside the trap cap reveals a point of the target cap with
conditional probability at most $2e^{-{c_{\mathrm{fail}}} d}$, so a direct jump has probability at most
$2H_ne^{-{c_{\mathrm{fail}}} d}=e^{-\Omega(d)}$, because $H_n=e^{o(d)}$. The conditional estimates bound the exploration under the original sampling law. Adding the probabilities that the trap, target, or progress properties fail therefore gives, for every fixed $q$, navigation from a uniform entry succeeds
with probability at most $e^{-4\zeta d}$ for a fixed $\zeta>0$, over the data and the entry. Let
$S_D(q)$ be the fraction of successful entries. Then $\E_D\int S_D\,d\sigma_{d-1}\le e^{-4\zeta d}$. By
Markov's inequality, with probability at least $1-e^{-\zeta d}$ we have
$\int S_D\,d\sigma_{d-1}\le e^{-3\zeta d}$, and the queries with $S_D(q)>e^{-\zeta d}$ have measure at most
$e^{-2\zeta d}$. This proves part~(b) for $G_{r_\gamma}$. In the logarithmic regime, $d=\lfloor c\log n\rfloor$ converts these bounds into part~(a) for $G_{\rho_\beta}$ after choosing $\eta>0$ sufficiently small.

\phantomsection\label{step:nn-rows}
Keep $\rho=\rho_\beta$ or $r_\gamma$, and let $\Delta=\lceil n^\beta\rceil$ or $\lceil e^{\gamma d}\rceil$ as before. The
neighbor list of $x$ in $\NN{\Delta}$ is the set of the $\Delta$ points nearest to $x$, and its radius $R(x)$ is
random. Let $\eps_n=(\log n/\Delta)^{1/4}$ and define $R_\pm$ by $nF(R_\pm)=\Delta(1\pm\eps_n)$. Given $x$, the
number of other points in $C(x,R_\pm)$ is binomial with mean $(1+O(1/n))\Delta(1\pm\eps_n)$, so by
\eqref{eq:chernoff} and a union bound, with probability at least $1-2ne^{-\eps_n^2\Delta/4}$ every neighbor-list radius
lies in $[R_-,R_+]$. By \cref{lem:cap}, $(\log F)'(t)=(d-1)\cot t\,(1+O(d^{-1}))$ near $\rho$, so
$R_+-R_-=O(\eps_n\tan\rho/d)=o(\rho/d)$.

Expanding $x_i$ now reveals the radius $R(x_i)$ and the unrevealed points of $C(x_i,R(x_i))$, with
their positions. Conditional on the revealed information, the remaining points are independent and uniform outside the revealed caps. Given $\mathcal H_i$ the revealed
points are known. So for a fixed set $S$ of unrevealed labels, a radius $r$, and positions of
the labels of $S$ in $C(x_i,r)$, the event that $R(x_i)=r$ and that the unrevealed points of
$C(x_i,r)$ are exactly the labels of $S$ at these positions is the intersection of a
constraint on $S$ alone with the event that every other unrevealed label lies outside
$C(x_i,r)$. Given it, the other unrevealed points are therefore independent and uniform on
$\mathbb S^{d-1}\setminus(E_i\cup C(x_i,R(x_i)))$. The argument that every move reaches a fresh point applies
verbatim.

Run the exploration with the additional rule that it stops if a revealed radius leaves
$[R_-,R_+]$. The stopping rule depends only on revealed information, and the probability of a radius violation is at most
$2ne^{-\eps_n^2\Delta/4}$. Before this stop, every revealed cap lies between $C(x_i,R_-)$ and
$C(x_i,R_+)$. Reconstruct $U$, $u$, $\Lambda$, $a$, $b$, $A_0$, $\theta_n$, $h_{\mathrm{near}}$, $h_{\mathrm{far}}$, $B_n$, and $H_n$ from the preceding formulas with $R_+$ in place of $\rho$. The lens, trap, strip, direct-jump, inner-progress, and revealed-mass estimates then use these reconstructed quantities. $R_+$ still satisfies
\eqref{eq:subcritical} with ${c_{\mathrm{sub}}}/2$ because $R_+/\rho\to1$. A slow move or a direct jump requires an
unrevealed point in a set defined with $R_+$, whose conditional probability is bounded
before the random radius is revealed. For the far phase, first set $\theta=\theta_n<\pi/2$ and $A=\min\{R_-,\theta_n\}/2$. Apply the map $\phi$ from
the strip estimate, with center at distance $A$ from $x$ toward $q$ and factor $1-1/(4d)$, to $C(x,R_+)\cap C(q,\theta_n)$.
The geodesics defining $\phi$ stay in $C(x,R_+)\cap C(q,\theta_n)$. Since both radii are below $\pi/2$, the functions $\delta(x,\cdot)$ and $\delta(q,\cdot)$ are convex along these geodesics. Every image point is within $(1-1/(4d))R_++A/(4d)\le R_+-R_+/(8d)\le R_-$ of $x$ for
large $n$, and within $\theta_n-A/(4d)\le\theta_n-h_{\mathrm{far}}$ of $q$. The image keeps at least $(1-1/(4d))^{d-1}\ge3/4$ of the lens mass. This gives the
same candidate mass at $\theta_n$ within radius $R_-$. The monotonicity in query distance used above to find a decrease of $h_{\mathrm{far}}$, now with
neighbor radius $R_-$, extends this lower bound to all $\theta\ge\theta_n$. The probability bound for this progress therefore holds for $\NN{\Delta}$ as well. Combining it with the transferred trap, target-containment, short-move, direct-jump, and entry estimates, and then adding the probability $2ne^{-\eps_n^2\Delta/4}$ that some data point has its neighbor radius outside $[R_-,R_+]$, gives the same fixed-query success bound. The two Markov arguments used for $G_\rho$ then give the stated data probability and exceptional surface measure for $\NN{\Delta}$. Intersecting the query sets and sample events for the two graphs, and decreasing $\zeta$ or $\eta$ if necessary, makes the guarantees simultaneous for both graphs in each regime.

\end{proof}

\subsection{Success probabilities and local exploration}\label{sec:local}

The bound in \cref{thm:random-entry} establishes that success is rare, but the number of fresh entries needed depends on the exact success exponent. We determine this exponent in the degree ranges specified below and bound the success probability of local exploration in terms of its numbers of expansions and entry requests. The data are uniform and the query $q$ is fixed. In the logarithmic regime let $\Delta=\lceil n^\beta\rceil$ with $0<\beta<\bstar(c)$ and $\rho=\rho_\beta$. In the dense regime let $\Delta=\lceil e^{\gamma d}\rceil$ with $0<\gamma<\gc$ and $\rho=r_\gamma$. Thus $nF(\rho)=\Delta(1+o(1))$, and we compare $G_\rho$ with $\NN{\Delta}$.

We use the local exploration model introduced before \cref{thm:informal-restarts}, with a budget $(T,E)$ of at most $T$ complete expansions and $E$ entry requests. A request evaluates its entry, whereas an expansion evaluates the complete neighbor list, including when it finds a target. Success is recorded after either operation if a point of $\TopSet_k(q)$ has been evaluated. Thus local exploration may stop at a target entry without expanding it. A greedy run still follows the convention for $\Work_G$ and completes that entry's expansion. In both cases, the cost counts distinct evaluated points across all operations.

To state the success exponent, we first identify the distance from the query at which greedy progress becomes scarce. Write $S=\sin^2\theta_1$. For a point at query distance $\theta$, the expected number $(n-1)u(\theta)$ of improving neighbors on $G_\rho$ increases from exponentially small to exponentially large as $\theta$ crosses the zero-exponent boundary in \eqref{eq:u-rate}. Denote this distance by $\theta_{\mathrm{trap}}$, defined by $\sin^2\theta_{\mathrm{trap}}=(1-\cos\rho)^2/(2(1-\cos\rho)-S)$. The inequalities in \eqref{eq:subcritical} place this distance strictly between $\theta_1$ and $\pi/2$.

For a point whose distance from $q$ is near $\theta_{\mathrm{trap}}$, we first estimate the probability that its neighbor list contains the nearest neighbor. For neighbor radius $0<\varrho<\pi/2$, consider a current point at query distance $\varphi\in(0,\pi)$ and another point at query distance $\psi\in(0,\pi)$. Their distance is at most $\varrho$ exactly when the inner product of their unit tangent directions at $q$ is at least $w^\varrho(\varphi,\psi)$. For independent uniform directions, $p^\varrho(\varphi,\psi)$ is the exponent of the probability of this edge in $G_\varrho$. The spherical cosine identity gives $w^\varrho$, and we define the probability exponent $p^\varrho$ by
\[
  \begin{gathered}
  w^\varrho(\varphi,\psi)=\frac{\cos\varrho-\cos\varphi\cos\psi}{\sin\varphi\sin\psi},\quad
  \text{and}\quad p^\varrho(\varphi,\psi)=\frac12\log\bigl(1-w^\varrho(\varphi,\psi)_+^2\bigr).
  \end{gathered}
\]
Here $t_+=\max\{t,0\}$ and $p^\varrho=-\infty$ when $w^\varrho\ge1$. For independent uniform directions about $q$, the one-coordinate cap estimate in the proof of \cref{lem:directions} gives edge probability $e^{dp^\varrho(\varphi,\psi)+o(d)}$ when the directional threshold stays a fixed distance below one. We omit the superscript when $\varrho=\rho$.

We determine the exact exponent of the single-run success probability in the \emph{near-threshold band}, where a direct jump from the trap boundary is optimal. In the logarithmic regime this means $\lim_{n\to\infty}\theta_{\mathrm{trap}}\le R$, where $R=\lim_{n\to\infty}\rho$. Equivalently, the limiting values of $\rho$ and $S$ satisfy $S\le \frac{(1-\cos\rho)(1+2\cos\rho)}{1+\cos\rho}$. In the dense regime the band is $\frac12\log(4/3)\le\gamma<\gc$. In either regime, the probability that points at query distances $\theta_{\mathrm{trap}}$ and $\theta_1$, with independent uniform directions about $q$, lie within distance $\rho$ has exponent $-\eta^*$, where
\begin{equation}\label{eq:eta-star}
  \eta^*=-\lim_{n\to\infty}p(\theta_{\mathrm{trap}},\theta_1).
\end{equation}
In the dense regime this is the function $\eta^*(\gamma)$ in \eqref{eq:dense-fresh-start-exponent}. The next theorem shows that allowing intermediate local moves does not improve this exponent.

\begin{theorem}[Single-start success and local exploration]\label{thm:exact-exponent}
For uniform data in either regime, assume that $\Delta$ lies in the near-threshold band and fix $k\ge1$. For each fixed query and each of $G_\rho$ and $\NN{\Delta}$, let $p_1$ be the probability, over the data and a uniform random entry, that one greedy run succeeds on the chosen graph. On each graph, $p_1=e^{-d(\eta^*+o(1))}$. Every local exploration algorithm on that graph with budget $(T,E)$, $T\le n/(8\Delta)$ and $E\ge1$, succeeds with probability at most $e^{o(d)}(T+E)p_1$ over the data and its internal randomness. The $o(d)$ term is independent of the algorithm and the budgets $T,E$.
\end{theorem}

To upper-bound both single-run and local-exploration success, we group revealed points by their distances from $q$ and bound the expected number of routes leading to a target. For the matching lower bound, we hide the nearest neighbor's direction and construct a greedy path that stops within a small enlargement of the trap radius. We then use the independent hidden direction to bound the probability that this stopping point lists the nearest neighbor. The geometric comparison needed for the upper bound comes first.

\phantomsection\label{sec:trap-geometry}

To bound success along a route with intermediate points, edge probability alone is insufficient because an expansion may reveal many candidates at each distance. Define $\nu(\psi)=\log(\sin\psi/\sin\theta_1)$. For $\psi$ in the angular range of \eqref{eq:cap-ratio}, $\nu(\psi)$ gives the exponent in $d$ of the expected number of points in $C(q,\psi)$. Adding $\nu(\psi)$ to the edge-probability exponent $p^\varrho(\varphi,\psi)$ gives the transition exponent $g^\varrho$ for the expected number of neighbors. Its symmetric version $\chi^\varrho$ assigns half the counting exponent to each endpoint. The difference between $g^\varrho$ and $\chi^\varrho$ then telescopes along a route. These rates satisfy
\begin{equation}\label{eq:g-kappa}
  \begin{aligned}
  g^\varrho(\varphi\to\psi)&=\nu(\psi)+p^\varrho(\varphi,\psi),\quad
  \chi^\varrho(\varphi,\psi)=\tfrac12\bigl(\nu(\varphi)+\nu(\psi)\bigr)+p^\varrho(\varphi,\psi),\\
   g^\varrho(\varphi\to\psi)&=\chi^\varrho(\varphi,\psi)+\tfrac12\log\frac{\sin\psi}{\sin\varphi}.
  \end{aligned}
\end{equation}
For a current point at distance $\varphi$, the supremum of $g^\varrho(\varphi\to\psi)$ over a distance interval bounds the exponent of the expected number of its neighbors whose distances to $q$ lie in that interval, as shown in the proof of \cref{thm:local}. 

For a route whose successive data points lie in each other's radius-$\rho$ caps, we describe its first crossing of the boundary and its subsequent distances by a \emph{chain} of distances $\varphi_0\ge\theta_{\mathrm{trap}}$ and $\varphi_1,\ldots,\varphi_\ell\in[\theta_1,\theta_{\mathrm{trap}}]$, with $\ell\ge0$, followed by an endpoint at the reference distance $\varphi_{\ell+1}=\theta_1$. Its value is the exponent obtained by multiplying the expected numbers of intermediate neighbors and the probability of the last edge to that endpoint. The logarithmic ratios in \eqref{eq:g-kappa} telescope, giving
\begin{equation}\label{eq:chain-value}
  \mathrm{val}(\varphi_0,\ldots,\varphi_\ell)
  =\sum_{i<\ell}g(\varphi_i\to\varphi_{i+1})+p(\varphi_\ell,\theta_1)
  =\sum_{i=0}^{\ell}\chi(\varphi_i,\varphi_{i+1})+\tfrac12\log\frac{\sin\theta_1}{\sin\varphi_0}.
\end{equation}
The rates and chain values just defined are finite-$n$ quantities. The cap formulas in \cref{lem:cap} and \eqref{eq:cap-ratio} retain their powers of $d-1$. Normalizing rates by $d$ adds an error covered below using $\log(1/\theta_1)=o(d)$. In the logarithmic regime the angles converge, with $\sin\theta_1\to e^{-1/c}$ and $\sin\rho\to e^{-(1-\beta)/c}$, so we take their angle limits. In the dense regime the angles tend to zero, so we instead use normalized distances $x=\varphi/\theta_1$ and $y=\psi/\theta_1$. Writing $M=e^\gamma$ for the limiting ratio $\rho/\theta_1$, we obtain $\nu(x)=\log x$ and $w(x,y)=(x^2+y^2-M^2)/(2xy)$. On the branch $w\ge0$, the symmetrized rate satisfies
\[
 e^{2\chi(x,y)}=\frac{(M^2-(x-y)^2)_+((x+y)^2-M^2)}{4xy}.
\]
When $w<0$, the probability exponent is zero and $e^{2\chi(x,y)}=xy$. The displayed quantities $\nu,w,e^{2\chi}$ converge uniformly on compact sets of positive normalized distances. Since $p,g,\chi$ can take the value $-\infty$, we use uniform convergence after replacing values below a fixed lower bound by that bound, as specified in the proof of \cref{thm:local}. Moreover, $\theta_{\mathrm{trap}}/\theta_1\to L$, where $L^2=M^4/(4(M^2-1))$. Thus the intermediate chain distances in $[\theta_1,\theta_{\mathrm{trap}}]$ become $[1,L]$, and the dense near-threshold condition is $L\le M$. For a chain with normalized distances $x_i=\varphi_i/\theta_1$, its limiting value is $\sum_{i=0}^{\ell}\chi(x_i,x_{i+1})-\frac12\log x_0$, with $x_{\ell+1}=1$. For finite $n$, the geometric formulas use actual angles $\varphi_i$, $\theta_1$, and $\theta_{\mathrm{trap}}$. When a calculation explicitly passes to limiting coordinates, the same function symbols $\nu,w,p,g,\chi$ refer to their limits, with normalized distances as arguments in the dense regime and limiting angles in the logarithmic regime.

To bound every route, let $\eta_{\mathrm{gen}}=-\sup_{\text{chains}}\mathrm{val}$, with chains and rates taken in these limiting coordinates. The direct-jump exponent is $\eta^*=-p(L,1)$ in the dense regime, where
\[
  w(L,1)=\frac{(2-M^2)(3M^2-2)}{8(M^2-1)L}.
\]
Substituting $L=M^2/(2\sqrt{M^2-1})$ gives \eqref{eq:dense-fresh-start-exponent}. In the logarithmic regime we must also account for an entry inside $C(q,\theta_{\mathrm{trap}})$. In the dense regime, landing in this cap has probability $e^{O(d)}/n=e^{-\omega(d)}$, so the entry term in \cref{thm:local} already makes it negligible. For routes starting inside $C(q,\theta_{\mathrm{trap}})$ in the logarithmic regime, allow all distances $\varphi_0,\ldots,\varphi_\ell$ in \eqref{eq:chain-value} to lie between the limiting nearest-neighbor and trap angles, with $\ell\ge0$. Including the probability of the entry location gives the exponent
\[
  \eta_{\mathrm{entry}}
  =\min\left\{\lambda,-\sup\bigl[\nu(\varphi_0)-\lambda+
        \mathrm{val}(\varphi_0,\ldots,\varphi_\ell)\bigr]\right\},
\]
where the supremum is over the routes just described. The term $\nu(\varphi_0)-\lambda$ is the probability exponent of the entry location, and the separate term $\lambda$ accounts for an entry that is itself a target. The following lemma shows that in the band, $\eta_{\mathrm{gen}}=\eta^*$ and the logarithmic probability exponent for routes from an interior entry is strictly smaller than $-\eta^*$. As the degree approaches the critical value, $\eta^*\to0$ in both regimes.

\begin{lemma}[Geometry of the trap]\label{lem:trap-geometry}
Let $\varrho$ satisfy \eqref{eq:subcritical}, $a=\cos\varrho$, and let
$\theta_{\mathrm{trap}}^\varrho$ be defined like $\theta_{\mathrm{trap}}$ with $\varrho$ in place of $\rho$.
\begin{enumerate}[label=\textup{(\alph*)},leftmargin=*]
\item For distances $\varphi,\psi$, with $u=\cos(\varphi-\psi)$ and $v=\cos(\varphi+\psi)$,
  \[
    e^{2\chi^\varrho(\varphi,\psi)}=\frac{2\,(u-a)_+\,(a-v)}{(u-v)\sin^2\theta_1}\qquad(w^\varrho\ge0).
  \]
  If $\varphi,\psi\in[\theta_1,\pi/2]$, then $\chi^\varrho(\varphi,\psi)\le\chi^\varrho(\bar\varphi,\bar\varphi)$ with
  $\bar\varphi=(\varphi+\psi)/2$. The function $\theta\mapsto\chi^\varrho(\theta,\theta)$ is increasing on
  $(0,\pi/2]$, with its unique zero at $\theta^\varrho_{\mathrm{trap}}$. In particular
  $\chi^\varrho\le\chi^\varrho(\theta,\theta)<0$ on $[\theta_1,\theta]^2$ for every $\theta<\theta_{\mathrm{trap}}^\varrho$.
\item For $\varphi\ge\psi$ and $\psi<\pi/2$, $\partial w^\varrho/\partial\varphi$ has the sign of
  $\cos\psi-a\cos\varphi>0$. So $p^\varrho(\varphi,\psi)$ and $g^\varrho(\varphi\to\psi)$ are nonincreasing
  in $\varphi\in[\psi,\pi]$, with the values at $\varphi=\pi$ interpreted as limits.
\item In the near-threshold band, the limiting function $\chi$ is nondecreasing in each argument between the limiting nearest-neighbor and trap distances, namely on $[1,L]^2$ in the dense regime. We have $\eta_{\mathrm{gen}}=\eta^*<\infty$. This exponent is smaller than the limiting exponent for the expected number of points within the trap radius, which is $\nu(L)=\log L$ in the dense regime. In the logarithmic regime moreover $\eta^*<\eta_{\mathrm{entry}}$ and
  $\eta^*<\lambda-\beta/c$.
\end{enumerate}
\end{lemma}

\begin{proof}

We have $\sin\varphi\sin\psi=(u-v)/2$ and $\cos\varphi\cos\psi=(u+v)/2$, so
$\sin^2\varphi\sin^2\psi-(a-\cos\varphi\cos\psi)^2=(u-a)(a-v)$, which gives the formula in part~(a) when $u>a$ and $w^\varrho\ge0$. When $u\le a$, we have $w^\varrho\ge1$, so both $e^{2\chi^\varrho}$ and the positive part $(u-a)_+$ in that formula are zero. If $\varphi,\psi\in[\theta_1,\pi/2]$, then $2\theta_1\le\varphi+\psi\le\pi$, where
the cosine decreases, so $a-v\ge\cos\varrho-\cos2\theta_1>0$, since $1-\cos\varrho<2\sin^2\theta_1=1-\cos2\theta_1$ by
\eqref{eq:subcritical}. At a fixed sum $\varphi+\psi$ the variable $v$ is fixed and $(u-a)/(u-v)$
increases with $u$, because its derivative is $(a-v)/(u-v)^2>0$. So the maximum over
pairs with a given sum is at $\varphi=\psi$. Finally, on the diagonal $e^{2\chi^\varrho(\theta,\theta)}=\Lambda_\varrho(\theta)^2/\sin^2\theta_1$ with
$\Lambda_\varrho(\theta)^2=\sin^2\theta\,(1-w^\varrho(\theta,\theta)_+^2)$. Let $\theta_w\in(0,\pi/2)$ be defined by
$\cos^2\theta_w=a$, so that $w^\varrho(\theta,\theta)=(a-\cos^2\theta)/\sin^2\theta$ is negative for $\theta<\theta_w$ and
nonnegative for $\theta\ge\theta_w$. Below the critical degree $a>\cos^2\theta_1$, so $\theta_w<\theta_1$. Hence
\[
  \Lambda_\varrho(\theta)^2=\begin{cases}\sin^2\theta,&0<\theta<\theta_w,\\ 2(1-a)-(1-a)^2/\sin^2\theta,&\theta_w\le\theta\le\pi/2.\end{cases}
\]
The expression in the second branch is nonnegative on its stated range. The two
branches agree at $\theta_w$, where both equal $(1-a)=\sin^2\theta_w$. Both branches increase, so
$\Lambda_\varrho^2$ increases on $(0,\pi/2]$, and it equals $\sin^2\theta_1$ exactly at $\theta^\varrho_{\mathrm{trap}}$, which lies in
the second range because $\Lambda_\varrho(\theta_1)^2<\sin^2\theta_1$.

For part~(b), moving the source farther from the query changes the directional threshold by $\partial_\varphi w^\varrho=(\cos\psi-a\cos\varphi)/(\sin^2\varphi\sin\psi)$. For $\varphi\ge\psi$ and
$\psi<\pi/2$ we have $\cos\psi-a\cos\varphi\ge(1-a)\cos\psi>0$, both when $\cos\varphi\ge0$ (then
$\cos\varphi\le\cos\psi$) and when $\cos\varphi<0$. Hence $w^\varrho$ increases and $p^\varrho$ does not
increase in $\varphi$, while $\nu(\psi)$ does not depend on $\varphi$.

For part~(c), first work with the limiting angles in the logarithmic regime. In this calculation, $\theta_1$ and $\theta_{\mathrm{trap}}$ denote their angle limits, with $a=\cos R$ and $S=\sin^2\theta_1$. Write $x=\cos\varphi$, $y=\cos\psi$ and $\Xi=(1-x^2)(1-y^2)-(a-xy)^2$. On
$[\theta_1,\theta_{\mathrm{trap}}]^2$ we have $0<w<1$: $w>0$ because $xy\le\cos^2\theta_1<a$ by
\eqref{eq:subcritical}, and $w<1$ because $|\varphi-\psi|<\theta_{\mathrm{trap}}\le R$. A direct
computation gives
\[
  2\,\Xi\,(1-x^2)\,\partial_x\chi=(1-x^2)(2ay-x-xy^2)-x(a-xy)^2 .
\]
In the band $x\ge\cos\theta_{\mathrm{trap}}\ge a$, hence $2ay\le2xy\le x(1+y^2)$. The first term is
nonpositive and the second negative. So $\chi$ increases with the distance $\varphi$, and by
symmetry with $\psi$.

The corresponding dense-regime calculation uses normalized distances $x,y$. On the branch $x^2+y^2\ge M^2$, where $w\ge0$, and for $|x-y|<M<x+y$, we have
\[
  \partial_x(2\chi)=\frac{4x^2(M^2-x^2)+(x^2+y^2-M^2)^2}{x\,(M^2-(x-y)^2)\,((x+y)^2-M^2)}.
\]
Multiplying $\partial_x(2\chi)=-2(x-y)/(M^2-(x-y)^2)+2(x+y)/((x+y)^2-M^2)-1/x$ by the positive
denominator and collecting terms gives the numerator. It is nonnegative for $x\le M$,
and on $[1,L]^2$ with $L\le M$ we have $|x-y|\le L-1<M<x+y$ and $x^2+y^2\ge2>M^2$, so the derivative formula applies throughout the required region.

These monotonicity bounds identify the best chain. First use the limiting angles in the logarithmic regime. The chain value does not decrease if $\varphi_0$ is
replaced by $\theta_{\mathrm{trap}}$, by part~(b), since only the first term in the $g,p$ expression of \eqref{eq:chain-value} depends on $\varphi_0$. Then all
distances lie in $[\theta_1,\theta_{\mathrm{trap}}]$, every term $\chi(\varphi_i,\varphi_{i+1})$ of
\eqref{eq:chain-value} is at most $\chi(\theta_{\mathrm{trap}},\theta_{\mathrm{trap}})=0$ by monotonicity, and
dropping all of them but the last,
\[
  \mathrm{val}\le\chi(\varphi_{\ell},\theta_1)+\tfrac12\log\frac{\sin\theta_1}{\sin\theta_{\mathrm{trap}}}
  \le\chi(\theta_{\mathrm{trap}},\theta_1)+\tfrac12\log\frac{\sin\theta_1}{\sin\theta_{\mathrm{trap}}}
  =p(\theta_{\mathrm{trap}},\theta_1)=-\eta^*,
\]
with equality for the chain $\ell=0$, $\varphi_0=\theta_{\mathrm{trap}}$. In the dense regime, the same telescoping argument in normalized coordinates gives $\mathrm{val}\le\chi(L,1)-\frac12\log L=p(L,1)$, with equality for $\ell=0$ and $x_0=L$. Hence $\eta_{\mathrm{gen}}=\eta^*$ in both regimes. It is finite because the threshold $w(\theta_{\mathrm{trap}},\theta_1)$ for the direct jump is less than one. In the dense regime
$w(L,1)=(L^2+1-M^2)/(2L)$, and $L^2+1-M^2=(2-M^2)(3M^2-2)/(4(M^2-1))$ gives
\eqref{eq:dense-fresh-start-exponent}.

In the logarithmic regime, an entry lands in $C(q,\theta_{\mathrm{trap}})$ with probability $e^{-d(\lambda-\nu(\theta_{\mathrm{trap}}))+o(d)}$, and
$\lambda-\nu(\theta_{\mathrm{trap}})=-\log\sin\theta_{\mathrm{trap}}$. The value of a chain that starts inside the trap
cap is at most $0$, because all its terms $\chi$ are nonpositive and
$\sin\theta_1\le\sin\varphi_0$. So $\eta_{\mathrm{entry}}\ge-\log\sin\theta_{\mathrm{trap}}$. By the derivative formula in part~(b), with the roles of the arguments exchanged, $\partial_\psi w(\theta_{\mathrm{trap}},\psi)$ has the sign of
$\cos\theta_{\mathrm{trap}}-a\cos\psi\ge a(1-\cos\psi)\ge0$, so
$w(\theta_{\mathrm{trap}},\theta_1)\le w(\theta_{\mathrm{trap}},\theta_{\mathrm{trap}})=1-(1-a)/\sin^2\theta_{\mathrm{trap}}=(S-(1-a))/(1-a)$. The
band condition gives $(S-(1-a))/(1-a)\le a/(1+a)<a\le\cos\theta_{\mathrm{trap}}$. Hence
$1-w(\theta_{\mathrm{trap}},\theta_1)^2>\sin^2\theta_{\mathrm{trap}}$, that is, $\eta^*<-\log\sin\theta_{\mathrm{trap}}\le\eta_{\mathrm{entry}}$. Also $1-w^2>1-a^2=\sin^2R$, that is, $\eta^*<-\log\sin R=\lambda-\beta/c$.

Finally, part~(a) gives $\chi(\theta_{\mathrm{trap}},\theta_{\mathrm{trap}})=0$. In the logarithmic regime, this yields $-p(\theta_{\mathrm{trap}},\theta_{\mathrm{trap}})=\nu(\theta_{\mathrm{trap}})$. The preceding entry-exponent comparison shows that $\partial_\psi w(\theta_{\mathrm{trap}},\psi)>0$ for $0<\psi\le\theta_{\mathrm{trap}}$, so $0<w(\theta_{\mathrm{trap}},\theta_1)<w(\theta_{\mathrm{trap}},\theta_{\mathrm{trap}})$ and $\eta^*<\nu(\theta_{\mathrm{trap}})$. In the dense regime, $\partial_yw(L,y)=(y^2-L^2+M^2)/(2Ly^2)>0$ because $L\le M$. Thus $0<w(L,1)<w(L,L)$, while the limiting diagonal identity gives $-p(L,L)=\nu(L)=\log L$. It follows that $\eta^*<\nu(L)$ as required.
\end{proof}

The geometric rates can now be applied to an adaptive exploration. To preserve independence of the unrevealed points, expose a cap containing the entire neighbor list at each expansion. Define $\rho^+$ by $nF(\rho^+)=2\Delta$, and retain $R_\pm$ from the proof of \cref{thm:random-entry}, where $nF(R_\pm)=\Delta(1\pm\eps_n)$ and $\eps_n=(\log n/\Delta)^{1/4}$. These radii differ from $\rho$ by a relative $O(1/d)$, so \eqref{eq:subcritical} still holds with a smaller fixed margin. For each data point $x$, its list in $G_\rho$ is determined by the points in $C(x,\rho^+)$. For $\NN{\Delta}$ this is true whenever every such cap contains at least $\Delta$ other data points. Denote that data event by $\mathcal C$. Chernoff's inequality gives $\Prob(\mathcal C^c)\le ne^{-\Delta/4}$, whereas for $G_\rho$ we take $\mathcal C$ to be the whole sample space. In the auxiliary exposure procedure, an entry request reveals that point's position, and an expansion reveals the complete cap $C(x,\rho^+)$. The algorithm still evaluates only its requested entry or scanned neighbor list. For $\NN{\Delta}$, form the list from the $\Delta$ nearest other exposed points in that cap, and scan all available points and stop if there are fewer than $\Delta$. Complete every cap exposure and list scan, including one that triggers a stop. On $\mathcal C$, these lists are the true lists. The next theorem bounds success using this auxiliary procedure.

\begin{theorem}[Local-exploration probability bound]\label{thm:local}
Fix $k\ge1$ and $\eps>0$, and let $\iota_0>0$ be small enough, depending on $\eps$. Let
$b_0=\theta_1+\iota_0$ in the logarithmic regime and $b_0=(1+\iota_0)\theta_1$ in the dense regime. Stop the auxiliary exposure procedure at the end of its first operation revealing a point of $C(q,b_0)$, or when the algorithm terminates earlier. Retain its exceptional stop if a nearest-neighbor list cannot be reconstructed. Every triggering expansion completes its scan and exposure. Let $T_{\mathrm{near}}\le T$ count the expansions at distances in $(b_0,\theta_{\mathrm{trap}}+\rho^+]$ through this stop.
There are $c'>0$, $n_0$, and, in the dense regime, $C=C(\gamma,\eps)>0$ such that for all $n\ge n_0$, every fixed query $q$, the graphs
$G_\rho$ and $\NN{\Delta}$, and every local exploration algorithm with budget $(T,E)$ and
$T\le n/(8\Delta)$,
\[
  \Prob\{\text{a point of }\TopSet_k(q)\text{ is evaluated}\}
  \le\E[T_{\mathrm{near}}]\,e^{-d(\eta_{\mathrm{gen}}-\eps)}+E\,\mathfrak e_n+\Prob(\mathcal C^c)+\exp(-e^{c'd}),
\]
where the probability is over the data and the algorithm's randomness, including random entry choices. The entry term is $\mathfrak e_n=e^{-d(\eta_{\mathrm{entry}}-\eps)}$ in the logarithmic regime and $\mathfrak e_n=e^{Cd}/n$ in the dense regime.
\end{theorem}

\begin{proof}

\phantomsection\label{step:shell}
We first bound how many new points one expansion can reveal in a distance interval. Let $x$ be a point of distance $\varphi$, let $0\le\psi_-<\psi_+\le\pi/2$, and let $K>0$ truncate very negative transition exponents at $-K$. We show
that, uniformly in $\varphi$,
\begin{multline*}
  n\,\sigma_{d-1}\bigl(C(x,\rho^+)\cap\{y:\psi_-<\delta(q,y)\le\psi_+\}\bigr)\\
  \le e^{4K+o(d)}\exp\Bigl\{d\max\Bigl(\sup_{\psi_-<\psi\le\psi_+}g^{\rho^+}(\varphi\to\psi),\,-K\Bigr)\Bigr\}.
\end{multline*}
If the set on the left is empty, the asserted bound is immediate, so assume
it is nonempty. Let $z$ be the projection of a uniform point onto the plane spanned by $x$ and $q$. By the density formula in the proof of \cref{lem:lens}, $z$ has density
$\frac{d-2}{2\pi}(1-|z|^2)^{(d-4)/2}$ on the unit disk.
At query distance $\psi$, membership in $C(x,\rho^+)$ requires the cosine between the directions of $x$ and $y$ about $q$ to be at least $w^{\rho^+}(\varphi,\psi)$. Consequently every feasible projection satisfies $1-|z|^2\le\sin^2\psi\,[1-(w^{\rho^+}(\varphi,\psi)_+)^2]=\sin^2\theta_1\exp\{2g^{\rho^+}(\varphi\to\psi)\}$. Taking the supremum over the distance interval and replacing rates below $-K$ by $-K$ bounds the maximum planar density. The planar image has area at most $\pi$. Applying this density bound and using $n=\sqrt{2\pi(d-1)}\cos\theta_1(\sin\theta_1)^{-(d-1)}(1+O(d^{-1}))$ gives the asserted estimate with prefactor $O(d^{3/2}(\sin\theta_1)^{-3}e^{4K})$. This prefactor is $e^{4K+o(d)}$ because $\log(1/\theta_1)=o(d)$.

We next make this geometric bound valid during an adaptive search. Include the algorithm's internal randomness, independent of the sample, in the history. Reveal the position of every entry when it is drawn, and at each expansion of a
label $v$ all labels and positions in $C(X_v,\rho^+)$. The next choice of the algorithm
is a function of the history, including the algorithm's internal randomness, because the neighbor lists are determined by the
revealed caps. For $\NN{\Delta}$ we use the auxiliary process. By
the conditional exposure argument in the proof of \cref{thm:random-entry}, conditionally on the
history the positions of the unrevealed labels are independent and uniform on the complement of the
union of the caps revealed in the completed expansions. An entry label chosen independently of the data
that is still unrevealed has a uniform position on that complement. Since these caps have total surface measure at most
$TF(\rho^+)=2T\Delta/n\le\frac14$, the conditional expected number of
unrevealed labels in a Borel set $U$ is at most $\frac43n\sigma_{d-1}(U)$.

To follow repeated expansions toward $C(q,b_0)$, partition the query distances from $b_0$ to a value slightly below $\theta_{\mathrm{trap}}$ into finitely many intervals. Fix a mesh $\iota\in(0,\iota_0)$, let $b_*=\theta_{\mathrm{trap}}-\iota_0$, respectively
$b_*=(L-\iota_0)\theta_1$, and choose $b_0<b_1<\cdots<b_J=b_*$ with steps at most $\iota$, in units of
$\theta_1$ in the dense regime. $J$ does not depend on $n$. Bin $j\ge1$ is the shell
$\{b_{j-1}<\delta(q,\cdot)\le b_j\}$, bin $0$ is the \emph{target cap} $C(q,b_0)$, and points of
distance above $b_*$ are \emph{outside}. By \eqref{eq:cap-ratio}, $nF(b_0)\ge e^{c'd}$ for some
$c'=c'(\iota_0)>0$, so $\TopSet_k(q)\subseteq C(q,b_0)$ except with probability
$\exp(-e^{c'd})$, after decreasing $c'$. Stop the auxiliary process at the end of the first operation that reveals a point of the target cap, or when the algorithm terminates without such a reveal. If an expansion triggers this stop, complete both the cap exposure and the neighbor-list scan. Then
\begin{equation}\label{eq:success-reveal}
  \Prob\{\text{success}\}\le\exp(-e^{c'd})+\Prob(\mathcal C^c)
  +\E\,\#\{\text{revealed points of }C(q,b_0)\}.
\end{equation}
For $j\ge1$, let $e_j\le T$ be the expected number of expanded labels in bin $j$. Define $e_0$ separately as the expected number of points in the target cap revealed through the stopping operation. Also let
$N_{\mathrm{near}}\le T_{\mathrm{near}}$ be the number of expanded labels of distance in
$(b_*,\theta_{\mathrm{trap}}+\rho^+]$. An expansion at a larger distance reveals no point of distance at most
$\theta_{\mathrm{trap}}$, by the triangle inequality. An expanded label of bin $j$ was revealed
earlier, as an entry or inside the cap of an earlier expansion. The shell-mass and conditional exposure bounds, applied before each expansion and then summed, give
\begin{equation}\label{eq:bin-recursion}
  e_j\le\sum_{i=1}^Je_i\,K_{ij}+\E[N_{\mathrm{near}}]\,K_{\mathrm{out},j}+E\,f_j,
  \qquad 0\le j\le J,
\end{equation}
where $K_{ij}=e^{4K+o(d)}\exp\{d\sup\max(g^{\rho^+}(\varphi\to\psi),-K)\}$, the supremum over $\varphi$
in bin $i$ and $\psi$ in bin $j$, $K_{\mathrm{out},j}$ is the same with the supremum over
$\varphi\ge b_*$, and $f_j=\frac43F(b_j)$ accounts for the entries. For $j=0$ the left side is
the expected number of revealed points of the target cap. In the dense regime
$f_j\le(L+1)^{d-1}e^{O(1)}/n\le e^{Cd}/n$.

To control the recursion, first use \cref{lem:trap-geometry}(b). The supremum over $\varphi\ge b_*$ in $K_{\mathrm{out},j}$ is attained at
$\varphi=b_*$, since all bins lie at distance at most $b_*$.

Write $g^{\rho^+}=\chi^{\rho^+}+t$ with $t(\varphi,\psi)=\frac12\log(\sin\psi/\sin\varphi)$. For $\varphi,\psi\in[b_0,b_*]$, we have $|t(\varphi,\psi)|\le\tau_*$, where one may take $\tau_*=\lambda/2$ in the logarithmic regime and $\tau_*=\frac12\log(2L)$ in the dense regime. Along a chain of distances in this interval, the sum of the $t$ terms equals $t$ evaluated at the first and last distances. By
\cref{lem:trap-geometry}(a) with $\varrho=\rho^+$, whose trap distance is $\theta_{\mathrm{trap}}-O(1/d)$,
respectively $(L-o(1))\theta_1$, there is ${c_{\mathrm{int}}}>0$ with $\chi^{\rho^+}\le-{c_{\mathrm{int}}}$ whenever both distances
lie in $[b_0,b_*]$ for large $n$. Fix a truncation level
$K>2\tau_*+\eta_{\mathrm{gen}}+\lim_{n\to\infty}\nu(\theta_{\mathrm{trap}})+2$, where $\nu$ inside the limit is evaluated at the actual angles for each $n$. The functions $\max(g^{\rho^+},-K)$ converge
uniformly for both distances in $[b_0,b_*]$, in normalized distances in the dense regime, to a continuous
limit. Changing each argument by at most $\iota$ therefore changes the truncated rate by $o_\iota(1)$, uniformly for all sufficiently large $n$. Put
${c_{\mathrm{path}}}=\min\{{c_{\mathrm{int}}},K-\tau_*\}$, so that $\max(g^{\rho^+},-K)\le t-{c_{\mathrm{path}}}$ for both distances in $[b_0,b_*]$.

The inequality $\max(g^{\rho^+},-K)\le t-{c_{\mathrm{path}}}$ bounds products of the coefficients $K_{ij}$ along transitions between bins $1,\ldots,J$. The final transition into bin $0$ instead has rate at most $\nu(b_0)$, as used in the final-step comparison below. Along a closed walk with $r$ transitions between the distance bins, the terms $t$ sum to zero, so the
product of the corresponding entries $K_{ij}$ is at most
$\exp\{d(-r{c_{\mathrm{path}}}+ro_\iota(1))+r(4K+o(d))\}\le e^{-r{c_{\mathrm{path}}}d/2}$ once $o_\iota(1)\le{c_{\mathrm{path}}}/4$ and $n$ is
large, with $o(d)$ uniform in $r$.

To bound the remaining paths within bins $1,\ldots,J$, the terms $t$ along a simple path with $r$ transitions sum to at most $\tau_*$. Including the discretization error and the prefactor of each coefficient gives a logarithmic bound $d(\tau_*-rc_{\mathrm{path}})+r(do_\iota(1)+4K+o(d))$. With the same choice of mesh and sufficiently large $n$, this is at most $d(\tau_*-rc_{\mathrm{path}}/2)$. Consequently, paths with $r>r_{\max}:=(2\tau_*+2K)/c_{\mathrm{path}}$ give products at most $e^{-Kd}$. If one step of a chain has $g^{\rho^+}<-K$ and its rate is replaced by $-K$,
the chain has value at most $-K+2\tau_*$, since the terms $t$ before and after
that step each sum to at most $\tau_*$. The coefficient
$K_{\mathrm{out},j}\le e^{d(\nu(\theta_{\mathrm{trap}})+o(1))}$ accounts for revealing the first point in the distance intervals through an expansion outside them. By $K>2\tau_*+\eta_{\mathrm{gen}}+\lim_{n\to\infty}\nu(\theta_{\mathrm{trap}})+2$, the preceding long-path bound $e^{-Kd}$ and the bound for chains containing a rate below $-K$ both remain smaller than
$e^{-d(\eta_{\mathrm{gen}}+1)}$ after multiplication by $K_{\mathrm{out},j}$ and the final-step factor $e^{4K+o(d)+d\nu(b_0)}$, for sufficiently small $\iota_0$. For $r\le r_{\max}$ the discretization error is at
most $r_{\max}(do_\iota(1)+4K)+o(d)$, which is $d\,o_\iota(1)$ at fixed $\iota_0$.

We now sum these path bounds to solve \eqref{eq:bin-recursion}. Set $K_{0j}=0$ for $0\le j\le J$, since the auxiliary process stops after its first operation revealing a point of the target cap, and view the coefficients $(K_{ij})_{0\le i,j\le J}$ as a matrix.
A walk is a simple path with cycles inserted. For fixed initial and final bins, there are at most $J^r$ walks with $r$ transitions. By the preceding path and cycle bounds, the series of powers of the transpose of this coefficient matrix converges for large $n$, and its $(0,i)$ entry is at most
$C_J\exp\{dW_i+d\,o_\iota(1)\}$, where $C_J$ is a constant depending only on the fixed number $J$ of bins and $W_i$ is the largest total of the truncated limiting transition rates $\max(g,-K)$ over chains from
bin $i$ into the target cap. All quantities in \eqref{eq:bin-recursion} are finite and
nonnegative, so iterating it and inserting the result into \eqref{eq:success-reveal}
gives, in the logarithmic regime,
\[
  \Prob\{\text{success}\}\le\exp(-e^{c'd})+\Prob(\mathcal C^c)
  +\E[N_{\mathrm{near}}]\,e^{-d(\eta_{\mathrm{gen}}(\iota_0)-o_\iota(1))}
  +E\,e^{-d(\eta_{\mathrm{entry}}(\iota_0)-o_\iota(1))},
\]
where the perturbed exponents use the same fixed cutoff $K$. For $\eta_{\mathrm{gen}}(\iota_0)$, take the negative supremum of sums of $\max\{g,-K\}$ along finite chains starting at $b_*$, with intermediate distances in $[b_0,b_*]$ and the
final distance in $(0,b_0]$. Use limiting angles in the logarithmic regime and distances divided by $\theta_1$ in the dense regime. In the logarithmic regime, $\eta_{\mathrm{entry}}(\iota_0)$ instead allows the initial distance $\varphi_0$
anywhere in $[b_0,b_*]$ and adds $\nu(\varphi_0)-\lambda$ to the chain sum before taking the negative supremum. It also includes direct entry into the target cap, with exponent $\nu(b_0)-\lambda$. In the dense regime the entry term is at most $Ee^{Cd}/n$, after increasing the same constant $C$ to absorb the path factors.

To compare the perturbed exponents with $\eta_{\mathrm{gen}}$ and $\eta_{\mathrm{entry}}$, first consider a chain with $\varphi_0=b_*$ and $\varphi_1,\ldots,\varphi_\ell\in[b_0,b_*]$ whose intermediate rates are all at least $-K$. Its intermediate rates are unchanged by truncation. By \cref{lem:trap-geometry}(a), $\chi\le0$ on $[\theta_1,\theta_{\mathrm{trap}}]^2$. The intermediate moves therefore satisfy
\[
 \sum_{i<\ell}g(\varphi_i\to\varphi_{i+1})
 =\sum_{i<\ell}\chi(\varphi_i,\varphi_{i+1})
  +\tfrac12\bigl(\nu(\varphi_\ell)-\nu(b_*)\bigr)\le0.
\]
Prepending the move from $\theta_{\mathrm{trap}}$ to $b_*$ adds an exponent tending to zero as $\iota_0\to0$. To obtain a chain of the form \eqref{eq:chain-value}, we must also compare the final step into $C(q,b_0)$ with the step to $\theta_1$. For angles, $g(\varphi\to\psi)$ increases with $\psi$ up to $\arccos(\cos\rho\cos\varphi)>\theta_1$. In the dense limit, $g(x\to y)$ increases with $y$ up to $\sqrt{x^2+M^2}>1$. Thus, for $\varphi\in[b_0,b_*]$ and sufficiently small $\iota_0$, the maximum over $\psi\le b_0$ is attained at $\psi=b_0$. At a fixed cutoff $K$, the function $\max\{g(\varphi\to\psi),-K\}$ is continuous on a fixed compact rectangle containing these $\varphi$ and all $\psi$ between $\theta_1$ and $b_0$, including where $g=-\infty$. In the dense regime this statement uses both distances divided by $\theta_1$. Uniform continuity and $g(\varphi\to\theta_1)=p(\varphi,\theta_1)$ give an error $\eps_K(\iota_0)\to0$ such that
\[
 \max\{g(\varphi\to b_0),-K\}
 \le\max\{p(\varphi,\theta_1),-K\}+\eps_K(\iota_0)
 \qquad(\varphi\in[b_0,b_*]).
\]
If $p(\varphi_\ell,\theta_1)\ge-K$, the chain's value is at most the value of its extension plus $\eps_K(\iota_0)+o_{\iota_0}(1)$. Otherwise its value is at most $-K+\eps_K(\iota_0)$, because the intermediate sum is nonpositive. This also covers an impossible final edge with rate $-\infty$. If instead an intermediate rate is truncated, the earlier telescoping bound gives at most $-K+2\tau_*$ for the intermediate sum. The final rate is at most $\nu(b_0)=o_{\iota_0}(1)$, even after truncation. Our choice of $K$ makes this total strictly smaller than $-\eta_{\mathrm{gen}}$ for sufficiently small $\iota_0$. Taking suprema over both types of chains gives
\[
 \eta_{\mathrm{gen}}(\iota_0)
 \ge\min\{\eta_{\mathrm{gen}},K\}-\eps_K(\iota_0)-o_{\iota_0}(1).
\]
Our fixed choice $K>\eta_{\mathrm{gen}}$ and sufficiently small $\iota_0$ prove the required comparison for expansions. For entry requests in the logarithmic regime, first take a chain with no truncated intermediate rate, and let $z_0,\ldots,z_r\in[b_0,b_*]$ be its intermediate distances. Including the probability of the entry location gives
\[
 \nu(z_0)-\lambda+\sum_{i<r}g(z_i\to z_{i+1})
 =\frac{\nu(z_0)+\nu(z_r)}2-\lambda+\sum_{i<r}\chi(z_i,z_{i+1})
 \le\nu(\theta_{\mathrm{trap}})-\lambda<0.
\]
Adding the final step and applying the same truncated comparison bounds this entry contribution by $-\eta_{\mathrm{entry}}+\eps_K(\iota_0)$ when $p(z_r,\theta_1)\ge-K$, and by $-K+\eps_K(\iota_0)$ otherwise. If an intermediate rate is truncated, the intermediate sum is at most $-K+2\tau_*$. Adding the entry exponent, at most $\nu(\theta_{\mathrm{trap}})-\lambda$, and the final rate, at most $\nu(b_0)=o_{\iota_0}(1)$, gives at most $-K+\nu(\theta_{\mathrm{trap}})+o_{\iota_0}(1)$ because $2\tau_*=\lambda$. The chosen $K$ makes this smaller than $-\lambda$. Direct entry into the target cap has exponent $\nu(b_0)-\lambda\to-\lambda$, with $\eta_{\mathrm{entry}}\le\lambda$. Taking $K>\eta_{\mathrm{entry}}$ therefore gives $\eta_{\mathrm{entry}}(\iota_0)\ge\eta_{\mathrm{entry}}-o_{\iota_0}(1)$.
The truncation level $K$ can be fixed before $\iota_0$, since the earlier requirement
$K>2\tau_*+\eta_{\mathrm{gen}}+\lim_{n\to\infty}\nu(\theta_{\mathrm{trap}})+2$, with
$\tau_*=\lambda/2$, already implies $K>\lambda\ge\eta_{\mathrm{entry}}$. The choices are made in the order $\eps$, $\iota_0$, $\iota$,
$n_0$. Since $N_{\mathrm{near}}\le T_{\mathrm{near}}$, this proves the theorem.
\end{proof}

\begin{proof}[Proof of \cref{thm:exact-exponent}]
We first prove a lower bound on the single-run success probability by constructing a greedy path without revealing the nearest neighbor's direction. We then bound the number of expansions and apply \cref{thm:local} to obtain the matching upper bound. We give the argument for $\NN{\Delta}$. For $G_\rho$, use $\rho$ in place of both the deterministic comparison radius $R_-$ and the random neighbor-list radius $R_{D'}(v)$ below. Fix $\eps>0$.

\phantomsection\label{step:condition-nn}
To keep the nearest neighbor's direction independent of the trajectory used to reach it, condition on the nearest neighbor's label and distance but not its direction. Let $y$ be the nearest neighbor of $q$, $j_*$ its label, and $\vartheta$ its distance.
Conditionally on $(j_*,\vartheta)$, the other $n-1$ points $D'$ are independent and uniform
on $\Omega_\vartheta=\mathbb S^{d-1}\setminus C(q,\vartheta)$, the direction $u_y$ of $y$ about $q$ is uniform on
$\mathbb S^{d-2}$ and independent of $D'$, and the entry label is uniform and independent of
both. This conditional independence follows because fixing one point's distance and requiring the other distances to exceed it imposes separate conditions on the independently sampled points. If the entry is $j_*$, its first complete expansion succeeds. Fix
small $\iota,\iota'>0$ with $e^{\iota'd}/n=o(\eps_n)$ and let $\theta_1^\pm$ be defined by $nF(\theta_1^\pm)=e^{\pm\iota'd}$. Then
$\Prob\{\vartheta>\theta_1^+\}=(1-F(\theta_1^+))^n\le\exp(-e^{\iota'd})$ and
$\Prob\{\vartheta<\theta_1^-\}\le nF(\theta_1^-)=e^{-\iota'd}$, so the event
$\mathcal G_0=\{\theta_1^-\le\vartheta\le\theta_1^+\}$ has probability at least $\frac34$ for large $n$.
By \eqref{eq:cap-ratio}, $\sin\theta_1^\pm/\sin\theta_1\to e^{\pm\iota'}$.

After removing the nearest neighbor, we need the remaining neighbor-list radii to be at least $R_-$. For $v\in D'$ let $R_{D'}(v)$ be the distance from $v$ to its $\Delta$-th nearest point of
$D'\setminus\{v\}$, and let $\mathcal G_1$ be the event that $R_{D'}(v)\ge R_-$ for every $v\in D'$.
Given $\vartheta$ and the position of $v$, the number of points of $D'\setminus\{v\}$ in $C(v,R_-)$
is binomial with mean at most $(n-2)F(R_-)/(1-F(\vartheta))\le\Delta(1-\eps_n)(1+2e^{\iota'd}/n)$,
which is at most $\Delta(1-\eps_n/2)$ on $\mathcal G_0$. By \eqref{eq:chernoff} it is at least $\Delta$
with probability at most $e^{-\eps_n^2\Delta/10}$, and a union bound gives
$\Prob(\mathcal G_0\setminus\mathcal G_1)\le ne^{-\sqrt{\Delta\log n}/10}\to0$, since $\Delta\ge400\log n$ by
$\log\log n=o(d)$.

We next show that a run on the data without the nearest neighbor $y$ cannot stop outside a small enlargement of $C(q,\theta_{\mathrm{trap}})$. Let $\theta^\iota=\theta_{\mathrm{trap}}+\iota$ in the logarithmic regime and $\theta^\iota=(L+\iota)\theta_1$ in the
dense regime, and let $\mathcal G_2$ be the event that every $v\in D'$ of distance at least $\theta^\iota$
has a point of $D'$ in $C(v,R_-)$ that is strictly closer to $q$. Let $u_-(\theta)$ be the
function $u$ of \cref{sec:random-entry} for the radius $R_-$, and $\Lambda_-$ its function
$\Lambda$. By \cref{lem:trap-geometry}(a) for $\varrho=R_-$, whose trap distance converges to
that of $\rho$, there is ${c_{\mathrm{prog}}}={c_{\mathrm{prog}}}(\iota)>0$ with $\Lambda_-(\theta^\iota)\ge e^{3{c_{\mathrm{prog}}}}\sin\theta_1$ for
large $n$. At $\theta=\theta^\iota$, \cref{lem:lens}(a) applies to the radius $R_-$, and
$nF(\theta)e^{-(d-1)J_\theta(R_-)}=(\Lambda_-(\theta)/\sin\theta_1)^{d-1}\cos\theta_1/\cos\theta\,(1+O(1/d))$ by
\eqref{eq:cap-ratio}, because $e^{-2J_\theta(R_-)}=\Lambda_-(\theta)^2/\sin^2\theta$. The error term in \cref{lem:lens}(a) is $o(d)$ by \eqref{eq:theta1-dense}. Hence $nu_-(\theta^\iota)\ge e^{2{c_{\mathrm{prog}}}d}$, and since $u_-$ is
nondecreasing (the lens-mass estimate in the proof of \cref{thm:random-entry}), the same
holds for every $\theta\ge\theta^\iota$. Further decrease $\iota'$ so that $\iota'<c_{\mathrm{prog}}$. On $\mathcal G_0$, the part of the lens of a
point $v$ of distance $\varphi_v\ge\theta^\iota$ that lies in $\Omega_\vartheta$ has mass at least
$u_-(\varphi_v)-F(\vartheta)\ge(e^{2{c_{\mathrm{prog}}}d}-e^{\iota'd})/n\ge e^{{c_{\mathrm{prog}}}d}/n$. Given $\vartheta$ and the position
of $v$, the other $n-2$ points of $D'$ are independent and uniform on $\Omega_\vartheta$, so this
part of the lens is empty with probability at most $\exp(-e^{{c_{\mathrm{prog}}}d}/2)$, and a union
bound gives $\Prob(\mathcal G_0\setminus\mathcal G_2)\le n\exp(-e^{{c_{\mathrm{prog}}}d}/2)\to0$.

We can now construct a trajectory independent of the nearest neighbor's direction. Assume that the entry is not $j_*$. The \emph{blind run} follows greedy successors on
$\NN{\Delta}(D')$ from the entry until no listed point is closer to $q$. It scans each neighbor list completely. Let $v_0,\ldots,v_{T_{\mathrm{path}}}$ be its vertices. Distances to $q$ strictly
decrease along it, so it is finite, and $v_{T_{\mathrm{path}}}$ has no point closer to $q$ in its neighbor list in
$\NN{\Delta}(D')$. When $\mathcal G_1$ and $\mathcal G_2$ hold, every vertex $v\in D'$ of distance at least $\theta^\iota$ has a closer point within distance $R_-\le R_{D'}(v)$ and hence in its neighbor list. The blind run therefore cannot stop at such a vertex, so $\varphi_{v_{T_{\mathrm{path}}}}<\theta^\iota$. Let
$\tau=\min\{i\le {T_{\mathrm{path}}}:\ \delta(v_i,y)<R_{D'}(v_i)\}$, with $\tau=\infty$ if there is no such $i$. We
claim that greedy navigation on $\NN{\Delta}(D_n)$ from the same entry follows
$v_0,\ldots,v_{\min\{\tau,{T_{\mathrm{path}}}\}}$ unless it succeeds earlier. If $\tau<\infty$ and the run has not already succeeded, its complete expansion at $v_\tau$ evaluates $y$.
Indeed $D_n=D'\cup\{y\}$. If $\delta(v,y)>R_{D'}(v)$, the $\Delta$ nearest points of $v$ in
$D_n\setminus\{v\}$ are its $\Delta$ nearest points in $D'\setminus\{v\}$, so the two neighbor lists of $v$
coincide, and so do the successors. If $\delta(v,y)<R_{D'}(v)$, then $y$ is among the $\Delta$
nearest points of $v$ in $D_n$, so $y$ is in its neighbor list and is evaluated. Induction on $i$
proves the claim. For
$G_\rho$, a neighbor list on $D_n$ is its list on $D'$ with $y$ added when $y$ lies within
distance $\rho$, so the same comparison holds.

The trajectory comparison turns the event that $y$ lies in the stopping point's neighbor cap into success on the original graph. On $\mathcal G_1$,
\[
  \Prob\{\text{success}\mid D',\text{entry},\vartheta\}\ge\Prob_{u_y}\{\delta(v_{T_{\mathrm{path}}},y)<R_-\}
  =C_{d-1}\bigl(w^{R_-}(\varphi_{v_{T_{\mathrm{path}}}},\vartheta)\bigr),
\]
using the direction probability $C_{d-1}$ from \cref{sec:tolerance}. The stopping point $v_{T_{\mathrm{path}}}$ depends only on $D'$ and the entry, whereas $u_y$ is uniform and independent of them. Write $\varphi=\delta(q,v_{T_{\mathrm{path}}})$ and let $u_v$ be the stopping point's unit tangent direction at $q$. The representations $v_{T_{\mathrm{path}}}=\cos\varphi\,q+\sin\varphi\,u_v$ and $y=\cos\vartheta\,q+\sin\vartheta\,u_y$ give $\delta(v_{T_{\mathrm{path}}},y)<R_-$ if and only if $u_y\cdot u_v>w^{R_-}(\varphi,\vartheta)$. All points of $D'$ are
farther from $q$ than $\vartheta$, so by \cref{lem:trap-geometry}(b), on $\mathcal G_1\cap\mathcal G_2$,
$w^{R_-}(\varphi_{v_{T_{\mathrm{path}}}},\vartheta)\le w^{R_-}(\theta^\iota,\vartheta)=:\bar w_n$. Here $\bar w_n$ depends on the conditioned distance $\vartheta$. Uniformly for $\vartheta\in[\theta_1^-,\theta_1^+]$, convergence of $w^{R_-}$ to the limiting threshold $w$ bounds $\bar w_n$ in the logarithmic regime by $o(1)$ plus the supremum of $w(\theta_{\mathrm{trap}}+\iota,t)$ over $\sin t/\sin\theta_1\in[e^{-\iota'},e^{\iota'}]$, with $\theta_1$ and $\theta_{\mathrm{trap}}$ denoting their angle limits.
In the dense regime, uniformly over the same allowed values of $\vartheta$, $\bar w_n$ is at most $o(1)$ plus the supremum of $w(L+\iota,t)$ over normalized distances $e^{-\iota'}\le t\le e^{\iota'}$.
These suprema tend respectively to $w(\theta_{\mathrm{trap}},\theta_1)$ and $w(L,1)$ as $\iota,\iota'\to0$, and both limits are less than one.
The lower bound of \cref{lem:directions} gives, for $\iota,\iota'$ small and $n$ large, uniformly on $\mathcal G_0$,
$C_{d-1}(\bar w_n)\ge c_0d^{-1/2}(1-(\bar w_n)_+^2)^{(d-2)/2}\ge e^{-d(\eta^*+\eps/2)}$. Therefore
$p_1\ge\Prob(\mathcal G_0\cap\mathcal G_1\cap\mathcal G_2)\,e^{-d(\eta^*+\eps/2)}\ge e^{-d(\eta^*+\eps)}$.

\phantomsection\label{step:short-run}
For the upper bound, apply \cref{thm:local} with $\eps/2$ in place of $\eps$, and use its resulting $b_0$. We first obtain a deterministic expansion budget. Retain $L_n=(\log n)^2$ and reconstruct $A_0,h_{\mathrm{far}},h_{\mathrm{near}},B_n,H_n$ from the proof of \cref{thm:random-entry} with $b_0$ in place of its trap-cap radius $a$. We show that, except with probability
$\varpi_n=ne^{-L_n/2}+4^{-B_n}+2ne^{-\eps_n^2\Delta/4}$, a run from a uniform entry makes at
most $O(dL_n)$ expansions of points of distance in $(b_0,\theta_{\mathrm{trap}}+\rho^+]$, and at most
$H_n=O(d/\theta_1+dL_n)=e^{o(d)}$ expansions in total, up to termination or the first completed expansion that evaluates a point of
the target cap $C(q,b_0)$ of \cref{thm:local}, whichever occurs first. Run the exploration of the proof of
\cref{thm:random-entry}, using its rule for revealing nearest-neighbor lists. We apply its strip and progress estimates with $b_0$ in place of the radius $a$, taking $A_0=\frac14\min\{\rho,b_0\}$. The inequality $b_0<\theta_n$ and the fact that $b_0$ is within fixed factors of $\theta_1$ preserve the radius conditions and the scale of $A_0$ used in those estimates. The property that every point beyond $\theta_n$ has a neighbor improving its query distance by $h_{\mathrm{far}}$ fails with
probability at most $ne^{-L_n/2}$. When it holds, every move from distance at least $\theta_n$ decreases the
distance by at least $h_{\mathrm{far}}=A_0/(4d)$, so at most $(\theta_{\mathrm{trap}}+\rho^+)/h_{\mathrm{far}}+1=O(d)$ such
moves start at distance at most $\theta_{\mathrm{trap}}+\rho^+$, because $A_0$, $\theta_1$, and $\rho^+$ are of the
same order. By the bound on slow moves in the proof of \cref{thm:random-entry}, the moves that start at distance in $(b_0,\theta_n]$
number at most $2B_n=O(dL_n)$ except with probability $4^{-B_n}$. For $\NN{\Delta}$, a neighbor list
radius leaves $[R_-,R_+]$ with probability at most $2ne^{-\eps_n^2\Delta/4}$. All three terms are
smaller than every fixed negative power of $n$, since $\Delta\ge(\log n)^K$ for every fixed positive power when
$\log\log n=o(d)$.

A run has one entry, but its expansion count is random. Use budget $(H_n,1)$ for the actual run from one uniform entry, truncating it after $H_n$ complete expansions or after the first complete expansion that evaluates a point of $C(q,b_0)$, whichever comes first. If the entry already lies in this cap, its full expansion is still completed. We have $H_n\Delta/n=e^{o(d)}\Delta/n\to0$, so $H_n\le n/(8\Delta)$ for all large $n$. The containing caps exposed during these expansions have total surface measure at most $H_nF(\rho^+)=o(1)$.

The auxiliary process in the proof of \cref{thm:local} stops at the first operation that reveals any point of $C(q,b_0)$. This can precede an actual evaluation in the cap, since exposure also reveals points outside the neighbor list. On $\mathcal C$, its scans and moves agree with those of the truncated run until this auxiliary stop. A stop triggered by an expansion completes that entire scan and exposure. A stop triggered by the entry request precedes the entry expansion of the actual run. In either case, the expansions included in $T_{\mathrm{near}}$ form a subset of those of the truncated run, so $\E T_{\mathrm{near}}\le O(dL_n)+n\varpi_n+H_n\Prob(\mathcal C^c)=e^{o(d)}$.

On $\mathcal C$ and $\TopSet_k(q)\subseteq C(q,b_0)$, success of the original run before the budget is exhausted requires the auxiliary process to reveal at least one point of $C(q,b_0)$. Thus the reduction in \eqref{eq:success-reveal} applies to the original run, with an additional error $\varpi_n$ for exhausting the budget before a target-cap evaluation. The proof of \cref{thm:local} bounds $e_0$, the expected number of target-cap points revealed through the auxiliary stop, by the expansion and entry terms. We use that bound with the preceding estimate on $\E T_{\mathrm{near}}$. By \cref{lem:trap-geometry}(c), $\eta_{\mathrm{gen}}=\eta^*$ and, in the logarithmic regime, $\eta_{\mathrm{entry}}>\eta^*$. In the dense regime, the entry term instead satisfies $e^{Cd}/n=e^{-\omega(d)}$.
The error terms $\Prob(\mathcal C^c)$, $\exp(-e^{c'd})$, and $\varpi_n$ are smaller than every fixed negative power of $n$, and hence than $e^{-d(\eta^*+1)}$. Therefore, for all sufficiently large $n$,
\[
  p_1\le e^{o(d)}e^{-d(\eta^*-\eps/2)}+e^{-d(\eta^*-\eps/2)}+\Prob(\mathcal C^c)+\exp(-e^{c'd})+\varpi_n
  \le e^{-d(\eta^*-\eps)}.
\]
Together with the preceding lower bound, this gives $p_1=e^{-d(\eta^*+o(1))}$. Finally,
expressing the terms in the bound of \cref{thm:local} as multiples of $p_1\ge e^{-d(\eta^*+\eps)}$, and absorbing its additive terms into $Ep_1$ since $E\ge1$, gives
a success probability of at most $e^{3\eps d}(T+E)\,p_1$. Since $\eps>0$ is arbitrary, this is
$e^{o(d)}(T+E)\,p_1$.
\end{proof}

\subsection{Independent restarts}\label{sec:restarts}

\Cref{thm:exact-exponent} gives a success probability averaged over the sample and the entry. Restarts reuse one sample, so that average alone does not determine how many attempts suffice. We now show that, with probability $1-e^{-\Omega(d)}$ over the sample, the fraction of successful entries on $G_\rho$ differs from the single-start probability by at most a factor $e^{\pm\eps d}$ for any fixed $\eps>0$. Conditional on the sample, independent attempts then succeed using a number of entries whose leading exponential scale is the reciprocal of the single-run success probability.

\begin{theorem}[Restarts]\label{thm:restarts}
For uniform data in either regime, take $k=1$ in the success rule of \cref{sec:model}, assume that $\Delta$ lies in the near-threshold band defined in \cref{sec:local}, and let $\eps>0$. Thus a run succeeds after its first completed expansion whose evaluated points include the nearest neighbor. There is $c_\eps>0$
such that for all large $n$, every query $q$ fixed before sampling, and every integer $B\ge e^{(\eta^*+\eps)d}$, $B$
independent runs of greedy navigation on $G_\rho$ from uniformly random entries, drawn
with replacement, all fail to evaluate the nearest neighbor of $q$ with probability at
most $e^{-c_\eps d}$ over the data and restart entries. Moreover, with probability at least $1-e^{-c_\eps d}$ over the data, the
fraction $S_D(q)$ of entries from which the run evaluates the nearest neighbor satisfies
$e^{-(\eta^*+\eps)d}\le S_D(q)\le e^{-(\eta^*-\eps)d}$. The constant $c_\eps$ depends only on $\eps$ and on $(c,\beta)$ in the logarithmic regime or $\gamma$ in the dense regime, and the bounds are uniform in $q$.
\end{theorem}

To obtain a lower bound on the fraction of successful entries on one sample, we compare two runs from independent random entries. We restrict the comparison to blind paths with at most $e^{o(d)}$ vertices that stay outside a cap centered at $q$ with radius slightly below $\theta_{\mathrm{trap}}$. In the sample with the nearest neighbor removed, points whose labels are not on a fixed path must avoid the part of each nonterminal point's neighbor cap closer to $q$ than the chosen successor, and all improving locations in the final point's neighbor cap. We bound the probability mass of the union of these regions. The bounds on path length, distance from $q$, and excluded probability mass let us show that the fraction of entries satisfying those bounds and succeeding on the original graph has small variance. Resampling after fixing the first path produces a second sample that is independent of that path. We then compare the second trajectory on the original and resampled samples. The resulting error in the joint success probability of the first and second runs requires both a change in the second trajectory and the hidden nearest neighbor to lie within radius $\rho$ of a vertex on the first path.

\begin{proof}
For the resampling comparison, we retain paths whose vertices stay outside the interior of a cap slightly smaller than $C(q,\theta_{\mathrm{trap}})$. Let $A=\theta_{\mathrm{trap}}-\iota$ in the logarithmic regime and $A=(L-\iota)\theta_1$ in the dense regime, where the positive constants $\iota,\iota',\iota_1$ will be chosen in that order. Write $\nu_{\mathrm{trap}}$ for the limiting exponent of the expected number of points within radius $\theta_{\mathrm{trap}}$, namely $\lim_{n\to\infty}\nu(\theta_{\mathrm{trap}})$ in the logarithmic regime and $\log L$ in the dense regime. The key comparison is $\eta^*<\nu_{\mathrm{trap}}$ from \cref{lem:trap-geometry}(c). For the paths retained in the comparison, we bound the probability that resampling changes the second path by a quantity with decay exponent approaching $\nu_{\mathrm{trap}}$. The joint success error also requires the hidden target direction to make the first path succeed. Conditional on both samples with the target removed and on the entry labels, this direction is independent and uniform, giving a further probability bound with decay exponent approaching $\eta^*$. The sum of the two decay exponents exceeds $2\eta^*$, allowing the joint success error to be smaller than the square of the single-run success probability.

We first delete the nearest neighbor and condition on its label and distance, using the conditioning argument in the proof of \cref{thm:exact-exponent}. Let $j_*$ and $\vartheta$ be the label and distance of the nearest neighbor $y$. To control this distance, define $\theta_1^-$ by $nF(\theta_1^-)=e^{-\iota'd}$ and restrict the analysis to the event $\mathcal T=\{nF(\vartheta)\le\log n,\ \vartheta\ge\theta_1^-\}$, whose complement has probability at most
$1/n+e^{-\iota'd}$. All statements below are conditional on $(j_*,\vartheta)$ satisfying the bounds defining $\mathcal T$, and
all bounds are uniform in such $\vartheta$. For sufficiently small $\iota$ and then $\iota'$, we have $A>\theta_1$ and $w(A,\vartheta)\ge0$ throughout $\mathcal T$. The other points $D'$ are independent with
the uniform law $\mu$ on $\Omega=\mathbb S^{d-1}\setminus C(q,\vartheta)$, the direction $u_y$ is uniform and
independent of $D'$, and the entry labels $e_1,e_2,\ldots$ are independent, uniform, and
independent of $(D',u_y)$. An entry equal to $j_*$ succeeds after its first complete expansion, with probability
$1/n$ per run. We keep uniform sampling over all $n$ labels but count this entry as unsuccessful in the
auxiliary comparison. For $e_i\ne j_*$, let its blind path be the vertex sequence of greedy navigation on $G_\rho(D')$, scanning every list completely and continuing until no listed neighbor is closer to $q$. The actual run $i$ on $G_\rho(D_n)$ follows this path until the
first vertex within distance $\rho$ of $y$, and then evaluates $y$. Its success event is
$S_i=\{u_y\in\bigcup_{x\in\mathrm{path}_i}A_x\}$, where $A_x=\{u:u\cdot u_x\ge w(\varphi_x,\vartheta)\}$, $u_x$ is the direction
of $x$ about $q$, and $\varphi_x$ its distance. For $e_i=j_*$, take the auxiliary path to be empty and $S_i$ to be the empty event.

\phantomsection\label{step:path-event}
We condition on the first blind path and identify the regions that all other points must avoid, up to null cap boundaries, for this to be the specified path. This conditional law will allow us to construct a sample for the second run that is independent of the first path. Let $(\ell_0,\ldots,\ell_{T_{\mathrm{path}}};x_0,\ldots,x_{T_{\mathrm{path}}})$ be the labels and positions of the blind path from $e_1$, with distances $\varphi_i$. Retain the lens $U(x)=C(x,\rho)\cap C(q,\varphi_x)$ from the proof of \cref{thm:random-entry}. The region excluded by the complete path is
\[
 E=\Bigl(\bigcup_{i<T_{\mathrm{path}}}[C(x_i,\rho)\cap C(q,\varphi_{i+1})]\cup U(x_{T_{\mathrm{path}}})\Bigr)\cap\Omega.
\]
For an empty auxiliary path, put $E=\varnothing$. For a nonempty path that can arise from the stated greedy rule, each specified move $i<{T_{\mathrm{path}}}$ selects the point of $D'$ in
$U(x_i)$ closest to $q$. The event that these are the complete blind path labels and positions is therefore $\{X_{\ell_i}=x_i\ \forall i\}\cap\{X_j\notin E\
\forall j\notin\{\ell_0,\ldots,\ell_{T_{\mathrm{path}}}\}\}$. The specified path coordinates must satisfy the greedy successor and stopping conditions among themselves. Conditional on such coordinates, interpreted through their joint density, the only remaining restriction on the other labels is avoidance of $E$. Their positions are therefore independent and uniform on $\Omega\setminus E$.

To bound the probability that resampling changes the second path, we need bounds on the first path's number of vertices and on the probability mass of its excluded region $E$. Let $H_n$ be the bound of the expansion-budget estimate in the proof of \cref{thm:exact-exponent}
and $K_n=3(H_n+1)\log n$. Both are $e^{o(d)}$. We restrict the entry distance to $[\pi/2-\iota_1,\pi/2+\iota_1]$ so that every later vertex has distance at most $\pi/2+\iota_1$, as needed for the direction comparison below. Let $N$ be the event that the path is nonempty, its
entry has distance in $[\pi/2-\iota_1,\pi/2+\iota_1]$, all its vertices have distance at least $A$, it
has at most $H_n$ vertices, and $n\mu(E)\le K_n$. The event $N$ is invariant under
rotations about $q$. Let $N_i$ impose these conditions on run $i$, using its own excluded region, defined by the formula for $E$ with that path's vertices and successors. Take $N_i$ to be the empty event when $e_i=j_*$, and put $S_i^*=S_i\cap N_i$, $p^*=\Prob(S_1^*)$, and $H^*=\Prob(S_1^*\mid D',u_y)$. Thus $p^*$ is the probability that a run both succeeds and satisfies $N$, while $H^*$ is the fraction of all $n$ entry labels whose runs have both properties on the fixed sample and target direction. The label $j_*$ contributes zero to this auxiliary fraction. We will bound the variance of $H^*$ to show that this fraction is close to its mean $p^*$ on most samples.
For the directional probabilities below, retain $\omega$ as normalized surface measure on $\mathbb S^{d-2}$, and interpret $C_{d-1}$ as its cap probability as defined in \cref{sec:tolerance}. On $N_1$, by \cref{lem:trap-geometry}(b) and \cref{lem:directions},
$\omega\bigl(\bigcup_xA_x\bigr)\le H_n\sqrt {d}\,(1-w(A,\vartheta)^2)^{(d-4)/2}\le e^{-(\eta^*-\eps_1)d}$, where
$\eps_1=\eps_1(\iota,\iota')\to0$ as $\iota,\iota'\to0$. We will also prove $p^*\ge e^{-(\eta^*+\eps_1)d}$, after enlarging
$\eps_1(\iota,\iota')$ by a quantity tending to zero if necessary. Choose the fixed margins small enough that $\eps_1<\eps/2$ and $3\eps_1<\nu_{\mathrm{trap}}-\eta^*$, leaving a positive gap for the resampling error estimated below. We give the
probability estimates for the event $N_1$ explicitly.
To control the probability mass of each part of $E$, fix a label $j\ne j_*$ and a position $x=X_j$. For $0\le t\le\delta(q,x)$, put $B_x(t)=C(x,\rho)\cap\Omega\cap\{z:\delta(q,z)<t\}$ and $b_x(t)=\mu(B_x(t))$.
The function $b_x$ is continuous. Set $u_n=3\log n/n$. If
$b_x(\delta(q,x))>u_n$, choose a deterministic $t_x<\delta(q,x)$ with
$b_x(t_x)=u_n$. If $x$ has no closer neighbor, or if its neighbor closest to $q$
has distance $t$ with $b_x(t)>u_n$, all the other $n-2$ labels avoid $B_x(t_x)$.
Conditionally on $X_j=x$ these labels are independent with law $\mu$, so this
has probability at most $(1-u_n)^{n-2}$. A union bound over $j$ gives at most
$(n-1)(1-u_n)^{n-2}\le2/n^2$ for large $n$. Off this event every constituent
set in the definition of $E$ has $\mu$-mass at most $u_n$, simultaneously for
every blind path. Boundaries have zero $\mu$-mass and do not affect this argument.
Consequently, a path with at most $H_n$ vertices has $n\mu(E)\le K_n$.

To control the path length, apply the expansion-budget estimate in the proof of \cref{thm:exact-exponent} up to the first complete scan that evaluates a point of distance below $A$. If that scan reveals such a point, the closest listed neighbor is also below $A$, so the next greedy vertex lies there. The triggering scan is included in the budget. Apply that expansion-budget argument first to $n-1$ independent uniform points. Reducing the number of labels by one preserves their strict margins, so under this uniform law the probability of exceeding the expansion budget is smaller than every fixed negative power of $n$. On $\mathcal T$ the density of $\mu^{\otimes(n-1)}$ relative to
$\sigma_{d-1}^{\otimes(n-1)}$ is at most
$(1-\log n/n)^{-(n-1)}\le n^2$. Multiplying the failure probability under uniform sampling by $n^2$ shows that the probability of exceeding the budget is still smaller than every fixed negative power of $n$ under $\mu$.
To show that the first blind path stays outside $C(q,A)$, expose the complete cap $C(x,\rho)$ at each expansion of its current vertex $x$ on $G_\rho(D')$. Stop after $H_n$ expansions or at its first vertex below $A$. Before a crossing, the expanded vertex has query distance at least $A$, and the revealed mass is at most $H_nF(\rho)=o(1)$.
Conditionally on this history, the expected number of fresh points of distance
below $A$ in the next neighbor list is at most
\[
 2n\sigma_{d-1}(C(x,\rho)\cap C(q,A))
 \le \exp\{d\chi(A,A)+o(d)\}.
\]
For this inequality, the lens mass is largest when the current point has distance $A$, among current distances at least $A$. At that distance, the shell estimate is maximized over candidate distances $\psi\le A$ at $\psi=A$. Where $w(A,\psi)\ge0$, differentiation of $e^{2g(A\to\psi)}$ has the sign of $\cos\psi-\cos\rho\cos A>0$ for $0<\psi\le A<\pi/2$. Where $w(A,\psi)<0$, we have $g(A\to\psi)=\nu(\psi)$, which is increasing as well. The resulting exponent is $g(A\to A)=\chi(A,A)<0$.
Every strict greedy successor is fresh, as in the conditional exposure argument in the proof of \cref{thm:random-entry}. A union
bound over $H_n=e^{o(d)}$ expansions shows that a crossing below $A$ has
probability $o(1)$. The entry itself lies in
$[\pi/2-\iota_1,\pi/2+\iota_1]$ with probability $1-o(1)$ under $\mu$.
Together these estimates give $\Prob(N_1^c)=o(1)$, uniformly on $\mathcal T$.
On the event $\mathcal G_2$ from the proof of \cref{thm:exact-exponent}, every point of $D'$ at distance at least $\theta^\iota$ has a closer neighbor within $R_-$.
Since $R_-<\rho$, the blind run on $G_\rho(D')$ stops at distance less than $\theta^\iota$. The event $\mathcal G_2$ has conditional probability $1-o(1)$ uniformly on $\mathcal T$.
Both $N_1$ and $\mathcal G_2$ depend on $D'$ and the entry, not on $u_y$. The hidden-target direction argument in the proof of \cref{thm:exact-exponent} therefore applies on their intersection. It gives
$p^*\ge(1-o(1))\exp\{-d(\eta^*+o_{\iota,\iota'}(1)+o_n(1))\}$,
which proves the lower bound on $p^*$.

The conditional law of the positions of labels not on path 1 permits a resampling whose law does not depend on that path. Given path 1, let $L_1$ be its label set. Draw fresh independent points $Y_j\sim\mu$ for
all labels $j\ne j_*$, and put $X_j''=Y_j$ if $j\in L_1$ or $Y_j\in E$, and $X_j''=X_j$ otherwise. Call $j\notin L_1$ with $Y_j\in E$ \emph{moved}. By the conditional law given path 1, the old positions $X_j$ for $j\notin L_1$ are independent with law $\mu$ when $\mu(E)=0$, and none of these labels is moved. When $\mu(E)>0$, each $X_j''$ with $j\notin L_1$ has law $\mu(E)\,\mathrm{Unif}(E)+(1-\mu(E))\,\mathrm{Unif}(\Omega\setminus E)=\mu$. In both cases, the positions $X_j''$ for $j\notin L_1$ are independent with law $\mu$. Thus $D''=(X_j'')$ has conditional law $\mu^{\otimes(n-1)}$ regardless of path 1. Consequently, path 1, the pair $(D'',e_2)$, and $u_y$ are mutually independent. Given path 1 and $D''$, the old positions
$X_j$ of the moved labels are independent and uniform on $\Omega\setminus E$. Let run $2''$ be
the blind run from $e_2$ on $G_\rho(D'')$. Define $S_2''^*$ by the same success-and-path-restriction rule as $S_2^*$, using $D''$ and the same $u_y$. We will show that
\[
  \Prob(S_1^*\cap S_2''^*)=p^{*2}\qquad\text{exactly.}
\]
For path 1, let $A_1^*=\bigcup_{x\in\mathrm{path}_1}A_x$ when $N_1$ holds and $A_1^*=\varnothing$ otherwise. Define $A_2''^*$ by the same rule for path $2''$ and its event $N$. Then $S_1^*=\{u_y\in A_1^*\}$ and $S_2''^*=\{u_y\in A_2''^*\}$.
Rotating $D''$ about $q$ rotates the second path and its direction set $A_2''^*$ by the same rotation, while leaving its entry label $e_2$ fixed. Since the law of $(D'',e_2)$ is rotation invariant, $\Prob\{u\in A_2''^*\}=p^*$ for
every direction $u$, and by independence
$\Prob\{u_y\in A_1^*\cap A_2''^*\}=\E\int_{A_1^*}\Prob\{u\in A_2''^*\}\,d\omega(u)=p^*\,\E\omega(A_1^*)=p^{*2}$.

It remains to bound how often this resampling changes the second trajectory. The data sets $D'$ and $D''$ agree outside the set $\mathfrak B=L_1\cup\{\text{moved labels}\}$ of labels whose positions were replaced. Let
$P_+''$ be the set of vertices of path $2''$ among its first $H_n+1$ that precede its first
vertex below $A$, and $P_+''=\emptyset$ unless the entry of path $2''$ has distance in
$[\pi/2-\iota_1,\pi/2+\iota_1]$. Rotating $D''$ about $q$ rotates this vertex set by the same rotation.
If $e_2\notin\mathfrak B$, the blind runs $2$ on $G_\rho(D')$ and $2''$ on $G_\rho(D'')$ start at the same point. At a common vertex $x$ whose label is outside $\mathfrak B$, the two lists have the same unchanged points.
Let $s'$ and $s''$ denote the successors on $D'$ and $D''$, respectively, whenever the corresponding run can move from $x$. If both exist and neither label is in $\mathfrak B$, then $s'=s''$.
If only one exists, its label must lie in $\mathfrak B$. A first divergence, including a difference in whether the runs stop, must therefore involve a replaced label. Before that divergence, $N_2$ keeps the common vertices at distance at least $A$ and within the first $H_n$ vertices, so the common expanded vertex belongs to $P_+''$. A replaced successor in the new sample lies either in $L_1$ or in $E$, and a replaced successor in the old sample is either a path-1 point or the old position of a moved label. On $N_1\cap N_2$, a first divergence is covered by one of the following five cases. (a) $e_2\in\mathfrak B$. (b1) Some $x\in P_+''$ has its $D''$-successor in $L_1$. (b2) Some $x\in P_+''$ has a moved $D''$-successor, hence a
successor in $E$. (c1) Some $x\in P_+''$ is within distance $\rho$ of a vertex of path 1 of distance at least $A$. (c2) Some $x\in P_+''$ is within distance $\rho$ of the old position of a moved label.

When path 1 satisfies the event $N_1$, its length and excluded mass bound all five ways the second trajectory can change. The total probability of cases (a) and (b1) is at most $(H_n+1)^2/(n-1)+K_n/n$, because
$L_1$ is independent of $(D'',e_2)$, labels are exchangeable, and a label is moved with
probability $\mu(E)\le K_n/n$. Case (c2) has probability at most $(H_n+1)K_n\cdot2F(\rho)$, because the old positions are uniform on $\Omega\setminus E$. To bound case (c1), use the independence of path 1 and $P_+''$. The rotation symmetry just described makes the direction of each vertex of $P_+''$ uniform conditional on its distance from $q$. The expected number of pairs within distance $\rho$ is
$\E\sum_i\sum_xC_{d-1}(w(\varphi_i,\varphi_x))\le(H_n+1)^2\sqrt {d}\,(1-w(A,A)^2)^{(d-4)/2}$.
To justify this bound, we show $w(\varphi,\psi)\ge w(A,A)$ for $A\le\psi\le\varphi\le\pi/2+\iota_1$. If $\psi<\pi/2$,
\cref{lem:trap-geometry}(b) gives $w(\varphi,\psi)\ge w(\psi,\psi)$, and $w(\psi,\psi)=1-(1-\cos\rho)/\sin^2\psi\ge
w(A,A)$. If $\pi/2\le\psi\le\varphi\le\pi/2+\iota_1$, then $0\le\cos\varphi\cos\psi\le\sin^2\iota_1$, so
$w(\varphi,\psi)\ge\cos\rho-\sin^2\iota_1$. The quantity $(1-\cos\rho)\cot^2A$ has a positive limit: in the dense regime it tends to $M^2/(2(L-\iota)^2)$, and in the logarithmic regime both angles have positive limits below $\pi/2$. We may therefore choose a fixed $\iota_1>0$ so that $\sin^2\iota_1\le(1-\cos\rho)\cot^2A$ for all sufficiently large $n$. This makes the preceding
bound at least $w(A,A)=\cos\rho-(1-\cos\rho)\cot^2A$. As $\iota\to0$, $-\frac12\log(1-w(A,A)^2)\to\nu_{\mathrm{trap}}$ by the last
calculation in the proof of \cref{lem:trap-geometry}. For case (b2), use $E\subseteq\bigcup_iC(x_i,\rho)$ and consider the moved
successor $s''$ that causes the first divergence. On $N_2$, the common expanded vertex $x$ belongs to path 2 among its at most $H_n$ vertices. Before this divergence the two paths coincide, so $s''$ is
among the first $H_n+1$ vertices of path $2''$. If $s''$ has
distance at least $A$, it belongs to $P_+''$ and this is case (c1). Otherwise it is
the first vertex of path $2''$ below $A$, reached within its first $H_n+2$
vertices. Write $Z$ for this first landing point, leaving $Z$ undefined if
no such landing occurs within that budget or if the entry distance of run $2''$ lies outside
$[\pi/2-\iota_1,\pi/2+\iota_1]$.

We now bound the probability that this first landing point $Z$ lies within distance $\rho$ of path 1, which is the remaining part of case (b2). Fix $\iota$ first and choose $\iota'$ small enough that $\chi(A,\psi)\le0$
for every $\psi\in[\theta_1^-,A]$, uniformly for large $n$. Such a choice is
possible because \cref{lem:trap-geometry}(a) gives the strict bound
$\chi(A,\psi)\le\chi(A,A)<0$ on $[\theta_1,A]$, and the functions are
continuous uniformly in normalized distances. Fix
$K>2\eta^*+\nu_{\mathrm{trap}}+2$.
Partition $[\vartheta,A)$ into intervals of mesh at most $1/d$ in the logarithmic regime, and $\theta_1/d$ in the dense regime. There are $O(d)$ intervals, uniformly on $\mathcal T$. For each interval $I$, set $G_I=\sup_{\psi\in I}\max\{g(A\to\psi),-K\}$ and $P_I=\sup_{\psi\in I}\max\{p(A,\psi),-K\}$. These control the probability of a landing in $I$ and its directional probability of lying near path 1.
The exploration of run $2''$ that reveals complete neighbor caps, restricted to its first $H_n+2$ vertices, leaves independent points uniform on the complement of revealed
caps in $\Omega$. That complement has $\sigma_{d-1}$-mass $1-o(1)$.
For each $\psi\in I$, $g(\varphi\to\psi)$ is nonincreasing in the current distance $\varphi\ge A$. The bound for one revealed distance interval in the proof of \cref{thm:local} therefore gives
\[
 \Prob\{Z\text{ is defined},\ \delta(q,Z)\in I\}
 \le 2(H_n+1)\exp\{dG_I+o(d)\}.
\]
This estimate is taken over $(D'',e_2)$, which is independent of path 1 and rotation invariant about $q$. Rotating $D''$ about $q$ rotates the path and its first landing point $Z$ by the same rotation. Consequently its direction is uniform
conditional on its distance and on being defined, and it is independent of
path 1. For every fixed vertex $x_i$ of path 1 of distance at least $A$, the
one-coordinate direction bound from the proof of \cref{lem:directions} applies with the distance of $x_i$ replaced by $A$, since $p(\varphi_i,\psi)\le p(A,\psi)$ for $\varphi_i\ge A$ and $\psi\in I$. Combining it with the preceding landing probability gives
\[
 \Prob\{Z\text{ is defined},\ \delta(q,Z)\in I,\ \delta(Z,x_i)\le\rho\mid
                  \text{path 1}\}
 \le 2(H_n+1)\exp\{d(G_I+P_I)+o(d)\}.
\]
All error terms here are uniform over the intervals and over
values of $\vartheta$ satisfying the bounds defining $\mathcal T$.
For the relevant landing distances $\psi\in[\vartheta,A)\subseteq[\theta_1^-,A]$, the choice of $\iota'$ gives $\chi(A,\psi)\le0$. Since $\nu(\psi)\le\nu(A)$, both $p(A,\psi)$ and $g(A\to\psi)$ are nonpositive, and
\[
 g(A\to\psi)+p(A,\psi)=2\chi(A,\psi)-\nu(A)\le-\nu(A).
\]
If either term is truncated, the sum of the truncated terms is at most
$-K< -\nu(A)$ for large $n$. Uniform continuity of the two truncated
functions for $\psi\in[\theta_1^-,A]$, using $\psi/\theta_1$ in the dense regime, now gives
$G_I+P_I\le-\nu(A)+o(1)$ for this shrinking mesh. Summing over the $O(d)$
intervals and at most $H_n$ vertices of path 1 yields
a bound $\exp\{-d\nu(A)+o(d)\}$ on the probability that $Z$ is defined and lies within distance $\rho$ of any vertex of path 1, since $H_n=e^{o(d)}$. The bounds for cases (a) and (b1) sum to at most $e^{o(d)}/n$, and case (c2) contributes at most $e^{o(d)}F(\rho)$, because $H_n=e^{o(d)}$ and $K_n=e^{o(d)}$. In the logarithmic regime, $w(\psi,\psi)\le\cos\rho$ gives $\nu_{\mathrm{trap}}\le-\log\sin\rho=\lambda-\beta/c$ in the limit. Together with $F(\rho)=\Delta(1+o(1))/n$, this bounds the contributions of (a), (b1), and (c2) by $e^{-(\nu_{\mathrm{trap}}-o(1))d}$. In the dense regime, both $e^{o(d)}/n$ and $e^{o(d)}F(\rho)$ are $e^{-\omega(d)}$. Combining these bounds with the estimates for (b2) and (c1) gives
\[
  \Prob\bigl(N_1\cap N_2\cap\{\text{path }2\ne\text{path }2''\}\bigr)\le e^{-(\nu_{\mathrm{trap}}-\eps_2)d},
\]
with $\eps_2=\eps_2(\iota)\to0$ as $\iota\to0$, for large $n$. We choose $\iota$ and then $\iota'$ sufficiently small that $3\eps_1+\eps_2<\nu_{\mathrm{trap}}-\eta^*$, in addition to the preceding requirement $\eps_1<\eps/2$. The fixed positive gap $\nu_{\mathrm{trap}}-\eta^*$ permits both conditions.

If path $2$ and path $2''$ coincide, then $S_2^*=S_2''^*$. The independent resampling comparison $\Prob(S_1^*\cap S_2''^*)=p^{*2}$ therefore gives $\Prob(S_1^*\cap S_2^*)\le p^{*2}+\Prob(S_1^*\cap N_2\cap\{\text{path }2\ne\text{path }2''\})$. To bound this error term, condition first on $(D',D'',e_1,e_2)$. The event $N_1\cap N_2\cap\{\text{path }2\ne\text{path }2''\}$ is then determined, while $u_y$ remains independent and uniform. On $N_1$, the bound on $\omega(\bigcup_x A_x)$ gives probability at most $e^{-(\eta^*-\eps_1)d}$ that this target direction makes path $1$ succeed. Combining this directional probability with the preceding bound on trajectory disagreement bounds the error term by $e^{-(\eta^*-\eps_1)d}e^{-(\nu_{\mathrm{trap}}-\eps_2)d}$. The two entry labels are independent given $(D',u_y)$, so
\[
  \mathrm{Var}(H^*)=\Prob(S_1^*\cap S_2^*)-p^{*2}\le e^{-(\eta^*+\nu_{\mathrm{trap}}-\eps_1-\eps_2)d}
  \le e^{-{c_{\mathrm{var}}} d}p^{*2},
\]
with ${c_{\mathrm{var}}}=\nu_{\mathrm{trap}}-\eta^*-3\eps_1-\eps_2>0$ once $\iota,\iota'$ are small, by the lower bound on $p^*$.
By Chebyshev's inequality, $\Prob\{H^*\le p^*/2\}\le4e^{-{c_{\mathrm{var}}} d}$. Given $(D',u_y)$ the runs are
independent and each succeeds with probability at least $H^*$, so all $B$ runs fail with
probability at most $\Prob(\mathcal T^c)+4e^{-{c_{\mathrm{var}}} d}+\exp(-Bp^*/2)$, and $Bp^*\ge e^{(\eps-\eps_1)d}$. This
proves the first statement with $c_\eps=\frac12\min\{{c_{\mathrm{var}}},\iota',\eps/2\}$. For the second statement, every entry counted by $H^*$ also succeeds in the original run, so
$S_D(q)\ge H^*\ge p^*/3\ge e^{-(\eta^*+\eps)d}$ off an event of probability at most $\Prob(\mathcal T^c)+4e^{-{c_{\mathrm{var}}} d}$. Applying \cref{thm:exact-exponent} with $k=1$ gives $\E S_D(q)\le e^{-(\eta^*-\eps/2)d}$. Hence
$S_D(q)\le e^{-(\eta^*-\eps)d}$ except with probability $e^{-\eps d/2}$, by Markov's inequality.
\end{proof}

\subsection{Evaluation cost and the benefit of fresh starts}\label{sec:tradeoff}

To complete \cref{thm:fresh-start-cost}, we must convert the success and expansion bounds into bounds on distinct evaluations. We first bound the total cost of independent restarts on $G_{r_\gamma}$, including failed attempts and the final complete expansion. For the lower bound, we combine the local exploration probability bound of \cref{thm:local} with a bound on overlaps between scanned neighbor lists. This gives a lower bound on distinct evaluations even when all attempts share a cache.

\begin{corollary}[Evaluation cost of independent restarts]\label{cor:tradeoff}
For uniform data in the dense regime, let $\Delta=\lceil e^{\gamma d}\rceil$ lie in the near-threshold band,
$\frac12\log\frac43\le\gamma<\gc$, and let $\eps>0$.
On $G_{r_\gamma}$, take $k=1$ in the success rule as in \cref{thm:restarts} and let
 $B=\lceil e^{(\eta^*+\eps)d}\rceil$ independent runs start from uniformly random
 entries. With probability $1-e^{-\Omega(d)}$ over the data and the $B$ entry labels, they evaluate the nearest
 neighbor of a fixed query, and the total number of distance evaluations
 performed by the runs is at most
 $e^{(\gamma+\eta^*(\gamma)+\eps+o(1))d}$.
 The cost includes every attempt and its final complete expansion, whether that attempt succeeds or stops unsuccessfully.

The achieved exponent $\gamma+\eta^*(\gamma)$ is minimized over this band at $e^{2\gamma}=(12+2\sqrt3)/11$, where its value is approximately $0.243$.
\end{corollary}

\begin{proof}
The success assertion follows from \cref{thm:restarts} with $\eps/2$.
For the evaluation cost, we bound the number of complete expansions made from every entry until success or termination. Fix the query, put $a=\theta_1/2$, and retain $L_n=(\log n)^2$. By \eqref{eq:cap-ratio}, $nF(a)=\exp\{(-\log2+o(1))d\}$.
Thus, except with probability $e^{-(\log2+o(1))d}$, the data contain no point
of $C(q,a)$. Apply \eqref{eq:strip} and the bounds on moves above $\theta_n$ and slow moves in $[a,\theta_n]$ from the proof of \cref{thm:random-entry}, using this $a$.
That proof requires $a$ to be within fixed factors of $\theta_1$ and to satisfy $nu(a)<2L_n$. Both conditions hold because $a$ is a fixed multiple of
$\theta_1$ and $nu(a)\le nF(a)<2L_n$. The monotonicity of $u$ and its growth above the trap boundary also give a fixed query distance below $\pi/2$
at which $nu$ exceeds $2L_n$ for all large $n$.
With $A_0=\frac14\min\{r_\gamma,a\}$, those arguments give an expansion budget $H_n=O(d/\theta_1+dL_n)=e^{o(d)}$ and a non-slow-move budget $B_n=\Theta(dL_n)$. A run from a uniform entry, stopped on entering $C(q,a)$, exceeds $H_n$ expansions with probability at most $\varpi_n=ne^{-L_n/2}+4^{-B_n}$.
The same bound holds for each fixed entry label by exchangeability.
A union bound over the $n$ labels gives failure probability at most $n\varpi_n=e^{-\omega(d)}$.
When $C(q,a)$ is empty, stopping at this cap is impossible. The expansion bound therefore applies to every complete run, including its final expansion, on a data event of probability $1-e^{-\Omega(d)}$.
We next bound the number of points evaluated by one complete expansion. Conditional on one source, its neighbor-list length is binomial with mean at
most $e^{\gamma d}$. Chernoff and a union over the labels give
$\max_x d_{G_{r_\gamma}}(x)\le2e^{\gamma d}$ outside an event of probability
$n\exp\{-c e^{\gamma d}\}=e^{-\omega(d)}$ for some fixed $c>0$.
Each run therefore performs at most
$1+2H_ne^{\gamma d}=e^{(\gamma+o(1))d}$ evaluations.
Summing over the $B$ runs proves the cost bound, even with separate caches. A shared cache can only reduce this total.

To optimize the achieved cost over the degree range, write $E_\gamma=e^{2\gamma}$ and express the cost exponent as
\[
 \begin{gathered}
 w(L,1)=\frac{(2-E_\gamma)(3E_\gamma-2)}{4E_\gamma\sqrt{E_\gamma-1}},\quad
 \text{and}\quad\mathcal F(E_\gamma)=\tfrac12\log\frac{E_\gamma}{1-w(L,1)^2}.
 \end{gathered}
\]
Here $w(L,1)$ is the direct-jump directional threshold from \cref{lem:trap-geometry}, so $\mathcal F(E_\gamma)=\gamma+\eta^*(\gamma)$.
Differentiation gives
\[
 \mathcal F'(E_\gamma)=\frac{(11E_\gamma^2-24E_\gamma+12)(5E_\gamma^2-8E_\gamma+4)}{32E_\gamma^3(E_\gamma-1)^2(1-w(L,1)^2)}.
\]
On $[4/3,2]$, the second numerator factor and the denominator are positive, so $\mathcal F'$ has the sign of $11E_\gamma^2-24E_\gamma+12$. The only root in the band is $(12+2\sqrt3)/11$, where
the sign changes from negative to positive. Substitution gives the exponent $0.243$ to three
decimal places.
\end{proof}

The operation lower bound from \cref{thm:exact-exponent} counts expansions and entry requests. To obtain a matching lower bound on distinct evaluations with a shared cache, we must show that the union of the scanned lists contains sufficiently many distinct points. The next lemma bounds this union simultaneously for every sufficiently small set of expanded points, including sets chosen adaptively.

\begin{lemma}[Union of scanned neighbor lists]\label{lem:small-row-union}
For uniform data in the dense regime, fix $\frac12\log(4/3)\le\gamma<\gc$ and let $\Delta=\lceil e^{\gamma d}\rceil$. Define $J_{\rm ov}(\gamma)=-\frac12\log(1-e^{-2\gamma}/4)$. For two caps of radius $r_\gamma$ whose centers are separated by $\theta_1$, $J_{\rm ov}(\gamma)$ is the limiting loss exponent for the surface measure of their intersection divided by $F(r_\gamma)$.
For every fixed $0<\tau<J_{\rm ov}(\gamma)$ there is $c>0$ such that,
for all sufficiently large $n$, a Borel data event $\mathcal E_n$ has
probability at least $1-e^{-cd^2}$. On this event, simultaneously for
$G=G_{r_\gamma}$ and $G=\NN{\Delta}$ and every set $A\subseteq D_n$ with
$1\le |A|\le\lfloor e^{\tau d}\rfloor$,
\begin{equation}\label{eq:small-row-union}
 \left|\bigcup_{x\in A}G(x)\right|\ge\frac{\Delta |A|}{4d}.
\end{equation}
The event $\mathcal E_n$ depends only on the sample. The constant depends only on $(\gamma,\tau)$. The sufficiently-large-$n$
threshold also depends on the given dense-regime sequence.
\end{lemma}

\begin{proof}

We separate the sources at a distance $a\theta_1$ small enough that at most $d$ sample points lie near any source. At this separation, the exponent $j$ governing pairwise cap intersections will satisfy $j>\tau$, where $|A|\le e^{\tau d}$. Put $E_\gamma=e^{2\gamma}$ and choose $j=(\tau+J_{\rm ov}(\gamma))/2$, $a=2e^\gamma\sqrt{1-e^{-2j}}$, and $\ell=-\log a$.
Then $0<a<1$, $\ell>0$, and
$-\frac12\log(1-a^2/(4E_\gamma))=j$.
Moreover, $\gamma-J_{\rm ov}(\gamma)=\tfrac12\log(E_\gamma-1/4)>0$, so the thresholds for neighbor-list intersections have positive fixed exponents. The gap $j-\tau$ leaves slack in those thresholds, while $\ell$ controls the expected number of points in the smaller caps. These constants depend only on $\gamma$ and $\tau$.

To enclose the lists of both graph types in caps of one radius, use $\rho^+$ defined by $nF(\rho^+)=2\Delta$ before \cref{thm:local}. Equations \eqref{eq:cap-ratio} and \eqref{eq:radii} give
$\rho^+/\theta_1\to e^\gamma$ and $r_\gamma/\theta_1\to e^\gamma$.
To compare these caps while retaining the exact trigonometric radii, we use the bounds
\begin{equation}\label{eq:row-union-cap-bounds}
 \frac{\alpha_{d-1}}{d-1}\sin^{d-1} t
 \le F(t)\le\frac{\alpha_{d-1}}{(d-1)\cos t}\sin^{d-1} t
 \qquad(0<t<\pi/2)
\end{equation}
obtained by integrating $(\sin^{d-1} t)'=(d-1)\sin^{d-2}t\cos t$.
In particular, $nF(a\theta_1)=\exp\{(-\ell+o(1))d\}$.
The errors in these radius limits and cap estimates depend only on $\gamma$, $\tau$, and the given dimension sequence, uniformly over the cap centers.

Conditional on $X_i$, the number $Z_i$ of other sample points in
$C(X_i,a\theta_1)$ is binomial and, eventually, has mean at most
$e^{-\ell d/2}$. For $d\ge5$, the elementary binomial tail bound gives
\[
 \Prob\{Z_i\ge d-1\}
 \le\left(\frac{e\,e^{-\ell d/2}}{d-1}\right)^{d-1}
 \le e^{-\ell d^2/4}.
\]
Since $\log n=o(d^2)$, a union bound over all labels gives failure probability at most $e^{-\ell d^2/8}$ for all sufficiently large $n$. Thus every such cap,
including its centre, contains at most $d$ sample points on a common
data event.

Chernoff bounds and a union over labels give, outside an event of
probability at most $2n e^{-\Delta/20}$ for all large $n$,
\[
 |G_{r_\gamma}(X_i)|\ge\Delta/2,
 \qquad\text{and}\qquad
 \#\bigl((D_n\setminus\{X_i\})\cap C(X_i,\rho^+)\bigr)\ge\Delta
 \quad\hbox{for every }i.
\]
Thus both graph families have neighbor-list lengths at least $\Delta/2$,
and all their neighbor lists are contained in the corresponding radius-$\rho^+$ caps.
For $\NN{\Delta}$, the lower bound on the number of points in $C(X_i,\rho^+)$ ensures that its $\Delta$ nearest neighbors lie in that cap.

For two centres $x,y$ with separation $s=\delta(x,y)\ge a\theta_1$, the
intersection of their radius-$\rho^+$ caps is empty if $s>2\rho^+$.
Otherwise use the midpoint direction $z=(x+y)/\|x+y\|$. Every point $u$ in the intersection obeys $u\cdot z\ge\cos \rho^+/\cos(s/2)$.
Consequently the intersection is contained in a cap about $z$ whose
radius is at most $v_n$, where
$v_n=\arccos(\cos\rho^+/\cos(a\theta_1/2))$. The ratio of squared sines satisfies
\[
 \frac{\sin^2v_n}{\sin^2\rho^+}=\frac{1-\sin^2(a\theta_1/2)/\sin^2\rho^+}{1-\sin^2(a\theta_1/2)}\longrightarrow1-\frac{a^2}{4E_\gamma}=e^{-2j}.
\]
For large $n$ this defines $0<v_n<\rho^+$. By
\eqref{eq:row-union-cap-bounds}, uniformly over all such pairs,
\[
 n\sigma_{d-1}(C(x,\rho^+)\cap C(y,\rho^+))
 \le nF(v_n)=\exp\{(\gamma-j+o(1))d\}.
\]
The same midpoint coordinates also verify the exact exponent $J_{\rm ov}(\gamma)$ stated in the lemma. For caps of radius $r_\gamma$ at separation $\theta_1$, their two center directions have coordinates $(\cos(\theta_1/2),\pm\sin(\theta_1/2))$. The point of minimum norm in the projected intersection is $(\cos r_\gamma/\cos(\theta_1/2),0)$. As in the proof of \cref{lem:lens}, place a parallelogram at this point with sides in the two center directions, each of length $(1-\cos^2r_\gamma/\cos^2(\theta_1/2))/(5d)$. Both cap constraints increase along either side, and throughout the parallelogram the remaining squared radius in the unit disk is at least $(1-1/d)(1-\cos^2r_\gamma/\cos^2(\theta_1/2))$. Integrating the two-coordinate density over this parallelogram and combining the result with the midpoint-cap upper bound gives
\[
 \log\sigma_{d-1}(C(x,r_\gamma)\cap C(y,r_\gamma))
 =\frac d2\log\left(1-\frac{\cos^2r_\gamma}{\cos^2(\theta_1/2)}\right)
 +O\bigl(\log d+\log(1/\theta_1)\bigr)
 \quad\bigl(\delta(x,y)=\theta_1\bigr).
\]
Here the parallelogram area is its squared side length times $\sin\theta_1$, so its logarithm contributes only $O(\log d+\log(1/\theta_1))$. Dividing by $F(r_\gamma)$ using \cref{lem:cap}, and using $r_\gamma/\theta_1\to e^\gamma$ and $\log(1/\theta_1)=o(d)$, yields
\[
 \frac1d\log\frac{\sigma_{d-1}(C(x,r_\gamma)\cap C(y,r_\gamma))}{F(r_\gamma)}
 \longrightarrow \frac12\log(1-e^{-2\gamma}/4)=-J_{\rm ov}(\gamma).
\]
We now use the uniform upper bound for pairs separated by at least $a\theta_1$ to control their sampled common neighbors.
After conditioning on $X_i,X_h$, the other $n-2$ points are independent
and uniform. Choose $n$ large enough that the expected number of remaining points in the intersection is at most $e^{(\gamma-\frac{3j+\tau}{4})d}$. We bound each pairwise overlap by the larger threshold $U_n=\lceil e^{(\gamma-\frac{j+\tau}{2})d}\rceil$.
The same elementary binomial upper-tail bound used for $Z_i$ bounds the probability that the number of such points exceeds $U_n$ by
\[
 \exp\{-\frac{j-\tau}{8}d\,e^{(\gamma-\frac{j+\tau}{2})d}\}
\]
for all sufficiently large $n$, uniformly in the conditioned centres.
The two centre labels cannot be in the intersection of their two neighbor lists,
since each neighbor list excludes its own source. On the event that every list lies in its radius-$\rho^+$ cap, a union over fewer than $n^2$ pairs therefore gives, simultaneously for both graph families,
\[
 |G(X_i)\cap G(X_h)|\le U_n
 \quad\hbox{whenever }\delta(X_i,X_h)\ge a\theta_1,
\]
outside an event of probability at most
$n^2\exp\{-\frac{j-\tau}{8}d e^{(\gamma-\frac{j+\tau}{2})d}\}$.
Let $\mathcal E_n$ be the intersection of the properties just proved for the number of nearby sample points, the lower bounds on list lengths and their containment in radius-$\rho^+$ caps, and the upper bounds on pairwise list intersections.

On this common data event, greedily select sources from $A$, discarding after
each selection all remaining sources within distance $a\theta_1$.
Each selection removes at most $d$ points in total, including the selected point. The resulting set $A'$ has
$K=|A'|\ge |A|/d$, has pairwise separation greater than $a\theta_1$,
and satisfies $K\le e^{\tau d}$. Inclusion--exclusion through pairs gives
\[
 \left|\bigcup_{x\in A}G(x)\right|
 \ge \frac{K\Delta}{2}-\binom K2U_n.
\]
Since $j>\tau$, we have $KU_n/\Delta\le2e^{-(j-\tau)d/2}\le1/2$
for all sufficiently large $n$. Substituting $KU_n/\Delta\le1/2$ into the inclusion--exclusion bound gives \eqref{eq:small-row-union} simultaneously for every set $A$, because the bounds on nearby points and pairwise intersections are already uniform over all sample labels. The exceptional probabilities sum to at most $e^{-cd^2}$. For example, one may take
$c=\min\{\ell/16,1/4\}$ after increasing the threshold for $n$.
All events are Borel because they involve only finitely many distances and numbers of points in caps.
\end{proof}

Combining \cref{thm:local} with the union estimate in \cref{lem:small-row-union} gives the following lower bound on distinct evaluations, even when all operations share a cache. \begin{theorem}[Distinct-evaluation lower bound in the near-threshold band]\label{thm:distinct-work-band}
For uniform data in the dense regime, fix $k\ge1$, $\frac12\log(4/3)\le\gamma<\gc$, and $\eps>0$, and put $\Delta=\lceil e^{\gamma d}\rceil$ and
$\eta=\eta^*(\gamma)$. On either $G_{r_\gamma}$ or $\NN{\Delta}$,
consider any measurable local exploration algorithm in the model of \cref{sec:local}, without an imposed expansion or entry-request budget.
For each fixed query $q$, let $V$ be the number of distinct query--data distance evaluations through the first completed expansion or entry request that evaluates a point of $\TopSet_k(q)$, using the local exploration convention in \cref{sec:local}. If the algorithm finds no target, set $V=\infty$.
Then for all sufficiently large $n$,
\begin{equation}\label{eq:distinct-work-lower}
 \Prob\{V\le e^{(\gamma+\eta-\eps)d}\}\le e^{-c_\eps d},
\end{equation}
where $c_\eps=\min\{\eps,\eta/2\}/16>0$. The probability is over the data and the algorithm's randomness.
Constants and the threshold for $n$ are uniform in the fixed query
and the local exploration algorithm. Caches may be shared throughout the search.
Every expansion scans the complete neighbor list, including an expansion that finds a target.
\end{theorem}

\begin{proof}

The union estimate applies to sets of expanded points of size $e^{\tau d}$ with $\tau<J_{\rm ov}(\gamma)$. To reach the single-start exponent, it suffices to check $\eta\le J_{\rm ov}(\gamma)$. Retain $E_\gamma=e^{2\gamma}\in[4/3,2)$ and the expression for $w(L,1)$ from the proof of \cref{cor:tradeoff}, so $\eta=-\tfrac12\log(1-w(L,1)^2)$.
The band identity of \cref{lem:trap-geometry}(c) identifies this
$\eta$ with $\eta_{\rm gen}$. Direct algebra gives
\[
 \frac1{4E_\gamma}-w(L,1)^2
 =\frac{(3E_\gamma-4)\{4-3E_\gamma(E_\gamma-2)^2\}}{16E_\gamma^2(E_\gamma-1)}\ge0.
\]
Indeed $3E_\gamma-4\ge0$ and $3E_\gamma(E_\gamma-2)^2\le 3\cdot2\cdot(2/3)^2=8/3<4$.
Since $t\mapsto-\tfrac12\log(1-t)$ is increasing, we obtain $\eta\le J_{\rm ov}(\gamma)$, with equality only at $E_\gamma=4/3$.

We choose an expansion limit small enough for \cref{lem:small-row-union} to apply, but large enough that completing that many expansions would exceed the evaluation budget. It suffices to use $\min\{\eps,\eta/2\}$ in place of $\eps$. Choose $c_{\mathrm{work}}=\min\{\eps,\eta/2\}/8$ and $\tau=\eta-2c_{\mathrm{work}}$, and set $M_n=\lfloor e^{\tau d}\rfloor$. Use the distinct-evaluation budget $\lfloor e^{(\gamma+\eta-\min\{\eps,\eta/2\})d}\rfloor$.
Here $M_n$ limits the number of distinct expansions. Since the evaluation cost is integer-valued, the event that it is at most $e^{(\gamma+\eta-\eps)d}$ is contained in the event that it is at most this integer budget. Since $0<\tau<J_{\rm ov}(\gamma)$, including at the lower band endpoint, apply \cref{lem:small-row-union} with this $\tau$. Its constants are
chosen next by the explicit formulas in its proof. Finally apply
\cref{thm:local} with exponent slack ${c_{\mathrm{work}}}$, choosing its constants in
the order specified there, and only then take $n$ sufficiently large.

Before applying the expansion bound, we remove entry requests whose distances are already in the actual algorithm's exploration history. Such requests can be simulated internally, preserving the original algorithm's random choices and subsequent decisions. The model permits each neighbor list to be expanded only once, so we count every expansion. Among entry requests, count only those that evaluate a previously unevaluated distance.
To apply the local-exploration estimate, we must also distinguish actual distance evaluations from the additional coordinate information exposed in its proof. Conditional on that exposed information, each label whose position has not yet been exposed, and which is selected by a data-independent entry rule from the local exploration model, is uniform on the unexposed region. Requesting an already exposed label adds no newly revealed point to the counts in \eqref{eq:bin-recursion}. If such a label lies in $C(q,b_0)$ from the proof of \cref{thm:local}, it was already included at first exposure in $e_0$, the expected number of revealed points in that cap. An entry whose position has not been exposed also has an unevaluated distance, so the number of requests contributing to the entry term is at most the number of first distance evaluations through entry requests.

On $\mathcal E_n$, completing $M_n$ distinct expansions requires at least
\[
 \frac{\Delta M_n}{4d}
 \ge\frac{e^{(\gamma+\eta-2{c_{\mathrm{work}}})d}}{8d}>\lfloor e^{(\gamma+\eta-\min\{\eps,\eta/2\})d}\rfloor
\]
distinct evaluations for all sufficiently large $n$. All points in the
union of these completed neighbor lists have actually been evaluated, regardless
of overlap with previous neighbor lists or entries. Hence an exploration whose first
target is found within this budget has fewer than $M_n$ completed expansions, including the final expansion, and at most $\lfloor e^{(\gamma+\eta-\min\{\eps,\eta/2\})d}\rfloor$ entry requests that evaluate a previously unevaluated distance.

Consequently, apart from $\mathcal E_n^c$, such success is contained in
the success event of the simulated algorithm stopped after $M_n$ complete
expansions or $\lfloor e^{(\gamma+\eta-\min\{\eps,\eta/2\})d}\rfloor$ entry requests that evaluate new distances. It obeys the required expansion bound $M_n\le n/(8\Delta)$ for all large $n$, since $\log n/d\to\infty$. The removal of redundant operations
above justifies applying the proof of \cref{thm:local} with these budgets,
including for an original algorithm that makes redundant requests.

Applying the conditional exposure estimate and recursion in the proof of \cref{thm:local}, together with \cref{lem:trap-geometry}(c) and $T_{\rm near}\le M_n$, bounds the stopped simulation's success probability by
\[
 M_n e^{-(\eta-{c_{\mathrm{work}}})d}
 +e^{(\gamma+\eta-\min\{\eps,\eta/2\})d}\frac{e^{Cd}}n
 +\Prob(\mathcal C^c)+\exp\{-e^{c'd}\}.
\]
Here $C,c'>0$ depend only on $\gamma$, $\eps$, and $k$, and $\mathcal C$ is the event that every exposure cap $C(x,\rho^+)$ contains at least $\Delta$ other data points, as in \cref{thm:local}. For $G_{r_\gamma}$ take $\mathcal C$ to be the whole sample space, so $\Prob(\mathcal C^c)=0$, while for $\NN{\Delta}$, $\Prob(\mathcal C^c)\le ne^{-\Delta/4}$.
The first term is at most $e^{-{c_{\mathrm{work}}} d}$. The entry term is
$e^{-\omega(d)}$, because the entry-request budget is $e^{O(d)}$ whereas $\log n/d\to\infty$.
The other terms in the displayed success bound are $e^{-\omega(d)}$. Adding $\Prob(\mathcal E_n^c)$ bounds the original event of success within the stated integer evaluation budget, and this additional term is also $e^{-\omega(d)}$. Increasing the threshold for $n$ gives
$e^{-{c_{\mathrm{work}}} d/2}=e^{-c_\eps d}$, as claimed. Rotation invariance and \cref{thm:local} give
uniformity in $q$.
\end{proof}

\begin{proof}[Proof of \cref{thm:fresh-start-cost}]
For each fixed $\gamma$ in the stated band, \cref{thm:distinct-work-band} gives the distinct-evaluation lower bound on both $G_{r_\gamma}$ and $\NN{\Delta}$. By \cref{cor:tradeoff}, independent restarts with $k=1$ attain the matching upper exponent on $G_{r_\gamma}$ with probability tending to one, counting all failed attempts and complete expansions. To obtain the upper bound for any fixed $k\ge1$, use the same sample and entry sequence for both procedures. Before the first completed expansion after which the evaluated labels include a point of $\TopSet_k(q)$, their failed attempts, successors, and shared caches coincide. Both procedures complete that expansion, including when its entry is already a target. The fixed-$k$ procedure therefore succeeds no later, and its evaluated labels form a subset of those evaluated by the $k=1$ procedure within the same number of attempts prescribed by \cref{cor:tradeoff}. This transfers both success and the distinct-evaluation bound to fixed $k$. The estimates hold for every fixed $\eps>0$ in \cref{thm:distinct-work-band,cor:tradeoff}. Letting $\eps$ tend to zero sufficiently slowly gives the $o(1)$ terms in \cref{thm:fresh-start-cost}. The minimization performed in the proof of \cref{cor:tradeoff} gives the value approximately $0.243$, while \cref{thm:critical,thm:uniform-cost} give the every-entry exponent $\gc\approx0.347$.
\end{proof}

\subsection{The rank of the stopping point}\label{sec:stopping-rank}

After determining the cost of exact recovery with fresh entries, we ask what rank greedy navigation can guarantee from every entry if exact recovery is relaxed. Throughout this subsection, a run continues until no listed neighbor is closer to the query than its current point. For uniform data below the critical degree, the lens estimates of \cref{sec:prelim} and the direction comparison in \cref{sec:random-entry} locate a cap outside which every data point has an improving neighbor. For nonuniform data, the progress bounds of \cref{sec:nonuniform} provide such a cap over a larger range of density variation than exact recovery allows. In each case, every full run stops inside the cap, and the number of sampled points in that cap bounds the stopping rank.

\begin{corollary}[Stopping rank for uniform data]\label{cor:stopping-rank}
For uniform data in the dense regime, fix $0<\gamma<\gc$ and $\eps>0$, and let $\Delta=\lceil e^{\gamma d}\rceil$. Define $\nu_\gamma=\frac12\log\frac{e^{4\gamma}}{4(e^{2\gamma}-1)}$. With probability $1-o(1)$ over the data, every query outside a set of surface measure $e^{-\Omega(d)}$ has the following property. From every entry, the full greedy run on either $G_{r_\gamma}$ or $\NN{\Delta}$ stops at a point of rank at most $e^{(\nu_\gamma+\eps)d}$.
\end{corollary}

\begin{proof}
To find a cap outside which every point has a closer neighbor, write $L_g=e^{2g}/(2\sqrt{e^{2g}-1})$, so that $\log L_\gamma=\nu_\gamma$. Choose $0<g<\gamma$ close enough to $\gamma$ that $\log L_g<\nu_\gamma+\eps/8$, and then choose $A>L_g$ with $\log A<\nu_\gamma+\eps/4$. Set $\rho=r_g$ and $\theta_0=A\theta_1$. The graph $G_\rho$ is contained in $G_{r_\gamma}$. It is also contained in $\NN{\Delta}$ with probability $1-\exp(-e^{\Omega(d)})$, since each radius-$\rho$ cap contains at most $e^{gd}$ points in expectation and $g<\gamma$. This follows from \eqref{eq:chernoff} and a union bound over the sources.

For a source $x$ at distance $\theta$ from a fixed query $q$, let $u_\rho(\theta)=\sigma_{d-1}(C(x,\rho)\cap C(q,\theta))$ be the mass of its improving region. Since $\rho/\theta_1\to e^g$ and $\theta_0/\theta_1=A$, \cref{lem:cap,lem:lens} give
\[
 \frac1{d}\log\bigl(nu_\rho(\theta_0)\bigr)=\frac12\log\left(e^{2g}-\frac{e^{4g}}{4A^2}\right)+o(1)>0.
\]
The strict inequality is exactly $A>L_g$. The error is $o(1)$ because $\log(1/\theta_1)=o(d)$. Moreover, $u_\rho(\theta)$ is nondecreasing in $\theta$. Indeed, for a point at distance $t\le\rho$ from $x$, the spherical cosine identity makes improvement equivalent to a directional cosine larger than $\cot\theta\tan(t/2)$, which decreases with $\theta$. Thus the improving region of every source at distance at least $\theta_0$ contains $e^{\Omega(d)}$ points in expectation. Conditional on the source, the remaining points are independent. The probability that any of the $n$ sources in this range lacks an improving neighbor is therefore at most $n\exp(-e^{\Omega(d)})=\exp(-e^{\Omega(d)})$.

Combining this estimate with retention of $G_\rho$ shows that every full run stops inside $C(q,A\theta_1)$. Its stopping rank is at most the number of data points in that cap. By \eqref{eq:cap-ratio}, this number has mean $e^{(\log A+o(1))d}$, and \eqref{eq:chernoff} bounds it by $e^{(\nu_\gamma+\eps)d}$ except with probability $\exp(-e^{\Omega(d)})$. The estimates are uniform in the fixed query. Integrating over surface measure and applying Markov's inequality gives one data event of probability $1-o(1)$ and an exceptional query set of measure $e^{-\Omega(d)}$. Distance ties are excluded as in the navigation conventions.
\end{proof}

For nonuniform data, the progress estimates can hold even when the inner cap contains false local minima. This gives a wider density range for bounded stopping rank than for exact recovery.

\begin{corollary}[Stopping rank for nonuniform data]\label{thm:approximate}
In the dense regime, fix $\gc<\gamma\le\log2$, $0\le\kappa<\gamma-\gc$, and $\eps>0$, and let $\Delta=\lceil e^{\gamma d}\rceil$. For every sufficiently large fixed $K$ depending on $\gamma$ and $\kappa$, let $f$ be $(\kappa,K)$-regular and let $Q$ be an absolutely continuous query distribution, both fixed before the data. With probability $1-e^{-\omega(d)}$ over the data, there is a query set of $Q$-probability at least $1-e^{-\omega(d)}$ on which every full greedy run on $\NN{\Delta}$ from every entry stops within $O(1+\rho_{\min}^{-1})=e^{o(d)}$ moves at a point of rank at most $2e^{(\kappa+\eps)d}$.
\end{corollary}

\begin{proof}
Regularity gives $f>0$ almost everywhere, as shown in the proof of \cref{thm:tolerance}(a), and every query satisfies the local density condition with parameter $\kappa$. Choose $g$ with $\kappa+\gc<g<\gamma$, then choose $a$ with $e^{\kappa-g}<a<\min\{1/\sqrt2,e^{\kappa+\eps-g}\}$. Put $A=g+\log a<\kappa+\eps$. Use the radius comparisons, retention, and outer progress estimates in the proof of \cref{loc:nn-sufficiency} with $\bar\kappa=\kappa_1=\kappa$ and with $K$ at least $K_0$ recomputed from the formula in that proof using the present $g$ and $\bar\kappa=\kappa_1=\kappa$. Their condition $g>\bar\kappa$ holds throughout the present range. The middle region estimate in that proof applies with the positive margins $A-\kappa$ and $g-\kappa-\gc$, yielding a factor $b<1$ and an expected number $e^{\Omega(d)}$ of neighbors that reduce the query distance by this factor. Thus, for each fixed query, every run reaches $C(q,\Theta)$, where $R=R_q$ is defined in that proof and $\Theta=\arcsin(a\sin R)$, except with probability $e^{-\omega(d)}$.

The path-length argument in that proof uses only retention of the radius-$R$ edges and $a<1/\sqrt2$, so it also bounds full runs that stop at a false local minimum. Every run stops inside this cap within $O(1+\rho_{\min}^{-1})$ moves. Using $M_q$ from the same proof, the number of sampled points in the cap is binomial with mean at most $nM_qF(\Theta)\le a^{-1}e^{Ad}$, and hence at most $2a^{-1}e^{Ad}\le2e^{(\kappa+\eps)d}$ except with doubly exponentially small probability, by \eqref{eq:chernoff}. Integration against $Q$ and Markov's inequality give the common query set and the stated probability bounds.
\end{proof}

The stopping rank guarantee can hold for the same sampling density and query distribution that force failing entries for every graph of degree at most $\Delta$. For any fixed $\eps>0$, $\gc<\gamma\le\log2$, and $h(\gamma)<\kappa<\gamma-\gc$, the density and query distribution $Q_n$ in \cref{thm:tolerance}(b) also meet the hypotheses of \cref{thm:approximate}, since the density is $(\kappa,K)$-regular for every fixed $K>0$. With probability $1-e^{-\Omega(d)}$ over the sample, every graph $G$ of degree at most $\Delta$ has a query set of $Q_n$-probability at least $1-e^{-\Omega(d)}$ obtained by intersecting the query sets from \cref{thm:tolerance}(b) and \cref{thm:approximate}. For every query in this intersection, greedy navigation on $G$ fails from $e^{\Omega(d)}$ entries and every full greedy run on $\NN{\Delta}$ from every entry has stopping rank at most $2e^{(\kappa+\eps)d}$.

\appendix

\section*{AI methodologies}

OpenAI ChatGPT and Codex were used during the mathematical development of this work. In particular, initial versions of the hard-instance constructions used in the proof of \cref{thm:tolerance}(b) and in \cref{enr:obstruction} were developed with ChatGPT. ChatGPT also assisted in checking the exponent identities and the compatibility of the parameter inequalities in these two constructions, as well as the minimization of the evaluation cost exponent for independent restarts in \cref{thm:fresh-start-cost} (see the proof of \cref{cor:tradeoff}). The author rewrote every argument retained in the paper, independently verified the correctness and originality of all content, including references, and takes full responsibility for the final manuscript.

\bibliographystyle{alphaurl}
\bibliography{bibliography}

@inproceedings{Kleinberg2000SmallWorld,
  author    = {Jon Kleinberg},
  title     = {The Small-World Phenomenon: An Algorithmic Perspective},
  booktitle = {Proceedings of the 32nd Annual ACM Symposium on Theory of Computing},
  pages     = {163--170},
  year      = {2000},
  doi       = {10.1145/335305.335325}
}

@article{MalkovYashunin2020HNSW,
  author  = {Yu. A. Malkov and D. A. Yashunin},
  title   = {Efficient and Robust Approximate Nearest Neighbor Search Using
             Hierarchical Navigable Small World Graphs},
  journal = {IEEE Transactions on Pattern Analysis and Machine Intelligence},
  volume  = {42},
  number  = {4},
  pages   = {824--836},
  year    = {2020},
  doi     = {10.1109/TPAMI.2018.2889473}
}

@inproceedings{SubramanyaEtAl2019DiskANN,
  author    = {Suhas Jayaram Subramanya and Fnu Devvrit and
               Harsha Vardhan Simhadri and Ravishankar Krishnaswamy and
               Rohan Kadekodi},
  title     = {{DiskANN}: Fast Accurate Billion-Point Nearest Neighbor Search
               on a Single Node},
  booktitle = {Advances in Neural Information Processing Systems 32},
  year      = {2019}
}

@inproceedings{Laarhoven2018GraphTradeoffs,
  author    = {Thijs Laarhoven},
  title     = {Graph-Based Time--Space Trade-Offs for Approximate Near Neighbors},
  booktitle = {34th International Symposium on Computational Geometry},
  series    = {LIPIcs},
  volume    = {99},
  pages     = {57:1--57:14},
  year      = {2018},
  doi       = {10.4230/LIPIcs.SoCG.2018.57},
  url       = {https://arxiv.org/abs/1712.03158}
}

@inproceedings{ProkhorenkovaShekhovtsov2020,
  author    = {Liudmila Prokhorenkova and Aleksandr Shekhovtsov},
  title     = {Graph-Based Nearest Neighbor Search: From Practice to Theory},
  booktitle = {Proceedings of the 37th International Conference on Machine Learning},
  series    = {Proceedings of Machine Learning Research},
  volume    = {119},
  pages     = {7803--7813},
  year      = {2020},
  url       = {https://proceedings.mlr.press/v119/prokhorenkova20a.html}
}

@inproceedings{IndykXu2023WorstCase,
  author    = {Piotr Indyk and Haike Xu},
  title     = {Worst-Case Performance of Popular Approximate Nearest Neighbor
               Search Implementations: Guarantees and Limitations},
  booktitle = {Advances in Neural Information Processing Systems 36},
  year      = {2023}
}

@misc{AviMusco2026AlmostNavigable,
  author        = {Pratyush Avi and Christopher Musco},
  title         = {Almost Navigable Graphs},
  year          = {2026},
  eprint        = {2607.14564},
  archivePrefix = {arXiv},
  primaryClass  = {cs.DS}
}

@inproceedings{AlJazzaziEtAl2025AdaptiveBeam,
  author    = {Yousef Al-Jazzazi and Haya Diwan and Jinrui Gou and
               Cameron Musco and Christopher Musco and Torsten Suel},
  title     = {Distance Adaptive Beam Search for Provably Accurate
               Graph-Based Nearest Neighbor Search},
  booktitle = {Advances in Neural Information Processing Systems 38},
  year      = {2025}
}

@inproceedings{DiwanEtAl2024Navigable,
  author    = {Haya Diwan and Jinrui Gou and Cameron Musco and
               Christopher Musco and Torsten Suel},
  title     = {Navigable Graphs for High-Dimensional Nearest Neighbor Search:
               Constructions and Limits},
  booktitle = {Advances in Neural Information Processing Systems 37},
  year      = {2024},
  eprint    = {2405.18680},
  archivePrefix = {arXiv}
}

@inproceedings{KhannaEtAl2026Sparse,
  author    = {Sanjeev Khanna and Ashwin Padaki and Erik Waingarten},
  title     = {Sparse Navigable Graphs for Nearest Neighbor Search:
               Algorithms and Hardness},
  booktitle = {Proceedings of the 37th Annual ACM--SIAM Symposium on
               Discrete Algorithms},
  pages     = {1606--1630},
  year      = {2026},
  doi       = {10.1137/1.9781611978971.58}
}

@inproceedings{ConwayEtAl2026Constructing,
  author    = {Alex Conway and Laxman Dhulipala and Martin Farach-Colton and
               Rob Johnson and Ben Landrum and Christopher Musco and
               Yarin Shechter and Torsten Suel and Richard Wen},
  title     = {Efficiently Constructing Sparse Navigable Graphs},
  booktitle = {Proceedings of the 37th Annual ACM--SIAM Symposium on
               Discrete Algorithms},
  pages     = {1631--1665},
  year      = {2026},
  doi       = {10.1137/1.9781611978971.59}
}

@article{FuEtAl2019NSG,
  author  = {Cong Fu and Chao Xiang and Changxu Wang and Deng Cai},
  title   = {Fast Approximate Nearest Neighbor Search with the Navigating
             Spreading-out Graph},
  journal = {Proceedings of the VLDB Endowment},
  volume  = {12},
  number  = {5},
  pages   = {461--474},
  year    = {2019}
}

@inproceedings{AndoniRazenshteyn2015,
  author    = {Alexandr Andoni and Ilya Razenshteyn},
  title     = {Optimal Data-Dependent Hashing for Approximate Near Neighbors},
  booktitle = {Proceedings of the 47th Annual ACM Symposium on Theory of Computing},
  pages     = {793--801},
  year      = {2015}
}

@inproceedings{AndoniEtAl2017Tradeoffs,
  author    = {Alexandr Andoni and Thijs Laarhoven and Ilya Razenshteyn and
               Erik Waingarten},
  title     = {Optimal Hashing-Based Time--Space Trade-offs for Approximate
               Near Neighbors},
  booktitle = {Proceedings of the 28th Annual ACM--SIAM Symposium on
               Discrete Algorithms},
  pages     = {47--66},
  year      = {2017}
}

@inproceedings{BeckerEtAl2016Sieving,
  author    = {Anja Becker and L{\'e}o Ducas and Nicolas Gama and
               Thijs Laarhoven},
  title     = {New Directions in Nearest Neighbor Searching with Applications
               to Lattice Sieving},
  booktitle = {Proceedings of the 27th Annual ACM--SIAM Symposium on
               Discrete Algorithms},
  pages     = {10--24},
  year      = {2016}
}

@misc{ShrivastavaSongXu2023,
  author       = {Anshumali Shrivastava and Zhao Song and Zhaozhuo Xu},
  title        = {A Theoretical Analysis of Nearest Neighbor Search on Approximate Near Neighbor Graph},
  year         = {2023},
  eprint       = {2303.06210},
  archivePrefix= {arXiv},
  primaryClass = {cs.LG},
}

@article{MaEtAl2025SNG,
  author  = {Xinran Ma and Zhaoqi Zhou and Chuan Zhou and Zaijiu Shang and
             Guoliang Li and Zhiming Ma},
  title   = {Sparse Neighborhood Graph-Based Approximate Nearest Neighbor Search
             Revisited: Theoretical Analysis and Optimization},
  journal = {Proceedings of the VLDB Endowment},
  volume  = {19},
  number  = {10},
  pages   = {2817--2830},
  year    = {2026},
  doi     = {10.14778/3828612.3828634}
}

@article{LuTao2025Proximity,
  author  = {Shangqi Lu and Yufei Tao},
  title   = {Proximity Graphs for Similarity Search: Fast Construction,
             Lower Bounds, and {Euclidean} Separation},
  journal = {Proceedings of the ACM on Management of Data},
  volume  = {3},
  number  = {5},
  pages   = {280:1--280:25},
  year    = {2025},
  note    = {Presented at PODS 2026},
  doi     = {10.1145/3767716}
}

@inproceedings{HarPeledRaichelRobson2025,
  author    = {Sariel Har-Peled and Benjamin Raichel and Eliot W. Robson},
  title     = {The Road to the Closest Point is Paved by Good Neighbors},
  booktitle = {Proceedings of the 2026 SIAM Symposium on Simplicity in Algorithms},
  pages     = {1--11},
  year      = {2026},
  doi       = {10.1137/1.9781611978964.1}
}

@inproceedings{GollapudiEtAl2025SortPrune,
  author    = {Siddharth Gollapudi and Ravishankar Krishnaswamy and Kirankumar Shiragur and Harsh Wardhan},
  title     = {Sort Before You Prune: Improved Worst-Case Guarantees of the {DiskANN} Family of Graphs},
  booktitle = {Proceedings of the 42nd International Conference on Machine Learning ({ICML})},
  year      = {2025}
}

@misc{OguriMatsui2024Entry,
  author       = {Yutaro Oguri and Yusuke Matsui},
  title        = {Theoretical and Empirical Analysis of Adaptive Entry Point Selection for Graph-based Approximate Nearest Neighbor Search},
  year         = {2024},
  eprint       = {2402.04713},
  archivePrefix= {arXiv},
  primaryClass = {cs.IR},
}

@inproceedings{AryaMount1993,
  author    = {Sunil Arya and David M. Mount},
  title     = {Approximate Nearest Neighbor Queries in Fixed Dimensions},
  booktitle = {Proceedings of the Fourth Annual {ACM-SIAM} Symposium on Discrete Algorithms ({SODA})},
  pages     = {271--280},
  year      = {1993}
}

@article{BaernsteinTaylor1976,
  author  = {Albert {Baernstein II} and B. A. Taylor},
  title   = {Spherical Rearrangements, Subharmonic Functions, and $*$-Functions in $n$-Space},
  journal = {Duke Mathematical Journal},
  volume  = {43},
  number  = {2},
  pages   = {245--268},
  year    = {1976}
}

@misc{WaizerDanieli2026Price,
  author        = {Tomer Waizer and Yoav Danieli},
  title         = {The Price of Almost Navigability},
  year          = {2026},
  eprint        = {2609.02498},
  archivePrefix = {arXiv},
  primaryClass  = {cs.DS},
  url           = {https://arxiv.org/abs/2609.02498}
}

@misc{MaghrebiEtAl2026PowerLaw,
  author        = {Sajad Faghfoor Maghrebi and Navid Eslami and Niv Dayan},
  title         = {A Power Law in Logarithm's Clothing: On the Scalability of Graph-Based Vector Search},
  year          = {2026},
  eprint        = {2609.02143},
  archivePrefix = {arXiv},
  url           = {https://arxiv.org/abs/2609.02143}
}

@misc{BhamidiEtAl2026Geometry,
  author        = {Shankar Bhamidi and Souvik Dhara and Lars Schroeder and Clara Stegehuis},
  title         = {The Geometry of Hierarchical Navigation: Accuracy and Query Cost for Point Process Input},
  year          = {2026},
  eprint        = {2610.12312},
  archivePrefix = {arXiv},
  primaryClass  = {cs.DS},
  url           = {https://arxiv.org/abs/2610.12312}
}

@misc{GilibertiEtAl2026WithoutSpread,
  author        = {Jeff Giliberti and Sariel Har-Peled and Jonas Sauer and Ali Vakilian},
  title         = {Graph-Based Nearest-Neighbor Search without the Spread},
  year          = {2026},
  eprint        = {2602.06633},
  archivePrefix = {arXiv},
  primaryClass  = {cs.CG},
  url           = {https://arxiv.org/abs/2602.06633}
}

@article{LiYanLu2026FastConvergent,
  author        = {Binhong Li and Xiao Yan and Shangqi Lu},
  title         = {Fast-Convergent Proximity Graphs for Approximate Nearest Neighbor Search},
  year          = {2026},
  journal       = {Proceedings of the ACM on Management of Data},
  volume        = {4},
  number        = {1},
  articleno     = {36},
  numpages      = {24},
  doi           = {10.1145/3786650},
  pages         = {36:1--36:24}
}

@article{WangEtAl2024SteinerHardness,
  author        = {Zeyu Wang and Qitong Wang and Xiaoxing Cheng and Peng Wang and Themis Palpanas and Wei Wang},
  title         = {{Steiner}-Hardness: A Query Hardness Measure for Graph-Based {ANN} Indexes},
  year          = {2024},
  journal       = {Proceedings of the VLDB Endowment},
  volume        = {17},
  number        = {13},
  pages         = {4668--4682},
  doi           = {10.14778/3704965.3704974}
}

@article{PengEtAl2023Tau,
  author        = {Yun Peng and Byron Choi and Tsz Nam Chan and Jianye Yang and Jianliang Xu},
  title         = {Efficient Approximate Nearest Neighbor Search in Multi-dimensional Databases},
  year          = {2023},
  journal       = {Proceedings of the ACM on Management of Data},
  volume        = {1},
  number        = {1},
  articleno     = {54},
  numpages      = {27},
  doi           = {10.1145/3588908},
  pages         = {54:1--54:27}
}

@inproceedings{HarwoodDrummond2016FANNG,
  author        = {Ben Harwood and Tom Drummond},
  title         = {{FANNG}: Fast Approximate Nearest Neighbour Graphs},
  year          = {2016},
  booktitle     = {Proceedings of the IEEE Conference on Computer Vision and Pattern Recognition},
  pages         = {5713--5722},
  doi           = {10.1109/CVPR.2016.616}
}

@article{FigielLindenstraussMilman1977,
  author = {Figiel, Tadeusz and Lindenstrauss, Joram and Milman, Vitali D.},
  title = {The Dimension of Almost Spherical Sections of Convex Bodies},
  journal = {Acta Mathematica},
  volume = {139},
  pages = {53--94},
  year = {1977},
  doi = {10.1007/BF02392234}
}

\end{document}